\pdfoutput=1

\documentclass[rmp,aps,twocolumn]{revtex4-1}

\usepackage{graphicx}
\usepackage{epsfig}
\usepackage{amsmath}
\usepackage{amssymb}
\usepackage{bm}
\usepackage{mathdots}

\usepackage[usenames,dvipsnames]{color}
\newcommand{\red}{\textcolor{Red}}

\newcommand{\beq}{\begin{equation}}
\newcommand{\eeq}{\end{equation}}
\newcommand{\bea}{\begin{eqnarray}}
\newcommand{\eea}{\end{eqnarray}}

\newcommand{\equ}[1]{Eq.~(\ref{eq:#1})}
\newcommand{\eqs}[2]{Eqs.~(\ref{eq:#1}) and (\ref{eq:#2})}
\newcommand{\equa}[1]{Equation~(\ref{eq:#1})}

\newcommand{\kk}{{\bf k}}
\newcommand{\qq}{{\bf q}}

\newcommand{\rr}{{\bf r}}
\newcommand{\vv}{{\bf v}}
\newcommand{\RR}{{\bf R}}
\newcommand{\PP}{{\bf P}}
\newcommand{\E}{{\cal E}}
\newcommand{\EE}{{\bm{{\cal E}}}}
\newcommand{\0}{{\bf 0}}

\newcommand{\enk}{\epsilon_{n\kk}}
\newcommand{\enkh}{\overline{\epsilon}_{n\kk}}

\newcommand{\hh}{{{\rm H}}}
\newcommand{\ww}{{{\rm W}}}

\newcommand{\la}{{\langle\kern-2.0pt\langle}}
\newcommand{\ra}{{\rangle\kern-2.0pt\rangle}}
\newcommand{\vt}{{\vert\kern-1.0pt\vert}}

\newcommand{\WW}{{\bf W}}
\newcommand{\HH}{{\bf H}}
\newcommand{\DD}{{\bf D}}
\newcommand{\BB}{{\bf B}}
\newcommand{\eps}{{\epsilon}}

\newcommand{\ket}[1]{\vert{#1}\rangle}
\newcommand{\bra}[1]{\langle#1\vert}
\newcommand{\inprod}[2]{\langle#1\vert{#2}\rangle}
\newcommand{\matel}[3]{\langle#1\vert{#2}\vert{#3}\rangle}

\newcommand{\bnk}{_{n\kk}}

\newcommand{\intr}{\int \! d{\bf r} \;}
\newcommand{\ibz}{\frac{V}{(2\pi)^3}\int_{\rm BZ}}

\newcommand{\nbw}{J}
\newcommand{\nwann}{J}
\newcommand{\nbands}{{\cal J}}

\begin{document}

\title{Maximally localized Wannier functions: Theory and applications}

\author{Nicola Marzari}
\affiliation{Theory and Simulation of Materials (THEOS),
\'Ecole Polytechnique F\'ed\'erale de Lausanne, Station 12, 1015 Lausanne, Switzerland}
\author{Arash A. Mostofi}
\affiliation{Departments of Materials and Physics, and the 
Thomas Young Centre for Theory and Simulation of Materials,
Imperial College London, London SW7 2AZ, UK}
\author{Jonathan R. Yates}
\affiliation{Department of Materials,
  University of Oxford, Parks Road, Oxford OX1 3PH, UK}
\author{Ivo Souza}
\affiliation{Centro de F\'{\i}sica de Materiales (CSIC) and DIPC, Universidad
del Pa\'\i s Vasco, 20018 San Sebasti\'an, Spain}
\affiliation{Ikerbasque, Basque Foundation for Science, 48011
Bilbao, Spain}
\author{David Vanderbilt}
\affiliation{Department of Physics and Astronomy, Rutgers University,
        Piscataway, NJ 08854-8019, USA}

\newcommand{\chapsum}[5]{\\
  \hbox to 1.0cm{Ch.~#1\hfil}
  \hbox to 3.5cm{#2\hfil}
  \hbox to 2.0cm{#3\hfil}
  \hbox to 2.0cm{#4\hfil}
  \hbox to 1.5cm{#5 pages\hfil}
}

\begin{abstract}  
The electronic ground state of a periodic system is usually
described in terms of extended Bloch orbitals, but
an alternative representation in terms of localized ``Wannier functions''
was introduced by Gregory Wannier in 1937.
The connection between the Bloch and Wannier representations
is realized by families of transformations in a continuous space of
unitary matrices, carrying a large degree of arbitrariness.
Since 1997, methods have been developed that allow one to iteratively
transform the extended Bloch orbitals of a first-principles calculation into
a unique set of {\it maximally localized} Wannier functions,
accomplishing the solid-state equivalent of constructing 
localized molecular orbitals, or ``Boys orbitals'' as previously
known from the chemistry literature.  These
developments are reviewed here, and a survey of the applications of these methods is presented.
This latter includes a description of their use in
analyzing the nature of chemical bonding, or as a local probe of
phenomena related to electric polarization and orbital magnetization.
Wannier interpolation schemes are also reviewed, by which quantities computed
on a coarse reciprocal-space mesh can be used to interpolate
onto much finer meshes at low cost, and applications in which
Wannier functions are used as efficient basis functions are discussed.
Finally the
construction and use of Wannier functions outside the context of
electronic-structure theory is presented, for cases that include phonon excitations,
photonic crystals, and cold-atom optical lattices.
\end{abstract}                                                                 

\maketitle

\marginparwidth 2.7in
\marginparsep 0.5in
\def\nmm#1{\marginpar{\small NM: #1}}
\def\amm#1{\marginpar{\small AM: #1}}
\def\jym#1{\marginpar{\small JY: #1}}
\def\ism#1{\marginpar{\small IS: #1}}
\def\dvm#1{\marginpar{\small DV: #1}}
\def\eqm#1{\marginpar{\small Label: #1}}
\def\exm#1{\marginpar{\small\red{#1}}}
\def\scr{\scriptsize}
\def\nmm#1{} \def\amm#1{} \def\jym#1{}
\def\ism#1{} \def\dvm#1{} \def\exm#1{}

\tableofcontents

\section{INTRODUCTION}
\label{sec:intro}


In the independent-particle approximation, the electronic ground state
of a system is determined by specifying a set of one-particle
orbitals and their occupations.  For the case of periodic system, these one-particle
orbitals are normally taken to be the Bloch functions
$\psi_{n\kk}(\rr)$ that are labeled, according to Bloch's theorem, by
a crystal momentum $\bf k$ lying inside the Brillouin zone and a band
index $n$.  Although this choice is by far the most widely used in
electronic-structure calculations, alternative representations are
possible.  In particular, to arrive at the Wannier representation
\cite{wannier-pr37,kohn-pr59,cloizeaux-pr63}, one carries out a
unitary transformation from the Bloch functions to a set of localized
``Wannier functions'' (WFs) labeled by a cell index $\bf R$ and a band-like
index $n$, such that in a crystal the WFs at different $\bf R$ are
translational images of one another.  Unlike Bloch functions, 
WFs are not eigenstates of the Hamiltonian; in selecting
them, one trades off localization in energy for localization in space.

In the earlier solid-state theory literature, WFs
were typically introduced in order to carry out some formal
derivation -- for example, of the effective-mass treatment
of electron dynamics, or of an effective spin Hamiltonian -- 
but actual calculations of the WFs were rarely performed.
The history is rather different in the chemistry literature, where
``localized molecular orbitals'' (LMO's)
\cite{boys-rmp60,foster-rmp60a,foster-rmp60b,edmiston-rmp63,boys-66}
have played a significant role in computational chemistry since
its early days.  Chemists have emphasized that such a representation
can provide an insightful picture of the nature of
the chemical bond in a material --- otherwise missing from the
picture of extended eigenstates --- or can serve as a compact
basis set for high-accuracy calculations.

The actual implementation of Wannier's vision in the
context of first-principles electronic-structure calculations,
such as those carried out in the Kohn-Sham framework of
density-functional theory \cite{kohn-pr65},
has instead been slower to unfold.
A major reason for this is that WFs are
strongly non-unique. This is a consequence of
the phase indeterminacy that Bloch orbitals
$\psi_{n{\bf k}}$ have at every wavevector {\bf k} -- or, more
generally, the ``gauge'' indeterminacy associated with the freedom
to apply any arbitrary unitary transformation to
the occupied Bloch states at each $\kk$.  This second
indeterminacy is all the more troublesome in the common case of
degeneracy for the occupied bands at certain high-symmetry
points in the Brillouin zone, making a partition into
separate ``bands'', that could separately be transformed in
Wannier functions, problematic.  Therefore, even 
before one could attempt to
compute the WFs for a given material, one had first
to resolve the question of which states to use to compute
WFs.

An important development in this regard was the introduction
by \textcite{marzari-prb97}
of a ``maximal localization'' criterion for identifying a unique
set of WFs for a given crystalline insulator.
The approach is similar in spirit to the construction of LMO's
in chemistry, but its implementation in the
solid-state context required significant developments, due to the
ill-conditioned nature of the position operator in periodic systems
\cite{nenciu-rmp91}, 
that was clarified in the context of the ``modern theory'' of polarization
\cite{king-smith-prb93,resta-rmp94}.
Marzari and Vanderbilt showed that the minimization of a 
localization functional corresponding to the sum of the
second-moment spread of each Wannier charge density about its own
center of charge was both formally attractive and computationally
tractable.  In a related
development, \textcite{souza-prb01} generalized
the method to handle the case in which one wants to construct a
set of WFs that spans a subspace containing, e.g.,
the partially occupied bands of a metal.

These developments touched off a transformational shift in which
the computational electronic-structure community started constructing
maximally-localized WFs (MLWFs) explicitly and
using these for different purposes.  The reasons are manifold: First,
WFs, akin to LMO's in molecules,
provide an insightful chemical analysis of the nature of bonding,
and its evolution during, say, a chemical reaction.  As such, they
have become an established tool in the post-processing of
electronic-structure calculations. More interestingly, there are
formal connections between the centers of charge of the WFs
and the Berry phases of the Bloch functions as they are
carried around the Brillouin zone.  This connection is embodied in
the microscopic modern theory of polarization, alluded to above,
and has led to important advances in the characterization
and understanding of dielectric response and polarization in
materials.
Of broader interest to the entire condensed matter community is the
use of WFs in the construction of model Hamiltonians
for, e.g., correlated-electron and magnetic systems.
An alternative use of WFs as localized, transferable
building blocks has taken place in the theory of ballistic
(Landauer) transport, where Green's functions and self-energies can
be constructed effectively in a Wannier basis, or that of
first-principles tight-binding Hamiltonians, where chemically-accurate
Hamiltonians are constructed directly on the Wannier
basis, rather than fitted or inferred from macroscopic
considerations.
Finally, the
ideas that were developed for electronic WFs have
also seen application in very different contexts.  For
example, MLWF's have been used in
the theoretical analysis of phonons, photonic crystals, cold atom
lattices, and the local dielectric responses of insulators.

Here we review these developments.  We first introduce the transformation
from Bloch functions to WFs in Sec.~\ref{sec:theory},
discussing their gauge freedom and the methods developed for constructing WFs
through projection or maximal localization.  A
``disentangling procedure'' for constructing WFs
for a non-isolated set of bands (e.g., in metals) is also
described.  In Sec.~\ref{sec:locorb} we discuss variants of these
procedures in which different localization
criteria or different algorithms are used, and discuss the relationship
to ``downfolding'' and linear-scaling methods.
Sec.~\ref{sec:bonding} describes how the calculation of WFs
has proved to be a useful tool for analyzing the nature
of the chemical bonding in crystalline, amorphous, and defective systems.
Of particular importance is the ability to use WFs as
a local probe of electric polarization, as
described in Sec.~\ref{sec:pol}.  There we also discuss how the Wannier
representation has been useful in describing 
orbital magnetization, NMR chemical shifts, orbital magnetoelectric
responses, and topological insulators.  Sec.~\ref{sec:winterp}
describes Wannier interpolation schemes, by which quantities computed
on a relatively coarse $\kk$-space mesh can be used to interpolate
faithfully onto an arbitrarily fine $\kk$-space mesh at relatively
low cost.  In Sec.~\ref{sec:basis} we discuss applications in which
the WFs are used as an efficient basis 
for the calculations of quantum transport properties, the
derivation of semiempirical potentials, and for describing
strongly-correlated systems.  Sec.~\ref{sec:other} contains a
brief discussion of the construction and use of WFs
in contexts other than electronic-structure theory, including
for phonons in ordinary crystals, photonic crystals, and cold
atoms in optical lattices.  Finally, Sec.~\ref{sec:summary}
provides a short summary and conclusions.

\section{REVIEW OF BASIC THEORY}
\label{sec:theory}

\subsection{Bloch functions and Wannier functions}
\label{sec:bloch-wannier}

Electronic structure calculations are often carried out using periodic
boundary conditions. This is the most natural choice for the study of
perfect crystals, and also applies to the common use of periodic
supercells for the study of non-periodic systems such as liquids,
interfaces, and defects. The one-particle effective Hamiltonian
$H$ 
then commutes with the lattice-translation operator
${T}_{\bf R}$, allowing one to choose as common eigenstates the
Bloch orbitals $\vert\,\psi_{n{\bf k}}\,\rangle$:
\begin{equation}
[\,{H},\,{T}_{\bf R}\,]\;=\;0 
\;\;\Rightarrow\;\; \psi_{n{\bf k}}({\bf r})\;
=\;u_{n{\bf k}}({\bf r})\,e^{i{\bf k\cdot r}}\;\;,
\label{eq:bloch}
\end{equation}
where $u_{n{\bf k}}({\bf r})$ has the periodicity of the Hamiltonian.

\begin{figure}
\begin{center}
\includegraphics[width=7cm]{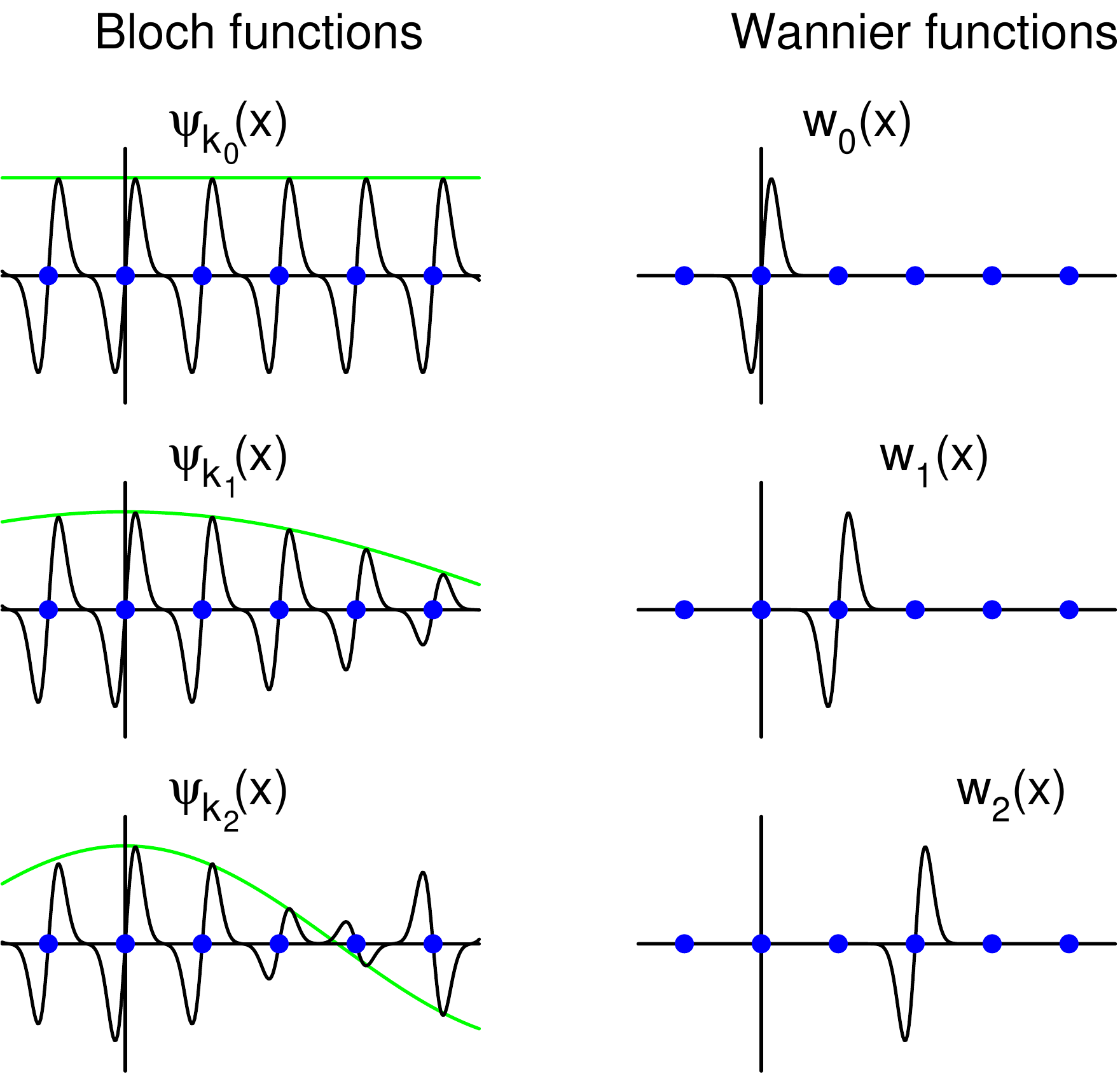}
\end{center}
\caption{\label{fig:wannier-bloch} (Color online)
Left: Bloch functions associated with a single band in 1D, at
three different values of wavevector $\kk$.  Right:
WFs associated with the same band, forming periodic
images of one another.  Blue dots indicate atoms; green curves
indicate envelopes $e^{ikx}$ of the Bloch functions.
Bloch and WFs span the same Hilbert space.}
\end{figure}

Several Bloch functions are sketched on the left-hand side of
Fig.~\ref{fig:wannier-bloch} for a toy model in which the
band of interest is composed of $p$-like orbitals centered on each atom.
For the moment, we suppose that this band is an isolated band,
i.e., it remains separated by a gap from the bands below and
above at all $\kk$.  Since Bloch functions at different $\kk$ have
different envelope functions $e^{i\kk\cdot\rr}$, one can expect
to be able to build a localized ``wave packet'' by superposing
Bloch functions of different $\kk$.  To get a very localized wave
packet in real space, we need to use a very broad superposition
in $\kk$ space.  But $\kk$ lives in the periodic Brillouin zone,
so the best we can do is to choose equal amplitudes all across
the Brillouin zone.  Thus, we can construct
\beq
w_0(\rr) \;=\;\ibz d\kk\; \psi_{n{\bf k}}(\rr)\ \;\;,
\label{eq:wannier-home}
\eeq
where $V$ is the real-space primitive cell volume and the integral is
carried over the Brillouin zone (BZ).
(See Sec.~\ref{sec:norm} for normalization conventions.)
\equa{wannier-home}
can be interpreted as the WF located in the
home unit cell, as sketched in the top-right panel of
Fig.~\ref{fig:wannier-bloch}.

More generally, we can insert a phase factor $e^{-i\kk\cdot\RR}$ into
the integrand of \equ{wannier-home}, where $\RR$ is a real-space
lattice vector; this has the effect of translating the real-space
WF by $\RR$, generating additional WFs
such as $w_1$ and $w_2$ sketched in Fig.~\ref{fig:wannier-bloch}.
Switching to the Dirac bra-ket notation and introducing the notation
that $\RR n$ refers to the WF $w_{n\RR}$ in cell $\RR$
associated with band $n$, WFs can be
constructed according to \cite{wannier-pr37}
\begin{equation}
\label{eq:wanniertransform}
\ket{\RR n}\;=\; \ibz d{\bf k}\;e^{-i{\bf k\cdot {\bf R}}} \,
\ket{\,\psi\bnk\,} \;\;.
\end{equation}
It is easily shown that the $\vert\,{{\bf R}n}\,\rangle$ form
an orthonormal set (see Sec.~\ref{sec:norm})
and that two WFs 
$\vert\,{{\bf R}n}\,\rangle$ and $\vert\,{{\bf R}^\prime}n\,\rangle$
transform into each other under translation by the lattice
vector ${\bf R}-{\bf R^\prime}$ \cite{blount-ssp62}.
\equ{wanniertransform} takes the form of a Fourier
transform, and its inverse transform is
\begin{equation}
\label{eq:wannier-inverse}
\ket{\,\psi\bnk\,} \;=\;\sum_\RR \,e^{i\kk\cdot\RR}\, \ket{\RR n}
\end{equation}
(see Sec.~\ref{sec:norm}).
Any of the Bloch functions on the left side of
Fig.~\ref{fig:wannier-bloch} can thus be built up by linearly superposing
the WFs shown on the right side, when the appropriate phases
$e^{i\kk\cdot\RR}$ are used.

The transformations of \eqs{wanniertransform}{wannier-inverse}
constitute a unitary transformation between Bloch and Wannier
states.  Thus, both sets of states
provide an equally valid description of the band subspace,
even if the WFs are not Hamiltonian eigenstates. 
For example, the charge density obtained by
summing the squares of the Bloch functions $\ket{\psi\bnk}$ or
the WFs $\ket{\RR n}$ is identical;
a similar reasoning applies to the trace of any one-particle operator.
The equivalence between the Bloch and the Wannier representations
can also be made manifest by expressing the band projection
operator ${P}$ in both representations, i.e., as
\begin{equation}
{P}=\ibz d{\bf k}\;\ket{\psi\bnk}\bra{\psi\bnk}=\sum_\RR \ket{\RR n}\bra{\RR n}\;.
\label{eq:bandproj}
\end{equation}

WFs thus provide an attractive option for representing the space
spanned by a Bloch band in a crystal, being localized
while still carrying the same information contained in
the Bloch functions.

\subsubsection{Gauge freedom}
\label{sec:gauge-freedom}

However, the theory of WFs is
made more complex by the presence of a ``gauge freedom'' that exists in
the definition of the $\psi_{n{\bf k}}$.  In fact, we can replace
\begin{equation}
\label{eq:gauge-psi}
\vert\,\tilde{\psi}_{n{\bf k}}\,\rangle\;=\;e^{i\varphi_n({\bf k})}\,\vert\,\psi_{n{\bf k}}\,\rangle\;\;,
\end{equation}
or equivalently,
\begin{equation}
\label{eq:gauge-u}
\vert\,\tilde{u}_{n{\bf k}}\,\rangle\;=\;e^{i\varphi_n({\bf k})}\,\vert\,u_{n{\bf k}}\,\rangle\;\;,
\end{equation}
without changing the physical description of the system, with
$\varphi_n({\bf k})$ being any real function that is periodic
in reciprocal space.\footnote{
  More precisely, the condition is that
  $\varphi_n({\bf k}+{\bf G})=\varphi_n({\bf k})+{\bf G}\cdot\Delta{\bf R}$
  for any reciprocal-lattice translation ${\bf G}$, where $\Delta{\bf R}$
  is a real-space lattice vector.  This allows for the possibility that
  $\varphi_n$ may shift by $2\pi$ times an integer upon translation by
  $\bf G$; the vector $\Delta{\bf R}$ expresses the corresponding
  shift in the position of the resulting WF.
}
A {\it smooth} gauge could, e.g., be defined such that
$\nabla_{\bf k}\vert u_{n{\bf k}}\rangle$ is well defined at all
$\bf k$. Henceforth we shall assume that the Bloch functions on
the right-hand side of \equ{wanniertransform} belong to a smooth
gauge, since we would not get well-localized
WFs on the left-hand side otherwise.
This is typical of Fourier transforms: the smoother the
reciprocal-space object, the more localized the resulting
real-space object, and vice versa.

One way to see this explicitly
is to consider the $\RR=0$ home-cell $w_{n0}(\rr)$
evaluated at a distant point $\rr$; using \equ{bloch}
in \equ{wanniertransform}, this is
given by $\int_{\rm BZ}\,\,u\bnk(\rr) e^{i\kk\cdot\rr}\,
d\kk$, which will be small due to cancellations arising from
the rapid variation of the exponential factor, provided that
$u\bnk$ is a smooth function of $\kk$ \cite{blount-ssp62}.

It is important to realize that the gauge freedom of
\eqs{gauge-psi}{gauge-u} propagates into the WFs.  That is, different
choices of smooth gauge correspond to different sets of WFs
having in general different shapes and spreads. In this sense, the WFs are ``more
non-unique'' than the Bloch functions, which only acquire a phase
change.  We also emphasize that there is no
``preferred gauge'' assigned by the Schr\"odinger equation to the
Bloch orbitals.  Thus, the non-uniqueness of the WFs resulting from 
\equ{wanniertransform} is unavoidable.

\subsubsection{Multiband case}
\label{sec:multi-band}

Before discussing how this non-uniqueness might be resolved, we first
relax the condition that band $n$ be a single isolated band,
and consider instead a manifold of $\nbw$ bands
that remain separated with respect to any lower or higher bands
outside the manifold.  Internal degeneracies and crossings among the
$\nbw$ bands may occur in general.  In the simplest case this manifold
corresponds to the occupied bands of an insulator, but more generally
it consists of any set of bands that is separated by a gap from both
lower and higher bands everywhere in the Brillouin zone.  Traces over
this band manifold are invariant with respect to any unitary
transformation among the $\nbw$ occupied Bloch orbitals at a given
wavevector, so it is natural to generalize the notion of a ``gauge
transformation'' to
\begin{equation}
\vert\,\tilde{\psi}_{n{\bf k}} \,\rangle=
\sum_{m=1}^\nbw\,U_{mn}^{({\bf k})}\,\vert\,\psi_{m{\bf k}} \,\rangle
\;\;.
\label{eq:blochlike}
\end{equation}
Here $U_{mn}^{({\bf k})}$ is a unitary matrix of dimension $\nbw$ that
is periodic in $\bf k$, with
\equ{gauge-psi} corresponding to the special case of a diagonal
$U$ matrix.  It follows that the projection operator onto this
band manifold at wavevector $\kk$,
\begin{equation}
{P}_{\,\kk}=\sum_{n=1}^\nbw\ket{\psi\bnk}\bra{\psi\bnk}
           =\sum_{n=1}^\nbw\ket{\tilde{\psi}\bnk}\bra{\tilde{\psi}\bnk}
\label{eq:proj-equiv}
\end{equation}
is invariant, even though the $\vert\tilde{\psi}_{n{\bf k}}
\rangle$ resulting from \equ{blochlike} are no longer generally
eigenstates of ${H}$, and $n$ is no longer a band index in
the usual sense.

\begin{figure}
\begin{center}
\includegraphics[angle=0,width=5cm]{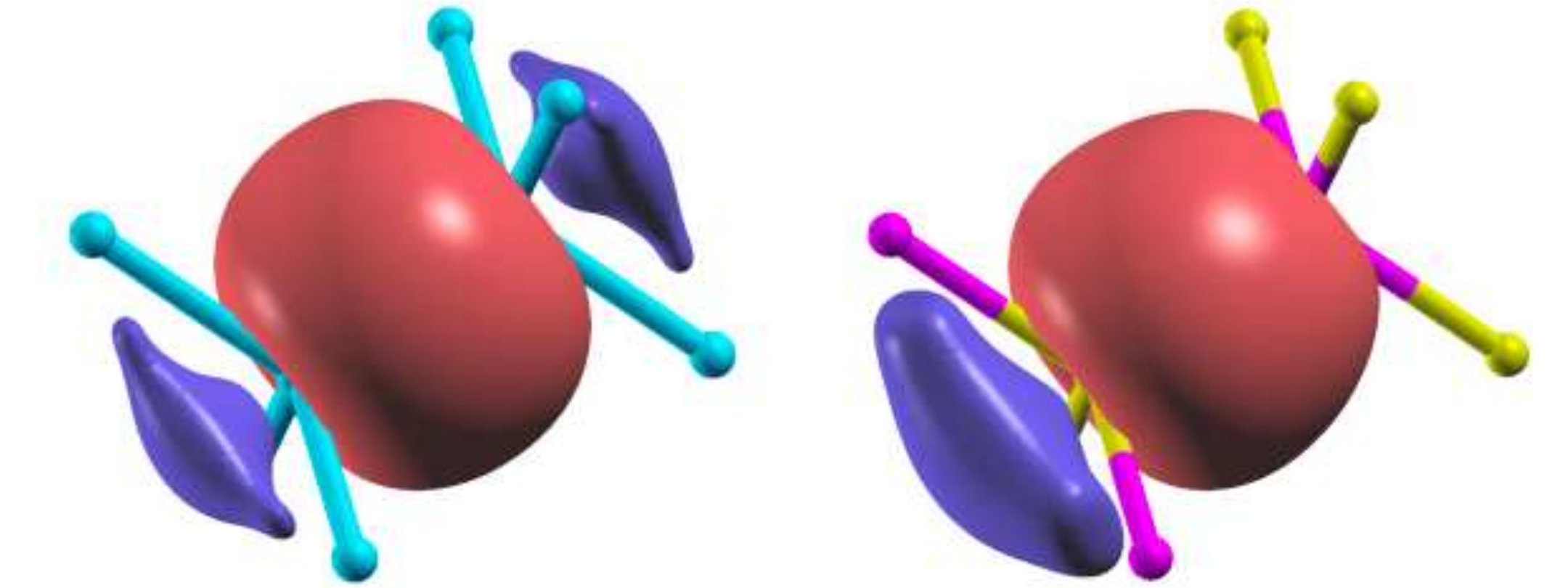}
\end{center}
\caption{\label{fig:si-gaas} (Color online) MLWFs constructed from the
four valence bands of Si (left) and GaAs (right; Ga at
upper right, As at lower left), having the
character of $\sigma$-bonded combinations of $sp^3$ hybrids. The red and
blue isosurfaces correspond to two opposite values for the amplitudes of the
real-valued MLWFs.}
\end{figure}

Our goal is again to construct WFs out of these
transformed Bloch functions using \equ{wanniertransform}.
Figs.~\ref{fig:si-gaas}(a) and (b) show, for example, what the result
might eventually look like for the case of the four occupied
valence bands of Si or GaAs, respectively.  From these four bands,
one obtains four equivalent WFs per unit cell, each
localized on one of the four nearest-neighbor Si-Si or Ga-As bonds.
The presence of a bond-centered inversion symmetry for Si, but not
GaAs, is clearly reflected in the shapes of the WFs.

\begin{figure}[b]
\begin{center}
\includegraphics[angle=0,width=9cm]{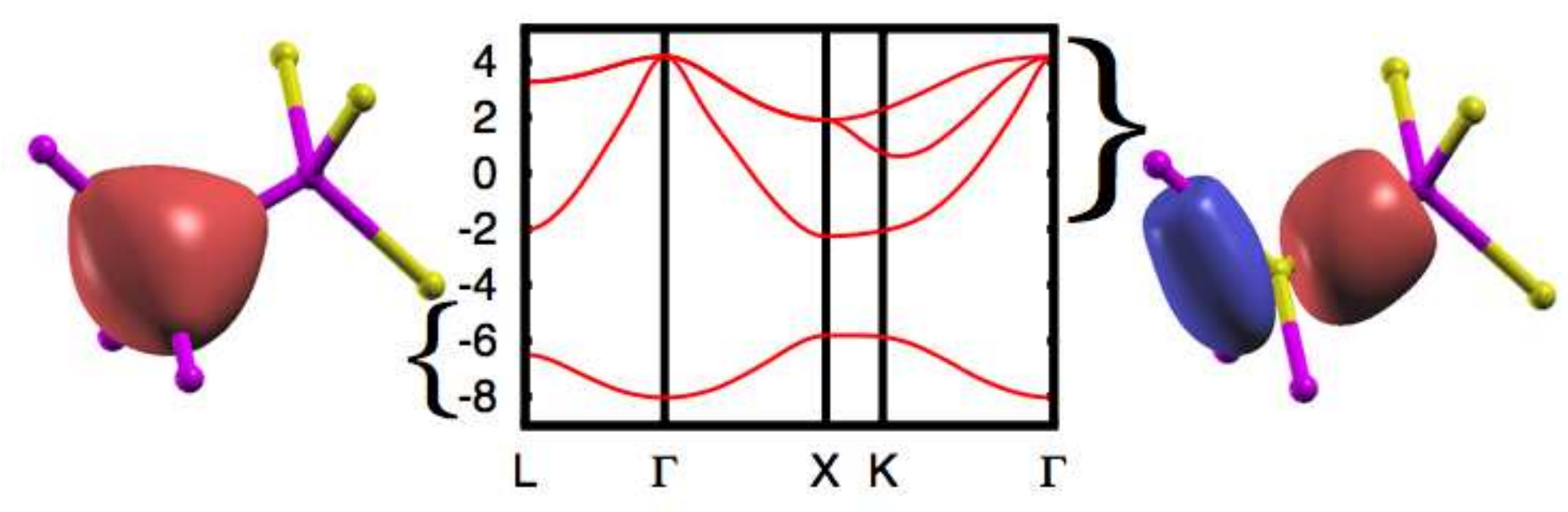}
\end{center}
\caption{\label{fig:gaas} (Color online) MLWFs constructed from
the $s$ band (left) or from the three $p$ bands (right) of GaAs.}
\end{figure}

Once again, we emphasize that the gauge freedom expressed in
\equ{blochlike} implies that the WFs are strongly
non-unique.  This is illustrated in Fig.~\ref{fig:gaas}, which
shows an alternative construction of WFs for GaAs.
The WF on the left was constructed from the lowest
valence band $n$=1, while the one on the right is one of three
constructed from bands $n$=2-4.  The former has primarily
As $s$ character and the latter has primarily As $p$ character,
although both (and especially the latter) contain some Ga $s$ and $p$
character as well.  The WFs of Figs.~\ref{fig:si-gaas}(b)
and Fig.~\ref{fig:gaas} are related to each other by a
certain manifold of 4$\times$4 unitary matrices $U_{nm}^{(\kk)}$
relating their Bloch transforms in the manner of \equ{blochlike}.

However, before we can arrive at well-localized WFs like
those shown in Figs.~\ref{fig:si-gaas} and \ref{fig:gaas}, we again
have to address questions of smoothness of the gauge choice expressed
in \equ{blochlike}.  This issue is even more profound in the
present multiband case, since this smoothness criterion
is generally {\it incompatible} with the usual construction of
Bloch functions.  That is, if we simply insert the usual Bloch
functions $\ket{\psi\bnk}$, defined to be eigenstates of ${H}$,
into the right-hand side of \equ{wanniertransform}, it is
generally {\it not} possible to produce well-localized WFs.
The problem arises when there are degeneracies among
the bands in question at certain locations in the Brillouin zone.
Consider, for example, what happens if we try to
construct a single WF from the highest occupied band $n=4$
in GaAs.  
This would be doomed to failure, since this band becomes
degenerate with bands two and three at the zone center, $\Gamma$, as shown
in Fig.~\ref{fig:gaas}. As a result, band four is non-analytic in
$\kk$ in the vicinity of $\Gamma$.  The Fourier transform of
\equ{wanniertransform} would then result in a poorly localized
object having power-law tails in real space.

In such cases, therefore, the extra unitary mixing expressed in
\equ{blochlike} is mandatory, even if it may be optional in the
case of a set of discrete bands that do not touch anywhere in the BZ.
So, generally speaking, our procedure must be that we
start from a set of Hamiltonian eigenstates
$\ket{\psi\bnk}$ that are not per se smooth in
$\kk$, and introduce unitary rotations $U_{mn}^{(\kk)}$ that
``cancel out'' the discontinuities in such a way that smoothness
is restored, i.e., that the resulting
$\ket{\tilde{\psi}\bnk}$ of \equ{blochlike} obey the smoothness
condition that $\nabla_\kk\ket{\tilde{\psi}\bnk}$ remains
regular at all $\kk$.  Then, when these $\ket{\tilde{\psi}\bnk}$ are
inserted into \equ{wanniertransform} in place of the $\ket{\psi\bnk}$,
well-localized WFs should result.  Explicitly,
this results in WFs constructed according to
\begin{equation}
\vert\,{{\bf R}n}\,\rangle\;= \;\ibz d{\bf k}
\,e^{-i{\bf k\cdot {\bf R}}}
\,\sum_{m=1}^\nbw\,U_{mn}^{({\bf k})}\,\vert\,\psi_{m{\bf k}} \,\rangle
\;\;.
\label{eq:wannier}
\end{equation}

The question remains how to choose the unitary rotations
$U_{mn}^{(\kk)}$ so as to accomplish this task.
We will see that one way to do this is to use a projection
technique, as outlined in the next section.  Ideally, however,
we would like the construction to result in
WFs that are ``maximally localized'' according
to some criterion.  Methods for accomplishing this are discussed
in Sec.~\ref{sec:maxloc}

\subsubsection{Normalization conventions}
\label{sec:norm}

In the above equations, {formulated for continuous $\kk$},
we have adopted the convention that
Bloch functions are normalized to one unit cell,
$\int_V d\rr\,|\psi_{n\kk}(\rr)|^2=1$, even though they extend
over the entire crystal.  {We also define $\inprod{f}{g}$ as
the integral of $f^*g$ over all space.} With this notation,
$\inprod{\psi_{n\kk}}{\psi_{n\kk}}$ is not unity; instead,
it diverges according to the rule
\beq
\inprod{\psi_{n\kk}}{\psi_{m\kk'}}=\frac{(2\pi)^3}{V}\,\delta_{nm}
\,\delta^3(\kk-\kk') \;\;.
\label{eq:inprodconv}
\eeq
{With these conventions it is easy to check that the WFs in
Eqs.~(\ref{eq:wanniertransform}-\ref{eq:wannier-inverse}) are
properly normalized, i.e., $\inprod{\RR n}{\RR' m}=\delta_{\RR\RR'}
\,\delta_{nm}$.}

It is often more convenient to work on a discrete uniform $\kk$ mesh
instead of continuous $\kk$ space.\footnote{The discretization of
  $\kk$-space amounts to imposing periodic boundary conditions on the
  Bloch wavefunctions over a supercell in real space.  Thus, it should
  be kept in mind that the WFs given by
  \eqs{discrete-abnormal}{discrete} are not truly localized, as they
  also display the supercell periodicity {(and are normalized to a
    supercell volume)}.  Under these circumstances the notion of
  ``Wannier localization'' refers to localization {\it within} one
  supercell, which is meaningful for supercells chosen large enough to
  ensure negligible overlap between a WF and its periodic images.
 \label{foot:discretization}}
Letting $N$ be the number of unit cells in the 
periodic supercell, or equivalently, the number of
mesh points in the BZ, it is possible to keep the conventions close to
the continuous case by defining the Fourier transform pair as
\bea
\label{eq:discrete-abnormal}
\ket{\psi_{n\kk}}&=&\sum_\RR\,e^{i\kk\cdot\RR}\,\ket{\RR n} \; \nonumber\\
&\Updownarrow& \\
\ket{\RR n}&=&\frac{1}{N}\sum_\kk \,e^{-i\kk\cdot\RR}\, \ket{\psi_{n\kk}}
\nonumber
\eea
with $\inprod{\psi_{n\kk}}{\psi_{m\kk'}}=N\delta_{nm}\,\delta_{\kk\kk'}$,
{so that \equ{bandproj} becomes, after generalizing to the multiband
case,
\begin{equation}
{P}=\frac{1}{N}\sum\bnk \ket{\psi\bnk}\bra{\psi\bnk}
       =\sum_{n\RR} \ket{\RR n}\bra{\RR n}\;.
\label{eq:mbandproja}
\end{equation}
Another commonly used convention is to write}
\bea
\label{eq:discrete}
\ket{\psi_{n\kk}}&=&
   \frac{1}{\sqrt{N}}\sum_\RR\,e^{i\kk\cdot\RR}\,\ket{\RR n} \; \nonumber\\
&\Updownarrow& \\
\ket{\RR n}&=&
   \frac{1}{\sqrt{N}}\sum_\kk \,e^{-i\kk\cdot\RR}\, \ket{\psi_{n\kk}} \nonumber
\eea
with $\inprod{\psi_{n\kk}}{\psi_{m\kk'}}=\delta_{nm}\,\delta_{\kk\kk'}$
{and \equ{mbandproja} replaced by}
\begin{equation}
{P}=\sum\bnk \ket{\psi\bnk}\bra{\psi\bnk}
       =\sum_{n\RR} \ket{\RR n}\bra{\RR n}\;.
\label{eq:mbandproj}
\end{equation}
{In either case, it is convenient to keep the $\ket{u_{n\kk}}$
functions normalized to the unit cell, with inner
products involving them, such as $\inprod{u_{m\kk}}{u_{n\kk}}$,
understood as integrals over one unit cell.  In the case
of \equ{discrete}, this means that
$u_{n\kk}(\rr)=\sqrt{N}e^{-i\kk\cdot\rr}\psi_{n\kk}(\rr)$.}

\subsection{Wannier functions via projection}
\label{sec:projection}

A simple yet often effective approach for constructing a smooth gauge
in $\kk$, and a corresponding set of well-localized WFs, is by 
projection - an approach that finds its roots in the analysis of 
\textcite{cloizeaux-pr64a}. Here, as discussed, e.g., in Sec. IV.G.1 of
\textcite{marzari-prb97}, one starts from a set of $\nbw$
{localized} trial
orbitals $g_n(\bf r)$ corresponding to some rough guess for the WFs in
the home unit cell.  {Returning to the continuous-$\kk$ formulation,}
these $g_n(\bf r)$ are projected onto the Bloch
manifold at wavevector $\bf k$ to obtain
\beq
\vert\phi_{n\bf k}\rangle=\sum_{m=1}^\nbw\,\vert\psi_{m\bf k} \rangle
\langle\psi_{m\bf k}\vert g_n\rangle\;,
\label{eq:proj}
\eeq
which are typically smooth in $\kk$-space, albeit not orthonormal.
{(The integral in $\langle\psi_{m\bf k}\vert g_n\rangle$ is over
all space as usual.)}
We note that in actual practice such projection is achieved by first
computing a matrix of inner products $(A_{\bf
  k})_{mn}=\langle\psi_{m\bf k}\vert g_n\rangle$ and then using these
in \equ{proj}.
{The overlap matrix $(S_{\bf k})_{mn}=\langle\phi_{m\bf
  k}\vert\phi_{n\bf k}\rangle_{V} =(A_\kk^\dagger
A_\kk^{\phantom{\dagger}})_{mn}$ (where subscript $V$ denotes
an integral over one cell)} is then computed and used to
construct the L\"owdin-orthonormalized Bloch-like states
\begin{equation}
\vert\tilde{\psi}_{n\bf k}\rangle = \sum_{m=1}^\nbw\, 
\vert\phi_{m\bf k}\rangle(S_{\bf k}^{-1/2})_{mn} \;.
\label{eq:symm}
\end{equation}
These $\vert\tilde{\psi}_{n\bf k}\rangle$ have now a smooth gauge in $\kk$, 
are related to the original
$\vert\psi_{n\bf k} \rangle$ by a unitary transformation,\footnote{One
  can prove that this transformation is unitary by performing the
  singular value decomposition $A=ZDW^\dagger$,
  with $Z$ and $W$ unitary and $D$ real and diagonal. It follows that
  $A(A^\dagger A)^{-1/2}$ is equal to $ZW^\dagger$, and thus
  unitary.}  and when substituted into \equ{wanniertransform} in place
of the $\vert\psi_{n\kk}\rangle$ result in a set of well-localized WFs.
We note in passing that the $\vert\tilde{\psi}_{n\bf k}\rangle$ 
 are uniquely defined by the trial orbitals $g_n(\bf r)$ and the
chosen (isolated) manifold, since any arbitrary unitary rotation among the
$\vert\psi_{n\bf k}\rangle$ orbitals cancels out exactly and does not
affect the outcome of Eq. \ref{eq:proj}, eliminating thus any gauge
freedom.

We emphasize that the trial functions do not need to resemble the
target WFs closely; it is often sufficient to choose simple analytic
functions (e.g., spherical harmonics times Gaussians) provided they
are roughly located on sites where WFs are expected to be centered and
have appropriate angular character.  The method is successful as long
as the inner-product matrix $A_\kk$ does not become singular (or
nearly so) for any $\kk$, which can be ensured by checking that the
ratio of maximum and minimum values of $\det S_\kk$ in the Brillouin
zone does not become too large.  For example, spherical ($s$-like)
Gaussians located at the bond centers will suffice for the
construction of well-localized WFs, akin to those shown in
Fig.~\ref{fig:si-gaas}, spanning the four occupied valence bands of
semiconductors such as Si and GaAs.

\subsection{Maximally localized Wannier functions}
\label{sec:maxloc}

The projection method described in the
previous subsection has been used by many authors
\cite{stephan-prb00,ku-prl02,lu-prb04,qian-prb08},
as has a related method involving downfolding of the band structure
onto a minimal basis \cite{andersen-prb00,zurek-c05}; some of these approaches
will also be discussed in Sec.~\ref{sec:proj-down}.  Other authors have made
use of symmetry considerations, analyticity requirements, and
variational procedures
\cite{sporkmann-prb94,sporkmann-jpcm97,smirnov-prb01}.
A very general and now widely used approach, however, has been that 
developed by \textcite{marzari-prb97} in which localization
is enforced by introducing a well-defined {\it localization
criterion}, and then refining the $U_{mn}^{({\bf k})}$ in order
to satisfy that criterion.  We first give an overview of this approach,
and then provide details in the following subsections.

First, the localization functional
\begin{eqnarray}
\Omega&=&\sum_n\left[    
\langle\,{{\bf 0}n}\,\vert\,r^2\,\vert\,{{\bf 0}n}\,\rangle\,-\,
\langle\,{{\bf 0}n}\,\vert\,{\bf r}\,\vert\,{{\bf 0}n}\,\rangle^2\,
\right]\nonumber\\
&=&\sum_n\left[    
\langle r^2\rangle_n-\bar{\bf r}_n^2
\right]
\label{eq:omega}
\end{eqnarray}
is defined, measuring of the sum of the quadratic spreads of the $\nwann$
WFs in the home unit cell around their centers.  This turns out to be
the solid-state equivalent of the Foster-Boys criterion of quantum
chemistry \cite{boys-rmp60,foster-rmp60a,foster-rmp60b,boys-66}.  The
next step is to express $\Omega$ in terms of the Bloch functions. This
requires some care, since expectation values of the position operator
are not well defined in the Bloch representation.  The needed
formalism will be discussed briefly in Sec.~\ref{sec:rsr} and more
extensively in Sec.~\ref{sec:bp}, with much of the conceptual work
stemming from the earlier development the modern theory of
polarization
\cite{blount-ssp62,resta-f92,king-smith-prb93,vanderbilt-prb93,resta-rmp94}.

Once a $\kk$-space expression for $\Omega$ has been derived, maximally
localized WFs are obtained by minimizing it with respect to the
$U_{mn}^{({\bf k})}$ appearing in \equ{wannier}.  This is done as a
post-processing step after a conventional electronic-structure
calculation has been self-consistently converged and a set of
ground-state Bloch orbitals $\vert\,\psi_{m{\bf k}}\,\rangle$ has been
chosen once and for all. The $U_{mn}^{({\bf k})}$ are then iteratively
refined in a direct minimization procedure of the localization
functional that is outlined in Sec.~\ref{sec:local} below.  This
procedure also provides the expectation values $\langle r^2\rangle_n$
and $\bar{\bf r}_n$; the latter, in particular, are the primary
quantities needed to compute many of the properties, such as the
electronic polarization, discussed in Sec.~\ref{sec:pol}.  If desired,
the resulting $U_{mn}^{({\bf k})}$ can also be used to construct
explicitly the maximally localized WFs via \equ{wannier}.  This step
is typically only necessary, however, if one wants to visualize the
resulting WFs or to use them as basis functions for some subsequent
analysis.

\subsubsection{Real-space representation}
\label{sec:rsr}

An interesting consequence stemming from the choice of
(\ref{eq:omega}) as the localization functional is that it allows
a natural decomposition of the functional into gauge-invariant
and gauge-dependent parts.  That is, we can write
\begin{equation}
\label{eq:spread-decomp}
\Omega\,=\,\Omega_{\rm\,I}\,+\,\widetilde{\Omega}
\end{equation}
where
\begin{equation}
\Omega_{\rm\,I} = \sum_n \left[ \langle 
\,{{\bf 0}n}\,\vert\,r^2\,\vert\,{{\bf 0}n}\,
\rangle\,-\,\sum_{{\bf R}m}
\,\bigl\vert \langle{{\bf R}m}\vert{\bf r}\vert
{\bf 0}n\rangle\bigr\vert^2 \right]
\label{eq:OmegaI}
\end{equation}
and
\begin{equation}
\widetilde{\Omega}=\sum_n \sum_{{\bf R}m\ne{\bf
0}n} \bigl\vert \langle{{\bf R}m}\vert{\bf r}\vert
{\bf 0}n\rangle\bigr\vert^2 \,.
\label{eq:Omegatil}
\end{equation}
It can be shown that not only $\widetilde{\Omega}$ but also
$\Omega_{\rm\,I}$ is {\it positive definite}, and moreover that
$\Omega_{\rm\,I}$ is {\it gauge-invariant}, i.e., invariant under any
arbitrary unitary transformation (\ref{eq:wannier}) of the Bloch
orbitals \cite{marzari-prb97}.  This follows straightforwardly from
recasting \equ{OmegaI} in terms of the band-group projection operator
$P$, as defined in \equ{mbandproj}, and its complement $Q=1-P$:
\begin{eqnarray}
\Omega_{\rm\,I} &=& \sum_{n\alpha}
\langle{{\bf 0}n}\vert r_\alpha Q r_\alpha \vert{\bf 0}n\rangle
\nonumber \\
&=& \sum_{\alpha} {\rm Tr}_{\rm c}\, [P r_\alpha Q r_\alpha ]
\;.
\label{eq:posdef}
\end{eqnarray}
The subscript `c' indicates trace per unit cell.  Clearly
$\Omega_{\rm\,I}$ is gauge invariant, since it is expressed in terms of
projection operators that are unaffected by any gauge transformation.
It can also be seen to be positive definite by using the idempotency
of $P$ and $Q$ to write
$\Omega_{\rm\,I}
= \sum_\alpha {\rm Tr}_{\rm c}\, [(P r_\alpha Q)(P r_\alpha Q)^\dagger ]
= \sum_\alpha ||Pr_\alpha Q||_{\rm c}^2$.

The minimization procedure of $\Omega$ thus actually corresponds to
the minimization of the 
non-invariant
part $\widetilde\Omega$
only.  At the minimum, the off-diagonal elements $\left\vert
\langle{{\bf R}m}\vert{\bf r}\vert {\bf 0}n\rangle\right\vert^2 $ are
as small as possible, realizing the best compromise in the
simultaneous diagonalization, within the subspace of
the Bloch bands considered, of the three position operators $x$, $y$
and $z$, which do not in general commute when projected 
onto this space.

\subsubsection{Reciprocal-space representation}
\label{sec:recipspace}

As shown by \textcite{blount-ssp62}, matrix elements of the position
operator between WFs take the form
\begin{equation}
\langle{\bf R}n\vert{\bf r}\vert{\bf 0}m\rangle = i\,{\frac{V}{(2\pi)^3}}
\int d{\bf k} \, e^{i{\bf k}\cdot{\bf R}}
\langle u_{n{\bf k}}\vert\nabla_{\bf k}\vert u_{m{\bf k}}\rangle 
\label{eq:rmatel}
\end{equation}
and
\begin{equation}
\langle{\bf R}n\vert r^2 \vert{\bf 0}m\rangle = -
{\frac{V}{(2\pi)^3}} \int d{\bf k}\,
e^{i{\bf k}\cdot{\bf R}}
\langle u_{n{\bf k}}\vert\nabla_{\bf k}^2\vert u_{m{\bf k}}\rangle
\;\;.
\label{eq:rrmatel}
\end{equation}
These expressions provide the needed connection with our underlying Bloch
formalism, since they allow to express the localization functional 
$\Omega$
in terms of the matrix elements of $\nabla_{\bf k}$ and $\nabla_{\bf k}^2$.
In addition, they allow to calculate the effects on the localization of
any unitary transformation of the $\vert u_{n{\bf k}}\rangle$  
without having to recalculate expensive (especially when
plane-wave basis sets are used) scalar products.
We thus determine the Bloch orbitals $\vert u_{m{\bf k}} \rangle$ 
on a regular mesh
of {\bf k}-points, and use finite differences to evaluate the above
derivatives. 
In particular, we make the assumption 
that the BZ has been discretized into a uniform 
Monkhorst-Pack mesh, and the Bloch orbitals have been determined on
that mesh.\footnote{Even the case of $\Gamma$-sampling --
where the Brillouin zone is sampled with a single {\bf k}-point -- is 
encompassed by the above formulation. 
In this case the neighboring {\bf k}-points
are given by reciprocal lattice vectors ${\bf G}$
and the Bloch orbitals differ only by phase factors
$\exp(i{\bf G\cdot r})$ from their counterparts at $\Gamma$. 
The algebra does become simpler, though, as will be discussed
in Sec.~\ref{sec:gpoint}.}

For any $f(\bf k)$ that is a smooth function of $\bf k$, it can be shown
that its gradient can
be expressed by finite differences as
\begin{equation}
\nabla f({\bf k})=\sum_{\bf b}\, w_b \, {\bf b}
 \,[f({\bf k+b})-f({\bf k})]
+{\cal O}(b^2)
 \;\;
\label{eq:grad}
\end{equation}
calculated on stars (``shells") of near-neighbor {\bf k}-points;
here $\bf b$ is a vector connecting a $\kk$-point to one of its 
neighbors, $w_b$ is an appropriate geometric factor that
depends on the number of points in the star and its
geometry (see Appendix B in \textcite{marzari-prb97} 
and \textcite{mostofi-cpc08} for a detailed
description). 
In a similar way,
\begin{equation}
\vert\nabla f({\bf k})\vert^2=\sum_{\bf b}\, w_b
 \,[f({\bf k+b})-f({\bf k})]^2 
+{\cal O}(b^3)
\;\;.
\label{eq:nabla}
\end{equation}

It now becomes straightforward to calculate the matrix elements in
\eqs{rmatel}{rrmatel}.  All the information needed for the
reciprocal-space derivatives is encoded in the overlaps
between Bloch orbitals at neighboring {\bf k}-points
\begin{equation}
M_{mn}^{(\bf k,b)} = \langle u_{m\bf k}\vert u_{n,\bf k+b}\rangle\;\;.
\label{eq:Mkb}
\end{equation}
These overlaps play a central role in the formalism, since all 
contributions to the localization functional can be expressed 
in terms of them. Thus, once the $M_{mn}^{(\bf k,b)}$ have been calculated,
no further interaction with the electronic-structure code that calculated
the ground state wavefunctions is necessary - making the entire
Wannierization procedure a code-independent post-processing step
\footnote{In particular, see \textcite{ferretti-jpcm07} for the
extension to ultrasoft pseudopotentials and the projector-augmented wave method,
and \textcite{posternak-prb02,freimuth-prb08,kunes-cpc10} for the full-potential linearized augmented
planewave method.}.
There is no unique form for the localization functional in terms of the overlap
elements, as it is possible to
write down many alternative finite-difference expressions for
$\bar{\bf r}_n$ and $\langle r^2 \rangle_n$ which agree numerically to
leading order in the mesh spacing $b$ (first and second order for
$\bar{\bf r}_n$ and $\langle r^2 \rangle_n$ respectively).  We give
here the expressions of \textcite{marzari-prb97}, which have the
desirable property of transforming correctly under gauge
transformations that shift $\ket{\0 n}$ by a lattice vector.  They are
\begin{equation}
\bar{\bf r}_n = - {\frac{1}{N}} \sum_{\bf k,b}\, w_b \, {\bf b} \,
{\rm Im}\,\ln M_{nn}^{(\bf k,b)} 
\label{eq:r}
\end{equation}
(where we use, as outlined in Sec.~\ref{sec:norm}, the convention of \equ{discrete}), and
\begin{equation}
\langle r^2 \rangle_n = {\frac{1}{N}} \sum_{\bf k,b}\, w_b \,
\left\{
  \left[
     1-\vert M_{nn}^{(\bf k,b)}\vert^2
  \right] +
  \left[
     {\rm Im}\,\ln M_{nn}^{(\bf k,b)}
  \right]^2 
\right\} 
\;\;.
\label{eq:rr}
\end{equation}
The corresponding expressions for the gauge-invariant and
gauge-dependent parts of the spread functional are
\begin{equation}
\label{eq:omega_i}
\Omega_{\rm I} = {\frac{1}{N}} \sum_{\bf k,b}\, w_b \,
\Big( \nbw-\sum_{mn} \vert M_{mn}^{(\bf k,b)} \vert^2 \Big)
\end{equation}
and 
\begin{eqnarray}
\label{eq:omtilde}
\widetilde\Omega &=& 
\frac{1}{N} \sum_{\bf k,b} w_b
\sum_{m\ne n} \vert M_{mn}^{\bf (k,b)} \vert^2 \\ \nonumber
&+& {\frac{1}{N}} \sum_{\bf k,b} w_b
\sum_n \left( - {\rm Im}\,\ln M_{nn}^{\bf (k,b)} 
 - {\bf b}\cdot\bar{\bf r}_n \right)^2 \;\;.
\end{eqnarray}
As mentioned, it is possible to write down alternative discretized
expressions which agree numerically with
Eqs.~(\ref{eq:r})--(\ref{eq:omtilde}) up to the orders indicated in
the mesh spacing $b$; at the same time, one needs to be careful in
realizing that certain quantities, such as the spreads, will display
slow convergence with respect to the BZ sampling (see \ref{sec:gpoint}
for a discussion), or that some exact results (e.g., that the sum of
the centers of the Wannier functions is invariant with respect to
unitary transformations) might acquire some numerical noise.  In particular,
\textcite{stengel-prb06} showed how to modify the above expressions in
a way that renders the spread functional strictly invariant under BZ
folding.

\subsection{Localization procedure}
\label{sec:local}

In order to minimize the localization functional, we 
consider the first-order change of the spread functional
$\Omega$ arising from an infinitesimal gauge transformation
$U_{mn}^{\bf(k)}=\delta_{mn}+dW_{mn}^{\bf(k)}\,,$
where $dW$ is an infinitesimal anti-Hermitian matrix,
$dW^\dagger=-dW$, so that
$\vert u_{n\bf k}\rangle \;\rightarrow\; \vert u_{n\bf k}\rangle
+ \sum_m dW_{mn}^{\bf(k)} \, \vert u_{m\bf k}\rangle\;.$
We use the convention
\begin{equation}
\left({\frac{d\Omega}{dW}}\right)_{nm}= {\frac{d\Omega}{dW_{mn}}}
\label{eq:dfdw}
\end{equation}
(note the reversal of indices)
and introduce $\cal A$ and $\cal S$ as the superoperators
${\cal A}[B]=(B-B^\dagger)/2$ and ${\cal S}[B]=(B+B^\dagger)/2i$.
Defining
\begin{equation}
q_n^{\bf(k,b)} = {\rm Im}\,\ln M_{nn}^{\bf(k,b)}
+ {\bf b\cdot\bar{r}}_n \,,
\label{eq:qdef}
\end{equation}
\begin{equation}
R_{mn}^{\bf(k,b)} = M_{mn}^{\bf(k,b)} M_{nn}^{\bf(k,b)*} \,,
\label{eq:Rdef}
\end{equation}
\begin{equation}
T_{mn}^{\bf(k,b)} = 
\frac{M_{mn}^{\bf(k,b)}}{M_{nn}^{\bf(k,b)}} \; q_n^{\bf(k,b)} \,,
\label{eq:Tdef}
\end{equation}
and referring to \textcite{marzari-prb97} for the details, we arrive
at the explicit expression for the gradient $G^{\bf(k)}=d\Omega/dW^{\bf(k)}$
of the spread functional $\Omega$ as
\begin{equation}
G^{\bf(k)} = 4 \sum_{\bf b}\, w_b
\left( {\cal A}[R^{\bf(k,b)}] - {\cal S}[T^{\bf(k,b)}]
\right) \;\;.
\label{eq:grado}
\end{equation}
This gradient is used to drive the evolution of the
$U_{mn}^{({\bf k})}$ (and, implicitly, of the
$\vert\,{{\bf R}n}\,\rangle$ of \equ{wannier}) towards the minimum
of $\Omega$. A simple steepest-descent implementation, for example,
takes small finite steps in the direction
opposite to the gradient $G$ until a minimum is reached.

For details of the minimization strategies and the enforcement of
unitarity during the search, the reader is referred to
\textcite{mostofi-cpc08}.
We should like to point out here, however, that most of the
operations can be performed using inexpensive matrix algebra on
small matrices.  The most computationally demanding parts of the
procedure are typically the calculation of the self-consistent
Bloch orbitals $\vert u_{n\bf k}^{(0)}\rangle$ in the first place,
and then the computation of a set of overlap matrices
\begin{equation}
\label{eq:overlapmv}
M_{mn}^{(0)\bf(k,b)} = \langle u_{m\bf k}^{(0)}
\vert u_{n,\bf k+b}^{(0)}\rangle
\end{equation}
that are constructed once and for all from the $\vert u_{n\bf
  k}^{(0)}\rangle$.  After every update of the unitary matrices
$U^{\bf(k)}$, the overlap matrices are updated with inexpensive matrix
algebra
\begin{equation}
M^{\bf(k,b)} = U^{\bf(k)\dagger} \, M^{(0)\bf(k,b)} \, U^{\bf(k+b)}
\label{eq:newM}
\end{equation}
without any need to access the Bloch wavefunctions themselves.
This not only makes the algorithm computationally fast and efficient,
but also makes it independent of the basis used to represent the
Bloch functions.  That is, any electronic-structure code package
capable of providing the set of overlap matrices $M^{\bf(k,b)}$
can easily be interfaced to a common Wannier maximal-localization code.

\subsection{{Local minima}}
\label{sec:local-minima}

{
It should be noted that the localization functional can
display, in addition to the desired global minimum,
multiple local minima that do not lead to the construction
of meaningful localized orbitals.
Heuristically, it is also found that the WFs corresponding to
these local minima are intrinsically complex, while they are 
found to be real, a part from a single complex phase,
at the desired global minimum (provided of course that
the calculations do not include spin-orbit coupling). 
Such observation in itself provides a useful diagnostic tool to 
weed out undesired solutions.
}

{
These false minima either correspond to the formation
of topological defects (e.g., ``vortices'')  in an otherwise smooth
gauge field in discrete $\kk$ space, or they can
arise when the branch cuts for the
complex logarithms in \equ{r} and \equ{rr} are inconsistent, i.e.,
when at any given {\bf k}-point the contributions from different {\bf
b}-vectors differ by random amounts of 2$\pi$ in the logarithm. Since a 
locally appropriate choice of branch cuts can always be performed, 
this problem is less severe than the topological problem.
The most straightforward way to avoid local minima altogether
is to initialize the minimization procedure with a gauge choice 
that is already fairly smooth.  For this purpose, the
projection method already described in Sec.~\ref{sec:projection} has been
found to be extremely effective.
Therefore, minimization is usually preceded by a
projection step, to generate a set of analytic Bloch
orbitals to be further optimized, as discussed more 
fully in \textcite{marzari-prb97} and \textcite{mostofi-cpc08}.
}

\subsection{The limit of isolated systems or large supercells}
\label{sec:isolated}

\begin{figure}
\begin{center}
\includegraphics[width=6.0cm]{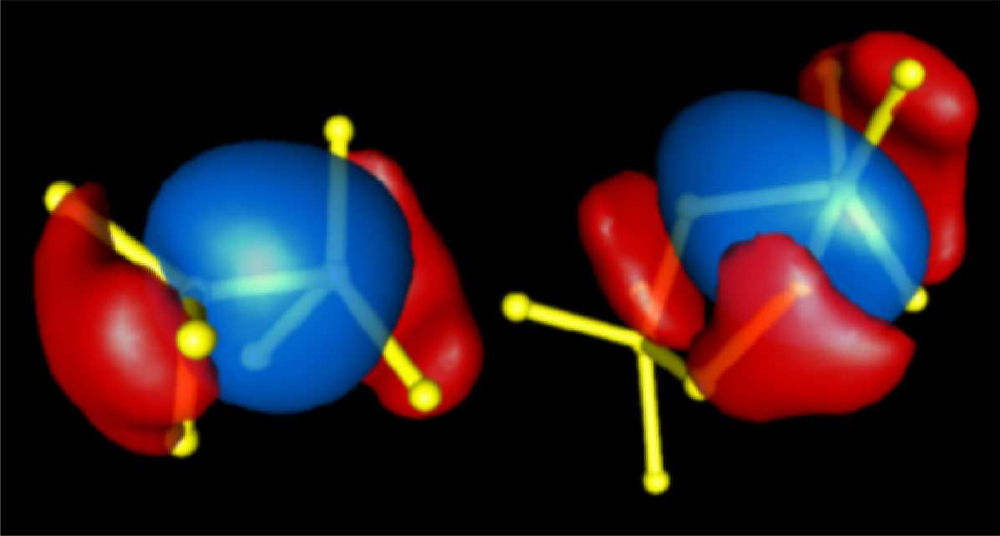}
\end{center}
\caption{\label{fig:a-si} (Color online) MLWFs
in amorphous Si, either around distorted but fourfold coordinated atoms,
or in the presence of a fivefold defect.
Adapted from \textcite{fornari-cms01}.}
\end{figure}

The formulation introduced above can be significantly simplified in
two important and related cases, which merit a separate
discussion. The first is the case of open boundary conditions: this is
the most appropriate choice for treating finite, isolated systems
(e.g., molecules and clusters) using localized basis sets, and is a
common approach in quantum chemistry.  The localization procedure can
then be entirely recast in real space, and corresponds exactly to
determining Foster-Boys localized orbitals.  The second is the case of
systems that can be described using very large periodic
supercells. This is the most appropriate strategy for non-periodic
bulk systems, such as amorphous solids or liquids (see
Fig.~\ref{fig:a-si} for a paradigmatic example), but obviously
includes also periodic systems with large unit cells.  In this
approach, the Brillouin zone is considered to be sufficiently small
that integrations over ${\bf k}$-vectors can be approximated with a
single ${\bf k}$-point (usually at the $\Gamma$ point, i.e., the
origin in reciprocal space).  The localization procedure in this
second case is based on the procedure for periodic boundary conditions
described above, but with some notable simplifications.  Isolated
systems can also be artificially repeated and treated using the
supercell approach, although care may be needed in dealing with the
long-range electrostatics if the isolated entities are charged or have
significant dipole or multipolar character \cite{makov-prb95,
  dabo-prb08}.

\subsubsection{Real-space formulation for isolated systems}

For an isolated system, described with open boundary conditions, all
orbitals are localized to begin with, and further localization 
is achieved via unitary transformations within this set.
We adopt a simplified notation
$\vert{\bf R}n\rangle\rightarrow\vert w_i\rangle$ to refer
to the localized orbitals of the isolated system
that will become maximally localized. 
We decompose again the localization functional
$\Omega=\sum_i [\langle r^2\rangle_i-\bar{\bf r}_i^2]$ into a
term $\Omega_{\rm I}=\sum_\alpha{\rm tr}\,[Pr_\alpha
Qr_\alpha]$ (where $P=\sum_i\vert w_i\rangle\langle w_i\vert$,
$Q={\bf 1}-P$, and `$\rm tr$' refers to a sum over all the states $w_i$) that is invariant under unitary rotations, and 
a remainder 
$\widetilde\Omega = \sum_\alpha \sum_{i\ne j} \, \vert\langle w_i
\vert r_\alpha \vert w_j \rangle \vert^2$ that needs to be minimized.
Defining the matrices $X_{ij}=\langle w_i \vert x \vert w_j \rangle$,
$X_{{\rm D},ij}=X_{ij}\,\delta_{ij}$, $X^\prime=X-X_{\rm D}$ (and
similarly for $Y$ and $Z$), $\widetilde\Omega$ can be rewritten as
\begin{equation}
\label{eq:omegatil}
\widetilde\Omega = 
{\rm tr}\,[X^{\prime\,2}+Y^{\prime\,2}+Z^{\prime\,2}] \;\;.
\end{equation}
If $X$, $Y$, and $Z$ could be simultaneously diagonalized, then
$\widetilde\Omega$ would be minimized to zero (leaving only
the invariant part). 
This is straightforward in one dimension, but is not generally possible
in higher dimensions.
The general solution to the three-dimensional problem consists instead
in the optimal, but approximate, simultaneous co-diagonalization of
the three Hermitian matrices $X$, $Y$, and $Z$ by a single unitary
transformation that minimizes the numerical value of the localization
functional.  Although a formal solution to this problem is missing,
implementing a numerical procedure (e.g., by steepest-descent or
conjugate-gradients minimization) is fairly straightforward. It should
be noted that the problem of simultaneous co-diagonalization arises
also in the context of multivariate analysis \cite{flury-sjssc86} and
signal processing \cite{cardoso-sjmaa96}, and has been recently
revisited in relation with the present localization approach
\cite{gygi-cpc03} (see also Sec.~IIIA in \textcite{berghold-prb00}).

To proceed further, we write
\begin{equation}
d\Omega = 2\,{\rm tr}\,[X^{\prime}dX+Y^{\prime}dY+Z^{\prime}dZ] \;\;,
\label{eq:dotr}
\end{equation}
exploiting the fact that ${\rm tr}\,[X^{\prime}X_{\rm D}]=0$, and similarly for
$Y$ and $Z$.
We then consider an infinitesimal unitary transformation $\vert
w_i\rangle \;\rightarrow\; \vert w_i\rangle +\sum_j dW_{ji} \vert
w_j\rangle$ (where $dW$ is antihermitian),
from which $dX=[X,dW]$, and similarly for $Y$, $Z$.  Inserting in \equ{dotr}
and using ${\rm tr}\,[A[B,C]]=\,{\rm tr}\,[C[A,B]]$ and
$[X^{\prime},X]=[X^{\prime},X_{\rm D}]$, we obtain
$d\Omega={\rm tr}\,[dW\,G]$ where the gradient of the spread functional
$G={d\Omega/dW}$ is given by
\begin{equation}
G= 2\,\Bigl\{ [X^{\prime},X_{\rm D}] + [Y^{\prime},Y_{\rm D}]
+ [Z^{\prime},Z_{\rm D}]\,dW \Bigr\} \;\;.
\label{eq:Greal}
\end{equation}
The minimization can then be carried out using a
procedure similar to that outlined above for
periodic boundary conditions.
Last, we note that minimizing 
$\widetilde\Omega$ is equivalent to maximizing 
${\rm tr}\,[X^2_{\rm D}+Y^2_{\rm D}+Z^2_{\rm D}]$, since
${\rm tr}\,[X^{\prime}X_{\rm D}]={\rm tr}\,[Y^{\prime}Y_{\rm D}]={\rm tr}\,[Y^{\prime}Y_{\rm D}]=0$.

\subsubsection{$\Gamma$-point formulation for large supercells}
\label{sec:gpoint}

A similar formulation applies in reciprocal space when 
dealing with isolated or very large systems
in periodic boundary conditions,
i.e., whenever it becomes
appropriate to sample the wavefunctions only at the $\Gamma$-point of 
the Brillouin zone.
For simplicity, we start with the case of a simple cubic lattice
of spacing $L$, and 
define the matrices $\cal{X}$, $\cal{Y}$, and $\cal{Z}$ as
\begin{equation}
{\cal{X}}_{mn}=\langle  w_m | e^{-2\pi ix/L} | w_n \rangle
\end{equation}
and its periodic
permutations (the extension to supercells of arbitrary symmetry has been derived by
\textcite{silvestrelli-prb99}).\footnote{We point out that the definition of the 
$\cal{X}, \cal{Y}, \cal{Z}$ matrices
for extended systems, now common in the literature, is different from the one used in the previous 
subsection for the $X, Y, Z$ matrices for isolated system.
We preserved these two notations for the sake of consistency with published work, 
at the cost of making less evident the close similarities that exist
between the two minimization algorithms.}
The coordinate $x_n$ of the $n$-th WF
center (WFC) can then be obtained from
\begin{equation}
x_n = -{\frac{L}{{2\pi}}}\,{\rm Im}\; {\rm ln} 
\langle  w_n | e^{-i{\frac{2\pi}{L}} x} | w_n \rangle 
= -{\frac{L}{{2\pi}}}\,{\rm Im}\; {\rm ln}\; {\cal{X}}_{nn} \;,
\label{eq:rcenter}
\end{equation}
and similarly for $y_n$ and $z_n$.
\equa{rcenter} has been shown by \textcite{resta-prl98}
to be the correct definition of the expectation value of 
the position operator for a system with periodic boundary conditions,
and had been introduced several years ago to deal with the 
problem of determining the average position of a single electronic 
orbital
in a periodic supercell \cite{selloni-prl87} (the
above definition becomes self-evident in the limit where $w_n$ tends
to a Dirac delta function).

Following the derivation of the previous subsection, or of
\textcite{silvestrelli-ssc98}, it can be shown that the
maximum-localization criterion is equivalent to maximizing the functional
\begin{equation}
\Xi = \sum_{n=1}^\nwann \left(|{\cal{X}}_{nn}|^2+|{\cal{Y}}_{nn}|^2+|{\cal{Z}}_{nn}|^2\right)\;.
\label{eq:ximax}
\end{equation}
The first term of the gradient $d\Xi/dA_{mn}$ is given by
$[{\cal{X}}_{nm}({\cal{X}}_{nn}^*-{\cal{X}}_{mm}^*)-{\cal{X}}_{mn}^*({\cal{X}}_{mm}-{\cal{X}}_{nn})]$, and 
similarly
for the second and third terms.  Again, once the gradient 
is determined, minimization can be performed using a steepest-descent
or conjugate-gradients algorithm; as always, the computational cost
of the localization procedure is small, given that it involves only
small matrices of dimension $\nbw \times \nbw$, 
once the scalar products needed to construct the initial
${\cal{X}}^{(0)}$, ${\cal{Y}}^{(0)}$ and ${\cal{Z}}^{(0)}$ have been calculated, which takes
an effort of order $\nbw^2*N_{\rm basis}$.
We note that in the 
limit of a single $\kk$ point the distinction between Bloch orbitals
and WFs becomes irrelevant, since no Fourier transform
from ${\bf k}$ to ${\bf R}$ is involved in the transformation
\equ{wannier}; rather, we want to find the optimal unitary
matrix that rotates the ground-state self-consistent orbitals
into their maximally localized representation, making this problem exactly equivalent to the one
of isolated systems.
{Interestingly, it should also be mentioned that the local minima alluded to
in the previous subsection are typically not 
found when using $\Gamma$ sampling in large supercells.}

Before concluding, we note that care should be taken when comparing
the spreads of MLWFs calculated in supercells of different sizes.
The Wannier centers and the general shape of the MLWFs often
converge rapidly as the cell size is increased, but
even for the ideal case of an isolated molecule,
the total spread $\Omega$ often displays much slower convergence.
This behavior derives from the finite-difference representation of the
invariant part $\Omega_{\,\rm I}$ of the localization functional
(essentially, a second derivative); while $\Omega_{\,\rm I}$
does not enter into the localization procedure, it does contribute
to the spread, and in fact usually represents the largest term.  This
slow convergence was noted by \textcite{marzari-prb97}
when commenting on the convergence properties of $\Omega$
with respect to the spacing of the Monkhorst-Pack mesh, and
has been studied in detail by others \cite{umari-prb03,stengel-prb06}.
For isolated systems in a supercell, this problem can be avoided
simply by using a very large $L$ in \equ{rcenter}, since in practice
the integrals only need to be calculated in the small region where
the orbitals are non-zero \cite{wu-phd04}. For extended bulk systems,
this convergence problem can be ameliorated significantly by
calculating the position operator using real-space integrals
\cite{lee-prl05,stengel-prb06,lee-phd06}.

\subsection{{Exponential localization}}
\label{sec:exp-loc}

{
The existence of exponentially localized WFs -- i.e., WFs whose
tails decay exponentially fast -- is a famous problem in the band
theory of solids, with close ties to the general properties of the
insulating state~\cite{kohn-pr64}.  
While the asymptotic decay of a Fourier transform 
can be related to the analyticity of the function and its
derivatives (see, e.g., \textcite{duffin-dmj53,duffin-dmj60} and
references therein), proofs of exponential
localization for the Wannier transform were obtained over the 
years only for limited cases,
starting with the work of \textcite{kohn-pr59}, who considered a
one-dimensional crystal with inversion symmetry.  Other milestones
include the work of \textcite{cloizeaux-pr64b}, who established the
exponential localization in 1D crystals without inversion symmetry
and in the centrosymmetric 3D case, and the subsequent removal of
the requirement of inversion symmetry in the latter case by
\textcite{nenciu-cmp83}.  
The asymptotic behavior of WFs was further clarified by
\textcite{he-prl01}, who found that the exponential decay is
modulated by a power law.
In dimensions $d>1$ the problem is further
complicated by the possible existence of band degeneracies, while
the proofs of des Cloizeaux and Nenciu were restricted to isolated
bands. The early work on composite bands in 3D only established the
exponential localization of the projection operator $P$,
\equ{mbandproja}, not of the WFs themselves~\cite{cloizeaux-pr64a}.
}

{
The question of exponential decay in 2D and 3D 
was finally settled by \textcite{brouder-prl07}
who showed, as a corollary to a
theorem by \textcite{panati-ahp07}, that a necessary and sufficient
condition for the existence of exponentially localized WFs in 2D and
3D is that the so-called ``Chern invariants'' for the composite
set of bands vanish identically.  \textcite{panati-ahp07} had
demonstrated that this condition ensures the possibility of carrying
out the gauge transformation of \equ{blochlike} in such a way that
the resulting cell-periodic states $\ket{\tilde{u}_{n{\bf k}}}$ are
analytic functions of ${\bf k}$ across the entire
BZ;\footnote{{Conversely, nonzero Chern numbers pose an
obstruction to finding a {\it globally} smooth gauge in ${\bf
k}$-space. The mathematical definition of a Chern number is
given in Sec.~\ref{sec:top}.}}
this in turn implies the
exponential falloff of the WFs given by \equ{wannier}.
}

{
It is natural to ask whether the MLWFs obtained by
minimizing the quadratic spread functional $\Omega$ are also
exponentially localized.  \textcite{marzari-prb97}
established this in 1D, by simply noting that the MLWF construction
then reduces to finding the eigenstates of the projected position
operator $PxP$, a case for which exponential localization had
already been proven~\cite{niu-mplb91}. The more complex
problem of exponential localization of MLWFs for composite bands in 
2D and 3D was finally proven by \textcite{panati-arxiv11}.
}

\subsection{Hybrid Wannier functions}
\label{sec:hybrid}

It is sometimes useful to carry out the Wannier transform in one
spatial dimension only, leaving wavefunctions that are still
delocalized and Bloch-periodic in the remaining directions
\cite{sgiarovello-prb01}.  Such orbitals are usually denoted as
``hermaphrodite'' or ``hybrid'' WFs.  Explicitly,
Eq.~(\ref{eq:wannier}) is replaced by the hybrid WF definition
\begin{equation}
\vert\,{l,n\kk_\parallel}\,\rangle\;=\;\frac{c}{2\pi}
\;\int_0^{2\pi/c} \ket{\,\psi\bnk\,}
\,e^{-ilk_\perp c}\,dk_\perp \;,
\label{eq:hybrid}
\end{equation}
where $\kk_\parallel$ is the wavevector in the plane (delocalized
directions) and $k_\perp$, $l$, and $c$ are the wavevector, cell
index, and cell dimension in the direction of localization.  The 1D
Wannier construction can be done independently for each
$\kk_\parallel$ using direct (i.e., non-iterative) methods as
described in Sec.~IV\,C\,1 of \textcite{marzari-prb97}.

Such a construction has proved useful for a variety of purposes, from
verifying numerically exponential localization in 1 dimension, to
treating electric polarization or applied electric fields along a
specific spatial direction
\cite{giustino-prl03,giustino-prb05,wu-prl06,stengel-prb06,murray-prb09}
or for analyzing aspects of topological insulators
\cite{coh-prl09,soluyanov-prb11a,soluyanov-prb11b}.  Examples will be
discussed in Secs.~\ref{sec:layer} and \ref{sec:surface-bands}.

\subsection{Entangled bands}
\label{sec:entangled}

The methods described in the previous sections were designed with {\it
  isolated} groups of bands in mind, separated from all other bands by
finite gaps throughout the entire Brillouin zone.  However, in many
applications the bands of interest are not isolated. This can happen,
for example, when studying electron transport, which is governed by
the partially filled bands close to the Fermi level, or when dealing
with empty bands, such as the four low-lying antibonding bands of
tetrahedral semiconductors, which are attached to higher conduction
bands. Another case of interest is when a partially filled manifold is
to be downfolded
into a basis of WFs - e.g.,  to construct model Hamiltonians. In all
these examples the desired bands lie within a limited energy range but
overlap and hybridize with other bands which extend further out in
energy.  We will refer to them as {\it entangled bands}.

The difficulty in treating entangled bands stems from the fact that it
is unclear exactly which states to choose to form $\nwann$ WFs,
particularly in those regions of $\kk$-space where the bands of interest
are hybridized with other unwanted bands.  Before a Wannier
localization procedure can be applied, some prescription is needed for
constructing $\nwann$ states per $\kk$-point from a linear combination
of the states in this larger manifold.  Once a suitable
$\nwann$-dimensional Bloch manifold has been identified at each $\kk$,
the same procedure described earlier for an isolated group of bands
can be used to generate localized WFs spanning that manifold.

The problem of computing well localized WFs starting from entangled
bands is thus broken down into two distinct steps, subspace selection
and gauge selection.  As we will see, the same guiding principle can
be used for both steps, namely, to achieve ``smoothness'' in
$\kk$-space.  In the subspace selection step a $\nwann$-dimensional
Bloch manifold which varies smoothly as function of $\kk$ is
constructed. In the gauge-selection step that subspace is represented
using a set of $\nwann$ Bloch functions which are themselves smooth
functions of $\kk$, such that the corresponding WFs are well
localized.

\subsubsection{Subspace selection via projection}
\label{sec:subspace_proj}

The projection technique discussed in Section \ref{sec:projection} can
be easily adapted to produce $\nwann$ smoothly-varying Bloch-like
states starting from a {\it larger} set of Bloch
bands~\cite{souza-prb01}.  The latter can be chosen, for example, as
the bands lying within a given energy window, or within a specified
range of band indices.  Their number $\nbands_\kk\geq \nwann $ is not
required to be constant throughout the BZ.

We start from a set of $\nwann$ localized trial orbitals $g_n(\bf r)$
and project each of them onto the space spanned by the chosen
eigenstates at each $\kk$,
\begin{equation}
  \vert\phi_{n\bf k}\rangle=\sum_{m=1}^{\nbands_\kk}\, \vert\psi_{m\bf k}\rangle\,\langle\psi_{m\bf k}\vert g_n\rangle\;.
\label{eq:disen_proj}
\end{equation}
This is identical to \equ{proj}, except for the fact that the overlap
matrix $(A_{\bf k})_{mn}=\langle\psi_{m\bf k}\vert g_n\rangle$ has
become rectangular with dimensions $\nbands_\kk \times\nwann$.  We
then orthonormalize the resulting $\nwann$ orbitals using \equ{symm},
to produce a set of $\nwann$ smoothly-varying Bloch-like states across
the BZ, 
\begin{equation}
\vert\tilde{\psi}_{n\bf k}\rangle = \sum_{m=1}^\nwann\, 
\vert\phi_{m\bf k}\rangle(S_{\bf k}^{-1/2})_{mn}\;.
\label{eq:disen_orth}
\end{equation}
As in \equ{symm}, 
$(S_{\bf k})_{mn}={\langle\phi_{m\bf
  k}\vert\phi_{n\bf k}\rangle_{V}} =(A_\kk^\dagger
A_\kk^{\phantom{\dagger}})_{mn}$, but with rectangular $A_\kk$
matrices.

The above procedure achieves simultaneously the two goals of subspace
selection and gauge selection, although neither of them is performed
optimally.  The gauge selection can be further refined by minimizing
$\widetilde{\Omega}$ within the projected subspace. It is also
possible to refine iteratively the subspace selection itself, as will
be described in the next section. However, for many applications this
one-shot procedure is perfectly adequate, and in some cases it may
even be preferable to more sophisticated iterative approaches
{(see also Sec.~\ref{sec:locorb-disc}.)}
For
example, it often results in ``symmetry-adapted'' WFs which inherit
the symmetry properties of the trial orbitals~\cite{ku-prl02}.

\begin{figure}
\begin{center}
\includegraphics[width=8.7cm]{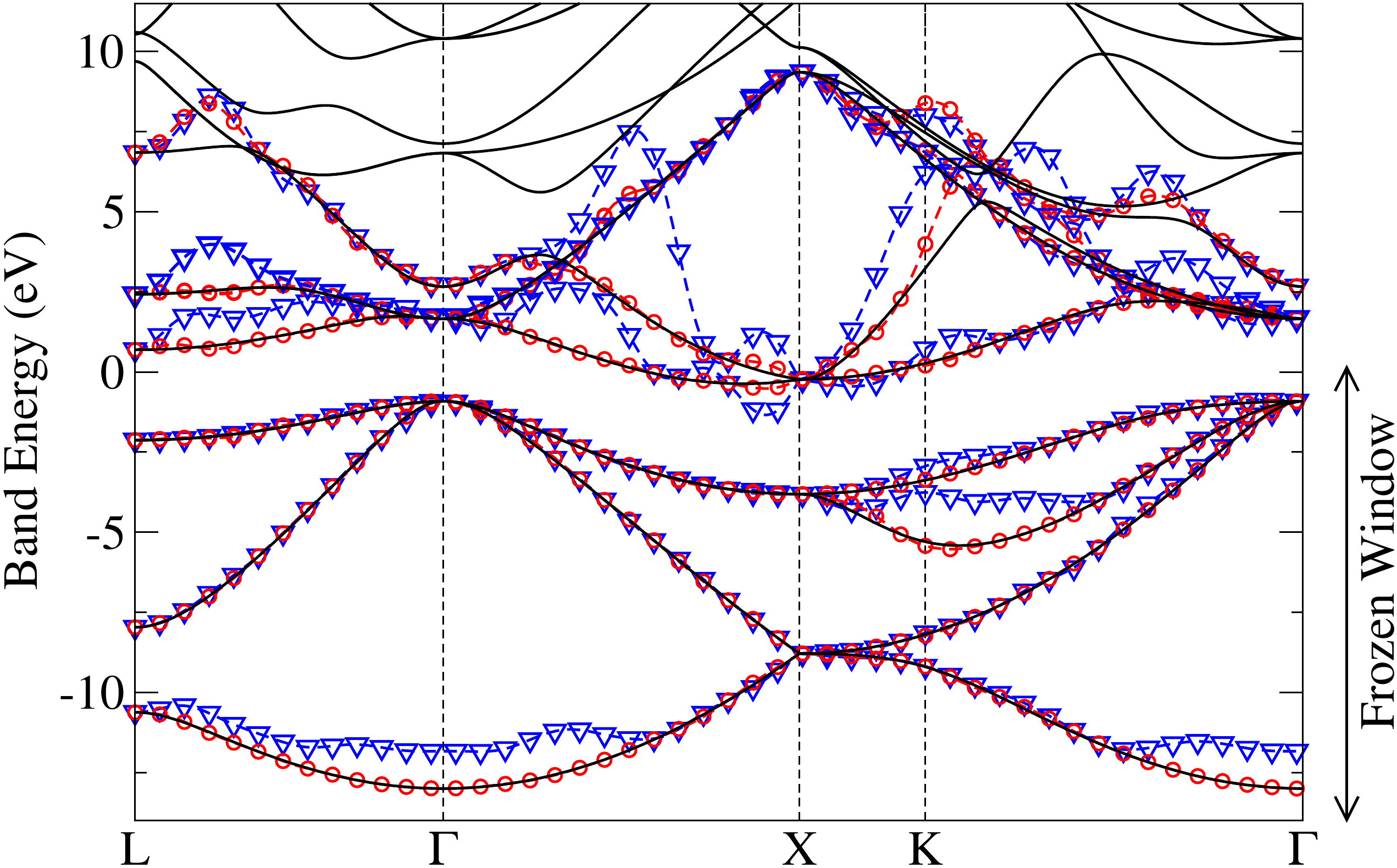}
\end{center}
\caption{\label{fig:si_proj} (Color online) Solid black lines: band
  structure of bulk crystalline Si. Blue triangles: band
  structure for the subspace selected by projection onto atomic $sp^3$
  orbitals. Red circles: band structure for the subspace selected by
  projection onto atomic $sp^3$ orbitals and forcing the valence
  manifold to be reproduced exactly using the frozen window
  indicated.}
\end{figure}

As an example, we plot in Fig.~\ref{fig:si_proj} the eight
disentangled bands obtained by projecting the band structure of
silicon, taken within an energy window that coincides with the entire
energy axis shown, onto eight atomic-like $sp^3$ hybrids. The
disentangled bands, generated using Wannier interpolation
(Sec.~\ref{sec:band-interpolation}), are shown as blue triangles,
along with the original first-principles bands (solid lines).  While
the overall agreement is quite good, significant deviations can be
seen wherever higher unoccupied and unwanted states possessing some
significant $sp^3$ character are admixed into the projected
manifold. This behavior can be avoided by forcing certain Bloch states
to be preserved identically in the projected manifold - we refer to
those as belonging to a frozen ``inner'' window, since this is often
the simplest procedure for selecting them.  The placement and range of
this inner window will depend on the problem at hand. For example, in
order to describe the low-energy physics for, e.g., transport
calculations, the inner window would typically include all states in a
desired range around the Fermi level.

We show as red circles in Fig.~\ref{fig:si_proj} the results obtained
by forcing the entire valence manifold to be preserved, leading to a
set of eight projected bands that reproduce exactly the four valence
bands, and follow quite closely the four low-lying conduction bands.
For the modifications to the projection algorithm required to enforce
this frozen inner window, we refer to Sec.~III.G of
\textcite{souza-prb01}.

\begin{figure}
\begin{center}
\includegraphics[width=8cm]{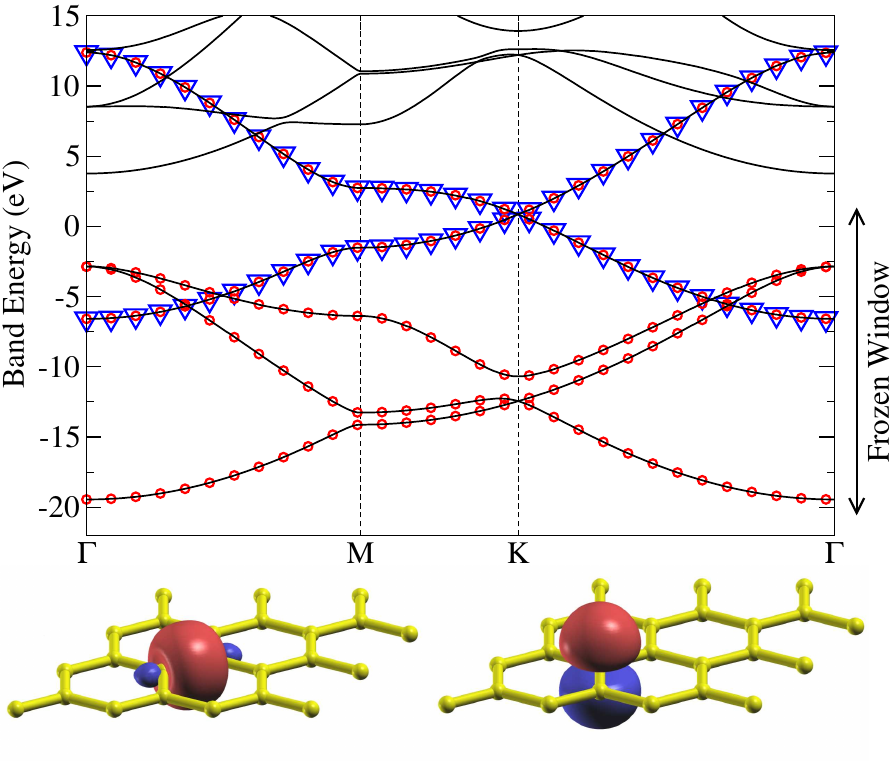}
\end{center}
\caption{\label{fig:graph_pz} (Color online) Solid black lines: band
  structure of graphene. Blue triangles: band structure for the
  subspace selected by projection onto atomic $p_z$ orbitals. Red
  circles: band structure for the subspace selected by projection onto
  atomic $p_z$ orbitals on each site and sp$^2$ orbitals on alternate
  sites, and using the frozen window indicated.  The lower panels show
  the MLWFs obtained from the standard localization procedure applied
  to these two projected manifolds.}
\end{figure}

Projection techniques can work very well, and an application of this
approach to graphene is shown in Fig.~\ref{fig:graph_pz}, where the
$\pi/\pi^\star$ manifold is disentangled with great accuracy by a
simple projection onto atomic $p_z$ orbitals, or the entire occupied
manifold together with $\pi/\pi^\star$ manifold is obtained by
projection onto atomic $p_z$ and $sp^2$ orbitals (one every other
atom, for the case of the $sp^2$ orbitals - albeit bond-centered $s$
orbitals would work equally well).

Projection methods have been extensively used to study
strongly-correlated systems \cite{ku-prl02,anisimov-prb05a,ku-prl11},
in particular to identify a ``correlated subspace'' for LDA+U or DMFT
calculations, as will be discussed in more detail in
Sec.~\ref{sec:basis}. It has also been argued~\cite{ku-prl10} that
projected WFs provide a more appropriate basis for the study of
defects, as the pursuit of better localization in a MLWF scheme
risks defining the gauge differently for the defect WF as compared to
the bulk. Instead, a straightforward projection approach ensures the
similarity between the WF in the defect (supercell) and in the
pristine (primitive cell) calculations, and this has been exploited to
develop a scheme to unfold the band-structure of disordered supercells
into the Brillouin zone of the underlying primitive cell, allowing a
direct comparison with angular resolved photoemission spectroscopy
experiments~\cite{ku-prl10}.

\subsubsection{Subspace selection via optimal smoothness}
\label{sec:disentanglement}

The projection onto trial orbitals provides a simple and effective way
of extracting a smooth Bloch subspace starting from a larger set of
entangled bands. The reason for its success is easily understood: the
localization of the trial orbitals in real space leads to smoothness
in $\kk$-space.  In order to further refine the subspace selection
procedure, it is useful to introduce a precise measure of the
smoothness in $\kk$-space of a manifold of Bloch states. The search for
an optimally-smooth subspace can then be formulated as a minimization
problem, similar to the search for an optimally-smooth gauge.

As it turns out, smoothness in $\kk$ of a Bloch space is precisely
what is measured by the functional $\Omega_{\rm I}$ introduced in
Sec.~\ref{sec:rsr}. We know from \equ{spread-decomp} that the
quadratic spread $\Omega$ of the WFs spanning a Bloch space of
dimension $\nwann$ comprises two positive-definite contributions, one
gauge-invariant ($\Omega_{\rm I}$), the other gauge-dependent
($\widetilde{\Omega}$).  Given such a Bloch space (e.g., an isolated
group of bands, or a group of bands previously disentangled via
projection), we have seen that the optimally smooth gauge can be found
by minimizing $\widetilde{\Omega}$ with respect to the unitary mixing
of states within that space.

From this perspective, the gauge-invariance of $\Omega_{\rm I}$ means
that this quantity is insensitive to the smoothness of the individual
Bloch states $\ket{\tilde{u}_{n\kk}}$ chosen to represent the Hilbert
space.  But considering that $\Omega_{\rm I}$ is a part of the spread
functional, it must describe smoothness in some other sense.  What
$\Omega_{\rm I}$ manages to capture is the {\it intrinsic} smoothness
of the underlying Hilbert space. This can be seen starting from the
discretized $\kk$-space expression for $\Omega_{\rm I}$, \equ{omega_i},
and noting that it can be written as
\beq
\label{eq:omega_i_spillage}
\Omega_{\rm I}=\frac{1}{N}\sum_{{\bf k},{\bf b}} \,
w_b \, T_{{\bf k},{\bf b}}
\eeq
with
\beq
\label{eq:spillage}
T_{{\bf k},{\bf b}}= {\rm Tr} [{P}_{\bf k}\,{Q}_{{\bf k}+{\bf b}}],
\eeq
where ${P}_{\bf k}=\sum_{n=1}^\nwann \, \ket{\tilde{u}_{n \bf
    k}}\bra{\tilde{u}_{n \bf k}}$ is the gauge-invariant projector
onto the Bloch subspace at $\kk$, ${Q}_{\bf k}={\bf 1}-{P}_{\bf k}$,
and ``Tr'' denotes the electronic trace over the full Hilbert
space. It is now evident that $T_{{\bf k},{\bf b}}$ measures the
degree of mismatch (or ``spillage'') between the neighboring Bloch
subspaces at $\kk$ and $\kk+{\bf b}$, vanishing when they are
identical, and that $\Omega_{\rm I}$ provides a BZ average of the
local subspace mismatch.

The optimized subspace selection procedure can now be formulated as
follows~\cite{souza-prb01}.  A set of $\nbands_\kk \geq \nwann$ Bloch
states is identified at each point on a uniform BZ grid, using, for
example, an energy window.  An iterative procedure is then used to
extract self-consistently across the BZ the $\nwann$-dimensional
subspaces having collectively the smallest possible value of
$\Omega_{\rm I}$ (typically the minimization starts from an initial
guess for the target subspaces given, e.g., by projection onto trial
orbitals).  Viewed as function of $\kk$, the Bloch subspace obtained
at the end of the minimization is ``optimally smooth'' in that it
changes as little as possible with $\kk$. The minimization algorithm
can be easily modified in order to preserve identically the Bloch
eigenstates within an inner energy window.

As in the case of the one-shot projection, the outcome of this
iterative procedure is a set of $\nwann$ Bloch-like states at each
$\kk$ which are linear combinations of the initial $\nbands_\kk$
eigenstates.  One important difference is that the resulting states
are not guaranteed to be individually smooth, and the minimization of
$\Omega_{\rm I}$ must therefore be followed by a gauge-selection step,
which is in every way identical to the one described earlier for
isolated groups of bands. Alternatively, it is possible to combine the
two steps, and minimize $\Omega=\Omega_{\rm I}+\widetilde{\Omega}$
simultaneously with respect to the choice of Hilbert subspace and the
choice of gauge~\cite{thygesen-prl05,thygesen-prb05}; this should lead
to the most-localized set of $\nwann$ WFs that can be constructed from
the initial $\nbands_\kk$ Bloch states.  In all three cases, the
entire process amounts to a linear transformation taking from
$\nbands_\kk$ initial eigenstates to $\nwann$ smooth Bloch-like
states,
\beq
\label{eq:wannierization-matrix}
\ket{\tilde{\psi}_{n\kk}}=\sum_{m=1}^{\nbands_\kk}\,\ket{\psi_{m\kk}}
V_{\kk,mn}.
\eeq
In the case of the projection procedure, the explicit expression for
the $\nbands_{\kk}\times\nwann$ matrix $V_\kk$ can be surmised from
\eqs{disen_proj}{disen_orth}.

\begin{figure}
\begin{center}
\includegraphics[angle=0,width=7cm]{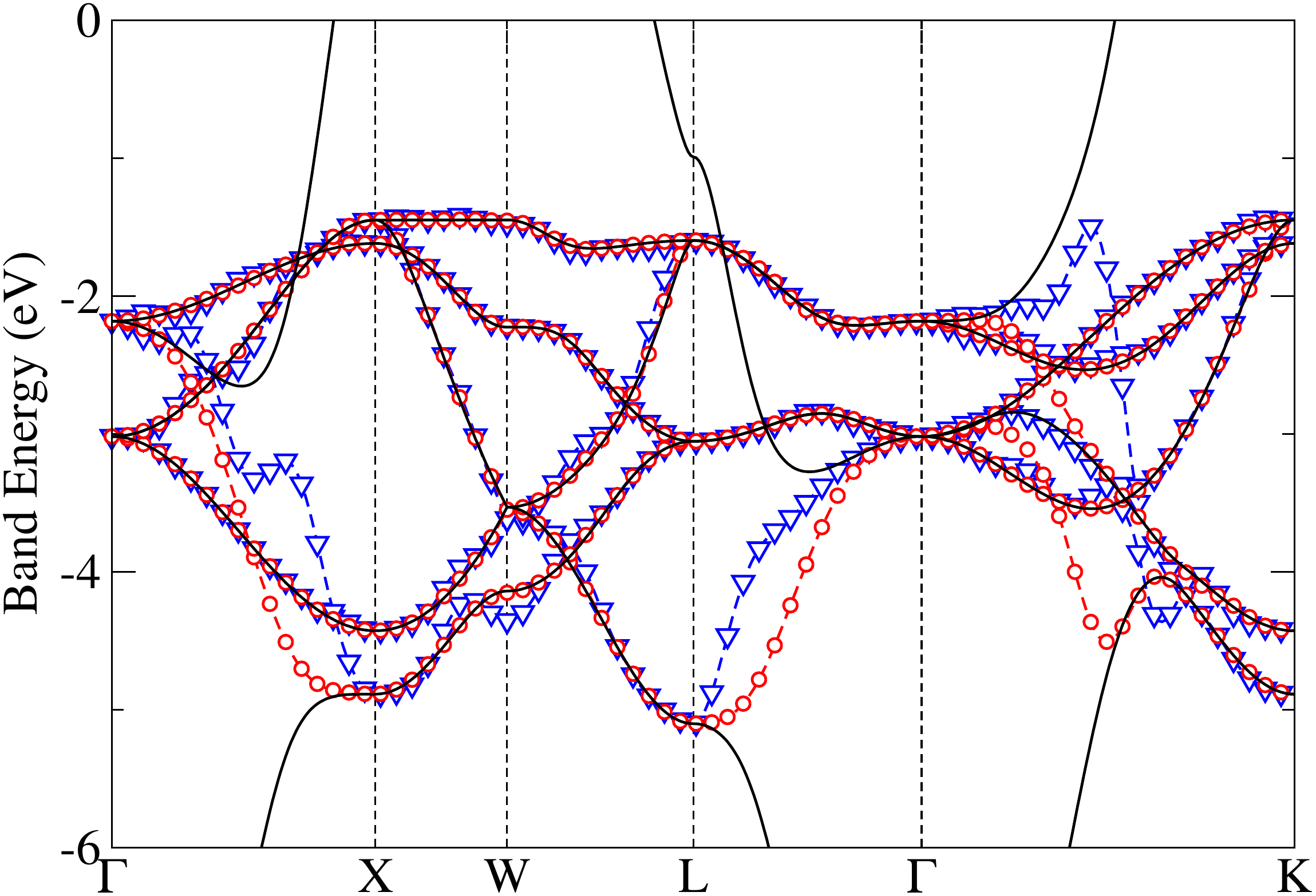}
\end{center}
\caption{\label{fig:cu_5} (Color online) Solid black lines: band
  structure of bulk crystalline Cu. Blue triangles: band structure
  for the subspace selected by projection onto atomic $3d$
  orbitals. Red circles: band structure for the subspace derived from
  the previous one, once the criterion of optimal smoothness has been
  applied.}
\end{figure}

Let us compare the one-shot projection and iterative procedures for
subspace selection, using crystalline copper as an example.  Suppose
we want to disentangle the five narrow $d$ bands from the wide $s$
band that crosses and hybridizes with them, to construct a set of
well-localized $d$-like WFs. The bands that result from projecting
onto five $d$-type atomic orbitals are shown as blue triangles in
Fig.~\ref{fig:cu_5}.  They follow very closely the first-principles
bands away from the $s$-$d$ hybridization regions, where the
interpolated bands remain narrow.

The red circles show the results obtained using the iterative scheme
to extract an optimally-smooth five-dimensional manifold.  The maximal
``global smoothness of connection'' is achieved by keeping the five
$d$-like states and excluding the $s$-like state. This happens because
the smoothness criterion embodied by \eqs{omega_i_spillage}{spillage}
implies that the orbital character is preserved as much as possible
while traversing the BZ.  Inspection of the resulting MLWFs confirms
their atomic $d$-like character.  They are also significantly
more localized than the ones obtained by projection and the
corresponding disentangled bands are somewhat better behaved,
displaying less spurious oscillations in the hybridization regions.

\begin{figure}
\begin{center}
\includegraphics[angle=0,width=7cm]{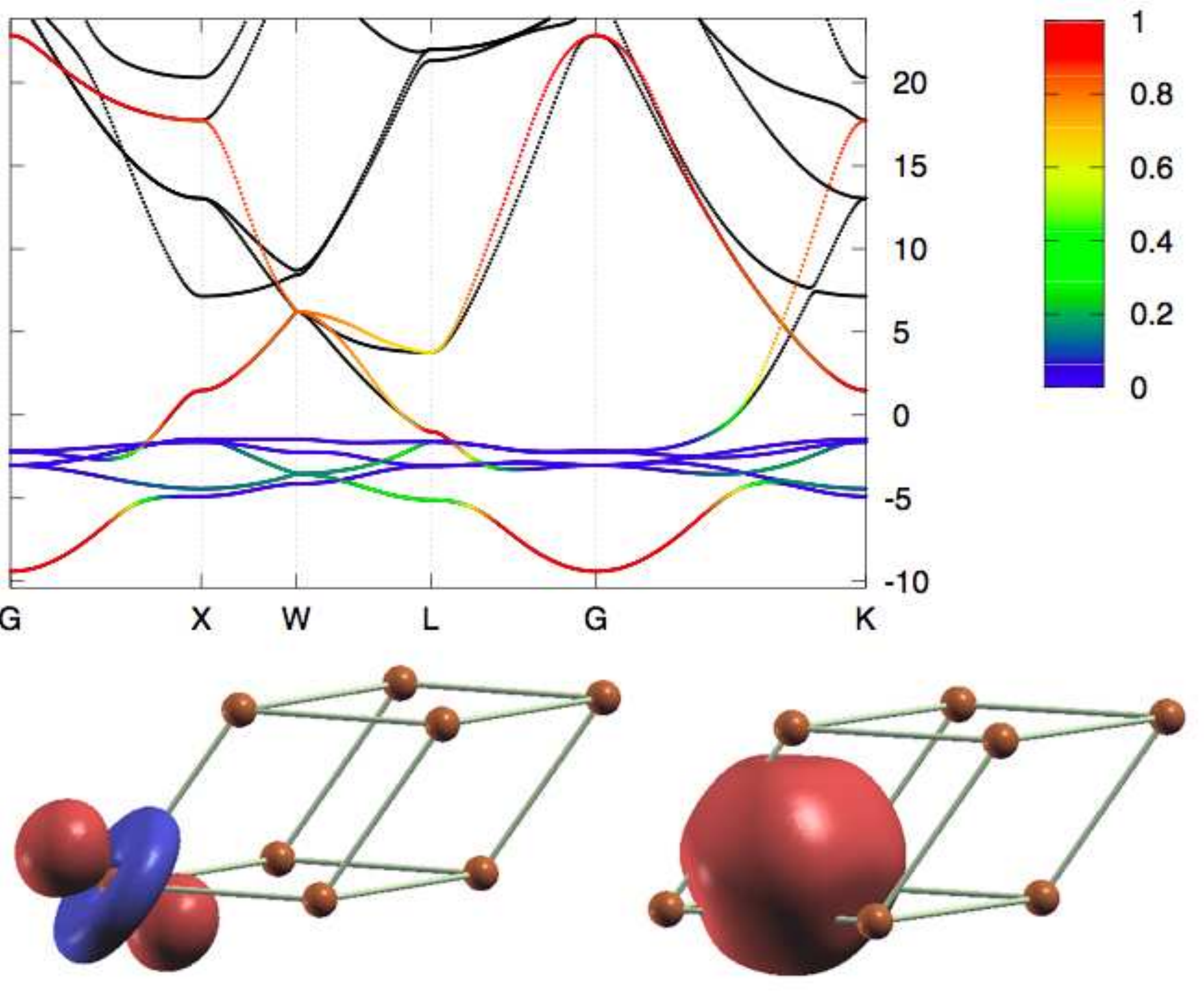}
\end{center}
\caption{\label{fig:cu_7} (Color online) Solid black lines: band
  structure of bulk crystalline Cu. Colored lines: band structure
  for the subspace selected by optimal smoothness, and a target
  dimensionality of 7, giving rise to 5 atom-centered $d$-like MLWFs
  and two $s$-like MLWFs in the tetrahedral interstitials, shown
  below.  The color coding represents the projection of the
  disentangled bands onto these MLWFs, smoothly varying from red
  (representing $s$-like interstitial MLWFs) to blue (atom-centered
  $d$-like MLWFs).}
\end{figure}

In addition, there are cases where the flexibility of the minimization
algorithm leads to surprising optimal states whose symmetries would
not have been self-evident  in advance.
One case is shown in Fig.~\ref{fig:cu_7}.
Here we wish to construct a minimal Wannier basis for copper,
describing both the narrow $d$-like bands and the wide
free-electron-like band with which they hybridize. By choosing
different dimensions for the disentangled subspace, it was found
that the composite set of bands is faithfully represented by seven
MLWFs, of which five are the standard $d$-like orbitals, and the
remaining two are $s$-like orbitals centered at the tetrahedral
interstitial sites of the fcc structure. The latter arise from the
constructive interference between $sp^3$ orbitals that would be part
of the ideal $sp^3d^5$ basis set; in this case, bands up to 20~eV
above the Fermi energy can be meaningfully described with a minimal
basis set of seven orbitals, that would have been difficult to
identify using only educated guesses for the projections.

\begin{figure}
\begin{center}
\includegraphics[angle=270,width=6cm]{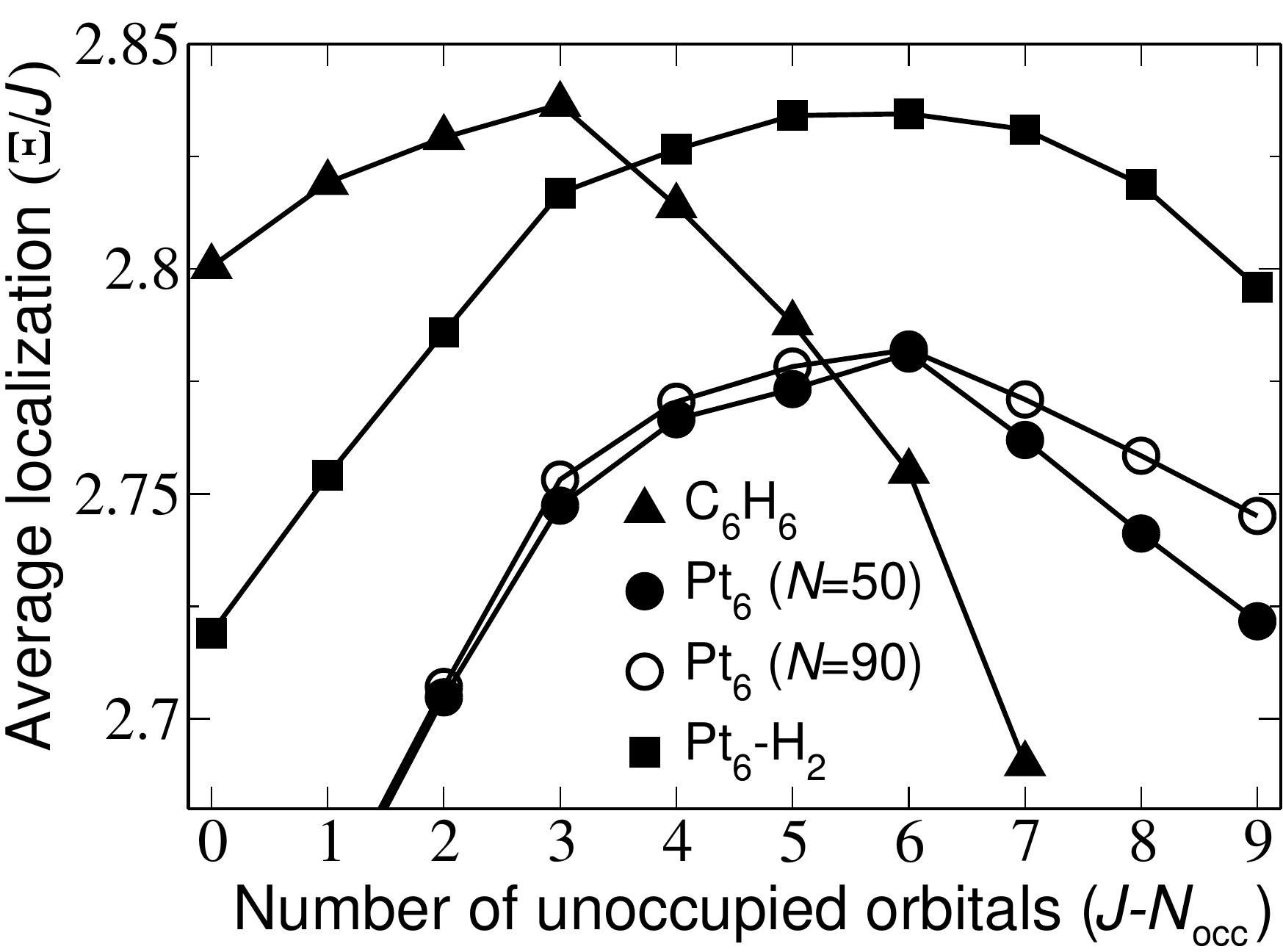}
\end{center}
\caption{\label{fig:thyg} 
  Plots of average $\Xi$ per Wannier function vs.~size of
  the Wannier space for several molecules.  $\Xi$
  is defined in \equ{ximax}; larger value indicates greater
  localization. $N_{\rm occ}$ is the number of occupied
  states, $J$ is the size of the Wannier space, and $N$ is
  the total number of states included in the DFT calculation.
  Adapted from \textcite{thygesen-prl05}.}
\end{figure}

  The concept of a natural dimension for the disentangled manifold has
  been explored further by
  \textcite{thygesen-prl05,thygesen-prb05}. As illustrated in
  Fig.~\ref{fig:thyg}, they showed that by
  minimizing $\Omega=\Omega_{\rm I}+\widetilde{\Omega}$ for different
  choices of $\nwann$, one can find an optimal $\nwann$
  such that the resulting ``partially-occupied''
  Wannier functions are most localized (provided enough bands are used
  to capture the bonding and anti-bonding combinations of those
  atomic-like WFs).

\subsubsection{Iterative minimization of $\Omega_{\rm I}$}

The minimization of $\Omega_{\rm I}$ inside an energy window is
conveniently done using an algebraic algorithm~\cite{souza-prb01}. The
stationarity condition $\delta \Omega_{\rm I}(\{\tilde{u}_{n \bf
  k}\})=0$, subject to orthonormality constraints, is equivalent to
solving the set of eigenvalue equations
\begin{equation}
\label{eq:stationary_cond}
\Big[
  \sum_{\bf b} \, w_b \, {P}_{{\bf k}+{\bf b}}
\Big]
\left| \left. \tilde{u}_{n{\bf k}} \right\rangle  \right.=
\lambda_{n \bf k} \left| \left. \tilde{u}_{n{\bf k}} \right\rangle  \right..
\end{equation}
Clearly these equations, one for each $\kk$-point, are coupled, so that
the problem has to be solved self-consistently throughout the
Brillouin zone.  This can be done using an iterative procedure: on the
$i$-th iteration go through all the $\kk$-points in the grid, and for
each of them find $\nwann$ orthonormal states
$\left|\left.\tilde{u}_{n \bf k}^{(i)}\right\rangle \right.$, defining
a subspace
whose spillage over the neighboring subspaces 
from the previous iteration is as small as possible.  At each step the
set of equations
\begin{equation}
\label{eq:stationary_cond_iter}
\Big[
  \sum_{\bf b} \, w_b \, {P}_{{\bf k}+{\bf b}}^{(i-1)}
\Big]
\left| \left. \tilde{u}_{n{\bf k}}^{(i)} \right\rangle  \right.=
\lambda_{n \bf k}^{(i)} \left| \left. 
\tilde{u}_{n{\bf k}}^{(i)} \right\rangle  \right.
\end{equation}
is solved, and the $\nwann$ eigenvectors 
with largest eigenvalues are selected. That choice ensures that at
self-consistency the stationary point corresponds to the absolute
minimum of $\Omega_{\rm I}$.

In practice \equ{stationary_cond_iter} is solved in the basis of
the original $\nbands_{\bf k}$ Bloch eigenstates $|u_{n \bf k}\rangle$
inside the energy window.  Each iteration then amounts to
diagonalizing the following $\nbands_{\bf k} \times \nbands_{\bf k}$
Hermitian matrix at every ${\bf k}$:
\begin{equation}
\label{eq:z_matrix}
Z_{mn}^{(i)}({\bf k})=
\Big<  u_{m \bf k}
\Big| \sum_{\bf b} w_b 
  {\left[ {P}^{(i-1)}_{{\bf k}+{\bf b}} \right]}_{\rm in}
\Big| u_{n \bf k} \Big> .
\end{equation}
Since these are small matrices,
each step of the iterative procedure is computationally inexpensive.

As a final comment, we mention that 
in the case of degeneracies or quasi degeneracies in the
spreads of orbitals centered on the same site, the localization
algorithm will perform rather arbitrary mixing of these (as can be the
case, e.g., for the $d$ or $f$ electrons of a transition-metal ion, or
for its $t_{2g}$ or $e_g$ groups). A solution to this problem is to
diagonalize an operator with the desired symmetry in the basis of the
Wannier functions that have been obtained (see
\textcite{posternak-prb02} for the example of the $d$ orbitals in
MnO).

\subsubsection{{Localization and local minima}}
\label{sec:loc-min-dis}

{
Empirical evidence and experience suggests that localized
Wannier functions can be readily constructed also in the case
of an entangled manifold of bands:
Even for metals, smooth manifolds
can be disentangled and ``wannierized'' to give MLWFs.
Such disentangled MLWFs, e.g., the $p_z$ MLWFs describing the
$\pi$/$\pi^{\ast}$ manifold of graphene shown in
Fig.~\ref{fig:graph_pz}, are found to be strongly localized. 
While exponential localization has not been proven, both 
numerical evidence and the analogy with the isolated composite case
suggest this might be the case.
}

{
Problems associated with reaching local minima of the spread functional,
and with obtaining Wannier functions that are not real-valued, are more 
  pronounced in the case of entangled bands.
They are usually alleviated by careful reconsideration
  of the energy windows used, in order to include higher energy states of the
  appropriate symmetry, and/or by using a better initial guess for the
  projections.
We infer, therefore, that such problems are associated not with
  the wannierization part of the procedure, but rather with the initial
  selection of the smooth subspace from the full manifold of entangled
  bands. 
}

{
It is worth noting that the $\Gamma$-point formulation
(Sec.~\ref{sec:gpoint}) appears to be less affected by these 
problems. 
In cases where it is not intuitive or obvious what
  the MLWFs should be, therefore, it can often be a
  fruitful strategy to use the $\Gamma$-point formulation to obtain
  approximate MLWFs that may then be used to inform the initial guess 
  for a subsequent calculation with a full k-point mesh. 
}

\subsection{Many-body generalizations}
\label{sec:manybody}

The concept of WFs is closely tied to the framework
of single-particle Hamiltonians.  Only in this case can we
define $\nbw$ occupied single-particle Bloch functions at each
wavevector $\kk$ and treat all $\nbw$ of them on an equal footing,
allowing for invariance with respect to unitary mixing among them.
Once the two-particle electron-electron interaction is formally
included in the Hamiltonian, the many-body wavefunction cannot be
reduced to any simple form allowing for the construction of
WFs in the usual sense.

One approach is to consider the reduced one-particle density matrix
\beq
n(\rr,\rr') =\int \Psi^*(\rr,\rr_2...)\,
 \Psi(\rr',\rr_2,...)\, d\rr_2\,d\rr_3 ...
\eeq
for a many-body insulator.  Since $n(\rr,\rr')$ is periodic in the
sense of $n(\rr+\RR,\rr'+\RR)=n(\rr,\rr')$, its eigenvectors -- the
so-called ``natural orbitals'' -- have the form of Bloch functions
carrying a label $n,\kk$.  If the insulator is essentially a
correlated version of a band insulator having $\nbw$ bands, then at
each $\kk$ there will typically be $\nbw$ occupation eigenvalues
$\nu_{n\kk}$ that are close to unity, as well as some small ones that
correspond to the quantum fluctuations into conduction-band states.
If one focuses just on the subspace of one-particle states spanned by
the $\nbw$ valence-like natural orbitals, one can use them to
construct one-particle WFs following the methods described earlier, as
suggested by \textcite{koch-ssc01}.  However, while such an approach
may provide useful qualitative information, it cannot provide the
basis for any exact theory.  For example, the charge density, or
expectation value of any other one-particle operator, obtained by
tracing over these WFs will not match their exact many-body
counterparts.

A somewhat related approach, adopted by \textcite{hamann-prb09},
is to construct WFs out
of the quasiparticle states that appear in the GW approximation
\cite{aryasetiawan-ropp98}.  Such an approach will be described
more fully in Sec.~\ref{sec:gw-bands}.
Here again, this approach may give useful physical and chemical
intuition, but the one-electron quasiparticle wavefunctions do not
have the physical interpretation of occupied states, and charge
densities and other ground-state properties cannot be computed
quantitatively from them.

Finally, a more fundamentally exact framework for a many-body
generalization of the WF concept, introduced
in \textcite{souza-prb00b}, is to consider
a many-body system with twisted boundary conditions applied to
the many-body wavefunction in a supercell.  For example, consider
$M$ electrons in a supercell consisting of $M_1\times M_2\times M_3$
primitive cells, and impose the periodic boundary condition
\beq
\Psi_\qq(..., \rr_j+\RR, ...)=e^{i\qq\cdot\RR}\,\Psi_\qq(..., \rr_j, ...)
\label{eq:superper}
\eeq
for $j=1,...,M$,
where $\RR$ is a lattice vector {\it of the superlattice}.
One can then construct a single ``many-body WF''
in a manner similar to \equ{wanniertransform}, but with
$\kk\rightarrow\qq$ and $\ket{\psi_{n\kk}}\rightarrow\ket{\Psi_q}$
on the right side.  The resulting many-body WF is
a complex function of $3M$ electron coordinates, and as such it
is unwieldy to use in practice.  However, it is closely related
to Kohn's theory of the insulating state \cite{kohn-pr64},
and in principle it can be used to formulate many-body versions of
the theory of electric polarization and related quantities, as shall
be mentioned in Sec.~\ref{sec:pol-manybody}.

\section{RELATION TO OTHER LOCALIZED ORBITALS}
\label{sec:locorb}

\subsection{Alternative localization criteria}
\label{sec:alt-loc}

As we have seen, WFs are inherently non-unique and, in practice, some
strategy is needed to determine the gauge. Two possible approaches
were already discussed in Sec.~\ref{sec:theory}, namely, projection
and maximal localization. The latter approach is conceptually more
satisfactory, as it does not depend on a particular choice of trial
orbitals. However, it still does not uniquely solve the problem of
choosing a gauge, as different localization criteria are possible and
there is, \emph{a priori}, no reason to choose one over another.

While MLWFs for extended
systems have been generated for the most part by minimizing the sum of
quadratic spreads, Eq. (17), a variety of other localization criteria
have been used over the years for molecular systems. We will briefly
survey and compare some of the best known schemes below. What they all
have in common is that the resulting localized molecular orbitals
(LMOs) $\phi_i(\rr)$ can be expressed as linear combinations of a set
of molecular eigenstates $\psi_i(\rr)$ (the ``canonical'' MOs),
typically chosen to be the occupied ones,
\beq
\phi_i({\rr})=\sum_{j=1}^{\nwann}U_{ji}\psi_j({\rr}) .
\label{eq:lmos}
\eeq
The choice of gauge then arises from minimizing or maximizing some
appropriate functional of the LMOs with respect to the coefficients
$U_{ij}$, under the constraint that the transformation (\ref{eq:lmos})
is unitary, which ensures the orthonormality of the resulting LMOs.

{\it The Foster-Boys criterion} (FB) \cite{boys-rmp60,foster-rmp60a,boys-66}.
This is essentially the same as the
Marzari-Vanderbilt criterion of minimizing the sum of the quadratic
spreads, except that the sum runs over the orbitals in the molecule,
rather than in one crystal cell,
\begin{equation}
\Omega_{\rm FB}=\sum_{i=1}^{\nwann} [\langle \phi_i|{\rr}^2|\phi_i\rangle -
\langle \phi_i|{\rr}|\phi_i\rangle ^2].
\label{eq:omega_boys}
\end{equation}
Interestingly, this criterion is equivalent to minimizing the
``self-extension'' of the orbitals \cite{boys-66},
\beq
\sum_{i=1}^{\nwann}\int d{\rr}_1 d{\rr}_2 \, |\phi_i({\rr}_1)|^2
 ({\rr}_1-{\rr}_2)^2 |\phi_i({\rr}_1)|^2
\label{eq:omega_boysa}
\eeq
and also to maximizing the sum of the squares of the distance between
the charge centers of the orbitals
\begin{equation}
\sum_{i,j=1}^{\nwann} |\langle \phi_i|{\rr}|\phi_i\rangle - \langle
\phi_j|{\rr}|\phi_j\rangle |^2 ,
\label{eq:omega_boys1}
\end{equation}
which is closely related to \equ{ximax} of Sec.~\ref{sec:gpoint}.
The relation between Eqs.~(\ref{eq:omega_boys})--(\ref{eq:omega_boys1}) is discussed in \textcite{boys-66}.

{\it The Edmiston-Ruedenberg criterion} (ER). Here localization is achieved by
maximizing the Coulomb self-interaction of the orbitals
\cite{edmiston-rmp63}
\begin{equation}
\Omega_{\rm ER}=
\sum_{i=1}^{\nwann}\int d{\rr}_1 d{\rr}_2 \, |\phi_i({\rr}_1)|^2
({\rr}_1-{\rr}_2)^{-1} |\phi_i({\rr}_2)|^2.
\label{eq:omega_er}
\end{equation}
From a computational point of view, the ER approach is more demanding 
(it scales as $\nwann^4$ versus $\nwann^2$ for FB),
but recently more efficient implementations have been developed
\cite{subotnik07}.

{\it The von Niessen criterion} (VN). The 
goal here is to maximize the density overlap of the
orbitals \cite{niessen-jcp72}
\begin{equation}
\Omega_{\rm VN}=
\sum_{i=1}^{\nwann}\int d{\rr}_1 d{\rr}_2 \, |\phi_i({\rr}_1)|^2
 \delta({\rr}_{1}-{\rr}_2) |\phi_i({\rr}_2)|^2.
\label{eq:omega_vn}
\end{equation}

{\it The Pipek-Mezey criterion} \cite{pipek-jcp89} (PM).
This approach differs from the
ones above in that it makes reference to some extrinsic objects. The
idea is to maximize the sum of the squares of the Mulliken atomic
charges \cite{mulliken-jcp55}. These are obtained with respect to a set
of atomic orbitals $\chi_{\mu}$ centered on atomic sites A. We define
\beq
\label{eq:PM}
P_{A}=\sum_{\mu\in A}\sum_{\nu} D_{\mu\nu}S_{\mu\nu}
\eeq
where the sum over $\mu$ involves all of the atomic states on atom
site A, $D_{\mu\nu}$ is the density matrix in the atomic basis, and
$S_{\mu\nu}=\inprod{\chi_{\mu}}{\chi_{\nu}}$ is the overlap operator.
The functional to be maximized is given by
\begin{equation}
\Omega_{\rm PM}=\sum_{i=1}^{\nwann} \sum_{A=1}^{N_A}
|\langle\phi_{i}|P_{A}|\phi_{i}\rangle |^{2}.
\label{eq:omega_pm}
\end{equation}
This is somewhat similar in spirit to the
projection scheme discussed in Sec.~\ref{sec:projection}, except that
it is not a one-shot procedure.

\begin{figure}
\begin{center}
\includegraphics[width=7cm]{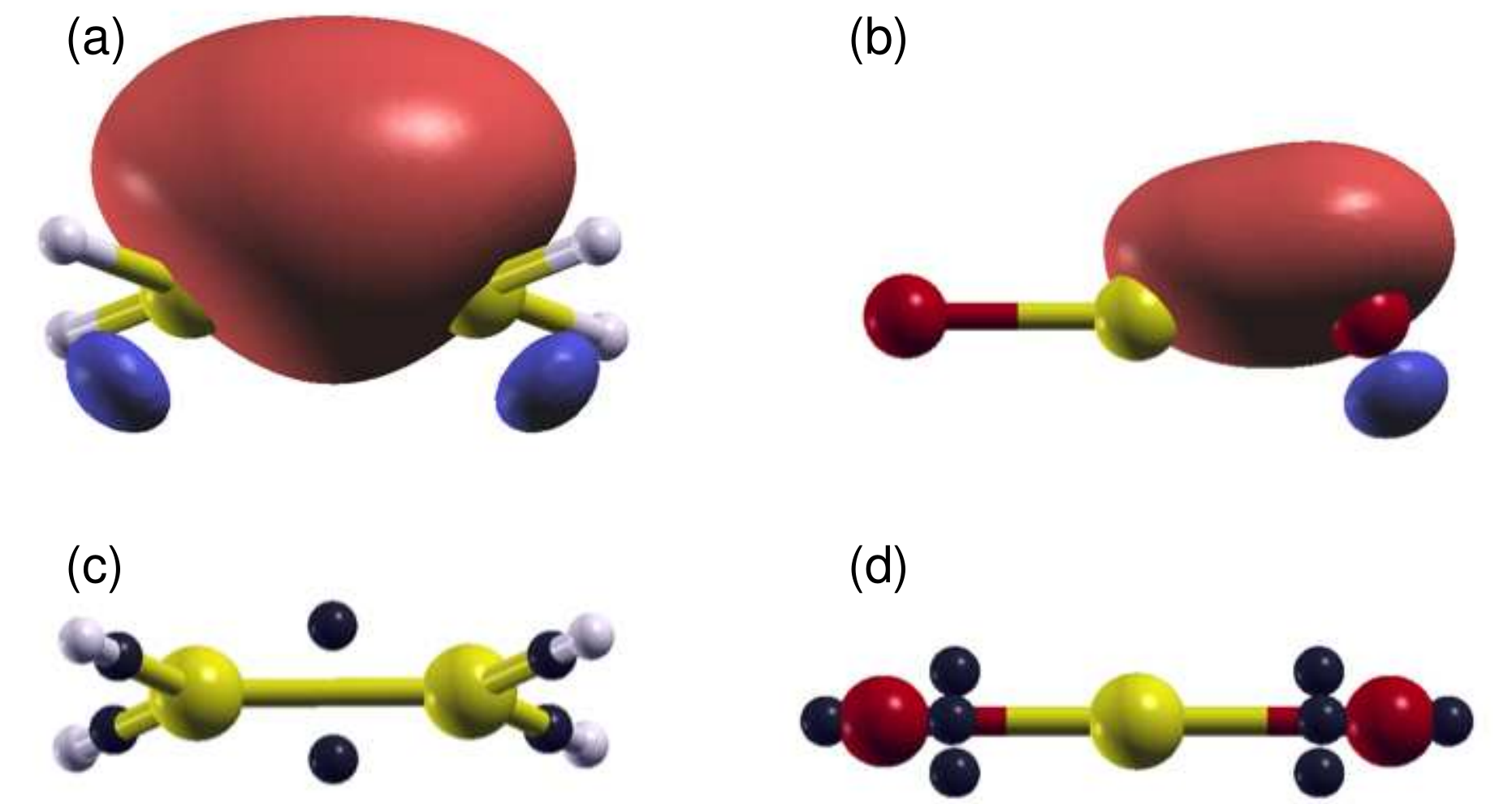}
\end{center}
\caption{\label{fig:banana_bond} (Color online) (a) One of the two
  ``banana'' or $\tau$ orbitals in ethene (b) Centers of the MLWF in
  ethene (c) $\tau$ MLWF in CO$_2$ (d) Centers of the MLWF in
  CO$_2$. Atoms colors are: hydrogen (white), carbon (yellow), oxygen
  (red). MLWF centers are shown in black. MLWF are computed using the
  scheme of Marzari and Vanderbilt \cite{marzari-prb97} with a large
  vacuum supercell and a $\Gamma$-point sampling of the BZ.}
\end{figure}

As indicated in \equ{lmos}, all of these schemes amount to unitary
transformations among the canonical MOs and, as such, they
give rise to representations of the electronic
  structure of the system that are equivalent to that provided by the
  original set of eigenstates.
For the purpose of providing chemical intuition, the usefulness of a
given scheme depends on how well it matches a particular
viewpoint of bonding. There have been few studies of the VN scheme, but
the FB, ER, and PM schemes have been extensively compared. In many
cases all three approaches lead to similar localized orbitals which
agree with the simple Lewis picture of bonding.
A notable
exception is for systems that exhibit both $\sigma$ and $\pi$
bonding. For a double bond, both the FB and ER schemes mix the
$\sigma$ and $\pi$ orbitals, giving rise to two bent ``banana bond''
(or $\tau$) orbitals \cite{pauling-book60} as shown in
Fig~\ref{fig:banana_bond}.  When applied to benzene,
both schemes give alternating $\sigma$ and $\tau$ orbitals. Trivially,
there are two equivalent sets of these orbitals, reminiscent of the
two Kekul\'{e} structures for benzene.  In contrast to the FB and ER
schemes, PM provides a clear separation of $\sigma$ and $\pi$
orbitals. For example, in benzene PM gives a framework of six $\sigma$
orbitals and three $\pi$ orbitals around the ring.

In some situations the FB and ER orbitals have been found to be quite
different. This has been observed when the bonding in a molecule can
be represented as two distinct resonant structures. The ER scheme
generally gives orbitals corresponding to one of the possible
structures, whilst the FB orbitals correspond a hybrid of the
structures.  An extreme example is CO$_2$ \cite{brown-jacs77}. In
agreement with the O=C=O Lewis picture, ER gives two lone pairs on
each oxygen and two $\tau$ orbitals between each carbon and oxygen. In
contrast, the FB scheme gives a single lone pair on each oxygen and
three highly polarized $\tau$ orbitals between the carbon and each
oxygen, as shown in Fig.~\ref{fig:banana_bond}.

MLWFs, which may be thought of
  as the solid-state equivalent of FB orbitals, have been
widely used to examine chemical bonding, as will be discussed in
detail in Sec.~\ref{sec:bonding}. The ER scheme has not been
extensively examined, an isolated exception being the work of
\textcite{miyake-prb08} who proposed a method to maximize the
Coulomb self-interaction of the orbitals in periodic systems (see also
Sec.~\ref{sec:mod-ham}).
They examined a range of bulk transition metals
and SrVO$_3$ and in each case found that the resulting WFs were
essentially identical to MLWFs. 
\subsection{Minimal-basis orbitals}
\label{sec:proj-down}

In the same way as was described for the Marzari-Vanderbilt scheme of
Sec.~\ref{sec:maxloc}, the alternative localization criteria
described above can be applied in a solid-state context to isolated
groups of bands. One is often interested in the more general situation
where the bands of interest are not isolated. The challenge then is to
generate a ``disentangled''
set of $\nwann$ localized WFs starting from some larger set
of bands.  Two procedures for doing so were discussed in
Sec.~\ref{sec:entangled}, namely, projection and iterative
minimization of the spread functional $\Omega$, not only with respect
to the choice of gauge, but also with respect to the choice of Hilbert
space.

 Here we discuss two further procedures which achieve the same goal by
different means. They have in common the fact that the resulting
orbitals constitute a minimal basis of atomic-like orbitals.

\subsubsection{Quasiatomic orbitals}

The quasiatomic orbitals (QOs)
scheme \cite{lu-prb04,chan-prb07,qian-prb08}
is a projection-based approach that
has the aim of extracting a minimal tight-binding basis
from an initial first-principles calculation.
In this regard it is similar in spirit to the Wannier interpolation
techniques discussion in Sec.~\ref{sec:winterp}.  Unlike WFs, however,
the quasiatomic orbitals form a {\it non-orthogonal} basis of {\it
  atom-centered} functions, each with specified angular
character. Their radial part, and hence their detailed local shape,
depends on the bonding environment.

As in the Wannier scheme, the first step is to construct a suitable
$\nwann$-dimensional subspace $\{\phi_{n}\}$ starting from a larger
set of $\nbands$ Bloch eigenstates $\{ \psi_{n}\}$. For simplicity
we consider a $\Gamma$-point only sampling of the BZ and hence omit the
$\kk$ index. 
The goal is to
identify a disentangled subspace
with
the desired atomic-orbital character, as specified by a
  given set of $J$ atomic orbitals (AOs) $\ket{A_i}$, where $i$ is a
composite site and angular-character index. One possible strategy
would be to employ the one-shot projection approach of
Sec.~\ref{sec:subspace_proj}, using the AOs as trial orbitals.
\textcite{lu-prb04} proposed a more optimized procedure, based on
maximizing the sum-over-square similarity measure\footnote{This is
  similar in spirit to the Pipek-Mazey procedure, \equ{PM}, but
  applied to subspace selection rather than gauge selection.}
\beq
\label{eq:similarity}
{\cal L}=\sum_{i=1}^\nwann\left|\left|
\sum_{n=1}^{\nwann}\ket{\phi_{n}}\inprod{\phi_{n}}{A_i}\right|\right|^2
\eeq
with respect to the orthonormal set $\{\phi_{n}\}$, expressed as
linear combinations of the original set $\{ \psi_{n}\}$. It is usually
required that a subset of $N\leq \nwann$ of the original
eigenstates are identically preserved (``frozen in'') in the
disentangled subspace, in which case the optimization is performed
with respect to the remaining $\nwann-N$ states $\phi_{n}$. $\nbands$
must be of sufficient size to capture all of the antibonding character
of the AOs.

In later work it was realized
\cite{qian-prb08} that the QOs can be constructed without explicit
calculation of the eigenstates outside the frozen window. The key insight
is to realize that the QOs will only have a contribution from the finite
subset of this basis spanned by the AOs (see \equ{qo} below).  
The component of the AOs projected onto
the $N$ states within the frozen window
$\ket{A_{i}^{\|}}$ is given by
\beq
 \ket{A_{i}^{\|}}=\sum_{n=1}^{N}  \ket{\psi_{n}}\inprod{\psi_{n}}{A_{i}}.
\eeq
The component of $\ket{A_{i}}$
projected onto the states outside the frozen window
can hence constructed directly using only the AOs and the states
within the frozen window, as
\beq
\ket{A_{i}^{\bot}}=\ket{A_{i}}-\ket{A_{i}^{\|}} .
\eeq
Using $\ket{A_{i}^{\bot}}$  as a basis, the set of $\{\phi_{n}\}$ which
maximize $\cal{L}$ can be obtained using linear algebra as
reported in \textcite{qian-prb08}. 

Once a subspace with the correct orbital character has been
identified, a basis of
quasiatomic orbitals 
can be
obtained by retaining the portion of the original AOs that ``lives''
on 
that
subspace,
\beq\label{eq:qo}
  \ket{Q_{i}}=\sum_{n=1}^\nwann\ket{\phi_{n}}\inprod{\phi_{n}}{A_{i}}.
  \eeq 
  In general the angular dependence of the resulting QOs will no
  longer that of pure spherical harmonics, but only approximately so.
In \textcite{qian-prb08}, QOs were obtained for simple metals and
semiconductors. Later applications have used the orbitals for the
study of quantum transport in nano-structures \cite{qian-prb10}.

\subsubsection{NMTO and Downfolding}
An alternative scheme for obtaining a minimal basis representation is the
perturbation approach introduced by \textcite{lowdin-jcp51}. Here the
general strategy is to partition a set of orbitals into an ``active'' set that is
intended to describe the states of interest, and a ``passive'' set that
will be integrated out.  Let us write the
Hamiltonian for the system in a block representation,
\beq \label{eq:lowdin_part}
H=\left(
  {\begin{array}{cc}
      H_{00} & 0  \\
      0 & H_{11} \\
 \end{array} } \right) +\left( {\begin{array}{cc}
 0 & V_{01}  \\
 V_{10} & 0  \\
\end{array} } \right) ,
\eeq
where $H_{00}$ ($H_{11}$) is the projection onto the active (passive)
subspace, and $V_{01}$ is the coupling between the two subspaces. An
eigenfunction can be written as the sum of its projections onto the
two subspaces $\ket{\psi}=\ket{\psi_0}+\ket{\psi_1}$.  This leads to
\beq
(H_{00}-\varepsilon) \ket{\psi_0} + V_{01}\ket{\psi_1}=0,
\eeq
\beq
V_{10}\ket{\psi_0}+(H_{11}-\varepsilon) \ket{\psi_1}=0,
\eeq
where $\varepsilon$ is the eigenvalue
  corresponding to $\ket{\psi}$.
Elimination of $\ket{\psi_1}$ gives an effective Hamiltonian for the
system which acts only on the active subspace:
\beq 
H^{\rm eff}_{00}(\varepsilon)=H_{00}-V_{01}(H_{11}-\varepsilon)^{-1}V_{10}.
\eeq
This apparent simplification has introduced an energy dependence
into the Hamiltonian. One practical way forward is to approximate this
as an energy-independent Hamiltonian $H^{\rm
  eff}_{00}(\varepsilon_0)$, choosing the reference energy
$\varepsilon_0$ to be the average energy of the states of
interest. This approach has been used to construct tight-binding
Hamiltonians from full electronic structure calculations \cite{solovyev_prb04}.

A form of L\"owdin partitioning has
been widely used in connection with
the linear-muffin-tin-orbital (LMTO) method, particularly in its most
recent formulation, the $N^{\rm th}$-order muffin-tin-orbital (NMTO)
approach \cite{andersen-prb00,zurek-c05}.
In this context it is usually referred to as ``downfolding'',
although we note that some authors use this term to refer to any scheme to produce a
minimal basis-set representation.

In studies of complex materials there may be a significant number of
MTOs, typically one for each angular momentum state ($s$, $p$, $d$) on
every atomic site. One may wish to construct a minimal basis to
describe states within a particular energy region, e.g., the occupied
states, or those crossing the Fermi level.  Let us assume we have of
basis orbitals (MTOs in this case) which we wish to partition into
``active'' and ``passive'' sets. Using the notation of
\equ{lowdin_part}, L\"owdin partitioning gives a set of
energy-dependent orbitals $\phi_0(\varepsilon,\rr)$ for the active
space according to~\cite{zurek-c05}
\beq
\phi_0(\varepsilon,\rr)=\phi_0(\rr)-\phi_1(\rr) (H_{11}-\varepsilon)^{-1}V_{10}.
\eeq
Taking into account the energy dependence, this reduced set of
orbitals spans the same space as the original full set of orbitals,
and can be seen to be the original orbitals of the
active set dressed by an energy-dependent linear combination of
orbitals from the passive set. In the NMTO scheme, the next step is
to form an energy-independent set of orbitals through an $n$th-order
polynomial fit to the energy dependence.

To give a specific example, accurate calculations on tetrahedral
semiconductors will require the inclusion of $d$ states in the MTO
basis, i.e., nine states ($s$, $p$, $d$) per site. However, it would
be desirable to construct a minimal basis to describe the valence and
lower conduction states with only four states ($s$, $p$) on each site
\cite{lambrecht_prb86}. We therefore designate $s$ and $p$ as
``active'' channels and $d$ as ``passive''.  Downfolding will result
in an MTO with either $s$ or $p$ character on a given site, with
inclusion of $d$ character on neighboring sites. In other words, the
tail of the MTO is modified to ``fold in'' the character of the
passive orbitals.

\begin{figure}
\begin{center}
\includegraphics[width=8.5cm]{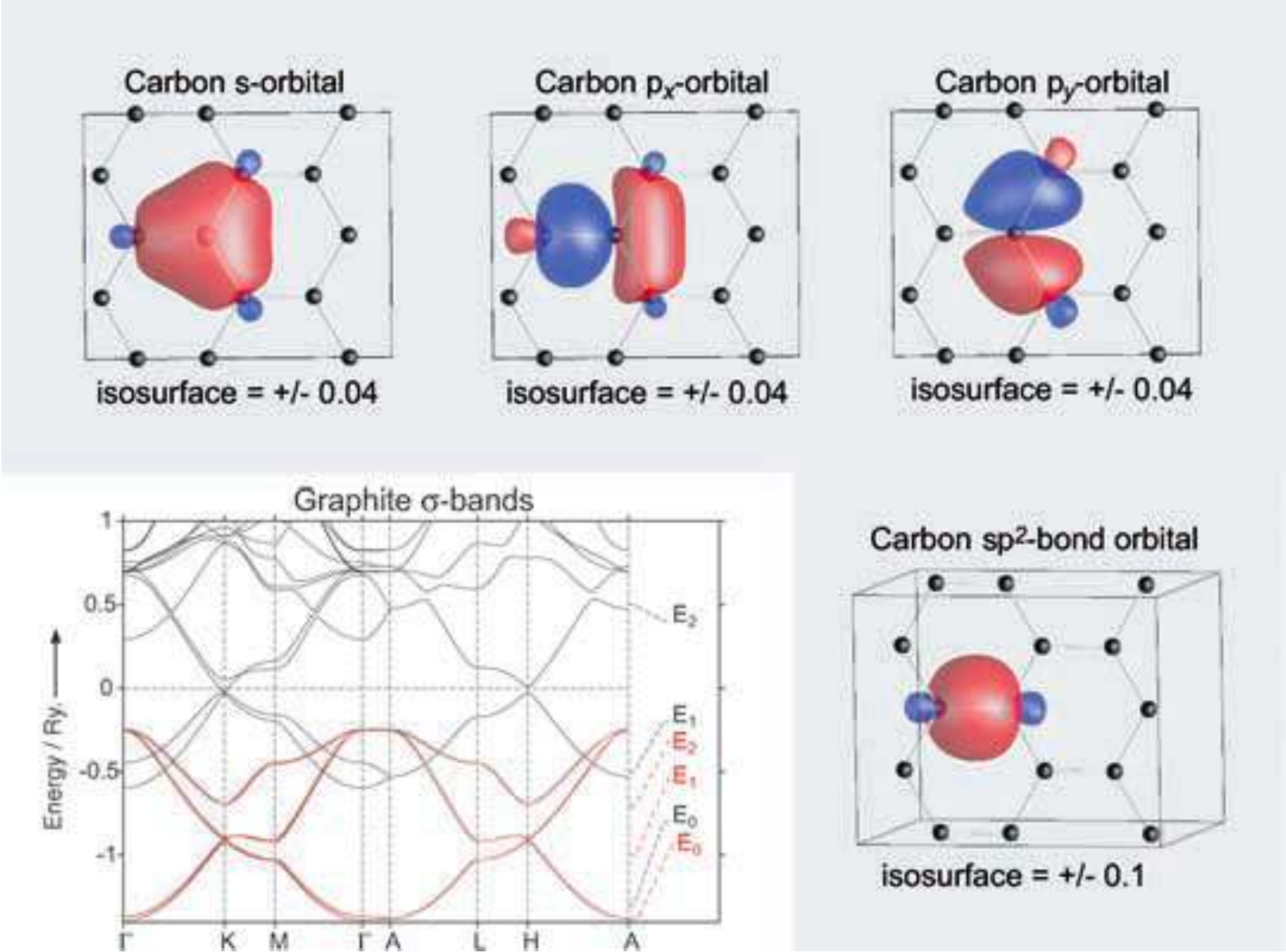}
\end{center}
\caption{\label{fig:nmto_graph} (Color online)
The band structure of graphite calculated with a full NMTO spd basis is
given in black. The red bands have been calculated with an $s$, $p_x$, and
$p_y$ orbital on every second carbon atom (shown above the band
structure). Also shown is one of the $sp^2$ bond
orbitals which arise by symmetrical orthonormalization of the $s$, $p_x$,
abd $p_y$ orbitals. The energy meshes used for each calculation
are given to the right of the band structure. From \textcite{zurek-c05}.}
\end{figure}

In Fig.~\ref{fig:nmto_graph}, the band structure of graphite
calculated using a full $s$, $p$, and $d$ basis on each carbon atom is shown in
black \cite{zurek-c05}. The red bands are obtained by choosing $s$,
$p_x$, and $p_y$ states on every second
carbon atom as the active channels and downfolding all other
states. The energy mesh spans the energy range of the $sp^2$ bonding
states (shown on the right-hand-side of the band-structure plot in
Fig.~\ref{fig:nmto_graph}). For these bands the agreement with the
full calculation is perfect to within the resolution of the
figure.  Symmetric orthonormalization of the three NMTOs gives the
familiar set of three symmetry related $\sigma$-bonding orbitals (compare
with the MLWF of graphene in Fig.~\ref{fig:graph_pz}).
\textcite{lechermann-prb06} have compared MLWFs and NMTO orbitals for
a set of t$_{2g}$ states located
around the Fermi level SrVO$_3$. It was found that both schemes gave
essentially identical orbitals.

\subsection{{Comparative discussion}}
\label{sec:locorb-disc}

{At this point it is worth commenting briefly on some of
the advantages and disadvantages of various choices of WFs.
We emphasize once again that no choice of WFs,
whether according to maximal localization or other
criteria, can be regarded as ``more correct'' than another.
Because WFs are intrinsically gauge-dependent, it is impossible,
even in principle, to determine the WFs experimentally.
By the same token, certain properties obtained from the WFs, such
as the dipole moments of molecules in condensed phases
(see Sec.~\ref{sec:mol}), must be interpreted with caution.
The measure of a WF construction procedure is, instead,
its \emph{usefulness} in connection with theoretical or
computational manipulations.}

{Where the WFs are to be used as basis functions for Wannier
interpolation (Sec.~\ref{sec:winterp}) or other purposes
(Sec.~\ref{sec:basis}), some variety of maximally localized WFs are
probably most natural both because the real-space matrix elements
can be restricted to relatively near neighbors, and because
Fourier-transformed quantities become relatively smooth in $\kk$ space.
However, in cases in which the set of MLWFs does not preserve
the space-group symmetry, it may be better to insist on
symmetry-preserving WFs even at the expense of some delocalization
(see also the discussion in Sec.~\ref{sec:subspace_proj}).
In this way, properties computed from the WFs, such as interpolated
bandstructures, will have the correct symmetry properties.
When using WFs to interpret the nature of chemical bonds, as in
Sec.~\ref{sec:bonding}, the results may depend to some degree on
the choice of WF construction method, and the optimal choice may
in the end be a matter of taste.}

\subsection{Non-orthogonal orbitals and linear scaling}
\label{sec:linear-scaling}
In recent years there has been much progress in the development
of practical linear-scaling methods for electronic structure
calculations, that is,
methods in which the computational cost of the calculation grows only
linearly with the size of the system. The fundamental principle behind
such approaches is the fact that electronic structure is inherently
local~\cite{heine-ssp80}, or `nearsighted'~\cite{kohn-prl96}. This
manifests itself in the exponential localization of the WFs
in insulators 
(discussed in Sec.~\ref{sec:exp-loc}) and, more generally, in the
localization properties of the single-particle density matrix
\beq
\rho(\rr,\rr') = \matel{\rr}{P}{\rr'}
= \ibz d\kk\; \sum_{n} f_{n\kk} \psi_{n\kk}(\rr)\psi^{\ast}_{n\kk}(\rr') , \label{eq:dm-bloch}
\eeq
where, following \textcite{janak-prb78}, the projection operator $P$ of
\equ{bandproj} has been generalized to the case of fractional
eigenstate occupancies $f_{n\kk}$. The quantity $\rho(\rr,\rr')$
has been shown to decay exponentially as $\exp(-\gamma|\rr-\rr'|)$
in insulators and semiconductors, where the exponent $\gamma$
can be heuristically related to the direct band gap of the
system~\cite{cloizeaux-pr64a, ismail-prl99, taraskin-prl02}.
It has also been shown to take the same form in metals at finite
temperature,\footnote{For metals at zero temperature, the
  discontinuity in occupancies as a function of $\kk$ results in the
  well-known \emph{algebraic} decay of the density matrix.} but with
$\gamma$ determined by
the ratio between the thermal energy $k_{\rm B}T$ and the Fermi
velocity~\cite{goedecker-prb98}.

Exponential localization may seem a surprising result given that the
Bloch eigenstates extend across the entire system.  Expressing the
density matrix in terms of WFs using \equ{wannier}, we find
\beq \rho(\rr,\rr') = \sum_{ij}\sum_{\RR\RR'}
w_{i\RR}(\rr)K_{ij}(\RR'-\RR)w^{\ast}_{j\RR'}(\rr'),
\label{eq:dm-wannier}
\eeq
where we have defined the 
\emph{density kernel}\footnote{This term
  was, to the best of our knowledge, first used by
  \textcite{mcweeny-romp60}.}
\beq K_{ij}(\RR) = \ibz d\kk\; e^{-i\kk\cdot\RR} 
\sum_{n} [U^{(\kk)\dagger}]_{in} f_{n\kk} [U^{(\kk)}]_{nj} ,
\label{eq:kij1}
\eeq
which reduces to $\delta_{ij}\delta_{\RR\0}$ in the case of a set of
fully occupied bands. 
We now see that the spatial localization of the density
matrix is closely linked to that of the Wannier functions
themselves. This locality is exploited in
linear-scaling methods by retaining an amount of information in the
density matrix that scales only linearly with system size.

Many different linear-scaling DFT approaches exist;
for comprehensive reviews the reader is referred to
\textcite{galli-cossms96},
\textcite{goedecker-romp99}, and \textcite{bowler-rpp12}. Many of
them are based on the variational minimization of an energy
functional expressed either in terms of localized Wannier-like
orbitals or the density operator itself.  The common point between
these variational methods is that the idempotency of the density
operator or the orthogonality of the Wannier orbitals is not
imposed explicitly during the minimization procedure. Instead,
the energy functionals are constructed such that these properties
are satisfied automatically at the minimum, which coincides with
the true ground state of the system.

Many of these methods also make use of non-orthogonal
localized orbitals, referred to as ``support
functions''~\cite{hernandez-prb95} or ``non-orthogonal generalized
Wannier functions'' (NGWFs)~\cite{skylaris-prb02}, in contrast to
canonical WFs, which are orthogonal. 
The density matrix in \equ{dm-wannier} 
can be generalized so as to be represented in terms of a set
of non-orthogonal localized orbitals $\{\phi_{\alpha\RR}(\rr)\}$ and a
corresponding non-unitary (and, in general, non-square) transformation
matrix $M^{(\kk)}$, which take the place of $\{w_{i\RR}(\rr)\}$ and
$U^{(\kk)}$, respectively. 
Two main benefits arise from allowing non-orthogonality. First, it is
no longer necessary to enforce explicit orthogonality constraints on
the orbitals during the energy minimization
procedure~\cite{galli-prl92, mauri-prb93, ordejon-prb93, hierse-prb94,
  hernandez-prb96}.
Second, a non-orthogonal representation can be
more localized than an essentially equivalent orthogonal
one~\cite{anderson-prl68, he-prl01}. 
In practice, linear-scaling methods target large systems, which means
that $\Gamma$-point only sampling of the BZ is usually sufficient. In
this case, the separable form for the density matrix simplifies to
\beq
\rho(\rr,\rr') = \sum_{\alpha\beta}
\phi_{\alpha}(\rr)K^{\alpha\beta}\phi^{\ast}_{\beta}(\rr'), 
\label{dm-ngwf2}
\eeq
where the density kernel is\footnote{It is worth noting that the
  non-orthogonality of the orbitals results in an important
  distinction between covariant and contravariant quantities, as
  denoted by
  raised and lowered
  Greek indices~\cite{artacho-pra91, oregan-prb11}.}
\beq
K^{\alpha\beta} = \sum_{n} [M^{\dagger}]^{\alpha}_{n} f_{n} [M]^{\beta}_{n}.
\eeq

\begin{figure}
\begin{center}
\includegraphics[width=5cm]{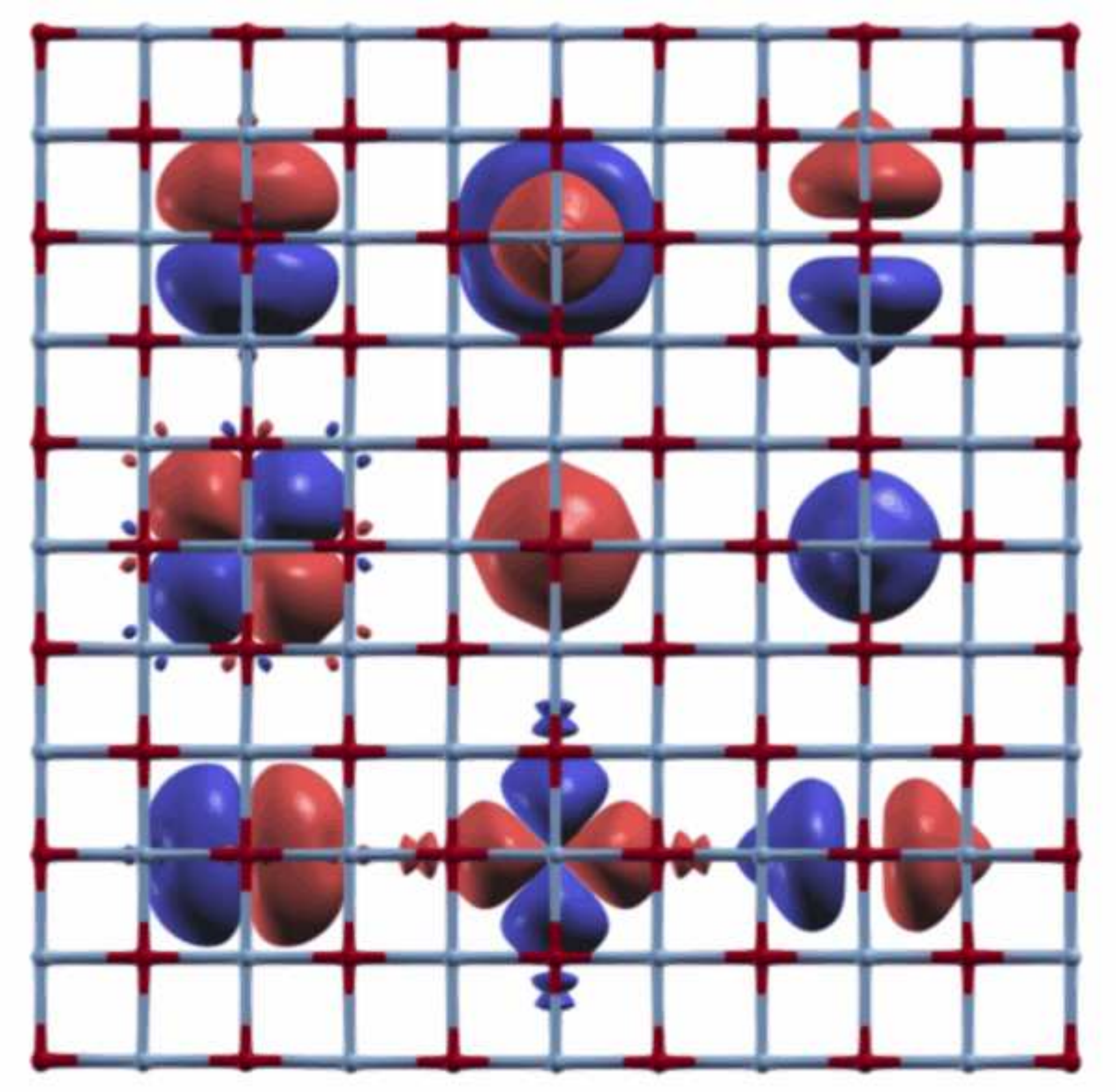}
\end{center}
\caption{\label{fig:ngwfs}
(Color online) Isosurfaces of the set of nine non-orthogonal
generalized Wannier functions (NGWFs) on a nickel atom in NiO (shown 
centered on different, symmetrically equivalent, Ni atoms in the
lattice). The isosurface is set to half of the maximum for the $s$ and
$p$-like NGWFs and $10^{-3}$ times the maximum for the $d$-like NGWFs. 
Adapted from \textcite{oregan-prb11}.}
\end{figure}

Minimization of an appropriate energy functional with respect to the
degrees of freedom present in the density matrix leads to
ground-state non-orthogonal orbitals that are very similar in
appearance to (orthogonal) MLWFs.
Fig.~\ref{fig:ngwfs} shows, for example, NGWFs on a Ni atom in bulk
NiO, obtained using the {\sc onetep} linear-scaling DFT
code~\cite{skylaris-jcp05}. A recent comparison of
  static polarizabilities for molecules, calculated using \equ{pol-real} with both
  MLWFs and NGWFs, demonstrates remarkable agreement between the
  two~\cite{oregan-arxiv12}.

\subsection{Other local representations}
Over the years, a number of other computational schemes have been
devised to provide a local analysis of the electronic structure in
molecules and solids. Here we briefly mention those most commonly used
in solid-state studies. The first choice is whether to work with the
electronic wavefunction or with the charge density.

One of the earliest and still most widely used wavefunction based
schemes is the ``Mulliken population analysis''
\cite{mulliken-jcp55}. This starts from a representation of the
density operator in an LCAO basis. If an extended basis, such as
planewaves, has been used, this can be obtained after first performing
a projection onto a suitable set of atomic orbitals
\cite{sanchez-portal-ssc95}.
Using the quantity $P_{A}$ introduced in \equ{PM} the Mulliken charge
on an atomic site A is given by
\beq
\label{eq:mull_charge}
Q_{A}=\sum_{i=1}^{\nwann} \langle\phi_{i}|P_{A}|\phi_{i}\rangle.
\eeq
The Mulliken scheme also provides a projection into local
angular-momentum eigenstates and an overlap (or bond) population between atom pairs.
The major disadvantage of the scheme is the fact that the absolute values
obtained have a marked dependence on the
LCAO basis. In fact, the results tend to become less meaningful as the
basis is expanded, as orbitals on one atomic site contribute to the
wavefunction on neighboring atoms. However, it is generally accepted
that so long as calculations using the same set of local orbitals are
compared, trends in the values can provide some chemical
intuition \cite{segall-prb96}. An early application was to the study of bonding
at grain boundaries of TiO$_2$ \cite{dawson-prb96}.

An alternative approach is to work directly with the charge density. The
scheme of \textcite{hirshfeld-tca77} attempts to partition the charge density by
first defining a so-called prodensity for the system, typically a
superposition of free atom charge densities $\rho^i({\rr})$.
The ground-state charge density
is then partitioned between atoms according to the proportions of the
procharge at each point in space. This can easily be integrated to
give, for example, a total charge
\begin{equation}
Q_{\rm H}^i=\int {\bf dr} \rho({\rr}) \frac{\rho^i({\rr})}{\Sigma_i \rho^i({\rr})}
\end{equation}
for each atomic site.  Hirschfield charges have
recently been used to parametrize dispersion corrections to local
density functionals \cite{tkatchenko-prl09}.

Partitioning schemes generally make reference to some arbitrary
auxiliary system; in the case of Hirschfield charges, this is the
free-atom charge density, which must be obtained within some
approximation. In contrast, the ``atoms in molecules'' (AIM) approach developed
by \textcite{bader-cr91} provides a uniquely defined way
 to partition the charge density.  It uses the vector field corresponding to the
 gradient of the charge density.  In many cases the only maxima in the
 charge density occur at the atomic sites. As all field lines must
 terminate on one of these atomic ``attractors'', each point in space can
 be assigned to a particular atom according to the basin of attraction that is
 reached by following the
 density gradient. Atomic regions are now separated by zero-flux
 surfaces ${\rm S}({\rr}_s)$
 defined by the set of points $({\rr}_s)$ at which
\begin{equation}
\nabla\rho({\rr}_s)\cdot{\bf n}({\rr}_s)=0 ,
\end{equation}
where ${\bf n}({\rr}_s)$ is the unit normal to ${\rm S}({\rr}_s)$. Having
made such a division it is straightforward to obtain values for the
atomic charges (and also dipoles, quadrupoles, and higher moments).
The AIM scheme has been widely used
to analyze bonding in both molecular and solid-state systems, as well as to give
a localized description of response properties such as infra-red absorption
intensities \cite{matta-book-07}.

A rather different scheme is the ``electron localization function'' (ELF)
introduced by \textcite{becke-jcp90} as a simple measure of electron
localization in physical systems. Their original definition is based
on the same-spin pair probability density $P({\rr},{\rr}')$, i.e., the
probability to find two like-spin electrons at positions ${\rr}$ and
${\bf r}'$. \textcite{savin-ac92} introduced a form for the ELF
$\epsilon({\rr})$ which can be applied to an independent-particle
model:
\begin{equation}
\epsilon({\rr})=\frac{1}{1+(D/D_h)^2}\,,
\end{equation}
\begin{equation}
D=\frac{1}{2}\sum^{\nbw}_{i=1}|\nabla\psi_i|^2-\frac{1}{8}\frac{|\nabla\rho|^2}{\rho},
\end{equation}
\begin{equation}
D_h=\frac{3}{10}(3\pi^2)^{2/3}\rho^{5/3}, \rho=\sum^{\nbw}_{i=1}|\psi_i|^2,
\end{equation}
where
the sum is over all occupied orbitals.  $D$ represents the
difference between the non-interacting kinetic energy and the kinetic
energy of an ideal Bose gas. $D_h$ is the kinetic energy of a
homogeneous electron gas with a density equal to the local density. As
defined, $\epsilon({\rr})$ is a scalar function which ranges from 0 to
1. Regions of large kinetic energy (i.e., electron delocalization) have
ELF values close to zero while larger values correspond to paired
electrons in a shared covalent bond or in a lone pair. In a uniform
electron gas of any density, $\epsilon({\rr})$ will take the value of
$1/2$. Early application of the ELF in condensed phases provided
insight into the nature of the bonding at surfaces of
Al \cite{santis-ss00} and Al$_2$O$_3$ \cite{jarvis-jpc01},
and a large number of other applications have appeared since.

\section{ANALYSIS OF CHEMICAL BONDING}
\label{sec:bonding}


%
As discussed in Sec.~\ref{sec:alt-loc}, there is a long tradition
in the chemistry literature of using localized molecular orbitals
\cite{boys-rmp60,foster-rmp60a,foster-rmp60b,edmiston-rmp63,boys-66}
as an appealing and intuitive avenue for investigating the nature
of chemical bonding in molecular systems.  The maximally-localized
Wannier functions (MLWFs) provide the natural generalization of this
concept to the
case of extended or solid-state systems. Since MLWFs are uniquely
defined for the case of insulators and semiconductors, they are
particularly suited to discuss hybridization, covalency, and
ionicity both in crystalline and disordered systems. In addition,
in the supercell approximation they can be used to describe any
disordered bulk, amorphous, or liquid system \cite{payne-rmp92},
providing a compact description of electronic states in terms of
their Wannier centers, their coordination with other atoms, and
the spatial distribution and symmetry of the MLWFs. As such, they
are often very useful for extracting chemical trends and for allowing
for statistics to be gathered on the nature of bonds (e.g covalent
bonds vs.\ lone pairs), be it in the presence of structural complexity,
as is the case of an amorphous solid, or following the intrinsic
dynamics of a liquid or an unfolding chemical reaction. They also
share the same strengths and weaknesses alluded to in section
\ref{sec:alt-loc}, whereby different localization criteria
can provide qualitatively different representations of chemical
bonds. This arbitrariness seems less common in extended system, and
often some of the most chemically meaningful information comes from
the statistics of bonds as obtained in large-scale simulations,
or in long first-principles molecular dynamics runs.  Finally, localized
orbitals can embody the chemical concept of transferable functional
groups, and thus be used to construct a good approximation for
the electronic-structure of complex systems starting from the
orbitals for the different fragments \cite{hierse-prb94,benoit-prl01,lee-prl05},
as will be discussed in Section \ref{sec:basis}.

\subsection{Crystalline solids}
\label{sec:bondcrystals}

One of the most notable, albeit qualitative, characteristics of MLWFs
is their ability to highlight the chemical signatures of different band manifolds.
This was realized early on, as is apparent from Fig. \ref{fig:si-gaas}, showing the
isosurfaces for one of the 4 MLWFs in crystalline silicon and gallium arsenide, respectively.
These are obtained from the closed manifold of four valence
bands, yielding four equivalent MLWFs that map into one another under
the space-group symmetry operations of the crystal.\footnote{It should
  be noted that the construction procedure does not necessarily lead to MLWFs that
  respect the space-group symmetry.  If desired, symmetries can be enforced by
  imposing co-diagonalization of appropriate operators \cite{posternak-prb02} or by using
  projection methods~\cite{ku-prl02,qian-prb08}.}
It is clearly apparent that these MLWFs represent the intuitive chemical
concept of a covalent bond, with each MLWF representing the bonding
orbital created by the constructive interference of two atomic $sp^3$
orbitals centered on neighboring atoms. In addition, it can be seen that
in gallium arsenide this covalent bond and its WFC are shifted towards
the more electronegative arsenic atom.  This has been explored further
by \textcite{abu-farsakh-prb07}, to introduce a formal definition of
electronegativity, or rather of a bond-ionicity scale, based on the
deviation of WFCs from their geometrical bond centers.  It is worth
mentioning that these qualitative features of Wannier functions tend
to be robust, and often independent of the details of the method used
to obtained them - maximally localized or not.  For example, similar
results are obtained for covalent conductors
whether one makes use of symmetry considerations
\cite{satpathy-pss88,smirnov-prb01,smirnov-ijqc02,usvyat-ijqc04},
finite-support regions in linear scaling methods \cite{fernandez-prb97,skylaris-prb02},
or projection approaches \cite{stephan-prb00}, or even if the MLWF
algorithm is applied within Hartree-Fock \cite{zicovich-wilson-jcp01}
rather than density-functional theory.

\begin{figure}
\begin{center}
\includegraphics[width=7.0cm]{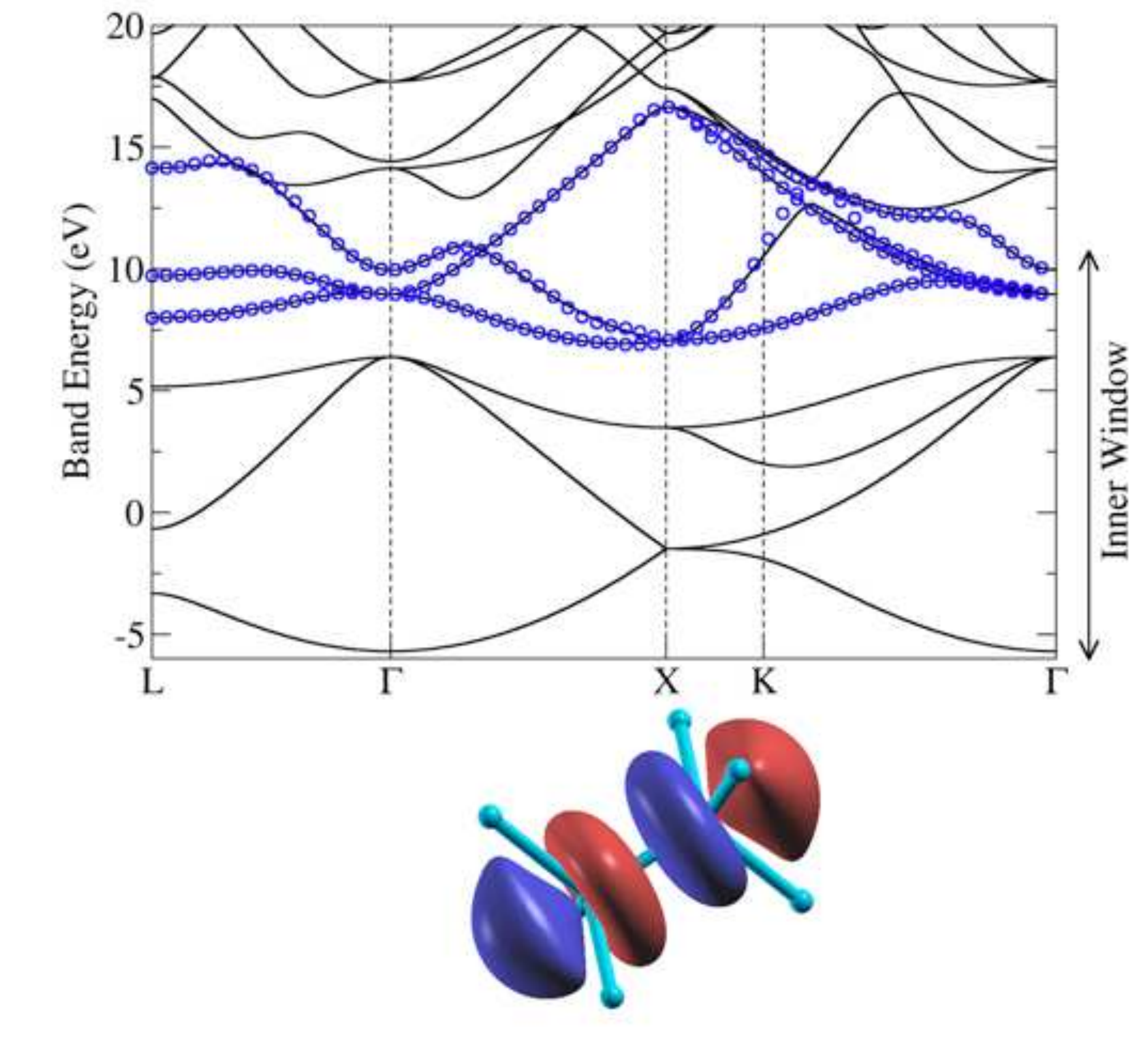}
\end{center}
\caption{\label{fig:si-anti} (Color online) Four-dimensional manifold
(blue circles) disentangled from the full band manifold (black lines) for Si,
together with one of the four antibonding MLWFs that are obtained after
wannierization of this four-dimensional manifold.  (The other three are
equivalent under space-group symmetry operations.)  The frozen inner
window is also indicated.}
\end{figure}

\begin{figure}[b]
\begin{center}
\includegraphics[width=7.0cm]{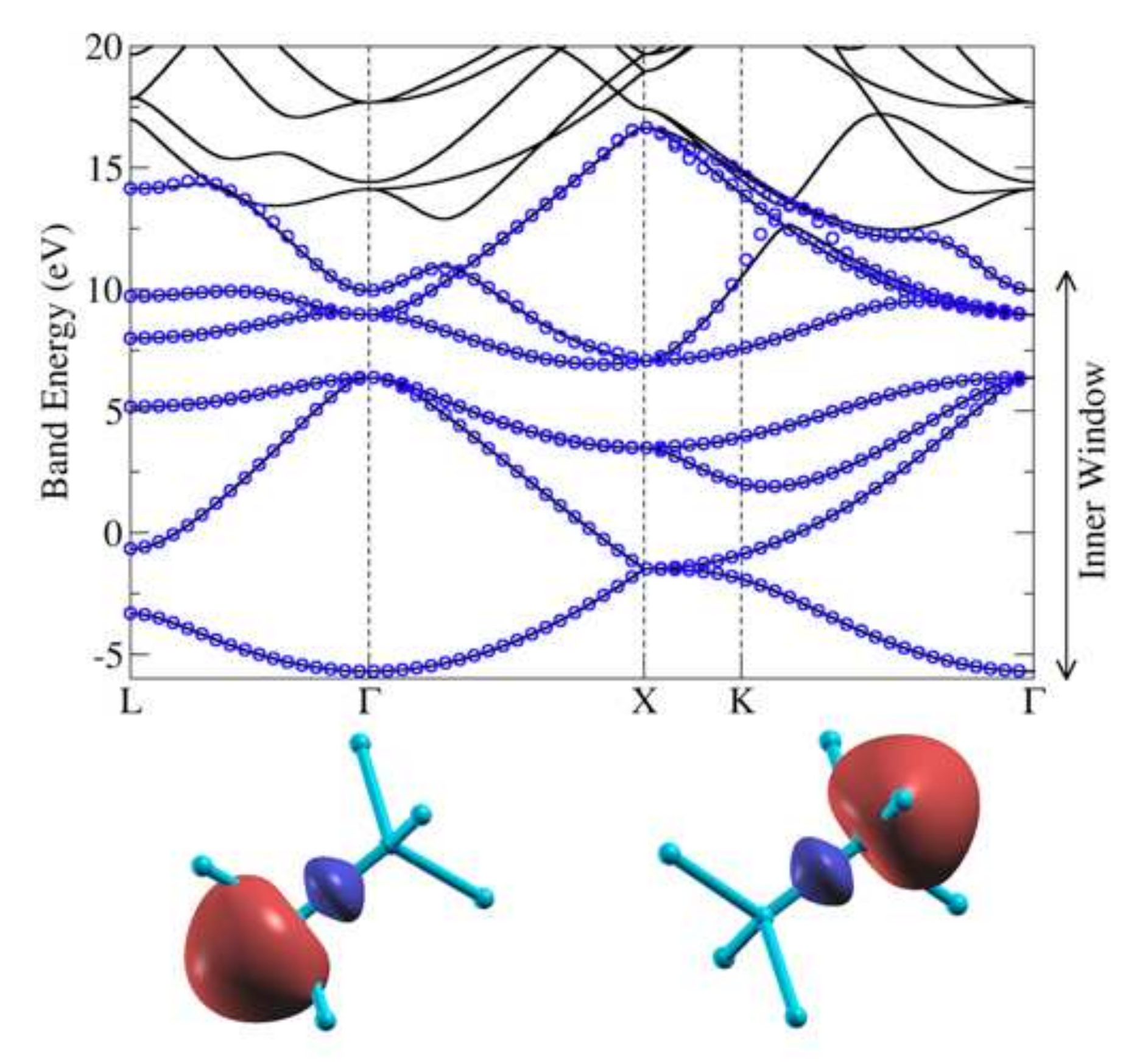}
\end{center}
\caption{\label{fig:si-sp3} (Color online)
Eight-dimensional manifold (blue circles)
disentangled from the full band manifold (black lines) for Si,
together with two of the eight atom-centered sp$^3$ MLWFs
that are obtained after wannierization of this manifold. (The other six are
equivalent under space-group symmetry operations.)  The frozen inner
window is also indicated.}
\end{figure}

Once conduction bands are included via the disentanglement procedure, results depend
on both the target dimensions of the disentangled manifold and on the states considered
in the procedure (e.g., the ``outer window'').  In this case, it becomes riskier to draw
conclusions from the qualitative features of the MLWFs.
Still, it is easy to see how MLWFs can make the connection between
atomic constituents and solid-state bands, representing a formal derivation of
``atoms in solids.'' That is, it can reveal the atomic-like orbitals
that conceptually lie behind
any tight-binding formulation, but that can now be obtained directly from first principles
according to a well-defined procedure. For crystalline silicon, the four-dimensional
manifold disentangled from the lowest part of the conduction bands gives rise
to four identical antibonding orbitals (see Fig.~\ref{fig:si-anti})
originating from the destructive interference of
two atomic $sp^3$ orbitals centered on neighboring atoms, to be contrasted with
the constructive interference shown in Fig.~\ref{fig:si-gaas} for the valence WFs.
In addition, an eight-dimensional manifold
disentangled from an energy window including both the valence bands and the lowest part of
the conduction bands gives rise to the atomic $sp^3$ tight-binding orbitals of crystalline
silicon (see Fig.~\ref{fig:si-sp3}). These can form the basis of
the construction of Hamiltonians for model systems
(e.g., strongly-correlated) or large-scale nanostructures, as will be discussed
in Chap. \ref{sec:basis}.

\begin{figure}
\begin{center}
\includegraphics[width=7.0cm]{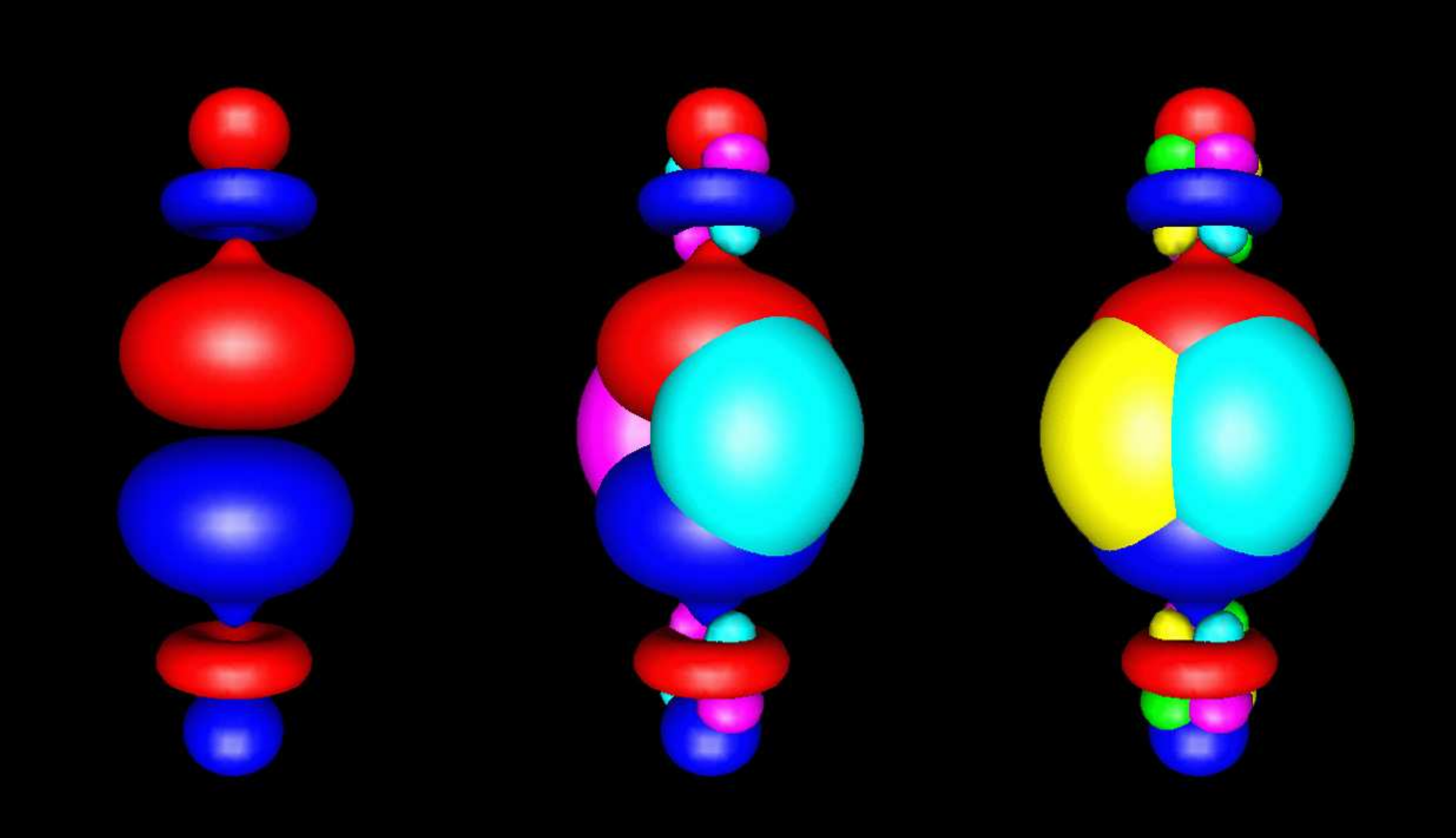}
\end{center}
\caption{\label{fig:bto3} (Color online) The three maximally-localized
Wannier functions derived from the O $2p$ bands of BaTiO$_{3}$,
showing the hybridization with the nominally empty Ti $3d$ orbitals.
The left panel shows the
O$[2p_z]$--Ti$[3d_{z^2}]$ MLWF, the central panel adds the
O$[2p_y]$--Ti$[3d_{yz}]$ MLWF, and the right panel adds to these the
O$[2p_x]$--Ti$[3d_{xz}]$ MLWF.  Adapted from \textcite{marzari-acp98}.}
\end{figure}

These considerations also extend to more complex systems. The case
of ferroelectric perovskites was studied relatively early, as by
\textcite{marzari-acp98,baranek-prb01,evarestov-ssc03}, thanks to the
presence of well-separated manifolds of bands \cite{king-smith-prb94}.
A classic example is shown in Fig.~\ref{fig:bto3}, showing for BaTiO$_3$
the three MLWFs per oxygen derived from the nine oxygen $2p$ bands.  While
in the classical ionic picture of perovskites the B-cation (here Ti) is
completely ionized in a 4+ state, the covalent nature of the bond becomes
clearly apparent here, with the MLWFs showing a clear hybridization in
the form of mixed $p-d$ orbitals. Such hybridization is at the origin
of the ferroelectric instability, as argued by \textcite{posternak-prb94}.
The analysis of the MLWF building blocks can also extend to quite different crystal types.
For example, \textcite{cangiani-prb04} discussed the case of TiO$_2$
polytypes, where bonding MLWFs associated with the O $2s/2p$
valence manifold are seen to be similar in the rutile and anatase
form, with the third polytype (brookite) an average between the two
\cite{posternak-prb06}.
Applications to other complex systems can be found, e.g.,
for antiferromagnetic MnO \cite{posternak-prb02} or for
silver halides \cite{evarestov-ijqc04}.

\subsection{Complex and amorphous phases}

Once the electronic ground state has been decomposed into well-localized
orbitals,
it becomes possible and meaningful to study their spatial
distribution or the distribution of their centers of charge
(the WFCs).  \textcite{silvestrelli-ssc98} were the first to argue
that the WFCs can be a powerful tool for
understanding bonding in low-symmetry cases, representing an insightful
and an economical mapping of the continuous electronic degrees of freedom into
a set of classical descriptors, i.e., the center positions and spreads
of the WFs.

The benefits of this approach become apparent when
studying the properties of disordered systems.
In amorphous solids the
analysis of the microscopic properties is usually based on the coordination
number, i.e., on the number of atoms lying inside a
sphere of a chosen radius $r_c$ centered on the selected atom
($r_c$ is typically inferred
from the first minimum in the pair correlation function).
This purely geometrical analysis is
completely blind to the actual electronic charge distribution,
which ought to be important in any description of chemical bonding.
An analysis of the full charge distribution and bonding in terms of the
Wannier functions, as for example in Fig.~\ref{fig:a-si} for the distorted
tetrahedral network of amorphous silicon,
would be rather complex, albeit useful to characterize
the most common defects \cite{fornari-cms01}.

Instead, just the knowledge of the positions of the WFCs and
their spreads can capture most of
the chemistry in the system and can identify the defects present.
In this approach, the WFCs are treated
as a second species of ``classical particles'' (representing electrons),
and the amorphous solid is treated as a statistical assembly of the two kinds of
particles (ions and WFCs).
Pair-correlation functions can thus be constructed for ions and classical electrons,
leading to the definition of novel bonding criteria based on the locations
of the WFCs.  For the case of amorphous silicon, for example,
the existence of a bond between two ions can be defined by their sharing
a common WFC within a distance that is smaller than
the first minimum of the silicon-WFC pair correlation function.
Following this definition, one can provide a more meaningful definition
of atomic coordination number,
argue for the presence (or absence) of bonds
in defective configurations, and propose specific electronic signatures for
identifying different defects \cite{silvestrelli-ssc98}.

\begin{figure}
\begin{center}
\includegraphics[width=8.2cm]{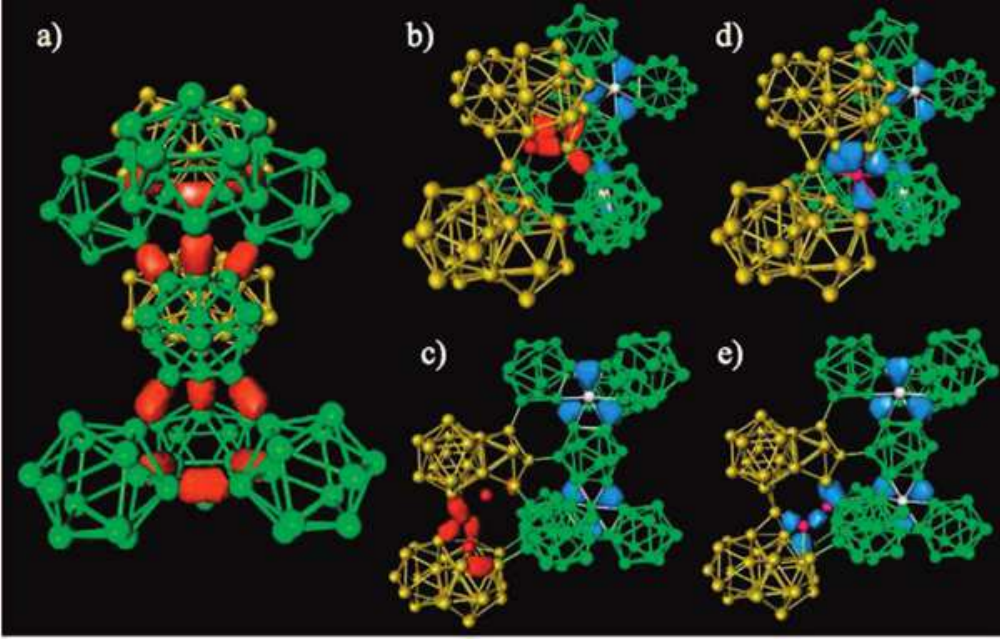}
\end{center}
\caption{\label{fig:boron} (Color online) Charge densities for
the MLWFs in $\beta$-rhombohedral boron. Red isosurfaces correspond
to electron-deficient bonds; blue correspond to fully
occupied bonds.  From \textcite{ogitsu-jacs09}.}
\end{figure}

The ability of Wannier functions to capture the electronic structure of
complex materials has also been demonstrated in the study of
boron allotropes. Boron is almost unique among the elements
in having at least four major crystalline phases -- all stable or
metastable at room temperature and with
complex unit cells of up to 320 atoms -- together with an amorphous phase.
In their study of $\beta$-rhombohedral boron, \textcite{ogitsu-jacs09}
were able to identify and study the relation between two-center and
three-center bonds and boron vacancies, identifying
the most electron-deficient bonds as the most chemically active.
Examples are shown in Fig.~\ref{fig:boron}.
\textcite{tang-prb10} were also able to study the evolution
of 2D boron sheets as they were made more compact (from hexagonal to
triangular), and showed that the in-plane bonding pattern
of the hexagonal system was preserved, with only minor changes in the
shape and position of the MLWFs.

\begin{figure}
\begin{center}
\includegraphics[width=8.7cm]{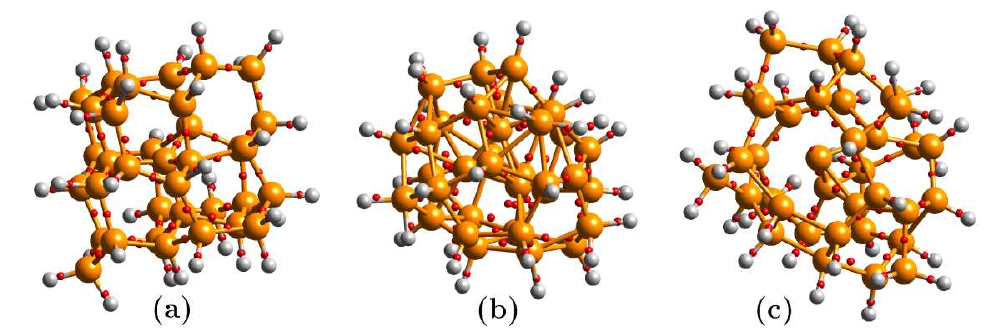}
\end{center}
\caption{\label{fig:nano1} (Color online)
Collapse and amorphization of a Si cluster under
pressure: increasing to 25 GPa (a) and then to 35 GPa (b), and then
back to 5 GPa (c).  Small red ``atoms'' are the Wannier centers.
From \textcite{martonak-cms01}.}
\end{figure}

Besides its application to the study of
disordered networks \cite{meregalli-ssc01,lim-ms02,fitzhenry-jpcm03}, the
above analysis can also be effectively employed
to elucidate the chemical and
electronic properties accompanying structural transformations.
In work on silicon nanoclusters under pressure
\cite{martonak-prl00,molteni-jcp01,martonak-cms01},
the location of the WFCs was monitored during compressive loading (up
to 35 GPa) and unloading.  Some resulting configurations are shown in
Fig.~\ref{fig:nano1}. The analysis of the
``bond angles'' formed by two WFCs and their common Si atom shows
considerable departure from the tetrahedral rule at the transition
pressure.  The MLWFs also become significantly more
delocalized at that pressure,
hinting at a metallization transition similar to the one that occurs in
Si in going from the diamond to the $\beta$-tin structure.

\subsection{Defects}

\begin{figure}[b]
\begin{center}
\includegraphics[width=4.5cm]{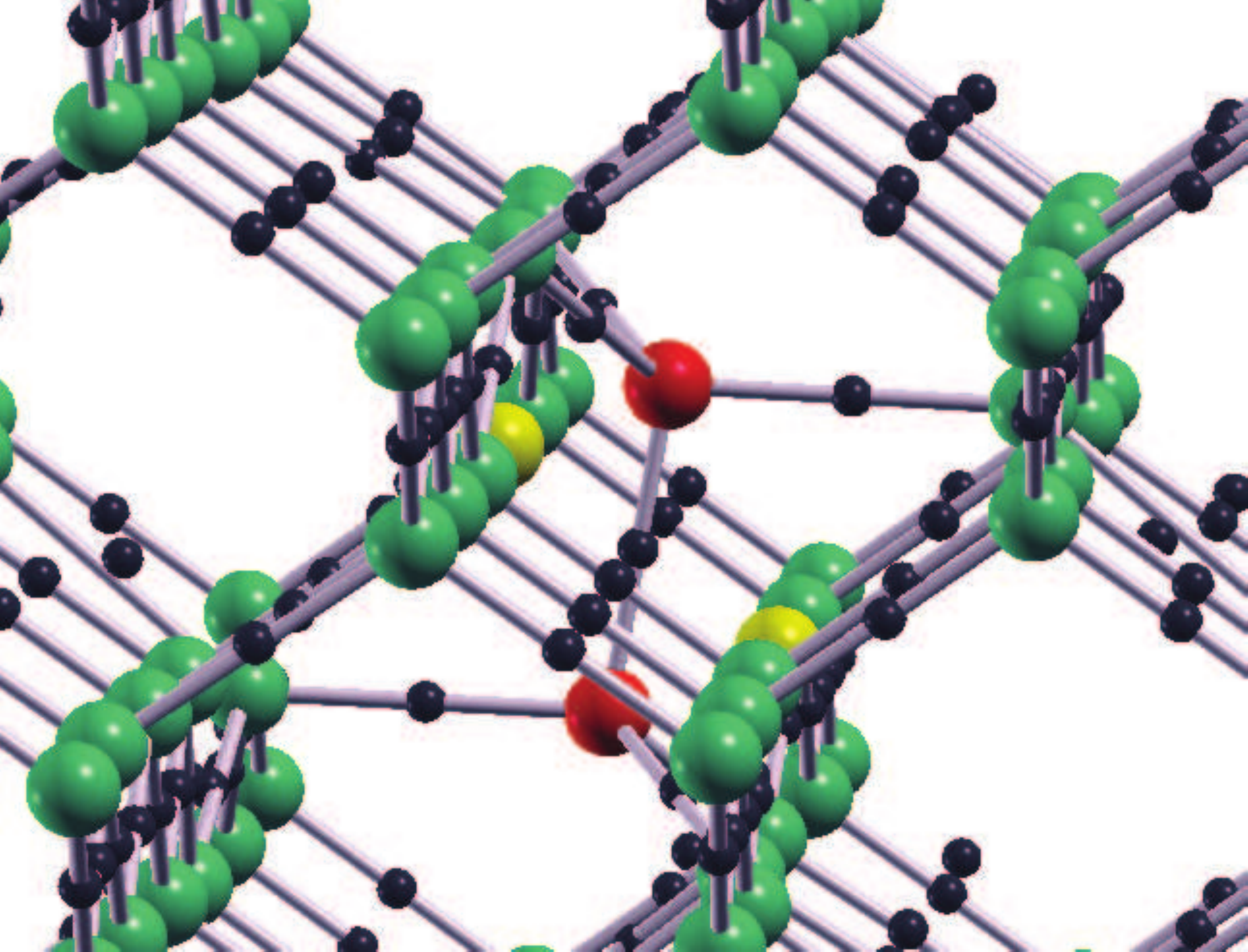}
\end{center}
\caption{\label{fig:fourfold} (Color online) The fourfold coordinated
defect in Si. Si atoms are in green, vacancies in black, and the centers
of the MLWFs in blue. Adapted from \textcite{goedecker-prl02}.}
\end{figure}

Interestingly, the MLWFs analysis can also point to structural
defects that do not otherwise exhibit any significant electronic signature.
\textcite{goedecker-prl02} have predicted
-- entirely from first-principles -- the existence of a new
fourfold-coordinated defect that is stable inside the
Si lattice (see Fig.~\ref{fig:fourfold}).
This defect had not been considered before, but displays by far the lowest
formation energy -- at the DFT
level -- among all native defects in silicon. Inspection of the relevant
``defective'' MLWFs
reveals that their spreads actually remain very close to those
typical of crystalline silicon,
and that the WFCs remain equally shared
between the atoms in a typical covalent arrangement.
These considerations suggest that the electronic
configuration is locally almost indistinguishable from that of
the perfect lattice, making this defect difficult to detect with standard
electronic probes.
Moreover, a low activation energy is required for the self-annihilation
of this defect; this consideration, in combination with the
``stealth'' electronic signature, hints at the reason why such
a defect could have
eluded experimental discovery
despite the fact that Si is one of the best studied materials in the
history of technology.

\begin{figure}
\begin{center}
\includegraphics[width=7.0cm]{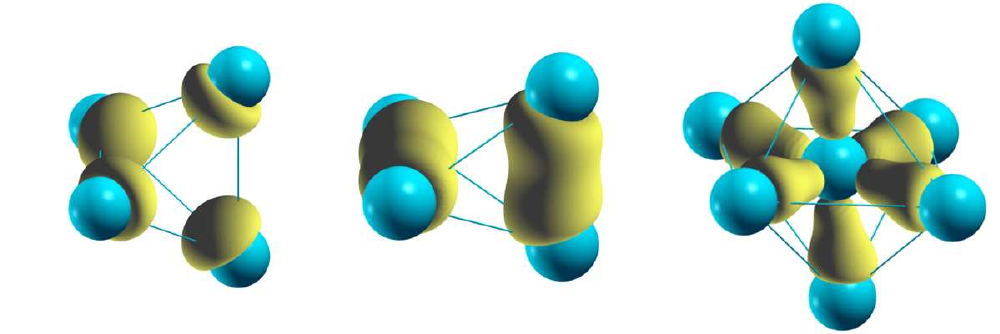}
\end{center}
\caption{\label{fig:corsetti} (Color online) Contour-surface plots of 
the MLWFs most strongly associated with a silicon vacancy in bulk
silicon, for different charge states of the vacancy (from left to
right: neutral unrelaxed, neutral relaxed, and doubly negative
relaxed). 
Adapted from \textcite{corsetti-prb11}.}
\end{figure}

For the case of the silicon vacancy, MLWFs have
been studied for all the charge states by
\cite{corsetti-prb11}, validating the canonical Watkins model
\cite{watkins-prl74}.  This work also demonstrated
the importance of including the occupied defect levels in the gap 
when constructing the relevant WFs, which are shown in the first
two panels of Fig.~\ref{fig:corsetti}.
%
%
For the doubly charged split-vacancy configuration, 
the ionic relaxation is such that one of the nearest neighbors of the
vacancy site moves halfway towards the vacancy, relocating to the
center of an octahedral cage of silicon atoms. This gives rise to
six defect WFs, each corresponding to a bond between $sp^3d^2$ hybrids on
the central atom and dangling $sp^3$ orbitals on the neighbors,
as shown in the last panel of Fig.~\ref{fig:corsetti}.

\subsection{Chemical interpretation}
\label{sec:chemint}

It should be stressed that a ``chemical'' interpretation of the
MLWFs is most appropriate when they are formed from a unitary
transformation of the occupied subspace.  Whenever unoccupied
states are included, MLWFs are more properly understood as forming
a minimal tight-binding basis, and not necessarily as descriptors
of the bonding.  Nevertheless, these tight-binding states sometimes
conform to our chemical intuition.  For example, referring back
to  Fig.~\ref{fig:graph_pz}) we recall that the band structure of
graphene can be described accurately by disentangling the partially
occupied $\pi$ manifold from the higher free-electron parabolic
bands and the antibonding $sp^2$ bands.  One can then construct
either a minimal basis of one $p_z$ MLWF per carbon, if interested
only in the $\pi/\pi^\star$ manifold around the Fermi energy,
or a slightly larger set with an additional MLWF per covalent
bond, if interested in describing both the partially occupied
$\pi/\pi^\star$ and the fully occupied $\sigma$ manifolds.  In this
latter case, the bond-centered MLWFs come from the constructive
superposition of two $sp^2$ orbitals pointing towards each other
from adjacent carbons \cite{lee-prl05}.

On the contrary, as discussed in Sec.~\ref{sec:disentanglement} and
shown in Fig.~\ref{fig:cu_7},
a very good tight-binding basis for $3d$ metals such as Cu can be
constructed \cite{souza-prb01} with 5 atom-centered $d$-like orbitals and
two $s$-like orbitals in the interstitial positions.  Rather than
reflecting a ``true'' chemical entity,
these represent linear combinations of $sp^3$ orbitals
that interfere constructively at the interstitial sites, thus
providing the additional variational freedom needed to describe
the entire occupied space.
%
%
Somewhere in between, it is worth pointing out that
the atom-centered $sp^3$ orbitals typical of group-IV or III-V semiconductors,
that can be obtained in the diamond/zincblende structure by disentangling the
lowest 4 conduction bands, can have a major lobe pointing either to the
center of the bond or in the opposite direction \cite{lee-phd06,wahn-pssbssp06}.
{
For a given spatial cutoff on the tight-binding Hamiltonian
constructed from these MLWFs, it is found that the former give a 
qualitatively much better description of the DFT band structure than the 
latter, despite the counter-intuitive result that the ``off-bond'' 
MLWFs are slightly more localized. 
The reason is that the ``on-bond'' MLWFs have a single dominant 
nearest neighbor interaction along a bond, whereas for the ``off-bond'' 
MLWFs there are a larger number of weaker nearest-neighbor interactions 
that are not directed along the bonds~\cite{corsetti-phd12}.
}

\subsection{MLWFs in first-principles molecular dynamics}
\label{sec:aimd}

The use of MLWFs to characterize electronic bonding in complex system
has been greatly aided by the implementation of efficient and robust
algorithms for maximal localization of the orbitals in the case of
large, and often disordered, supercells in which the Brillouin zone
can be sampled at a single point, usually the zone-center $\Gamma$ point
\cite{silvestrelli-ssc98,silvestrelli-prb99,berghold-prb00,bernasconi-jms01}.
Such efforts and the implementation in widely-available computer codes
have given rise to an extensive literature
dedicated to understanding and monitoring the nature of
bonding in complex and realistic systems under varying thermodynamical
conditions or during a chemical reaction. Such approaches are particularly
useful when combined with molecular-dynamics simulations, and
most applications have taken place within the framework first proposed
by Car and Parrinello \cite{car-prl85}. In fact, specialized algorithms have
been developed to perform on-the-fly Car-Parrinello molecular dynamics in a
Wannier representation \cite{sharma-ijqc03,iftimie-jcp04,wu-prb09}.

First applications
were to systems as diverse as high-pressure ice \cite{bernasconi-prl98},
doped fullerenes \cite{billas-jcp99}, adsorbed organic molecules
\cite{silvestrelli-prb00}, ionic
solids \cite{bernasconi-p02,posternak-prb02} and
the Ziegler-Natta polymerization \cite{boero-jacs00}.
This latter case is a paradigmatic
example of the chemical insight that can be gleaned by following
the WFCs in the course of an first-principles simulation. In the Ziegler-Natta
reaction we have an interconversion of a double carbon bond into
a single bond, and a characteristic agostic interaction between the C-H
bond and the activated metal center. Both become immediately visible once
the WFCs are monitored, greatly aiding the interpretation of the
complex chemical pathways.

Car-Parrinello molecular dynamics is particularly suited to the study of
liquid systems, and applications have been numerous in all areas of physical
chemistry.  Examples include the work of \textcite{
boero-jcp00,
raugei-jcp99,
sullivan-jpca99,
bako-jcp02, bernasconi-jcp04, blumberger-jacs04,
boero-jacs00, bucher-jpcb08, costanzo-jpcb08, dauria-jpca08,
faralli-jpcb06, heuft-jcp05,
ikeda-jcp05, jungwirth-jpca02,
kirchner-jcp04,
krekeler-jcp06, leung-jacs04, lightstone-c05,
lightstone-cpl01, odelius-jpca04, raugei-jcp02a,
salanne-jpcm08, schwegler-cpl01, suzuki-pccp08, tobias-jcp01, vanerp-jcp03,
vuilleumier-jcp01,
kreitmeir-jcp03, saharay-c04, todorova-jctc08}.

\begin{figure}
\begin{center}
\includegraphics[width=8.7cm]{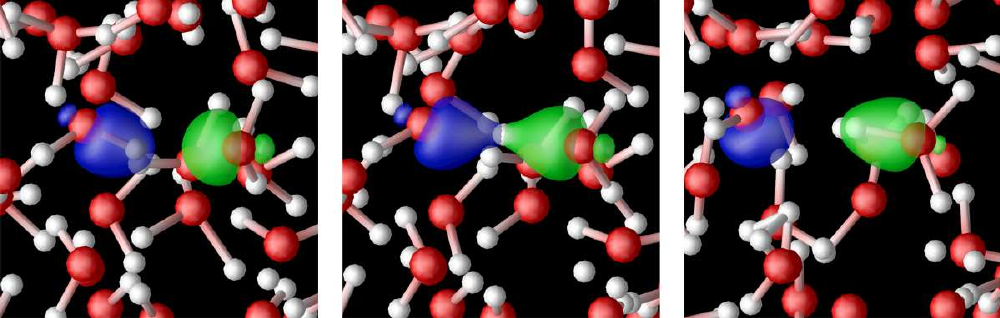}
\end{center}
\caption{\label{fig:hip_water} (Color online) Snapshots of a rapid
water-molecule dissociation
under high-temperature (1390 K) and high-pressure (27 GPa) conditions;
one of the MLWFs in the proton-donor molecule is highlighted in blue, and
one of the MLWFs in the proton-acceptor molecule is highlighted in green.
From \textcite{schwegler-prl01}.}
\end{figure}

Water in particular has been studied extensively, both
at normal conditions \cite{grossman-jcp04,sit-prb07} and in
high- and low-pressure phases at high temperature
\cite{silvestrelli-prl99,silvestrelli-jcp99,
boero-potps00,boero-prl00,boero-jcp01,schwegler-prl01,romero-jcp01}
(a fast dissociation event from one of these simulations is shown in
Fig.~\ref{fig:hip_water}).  Behavior in the presence
of solvated ions \cite{lightstone-cpl01,schwegler-cpl01,tobias-jcp01,
raugei-jcp02a,bako-jcp02} or a hydrated electron
\cite{boero-prl03, boero-jpca07}, or at surfaces and interfaces
\cite{kudin-jacs08,kuo-s04,kuo-jpcb06,kuo-jpcc08,mundy-cr06,salvador-pccp03},
has also been studied.
Moreover, MLWFs have been used to calculate the electronic momentum
density that can be measured in Compton scattering \cite{romero-pssbr00}.
This work elucidated the relation between the anisotropy of the Compton profiles for
water and the nature of hydrogen bonding \cite{romero-jcp01}, and led to
the suggestion that the number of hydrogen bonds present can be directly
extracted from the Compton profiles \cite{sit-prb07}.
{
The population of covalent bond pairs in liquid silicon and the
Compton signature of covalent bonding has also recently been studied using 
MLWFs~\cite{okada-prl12}.}

Even more complex biochemical systems have been investigated,
including wet DNA \cite{gervasio-prl02}, HIV-1 protease
\cite{piana-jacs01}, reverse transcriptase \cite{sulpizi-jpcb00},
phosphate groups (ATP, GTP and ribosomal units) in different environments
\cite{alber-jpcb99,minehardt-bj02,spiegel-jpcb03},
drug-DNA complexes \cite{spiegel-obc06}, and caspases and kinases
\cite{sulpizi-pfag03,sulpizi-jbc01}.  Extensive reviews of first-principles
quantum simulations and molecular dynamics, with discussions of MLWFs in
these contexts, have appeared reviews by \textcite{dovesi-rcc05, kirchner-prsopl07,
tuckerman-jpcb00, tuckerman-jpcm02, vuilleumier-incoll06,
tse-aropc02}, with \textcite{marx-book09} providing a very
comprehensive methodological overview.

Further applications of first-principles molecular dynamics oriented specifically
to extracting information about dipolar properties and dielectric
responses are discussed later in Sec.~\ref{sec:mol}.

\section{ELECTRIC POLARIZATION AND ORBITAL MAGNETIZATION}
\label{sec:pol}

First-principles calculations of electric dipoles and orbital magnetic
moments of molecular systems are straightforward.  The electric dipole
is
\beq
{\bf d}=-e\sum_j \matel{\psi_j}{\rr}{\psi_j}
\label{eq:finite-p}
\eeq
and the orbital moment is
\beq
{\bf m}= -\frac{e}{2c}\sum_j\matel{\psi_j}{\rr\times\vv}{\psi_j} \,,
\label{eq:finite-m}
\eeq
where the sum is over occupied Hamiltonian eigenstates $\ket{\psi_j}$,
$\rr$ is the position operator, $\vv=(i/\hbar)[H,\rr]$ is the
velocity operator, and Gaussian units are used.
However, these formulas cannot easily be generalized to the case of
crystalline systems, because the Hamiltonian
eigenstates take the form of Bloch functions $\ket{\psi\bnk}$ that
extend over all space.  The problem is that matrix
elements such as $\matel{\psi\bnk}{\rr}{\psi\bnk}$ and
$\matel{\psi\bnk}{\rr\times\vv}{\psi\bnk}$ are ill-defined for such
extended states \cite{nenciu-rmp91}.

To deal with this problem, the so-called ``modern theory of polarization''
\cite{resta-f92,king-smith-prb93,vanderbilt-prb93,resta-rmp94}
was developed in the 1990's, and a corresponding ``modern theory of
magnetization'' in the 2000's
\cite{xiao-prl05,thonhauser-prl05,ceresoli-prb06,shi-prl07,souza-prb08}.
Useful reviews of these topics have appeared \cite{resta-jpcm00,
vanderbilt-cccms06,resta-pf07,resta-jpcm10}.

These theories can be formulated either in terms of Berry phases and
curvatures, or equivalently, by working in the Wannier representation.
The basic idea of the latter is to consider a large but finite sample
surrounded by vacuum and carry out a unitary transformation from the
set of delocalized Hamiltonian eigenstates $\psi_j$ to a set of Wannier-like
localized molecular orbitals $\phi_j$.  Then one can use
\equ{finite-p} or \equ{finite-m}, with the $\psi_j$ replaced
by the $\phi_j$, to evaluate the electric or orbital magnetic
dipole moment per unit volume in the thermodynamic limit.  In doing
so, care must be taken to consider whether any surface contributions
survive in this limit.

In this section, we briefly review the modern theories of electric
polarization and orbital magnetization and related topics.
The results given in this section are 
valid for any set of localized WFs; maximally localized
ones do not play any special role.  Nevertheless, the
close connection to the theory of polarization has been one of
the major factors behind the resurgence of interest in WFs.
Furthermore, we shall see that the use of
MLWFs can provide a very useful,
if heuristic, local decomposition of polar properties in a an
extended system.  For these reasons, it is appropriate to
review the subject here.

\subsection{Wannier functions, electric polarization, and localization}
\label{sec:pol-theo}

\subsubsection{Relation to Berry-phase theory of polarization}
\label{sec:bp}

Here we briefly review the connection between the Wannier
representation and the Berry-phase theory of polarization
\cite{king-smith-prb93,vanderbilt-prb93,resta-rmp94}.
Suppose that we have constructed via \equ{blochlike} a set of
Bloch-like functions $\ket{\tilde{\psi}\bnk}$ that are smooth
functions of $\kk$.  Inserting these in place of $\ket{\psi\bnk}$
on the right side of \equ{wanniertransform}, the WFs in
the home unit cell $\RR$=$\0$ are simply
\beq
\ket{\0 n}=\ibz d\kk\; \ket{\tilde{\psi}\bnk} \,.
\label{eq:homewf}
\eeq
To find their centers of charge, we note that
\beq
\rr\,\ket{\0 n}=\ibz d\kk\; (-i\nabla_\kk e^{i\kk\cdot\rr}) \,\ket{\tilde{u}\bnk}\,.
\label{eq:rhomewf}
\eeq
Performing an integration by parts and applying $\bra{\0 n}$ on the left,
the center of charge is given by
\beq
\rr_n = \matel{\0 n}{\rr}{\0 n}=
  \ibz d\kk\; \matel{\tilde{u}\bnk}{i\nabla_\kk}{\tilde{u}\bnk} \,,
\label{eq:wfc}
\eeq
which is a special case of \equ{rmatel}.  Then, in the home unit cell, in
addition to the ionic charges $+eZ_\tau$ located at positions $\rr_\tau$,
we can imagine electronic charges $-e$ located at positions
$\rr_n$.\footnote{In these formulas, the sum over $n$ includes a
   sum over spin.  Alternatively a factor of 2 can be inserted to
   account explicitly for spin.}
Taking the dipole moment of this imaginary cell and dividing by the
cell volume, we obtain, heuristically 
\beq
\PP= \frac{e}{V} \Big( \sum_\tau Z_\tau \rr_\tau - \sum_n \rr_n \Big)
\label{eq:pol-real}
\eeq
for the polarization.

This argument can be put on somewhat firmer ground by imagining a
large but finite crystallite cut from the insulator of interest,
surrounded by vacuum.  The crystallite is divided into an ``interior''
bulk-like region and a ``skin'' whose volume fraction vanishes
in the thermodynamic limit.  The dipole moment is computed from
\equ{finite-p}, but using LMOs $\phi_j$ in place of the
Hamiltonian eigenfunctions $\psi_j$ on the right-hand side.
Arguing that the contribution of the skin to $\bf d$ is negligible
in the thermodynamic limit and that the interior LMOs
become bulk WFs, one can construct an
argument that arrives again at \equ{pol-real}.

If these arguments still seem sketchy, \equ{pol-real} can be
rigorously justified by noting that its second term
\beq
\PP_{\rm el}=-\frac{e}{(2\pi)^3}\sum_n \int_{\rm BZ} d\kk\;
   \matel{\tilde{u}\bnk}{i\nabla_\kk}{\tilde{u}\bnk} \,,
\label{eq:Pel}
\eeq
is precisely the expression for the electronic
contribution to the polarization in the Berry-phase theory
\cite{king-smith-prb93,vanderbilt-prb93,resta-rmp94},
which was derived by considering the flow of
charge during an arbitrary adiabatic change of the crystalline
Hamiltonian.

\begin{figure}
\begin{center}
\includegraphics[width=7cm]{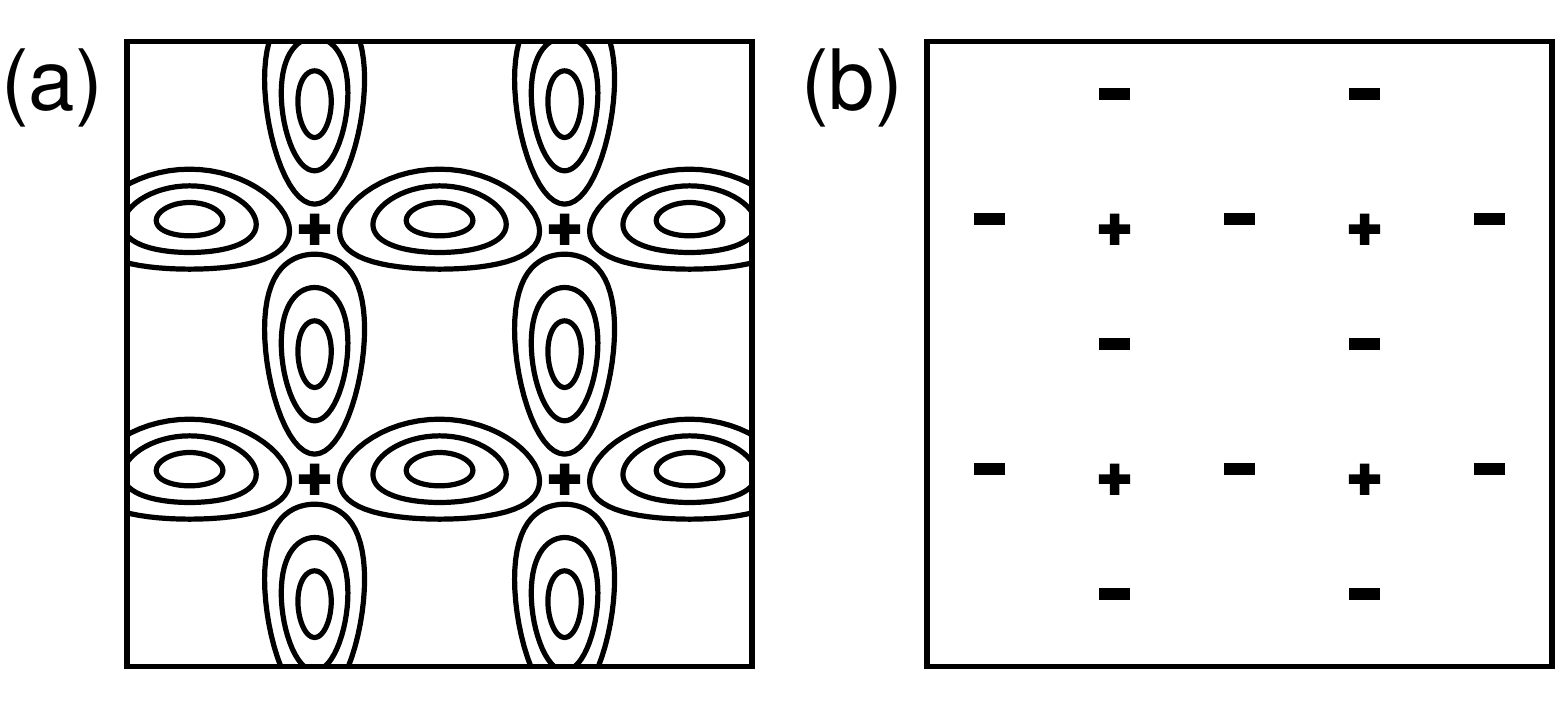}
\caption{Illustration of mapping from physical crystal
  onto equivalent point-charge system with correct dipolar
  properties.  (a) True system composed of point ions (+) and
  charge cloud (contours).  (b) Mapped system in which charge
  cloud is replaced by quantized electronic charges ($-$).
  In the illustrated model there are two occupied bands, i.e.,
  two Wannier functions per cell.}
\label{fig:mapping}
\end{center}
\end{figure}

The Berry-phase theory can be regarded
as providing a mapping of the distributed quantum-mechanical
electronic charge density onto a lattice of negative point charges
of charge $-e$, as illustrated in Fig.~\ref{fig:mapping}.
Then, the change of polarization
resulting from any physical change, such as the displacement of one
atomic sublattice or the application of an electric field, can
be related in a simple way to the displacements of the Wannier centers
$\rr_n$ occurring as a result of this change.

A well-known feature of the Berry-phase theory is that the polarization
is only well-defined modulo a quantum $e\RR/V$, where $\RR$ is a
real-space lattice vector.  Such an indeterminacy is immediately
obvious from \equ{pol-real}, since the choice of which WFs
are assigned to the home unit cell ($\RR$=$\0$) --
or, for that matter, which ions are assigned to it -- is arbitrary.
Shifting one of these objects by a lattice vector $\RR$ merely
changes $\PP$ by the quantum.
Correspondingly, it can be shown that an arbitrary change of
gauge can shift individual Wannier centers $\rr_n$ in arbitrary
ways, except that the sum $\sum_n\rr_n$ is guaranteed to remain
invariant (modulo a lattice vector).
{The same $e\RR/V$ describes the quantization of charge
transport under an adiabatic cycle \cite{thouless-prb83}, and indeed
the shifts of Wannier charge centers under such a cycle were
recently proposed as a signature of formal oxidation state in
crystalline solids \cite{jiang-prl12}.}

\subsubsection{Insulators in finite electric field}
\label{sec:efields}

The theory of crystalline
insulators in finite electric field $\EE$ is a subtle one;
the electric potential $-\EE\cdot\rr$ does not obey the conditions
of Bloch's theorem, and moreover is unbounded from below,
so that there is no well-defined ground state.  In practice one
wants to solve for a long-lived resonance in which the charge
density and other properties of the insulator remain periodic,
corresponding to what is meant experimentally by an insulator
in a finite field.  This is done by searching for
local minima of the electric enthalpy per cell
\beq
F=E_{\rm KS}-V\EE\cdot\PP
\label{eq:enthalpy}
\eeq
with respect to both the electronic and the ionic degrees of freedom.
$E_{\rm KS}$ is the ordinary Kohn-Sham energy as it would
be calculated at zero field, and the second term is the coupling of
the field to the polarization
as given in \equ{pol-real} \cite{nunes-prl94} or via the
equivalent Berry-phase expression \cite{nunes-prb01,souza-prl02,
umari-prl02}.  This approach is now standard for
treating periodic insulators in finite electric fields in 
density-functional theory.

\subsubsection{Wannier spread and localization in insulators}
\label{sec:wanloc}

We touch briefly here on another interesting connection to the
theory of polarization.  Resta and coworkers have defined a measure
of localization \cite{resta-prl99,sgiarovello-prb01,resta-jpcm02,
resta-jcp06} that distinguishes insulators from metals quite
generally, and have shown that this localization measure reduces,
in the absence of two-particle interactions or disorder, to the
invariant part of the spread functional $\Omega_{\rm I}$ given
in \equ{OmegaI}.  Moreover, \textcite{souza-prb00b}
have shown that this same
quantity characterizes the root-mean-square quantum fluctuations
of the macroscopic polarization. 
Thus,
while the Wannier charge centers are related to the {\it mean} value of
$\PP$ under quantum fluctuations, their invariant quadratic spread
$\Omega_{\rm I}$ is related to the corresponding {\it variance} of $\PP$.

\subsubsection{Many-body generalizations}
\label{sec:pol-manybody}

In the same spirit as for the many-body WFs discussed at
the end of Sec.~\ref{sec:manybody}, it is possible to generalize the
formulation of electric polarization and electron localization to the
many-body context.  One again considers $N$ electrons in a supercell;
for the present discussion we work in 1D and let the supercell have
size $L$.  The many-body theory of electric polarization was
formulated in this context by \textcite{ortiz-prb94}, and later
reformulated by \textcite{resta-prl98}, who introduced a ``many-body
position operator'' $\hat{X}=\exp(i2\pi\hat{x}/L)$ defined in terms of
the ordinary position operator $\hat{x}=\sum_{i=1}^N \hat{x}_i$.
While $\matel{\Psi}{\hat{x}}{\Psi}$ is ill-defined in the extended
many-body ground state $\ket{\Psi}$, the matrix element
$\matel{\Psi}{\hat{X}}{\Psi}$ is well-defined and can be used to
obtain the electric polarization, up to the usual quantum.  These
considerations were extended to the localization functional, and the
relation between localization and polarization fluctuations, by
\textcite{souza-prb00b}.  The variation of the many-body localization
length near an insulator-to-metal transition in 1D and 2D model
systems was studied using quantum Monte Carlo methods by
\textcite{hine-jpcm07}. Finally, the concept of electric enthalpy
was generalized to the many-body case by \textcite{umari_prl05},
allowing to calculate for the first time dielectric properties with
quantum Monte Carlo, and applied to the case of the polarizabilities
\cite{umari_prl05} and
hyperpolarizabilities \cite{umari_jcp09} of periodic hydrogen chains.

\subsection{Local polar properties and dielectric response}
\label{sec:pol-diel}

Is Sec.\ref{sec:bp} we emphasized the equivalence of the $\kk$-space
Berry-phase expression for the electric polarization, \equ{Pel},
and the expression written in terms of the locations of the
Wannier centers $\rr_n$, \equ{pol-real}.  The latter
has the advantage of being a real-space expression, thereby
opening up opportunities for localized descriptions and
decompositions of polar properties and dielectric responses.
We emphasize again that MLWFs have no
privileged role in \equ{pol-real}; the expression remains correct
for any WFs that are sufficiently well localized that the
centers $\rr_n$ are well defined.  Nevertheless, one may
argue heuristically that MLWFs provide the most natural
local real-space description of dipolar properties in crystals
and other condensed phases.

\subsubsection{Polar properties and dynamical charges of crystals}
\label{sec:pbulk}

Many dielectric properties of crystalline solids are
most easily computed directly in the $\kk$-space Bloch
representation.  Even before it was understood how to compute the
polarization $\bf P$ via the Berry-phase theory of \equ{Pel}, it was
well known how to compute derivatives of $\bf P$ using linear-response
methods \cite{gironcoli-prl89,resta-f92,baroni-rmp01}.  Useful derivatives include the electric
susceptibility $\chi_{ij}=dP_i/d\EE_j$ and the Born (or dynamical)
effective charges $Z^*_{i,\tau j}=VdP_i/dR_{\tau j}$,
where $i$ and $j$ are Cartesian labels and $R_{\tau j}$ is the
displacement of sublattice $\tau$ in direction $j$.  With the
development of the Berry-phase theory,
it also became possible to compute
effective charges by finite differences.  Similarly, with the
electric-enthalpy approach of \equ{enthalpy} it became possible to
compute electric susceptibilities by finite differences as well
\cite{souza-prl02,umari-prl02}.

The Wannier representation provides an alternative method for
computing such dielectric quantities by finite differences.  One computes
the derivatives $dr_{n,i}/d\EE_j$ or $dr_{n,i}/dR_{\tau j}$ of the
Wannier centers by finite differences, then sums these to get the desired
$\chi_{ij}$ or $Z^*_{i,\tau j}$.  An example of such a calculation
for $Z^*$ in GaAs was presented already in Sec.~VII of
\textcite{marzari-prb97},
and an application of the Wannier approach of \textcite{nunes-prl94}
in the density-functional context was used to compute $\chi$ by finite
differences for Si and GaAs \cite{fernandez-prb98}.
Dynamical charges were computed for several TiO$_2$ phases by
\textcite{cangiani-prb04} and \textcite{posternak-prb06}, and, as mentioned in 
Sec.~\ref{sec:bondcrystals},
observed differences between polymorphs were correlated with changes
in the chemical nature of the WFs associated with
OTi$_3$ structural units.
Piezoelectric coefficients, which are derivatives of $\bf P$ with
respect to strain, have also been carried out in the Wannier
representation for ZnO and BeO by \textcite{noel-prb02}.

Some of the most extensive applications of this kind have been to
ferroelectric perovskites, for which the dynamical charges have been
computed in density-functional and/or Hartree-Fock contexts for
BaTiO$_3$, KNbO$_3$, SrTiO$_3$, and PbTiO$_3$
\cite{marzari-acp98,baranek-prb01,evarestov-ssc03,usvyat-ijqc04}.  In
these materials, partially covalent bonding associated with
hybridization between O $2p$ and Ti $3d$ states plays a crucial role
in stabilizing the ferroelectric state and generating anomalous
dynamical effective charges \cite{posternak-prb94,zhong-prl94}.
Recall that the dynamical, or Born, effective charge $Z^*$
is defined as $Z^*_{i,\tau j}=VdP_i/dR_{\tau j}$
and carries units of charge.  Naively, one might expect values around
$+4e$ for Ti ions and $-2e$ for oxygen ions in BaTiO$_3$ based on
nominal oxidation states, but instead one finds ``anomalous'' values
that are much larger.  For example, \textcite{zhong-prl94} reported
values of $+7.2e$ for Ti displacements, and $-5.7e$ for O
displacements along the Ti--O--Ti chains.

\begin{figure}
\begin{center}
\includegraphics[width=2.5in]{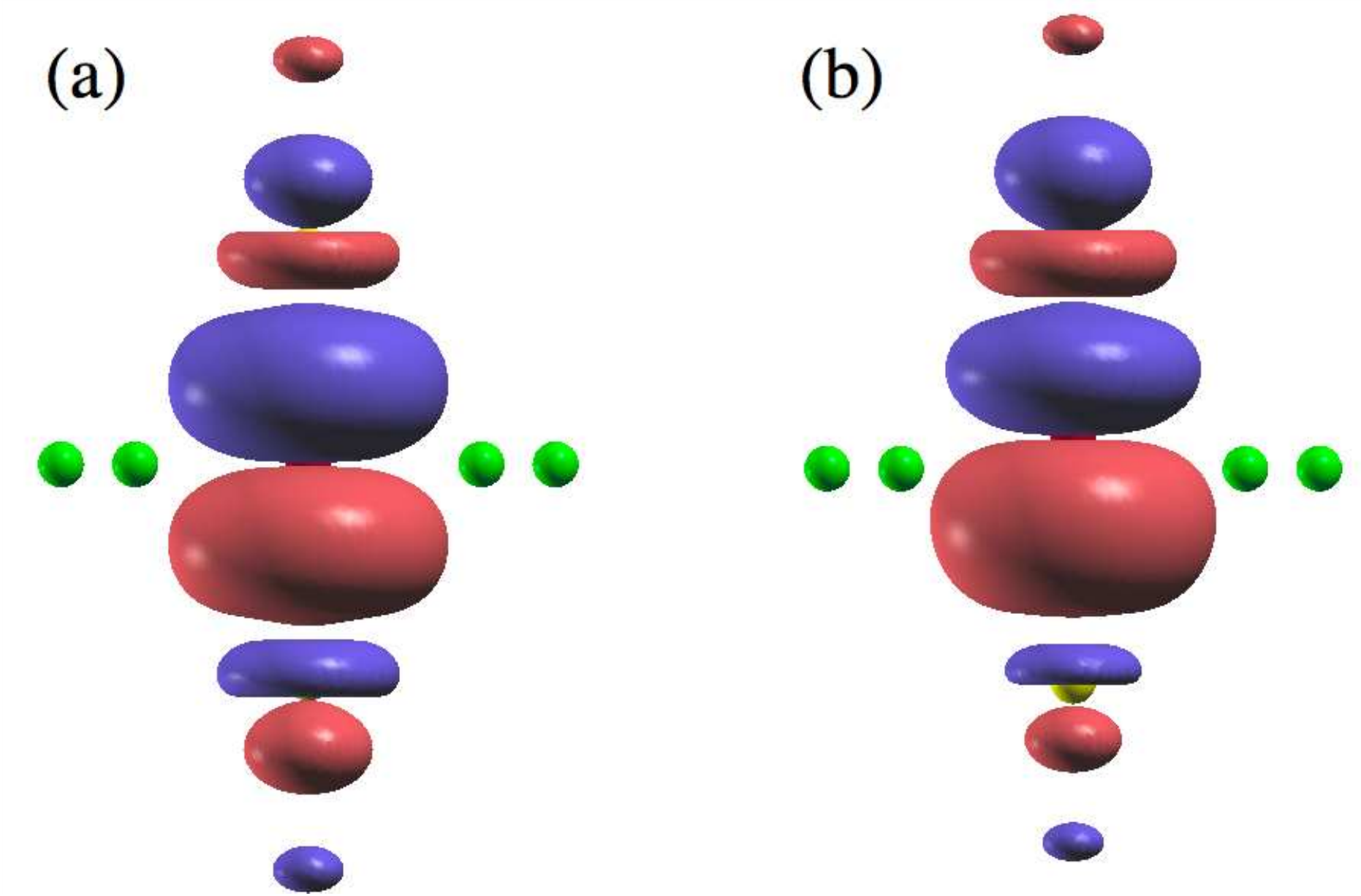}
\caption{(Color online) Amplitude isosurface plots of the maximally-localized
O$[2p_z]$--Ti$[3d_{z^2}]$ Wannier functions in
BaTiO$_3$. O is at center, surrounded by a plaquette of four
Ba atoms (green); Ti atoms (yellow, almost hidden) are above and below.
(a) Centrosymmetric structure.  (b) Ferroelectric structure
in which central O is displace upward relative to neighboring
Ti atoms.}
\label{fig:bto}
\end{center}
\end{figure}

This behavior can be
understood \cite{posternak-prb94,zhong-prl94} as arising from
hybridization between neighboring O $2p$ and Ti $3d$ orbitals that
dominate the valence and conduction bands, respectively.  This
hybridization, and the manner in which it contributes to an anomalous
$Z^*$, can be visualized by inspecting the changes in the MLWFs induced
by the atomic displacements.  Fig.~\ref{fig:bto}(a)
shows an 
O$[2p_z]$--Ti$[3d_{z^2}]$ MLWF in centrosymmetric
BaTiO$_3$ \cite{marzari-acp98}.
The hybridization to Ti $3d_{z^2}$ states appears in the form of
the ``donuts'' surrounding the
neighboring Ti atoms.  When the O atom moves upward relative to the
geometric center of the two neighboring Ti atoms as
shown in Fig.~\ref{fig:bto}(b), as it does in
ferroelectrically distorted BaTiO$_3$, the hybridization strengthens
for the upper O--Ti bond and weakens for the lower one, endowing the
WF with more Ti $3d$ character on the top than on the bottom.  As a
result, the center of charge of the WF shifts upward, and since
electrons carry negative charge, this results in a negative anomalous
contribution to the $Z^*$ of the oxygen atom.  The figure illustrates
this process for $\sigma$-oriented oxygen WFs, but a similar effect
occurs for the $\pi$-oriented oxygen WFs, and the total anomalous
dynamical charge can be accounted for quantitatively on the basis of
the distortion-induced changes of each kind of WF in the crystal
\cite{marzari-acp98}.

The above illustrates the utility of the MLWFs in providing a
{\it local} description of dielectric and polar responses in crystals.
This strategy can be carried further in many ways.  For example,
it is possible to decompose the $Z^*$ value for a given atom in
a crystal into contributions coming from various different
neighboring WFs, as was done for GaAs in Sec.~VII of
\textcite{marzari-prb97} and for BaTiO$_3$ by \textcite{marzari-acp98}.
Some chemical intuition is already gained by carrying out a
band-by-band decomposition of the $Z^*$ contributions
\cite{ghosez-prb95,ghosez-jpcm00}, but the WF analysis allows
a further spatial decomposition into individual WF contributions
within a band.  A deeper analysis that also involves the
decomposition of the WFs into atomic orbitals
has been shown to provide further insight into the anomalous
$Z^*$ values in perovskites \cite{bhattacharjee-pccp10}.

\begin{figure}
\begin{center}
\includegraphics[width=3.2in]{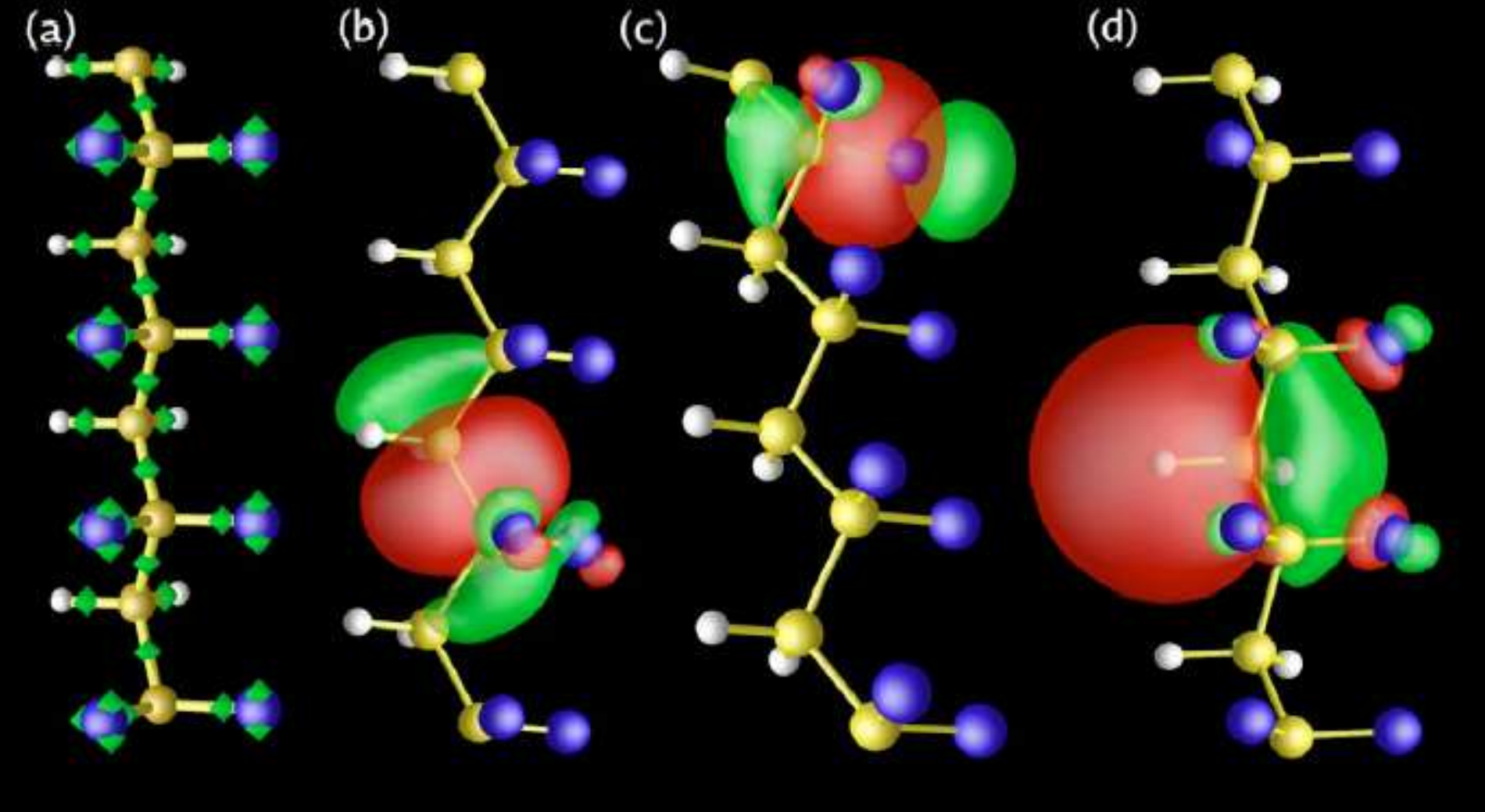}
\caption{(Color online) MLWFs for $\beta$-PVDF polymer chain.
(a) MLWF charge centers, indicated by diamonds.
(b)-(d) MLWFs localized on C--C, C--F, and C--H bonds,
respectively.  From \textcite{nakhmanson-prb05}.}
\label{fig:pvdf}
\end{center}
\end{figure}

Some insightful studies of the polar properties of polymer systems
in terms of MLWFs have also been carried out.  Figure~\ref{fig:pvdf},
for example, shows the WF centers and characters for the $\beta$
conformation of polyvinylidene fluoride ($\beta$-PVDF)
\cite{nakhmanson-prb05}, one of the more promising ferroelectric 
polymer systems.  An inspection of WF centers has also been
invoked to explain the polar properties of so-call ``push-pull''
polymers by \textcite{kudin-jcp07b} and of H$_2$O ice by
\textcite{lu-prl08}.  Finally, we note an interesting recent study
in which changes in polarization induced by corrugations in BN sheets
were analyzed in terms of WFs \cite{naumov-prl09}.

\subsubsection{Local dielectric response in layered systems}
\label{sec:layer}

In a similar way, the theoretical study of dielectric properties of
ultrathin films and superlattices can also be enriched by a local
analysis.  Two approaches have been introduced in the literature.
In one, the local $x$-$y$-averaged electric field $\bar{\E}_z(z)$
is calculated along the stacking direction $z$, and then the local
dielectric permittivity profile $\varepsilon(z)=\bar{\E}_z(z)/\bar{D}_z$ or
inverse permittivity profile $\varepsilon^{-1}(z)=\bar{D}_z/\bar{\E}_z(z)$
is plotted, where $\bar{D}_z$ is the $x$-$y$-averaged electric displacement
field (constant along $z$ in the absence of free charge) determined
via a Berry-phase calculation of $P_z$ or by inspection of $\bar{\E}_z$
in a vacuum region.  Such an approach has been applied to study
dielectric materials such as SiO$_2$ and HfO$_2$ interfaced to Si
\cite{giustino-prl03,giustino-prb05,shi-prb06,shi-apl07}
and perovskite films and superlattices \cite{stengel-n06,stengel-prb06}.

\begin{figure}
\begin{center}
\includegraphics[width=2.6in]{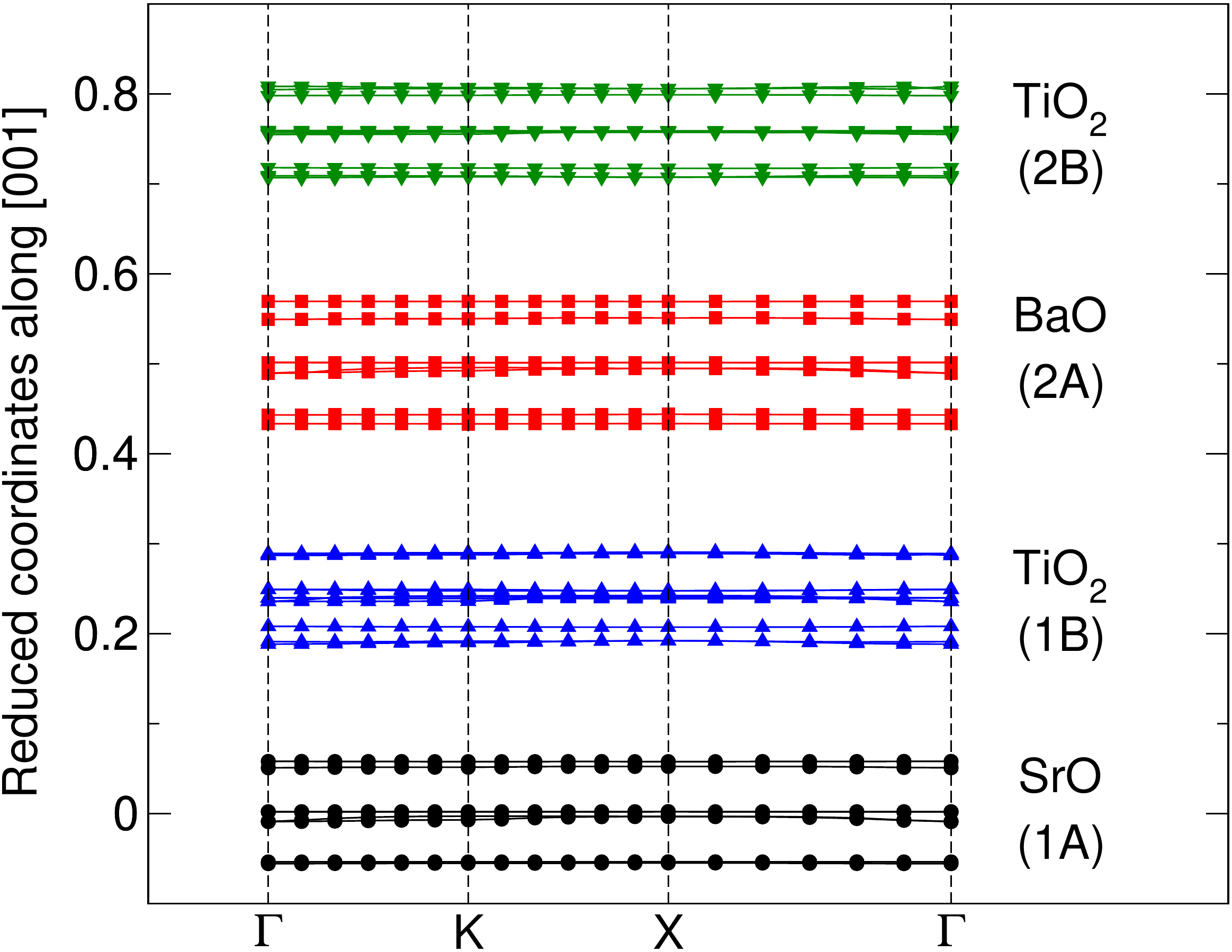}
\caption{(Color online) Dispersion of WF center positions along $z$ as a function
of $(k_x,k_y)$ for a superlattice composed of alternating layers of
SrTiO$_3$ (sublayers 1A and 1B) and BaTiO$_3$ (sublayers 2A
and 2B).  From \textcite{wu-prl06}.}
\label{fig:wcband}
\end{center}
\end{figure}

The second approach is to use a Wannier analysis to assign a
dipole moment to each layer.  This approach, based on the concept
of hybrid WFs discussed in Sec.~\ref{sec:hybrid}, was pioneered by
\textcite{giustino-prl03,giustino-prb05} and used by them to study
Si/SiO$_2$ interfaces and related systems.  Later applications
to perovskite oxide films and superlattices have been fairly
extensive.  The essential observation is that, when studying a
system that is layered or stacked along the $z$ direction, one
can still work with Bloch functions labeled by $k_x$ and $k_y$ while
carrying out a Wannier construction or analysis only along $z$.  Since
the extraction of Wannier centers in 1D is rather trivial, even in the
multiband case \cite{marzari-prb97,sgiarovello-prb01,bhattacharjee-prb05}, it is not
difficult to construct a ``Wannier center band structure''
$\bar{z}(k_x,k_y)$, and use the planar-averaged values to assign
dipole moments to layers.  This approach was demonstrated by
\textcite{wu-prl06}, as shown in Fig.~\ref{fig:wcband}, and has since
been used to study perovskite superlattices and artificial
nanostructures \cite{wu-prl08,murray-prb09}.  In a related
development, \textcite{stengel-prb06} introduced a Wannier-based
method for computing polarizations along $z$ and studying electric
fields along $z$ that work even in the case that the stacking includes
metallic layers, as long as the system is effectively insulating along
$z$ \cite{stengel-prb07}.  This allows for first-principles
calculations of the nonlinear dielectric response of ultrathin
capacitors and related structures under finite bias, providing an
insightful avenue to the study of finite-size and dead-layer effects
in such systems \cite{stengel-nm09,stengel-np09,stengel-prb09}.

\subsubsection{Condensed molecular phases and solvation}
\label{sec:mol}

Wannier-function methods have also played a prominent role in the
analysis of polar and dielectric properties of dipolar liquids, mainly
H$_2$O and other H-bonded liquids.  While the dipole moment of an
isolated H$_2$O molecule is obviously well defined, a corresponding
definition is not easy to arrive at in the liquid phase where
molecules are in close contact with each other.  An influential
development was the proposal made by
\textcite{silvestrelli-prl99, silvestrelli-jcp99}
that the dipole moment of a solvated molecule could be defined
in terms of positive charges on ionic cores and negative charges
located at the centers of the MLWFs.  Using this definition,
these authors found that the water molecule dipole is somewhat
broadly distributed, but has an average value of about 3.0\,D,
about 60\% higher than in the gas phase.
These features were shown to be in conflict with the behavior
of widely-used empirical models.

Of course, such a definition in terms of the dipole of the
molecular WF-center configuration remains heuristic
at some level.  For example, this local measure of the dipole
does not appear to be experimentally measurable
even in principle, and clearly the use of one of the alternative
measures of maximal localization discussed in Sec.~\ref{sec:alt-loc}
would give rise to slightly different values.  Nevertheless,
the approach has been widely adopted.  For example,
subsequent work elaborated \cite{sagui-jcp04,dyer-jcp06} and
extended this analysis to water in
supercritical conditions \cite{boero-prl00,boero-potps00},
confined geometries \cite{coudert-c06,dellago-cpc05},
and with solvated ions present \cite{scipioni-jcp09}, and compared
the results obtained with different exchange-correlation
functionals~\cite{todorova-jpcb06}.

It should be noted that the decomposition into Wannier dipoles
is closer to the decomposition of the charge density into static
(Szigeti) charges than to a decomposition into dynamical (Born)
charges.  The first one corresponds to a spatial decomposition of
the total electronic charge density, while the second is connected
with the force that appears on an atom in response to an applied
electric field.  As a counterpoint to the WF-based definition,
therefore, \textcite{pasquarello-prb03} have argued the a
definition based on these forces provides a more fundamental
basis for inspecting molecular dipoles in liquids.  In particular,
they define a second-rank tensor as the derivative of the
torque on the molecule with respect to an applied electric field,
and finding that this is typically dominated by its antisymmetric part,
they identify the latter (rewritten as an axial vector) as the
molecular dipole.  Surprisingly, they find that the magnitude of
this dipole vector is typically only about 2.1\,D in liquid water,
much less than the value obtained from the WF analysis.

Clearly the WF-based and force-based approaches to defining
molecular dipoles provide complementary perspectives, and
a more complete reconciliation of these viewpoints is the subject
of ongoing work.

Finally, we note that there is an extensive literature in which
Car-Parrinello molecular-dynamics simulations are carried out for H$_2$O
and other liquids, as already mentioned in Sec.~\ref{sec:aimd}
and surveyed in several reviews
\cite{tuckerman-jpcb00,tse-aropc02,tuckerman-jpcm02,kirchner-prsopl07}.
Using such approaches, it is possible to compute the dynamical
dipole-dipole correlations of polar liquids and compare the results
with experimental infrared absorption spectra.  While it is possible
to extract the needed information from the time-time correlation
function of the total polarization ${\bf P}(t)$ of the entire
supercell as calculated using the Berry-phase approach, methods which
follow the time-evolution of local dipoles, as defined via WF-based
methods, provide additional insight and efficiency
\cite{bernasconi-prl98,pasquarello-prb03,sharma-prl05,
  sharma-prl07,mcgrath-mp07,chen-prb08} and can easily be extended to
other kinds of molecular systems
\cite{gaigeot-mp07,gaigeot-jpcb03,gaigeot-jctc05,pagliai-jcp08}.  The
applicability of this kind of approach has benefited greatly from the
development of methods for computing WFs and their centers ``on the
fly'' during Car-Parrinello molecular dynamics simulations
\cite{sharma-ijqc03,iftimie-jcp04,wu-prb09}.

\subsection{Magnetism and orbital currents}
\label{sec:magnetism}

\subsubsection{Magnetic insulators}
\label{sec:mag}

Just as an analysis in terms of WFs can help clarify
the chemical nature of the occupied states in an ordinary insulator,
they can also help describe the orbital and magnetic ordering in
a magnetic insulator.

In the magnetic case, the maximal localization proceeds in the same
way as outlined in Sec.~\ref{sec:theory}, with trivial extensions
needed to handle the magnetic degrees of freedom.  In the case of
the local (or gradient-corrected) spin-density approximation,
in which spin-up and spin-down electrons are treated independently,
one simply carries out the maximal localization procedure
independently for each manifold.  In the case of a spinor calculation
in the presence of spin-orbit interaction, one instead implements
the formalism of Sec.~\ref{sec:theory} treating all wavefunctions as
spinors.  For example, each matrix element on the right-hand side
of \equ{Mkb} is computed as an inner product between spinors, and
the dimension of the resulting matrix is the number of occupied
spin bands in the insulator.

Several examples of such an analysis have appeared in the literature.
For example, applications to
simple antiferromagnets such as MnO \cite{posternak-prb02},
novel insulating ferromagnets and antiferromagnets \cite{ku-prl02,ku-jssc03},
and complex magnetic ordering in rare-earth manganates
\cite{yamauchi-prb08,picozzi-jpcm08} have proven to be illuminating.

\subsubsection{Orbital magnetization and NMR}
\label{sec:om}

In a ferromagnetic (or ferrimagnetic) material, the total
magnetization has two components. One arises from electron spin and is
proportional to the excess population of spin-up over spin-down
electrons; a second corresponds to circulating orbital currents.  The
spin contribution is typically dominant over the orbital one (e.g., by
a factor of 10 or more in simple ferromagnets such as Fe, Ni and Co
\cite{ceresoli-prb10a}), but the orbital component is also of
interest, especially in unusual cases in which it can dominate, or in
the context of experimental probes, such as the anomalous Hall
conductivity, that depend on orbital effects.  Note that inclusion of
the spin-orbit interaction is essential for any description of orbital
magnetic effects.

Naively one might imagine computing the orbital magnetization
${\bf M}_{\rm orb}$ as the thermodynamic limit of \equ{finite-m}
per unit volume for a large crystallite.  However, as we discussed
at the beginning of Sec.~\ref{sec:pol}, Bloch matrix elements
of the position operator $\rr$ and the circulation operator
$\rr\times\vv$ are ill-defined.  Therefore, such an approach is
not suitable.  Unlike for the case of electric polarization,
however, there is a simple and fairly accurate approximation
that has long been used to compute ${\bf M}_{\rm orb}$: one
divides space into muffin-tin spheres and interstitial regions,
computes the orbital circulation inside each sphere as a spatial
integral of $\rr\times{\bf J}$, and sums these contributions.
Since most magnetic moments are fairly local, such an approach
is generally expected to be reasonably accurate.

Nevertheless, it is clearly of interest to have available an
exact expression for ${\bf M}_{\rm orb}$ that can be used to
test the approximate muffin-tin
approach and to treat cases in which itinerant
contributions are not small.  The solution to this problem has
been developed recently, leading to a closed-form expression for
${\bf M}_{\rm orb}$ as a bulk Brillouin-zone integral.
Derivations of this formula via a semiclassical
\cite{xiao-prl05} or long-wave quantum \cite{shi-prl07} approach
are possible, but here we emphasize the derivation carried out
in the Wannier representation
\cite{thonhauser-prl05,ceresoli-prb06,souza-prb08}.
For this purpose, we restrict our interest to insulating
ferromagnets.  For the case of electric polarization, the
solution to the problem of $\rr$ matrix elements was sketched in
Sec.~\ref{sec:bp}, and a heuristic derivation of \equ{pol-real}
was given in the paragraph following that equation.  A similar
analysis was given in the above references for the case of
orbital magnetization, as follows.

Briefly, one again considers a
large but finite crystallite cut from the insulator of interest,
divides it into ``interior'' and ``skin'' regions, and transforms
from extended eigenstates to LMOs $\phi_j$.
For simplicity we consider the case of a two-dimensional
insulator with a single occupied band.  The interior gives a
rather intuitive ``local circulation'' (LC) contribution to the
orbital magnetization of the form
\beq
M_{\rm LC} = -\frac{e}{2A_0c}\matel{\0}{\rr\times\vv}{\0}
\label{eq:MLC}
\eeq
where $A_0$ is the unit cell area, since in the interior the
LMOs $\phi_j$ are really just bulk WFs.
This time, however, the skin contribution does {\it not} vanish.
The problem is that $\matel{\phi_j}{\vv}{\phi_j}$ is nonzero
for LMOs in the skin region, and the pattern of these velocity
vectors is such as to describe a current circulating around the
boundary of the sample, giving a second ``itinerant circulation''
contribution that can, after some manipulations, be written
in terms of bulk WFs as
\beq
M_{\rm IC} = -\frac{e}{2A_0c\hbar}\sum_{\RR}
[R_x\matel{\RR}{y}{\0}-R_y\matel{\RR}{x}{\0}] \,.
\label{eq:MIC}
\eeq
When these contributions are converted back to the Bloch
representation and added together, one finally obtains
\beq
{\bf M}_{\rm orb}=\frac{e}{2\hbar c} \, {\rm Im}
\int\frac{d^2k}{(2\pi)^2}\,\matel{\partial_\kk u_\kk}
{\times\,(H_\kk+E_\kk)}{\partial_\kk u_\kk},
\label{eq:Mkspace}
\eeq
which is the desired $\kk$-space bulk expression for the orbital
magnetization~\cite{thonhauser-prl05}.\footnote{In the case of metals
  \equ{Mkspace} must be modified by adding a $-2\mu$ term inside the
  parenthesis, with $\mu$ the chemical potential \cite{xiao-prl05,
    ceresoli-prb06}.  Furthermore, the integration is now restricted
  to the {\it occupied} portions of the Brillouin zone.}  The
corresponding argument for multiple occupied bands in three dimensions
follows similar lines \cite{ceresoli-prb06,souza-prb08}, and the
resulting formula has recently been implemented in the context of
pseudopotential plane-wave calculations \cite{ceresoli-prb10a}.
Interestingly, it was found that the interstitial contribution
--~defined as the difference between the total orbital magnetization,
\equ{Mkspace}, and the muffin-tin result~-- is not always
negligible. In bcc Fe, for example, it amounts to more than 30\% of
the spontaneous orbital magnetization, and its inclusion improves the
agreement with gyromagnetic measurements.

The ability to compute the orbital magnetization is also of use
in obtaining the magnetic shielding of nuclei. This is responsible for
the chemical shift effect observed in nuclear magnetic resonance (NMR)
experiments. A first principles theory for magnetic shielding in
solids was established by examining the perturbative response to a
periodic magnetic field in the long wavelength limit
\cite{mauri-prl96,pickard-prb01}. An alternative perturbative approach
used a WF representation of the electronic structure
together with a periodic position operator
\cite{sebastiani-jpca01,sebastiani-cpc02,sebastiani-mplb03}. However,
magnetic shieldings can also be computed using a ``converse'' approach
in which one uses Eq.~(\ref{eq:Mkspace}) to compute the orbital
magnetization induced by a fictitious point magnetic dipole on the
nucleus of interest~\cite{thonhauser-jcp09,ceresoli-prb10b}.
The advantage of such approach is that it does not require 
linear-response theory, and so it is amenable to large-scale
calculations or complex exchange-correlation functionals (e.g., including
Hubbard U corrections, or Hartree-Fock exchange), albeit at the cost
of typically one self-consistent iteration for every 
nucleus considered. Such converse approach has then been extended also to the
calculation of the EPR $g$-tensor by \textcite{ceresoli-prb10a}.

\subsubsection{Berry connection and curvature}
\label{sec:berrystuff}

Some of the concepts touched on in the previous section can be
expressed in terms of the $\kk$-space Berry connection
\beq
\label{eq:berry-connection}
{\bf A}_{n\kk}=\bra{u_{n\kk}}i{\nabla}_\kk \ket{u_{n\kk}}
\eeq
and Berry curvature
\beq
\label{eq:curv-pot}
\bm{\mathcal{F}}_{n\kk}={\nabla}_\kk\times {\bf A}_{n\kk}
\eeq
of band $n$.  In particular, the contribution of this band to the
electric polarization of Eq.~(\ref{eq:Pel}), and to the second term
in the orbital magnetization expression of Eq.~(\ref{eq:Mkspace}), are
proportional to the Brillouin-zone integrals of
${\bf A}_{n\kk}$ and $E_{n\kk}{\bm{\mathcal{F}}}_{n\kk}$,
respectively.  These quantities will also play a role in the
next subsection and in the discussion of the anomalous Hall
conductivity and related issues in Sec.~\ref{sec:curv-ahc}.

\subsubsection{Topological insulators and orbital magnetoelectric
response}
\label{sec:top}

There has recently been a blossoming of interest in so-called
topological insulators, i.e., insulators that cannot be adiabatically
connected to ordinary insulators without a gap closure.
\textcite{hasan-rmp10,hasan-arcmp11} provide excellent reviews
of the background, current status of this field, and provide
references into the literature.

One can distinguish two kinds of topological insulators.  First,
insulators having broken time-reversal ($T$) symmetry (e.g.,
insulating ferromagnets and ferrimagnets) can be classified by
an integer ``Chern invariant'' that is proportional to the
Brillouin-zone integral of the Berry curvature ${\bm{\mathcal{F}}}_{n\kk}$
summed over occupied bands $n$. Ordinary insulators are characterized
by a zero value of the invariant. An insulator with a non-zero
value would behave like an integer quantum Hall system, but without
the need for an external magnetic field; such systems are usually
denoted as ``quantum anomalous Hall'' (QAH) insulators.  While no
examples are known to occur in nature,
tight-binding models exhibiting such a
behavior are not hard to construct \cite{haldane-prl88}.  It can
be shown that a Wannier representation is not possible for a QAH
insulator, and \textcite{thonhauser-prb06} have explored the
way in which the usual Wannier construction
breaks down for model systems.

Second, depending on how their Bloch functions wrap the Brillouin
zone, nonmagnetic ($T$-invariant) insulators can be sorted into
two classes denoted as ``${\cal Z}_2$-even'' and ``${\cal
Z}_2$-odd'' (after the name ${\cal Z}_2$ of the group \{0,\,1\}
under addition modulo 2).  Most (i.e., ``normal'') insulators are
${\cal Z}_2$-even, but strong spin-orbit effects can lead to the
${\cal Z}_2$-odd state, for which the surface-state
dispersions are topologically required to display characteristic
features that are amenable to experimental verification.
Several materials realizations of ${\cal Z}_2$-odd insulators
have now been confirmed both theoretically and experimentally
\cite{hasan-rmp10,hasan-arcmp11}.

In a related development, the orbital magnetoelectric coefficient
$\alpha_{ij}=\partial M_{{\rm orb},j}/\partial \E_i$ was found to
contain an isotropic contribution having a topological character
(the ``axion'' contribution, corresponding to an $\EE\cdot\bm{B}$
term in the effective Lagrangian).  This term 
can be written as a Brillouin-zone integral of the Chern-Simons
3-form, defined in terms of multiband generalizations of the Berry
connection ${\bf A}_\kk$ and curvature ${\bm{\mathcal{F}}}_\kk$
introduced in the previous subsection \cite{qi-prb08,essin-prl09}.
The Chern-Simons magnetoelectric coupling has been evaluated from
first-principles with the help of WFs for both
topological and ordinary insulators~\cite{coh-prb11}.

A careful generalization of
Eq.~(\ref{eq:Mkspace}) to the case in which a finite electric field is
present has been carried out by \textcite{malashevich-njp10} in the
Wannier representation using arguments similar to those in
Secs.~\ref{sec:bp} and \ref{sec:om}, and used to derive a complete
expression for the orbital magnetoelectric response,
of which the topological Chern-Simons term is only one contribution
\cite{malashevich-njp10,essin-prb10}.

\section{WANNIER INTERPOLATION}
\label{sec:winterp}




\begin{figure*}
\begin{center}
\includegraphics[width=13.5cm]{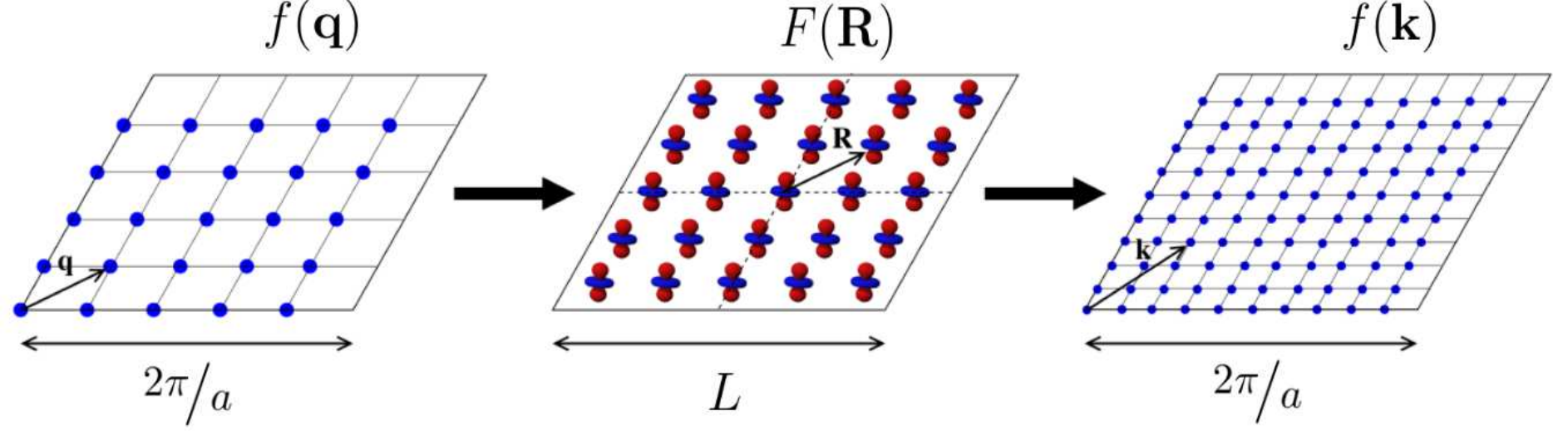}
\caption{(Color online) Schematic overview of the Wannier
  interpolation procedure.  The left panel shows the 
  BZ mesh $\qq$ used in the first-principles calculation, where the quantity of interest $f(\qq)$ is explicitly
  calculated. The Wannier-transformed quantity $F(\RR)$ is strongly
  localized near the origin of the equivalent supercell, shown in the
  middle panel with WFs at the lattice points. The right panel shows a
  dense mesh of interpolation points $\kk$ in the BZ, where the
  quantity $f(\kk)$ is evaluated at low cost starting from $F(\RR)$.}
\label{fig:interp_schem}
\end{center}
\end{figure*}

Localized Wannier functions are often introduced in textbooks as a
formally exact localized basis spanning a band, or a group of bands,
and their existence provides a rigorous justification for the
tight-binding (TB) interpolation
method~\cite{ashcroft-book76,harrison-book80}.

In this section we explore the ways in which WFs can be used as an
exact or very accurate TB basis, allowing to perform, very efficiently and accurately,
a number of operations on top of a conventional first-principles
calculation. The applications of this ``Wannier interpolation''
technique range from simple band-structure plots to the evaluation of
various physical quantities as BZ integrals.  The method is
particularly useful in situations where a very fine sampling of the BZ
is required to converge the quantity of interest. This is often the
case for metals, as the presence of a Fermi surface introduces sharp
discontinuities in $\kk$-space.

The Wannier interpolation procedure is depicted schematically in
Fig.~\ref{fig:interp_schem}.  The actual first-principles calculation
is carried out on a relatively coarse uniform reciprocal-space mesh
$\qq$ (left panel), where the quantity of interest $f(\qq)$ is
calculated from the Bloch eigenstates. The states in the selected
bands are then transformed into WFs, and $f(\qq)$ is transformed
accordingly into $F(\RR)$ in the Wannier representation (middle
panel). By virtue of the spatial localization of the WFs, $F(\RR)$
decays rapidly with $|\RR|$. Starting from this short-range real-space
representation, the quantity $f$ can now be accurately interpolated
onto an arbitrary point $\kk$ in reciprocal space by carrying out an
inverse transformation (right panel).  This procedure will succeed in
capturing variations in $f(\kk)$ over reciprocal lengths smaller than
the first-principles mesh spacing $\Delta q$, provided that the linear
dimensions $L=2\pi/\Delta q$ of the equivalent supercell are large
compared to the decay length of the WFs.

\subsection{Band-structure interpolation}
\label{sec:band-interpolation}

The simplest application of Wannier interpolation is to generate
band-structure plots. We shall describe the procedure in some detail, as
the same concepts and notations will reappear in the more advanced
applications to follow. 

From the WFs spanning a group of $\nwann$ bands, a set of Bloch-like states
can be constructed using \equ{wannier-inverse}, which we repeat here
with a slightly different notation, 
\beq
\label{eq:bloch-w-gauge}
\ket{\psi_{n\kk}^\ww}=\sum_\RR\,e^{i\kk\cdot\RR}\ket{\RR n}
\hspace{1cm} (n=1,\ldots, \nwann),
\eeq
{where the conventions of
  Eqs.~(\ref{eq:discrete-abnormal}-\ref{eq:mbandproja}) have been
  adopted.}  This has the same form as the Bloch-sum formula in
tight-binding theory, with the WFs playing the role of the atomic
orbitals.  The superscript W serves as a reminder that the states
$\ket{\psi_{n\kk}^\ww}$ are generally not eigenstates of the
Hamiltonian.\footnote{In Ch.~\ref{sec:theory} the rotated Bloch states
  $\ket{\psi_{n\kk}^\ww}$ were denoted by $\ket{\tilde{\psi}_{n\kk}}$,
  see \equ{blochlike}.}  We shall say that they belong to the {\it
  Wannier gauge}.

At a given $\kk$, the Hamiltonian matrix elements in the space of the
$\nwann$ bands is represented in the Wannier gauge by the matrix
\beq
\label{eq:h_kw}
H_{\kk,nm}^\ww
=\bra{\psi_{\kk n}^\ww}H\ket{\psi_{\kk m}^\ww}
=\sum_\RR\,e^{i\kk\cdot\RR}\bra{\0 n}H\ket{\RR m}.
\eeq
In general this is a non-diagonal matrix in the band-like indices,
and the interpolated eigenenergies are obtained by diagonalization,
\beq
\label{eq:h_k-diag}
H^\hh_{\kk,nm}
=\left[ U^\dagger_\kk H^\ww_\kk U_\kk\right]_{nm}=
\delta_{nm}\enkh.
\eeq

In the following, it will be useful to view the unitary matrices
$U_\kk$ as transforming between the Wannier gauge on the one hand,
and the Hamiltonian (H) gauge (in which the projected Hamiltonian
is diagonal) on the other.\footnote{The
  unitary matrices $U_\kk$ are related to, but not the same as, the
  matrices $U^{(\kk)}$ introduced in \equ{blochlike}. The latter are
  obtained as described in Secs.~\ref{sec:projection} and
  \ref{sec:maxloc}. In the present terminology, they transform from the
  Hamiltonian to the Wannier gauge on the 
  mesh used in the first-principles calculation. Instead, $U_\kk$ transforms from the Wannier to the Hamiltonian
  gauge on the interpolation mesh. That is, the matrix $U_\kk$
  is essentially an interpolation of the
  matrix $\left[ U^{(\kk)}\right]^\dagger$.}
From this point forward we adopt a condensed notation in
which band indices are no longer written explicitly, so that,
for example,
$H_{\kk,nm}^\hh=\bra{\psi^\hh_{\kk n}}H\ket{\psi^\hh_{\kk m}}$
is now written as $H_\kk^\hh=\bra{\psi^\hh_\kk}H\ket{\psi^\hh_\kk}$,
and matrix multiplications are implicit.  Then \equ{h_k-diag}
implies that the transformation law for the Bloch states is
\beq
\label{eq:psik-h}
\ket{\psi_\kk^\hh}=\ket{\psi_\kk^\ww}U_{\kk}.
\eeq

If we insert into Eqs.~(\ref{eq:h_kw}) and (\ref{eq:h_k-diag}) a
wavevector belonging to the first-principles grid, we simply recover
the first-principles eigenvalues $\enk$, while for arbitrary $\kk$ the
resulting $\enkh$ interpolate smoothly between the values on the
grid. (This is strictly true only for an isolated group of bands. When
using disentanglement, the interpolated bands can deviate from the
first-principles ones outside the inner energy window, as discussed in
Sec.~\ref{sec:entangled} in connection with Fig.~\ref{fig:si_proj}.)

Once the matrices $\bra{\0}H\ket{\RR}$ have been tabulated, the
band structure can be calculated very inexpensively by Fourier transforming
[\equ{h_kw}] and diagonalizing [\equ{h_k-diag}] matrices of rank
$\nwann$. Note that $\nwann$, the number of WFs per cell, is typically
much smaller than the number of basis functions (e.g., plane waves)
used in the first-principles calculation.

In practice the required matrix elements are obtained by inverting
\equ{h_kw} over the first-principles grid,
\beq
\label{eq:ham-R}
\begin{split}
\bra{\0}H\ket{\RR}&=\frac{1}{N}\sum_\qq\,e^{-i\qq\cdot\RR}
\bra{\psi_{\qq}^\ww}H\ket{\psi_{\qq}^\ww}\\
&=\frac{1}{N}\sum_\qq\,e^{-i\qq\cdot\RR}
        V_\qq^\dagger E_\qq V_\qq.
\end{split}
\eeq
Here $N$ is the number of grid points, $E_\qq$ is the diagonal matrix
of first-principles eigenenergies, and $V_\qq$ is the matrix defined in
\equ{wannierization-matrix}, which converts the $\nbands_\qq$ {\it ab
  initio} eigenstates into the $\nwann\leq \nbands_\qq$ Wannier-gauge
Bloch states,
\beq
\label{eq:psik-w}
\ket{\psi_\qq^\ww}=
\ket{\psi_\qq}V_{\qq}.
\eeq

%
%
%
%

The strategy outlined above~\cite{souza-prb01} is essentially the
Slater-Koster interpolation method.  However, while the
Hamiltonian matrix elements in the localized basis are treated as
adjustable parameters in empirical TB, they are calculated
from first-principles here.  A similar interpolation strategy is widely
used to obtain phonon dispersions starting from the interatomic force
constants calculated with density-functional perturbation
theory~\cite{baroni-rmp01}.  We shall return to this analogy between
phonons and tight-binding electrons \cite{martin-book04} when
describing the interpolation of the electron-phonon matrix elements in
Sec.~\ref{sec:el-ph}.

Wannier band-structure interpolation is extremely accurate. By virtue
of the exponential localization of the WFs within the periodic
supercell (see Footnote \ref{foot:discretization}), the magnitude of the
matrix elements $\bra{\0}H\ket{\RR}$ decreases rapidly with
$|\RR|$, and this exponential localization is preserved even in the
case of metals, provided a smooth subspace has been disentangled.
As the number of lattice vectors included in the summation in
\equ{h_kw} equals the number of first-principles mesh points, beyond a
certain mesh density the error incurred decreases
exponentially~\cite{yates-prb07}.
%
%
In the following we will illustrate the method with a few
representative examples selected from the literature.

\subsubsection{Spin-orbit-coupled bands of bcc Fe}
\label{sec:fe-bands}

As a first application, we consider the relativistic band structure of
bcc Fe.  Because of the spin-orbit interaction, the spin density is
not perfectly collinear, and the Bloch eigenstates are spinors.  As
mentioned in Sec.~\ref{sec:mag}, spinor WFs can be constructed via a
trivial extension of the procedure described in Sec.~\ref{sec:theory}
for the non-magnetic (spinless) case.  It is also possible to further
modify the wannierization procedure so as to produce two separate
subsets of spinor WFs: one with a predominantly spin-up character, and
the other with a predominantly spin-down character~\cite{wang-prb06}.

\begin{figure}
\begin{center}
\includegraphics[width=3.2in]{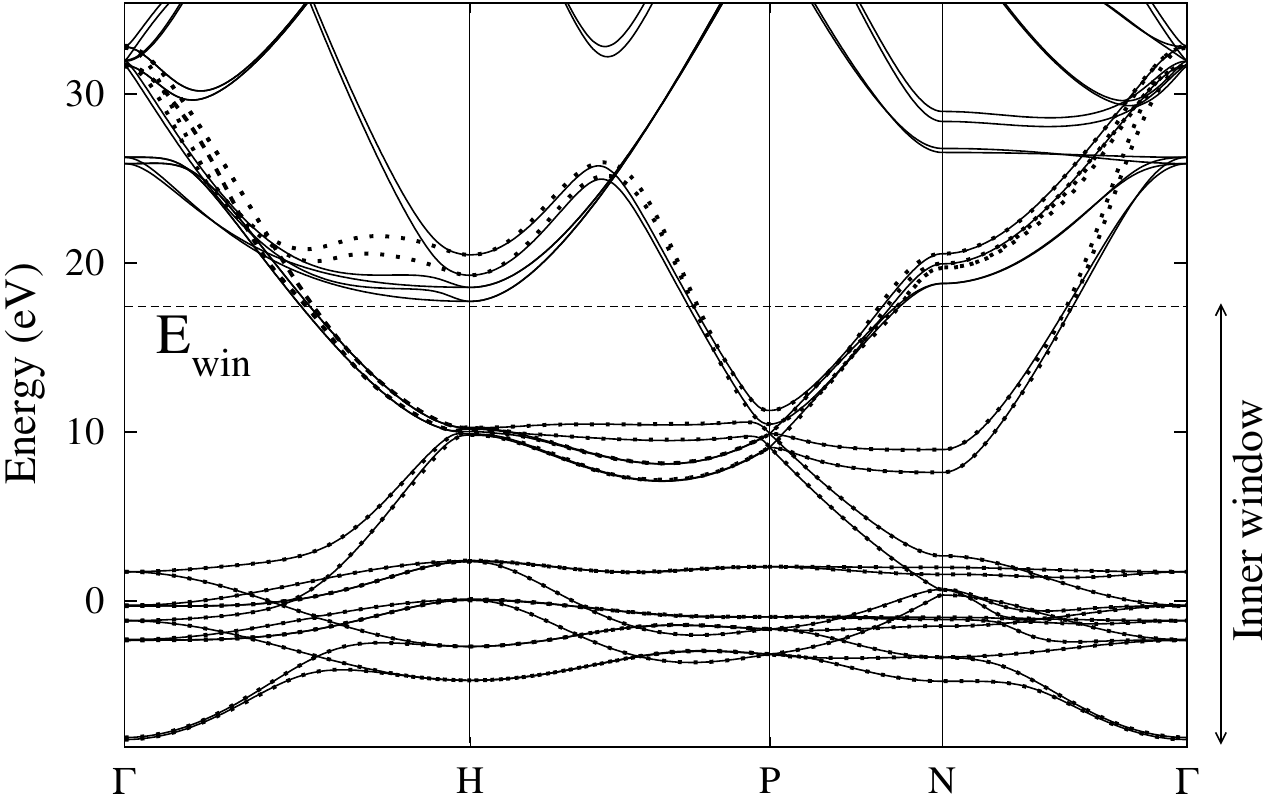}
\caption{Band structure of bcc Fe with spin-orbit coupling
  included. Solid lines: original band structure from a conventional
  first-principles calculation.  Dotted lines: Wannier-interpolated
  band structure. The zero of energy is the Fermi level. From
  \textcite{wang-prb06}.}
\label{fig:Fe-bands}
\end{center}
\end{figure}

Using this modified procedure, a set of nine disentangled WFs per spin
channel was obtained for bcc Fe by \textcite{wang-prb06}, consisting
of three $t_{2g}$ $d$-like atom-centered WFs and six $sp^3d^2$-like
hybrids pointing along the cubic directions.  An inner energy window
was chosen as indicated in Fig.~\ref{fig:Fe-bands}, so that these 18
WFs describe exactly all the the occupied valence states, as well as
the empty states up to $E_{\rm win}$, which was set at approximately
18~eV above the Fermi level.

\begin{figure}
\begin{center}
\includegraphics[width=2.6in]{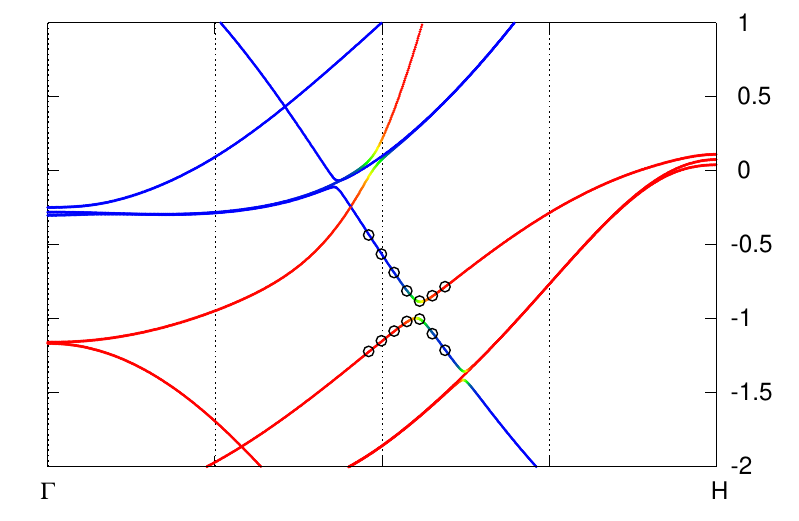}
\caption{(Color online)  Wannier-interpolated relativistic band
  structure of ferromagnetic bcc Fe along $\Gamma$--H. The bands are
  color-coded according to the expectation value of $S_z$: red for
  spin up and blue for spin down. The energies in eV are referred to
  the Fermi level. The vertical dashed lines indicate $\kk$-points on
  the mesh used in the first-principles calculation for constructing the WFs. From
  \textcite{yates-prb07}.}
\label{fig:Fe-avoided-crossing}
\end{center}
\end{figure}

The interpolated bands obtained using an $8\times 8\times 8$
$\qq$-grid in the full BZ are shown as dashed lines in
Fig.~\ref{fig:Fe-bands}.  The comparison with the first-principles
bands (solid lines), reveals essentially perfect agreement within the
inner window.  This is even more evident in
Fig.~\ref{fig:Fe-avoided-crossing}, where we zoom in on the
interpolated band structure near the Fermi level along $\Gamma$--H, and
color-code it according to the spin-projection along the quantization
axis.  The vertical dotted lines indicate points on the $\qq$-mesh.
For comparison, we show as open circles the eigenvalues calculated
directly from first-principles around a weak spin-orbit-induced
avoided crossing between two bands of opposite spin. It is apparent
that the interpolation procedure succeeds in resolving details of the
true band structure on a scale much smaller than the spacing between
$\qq$-points.


\subsubsection{Band structure of a metallic carbon nanotube}
\label{sec:nanotube-bands}

\begin{figure}
\begin{center}
\includegraphics[width=3.2in]{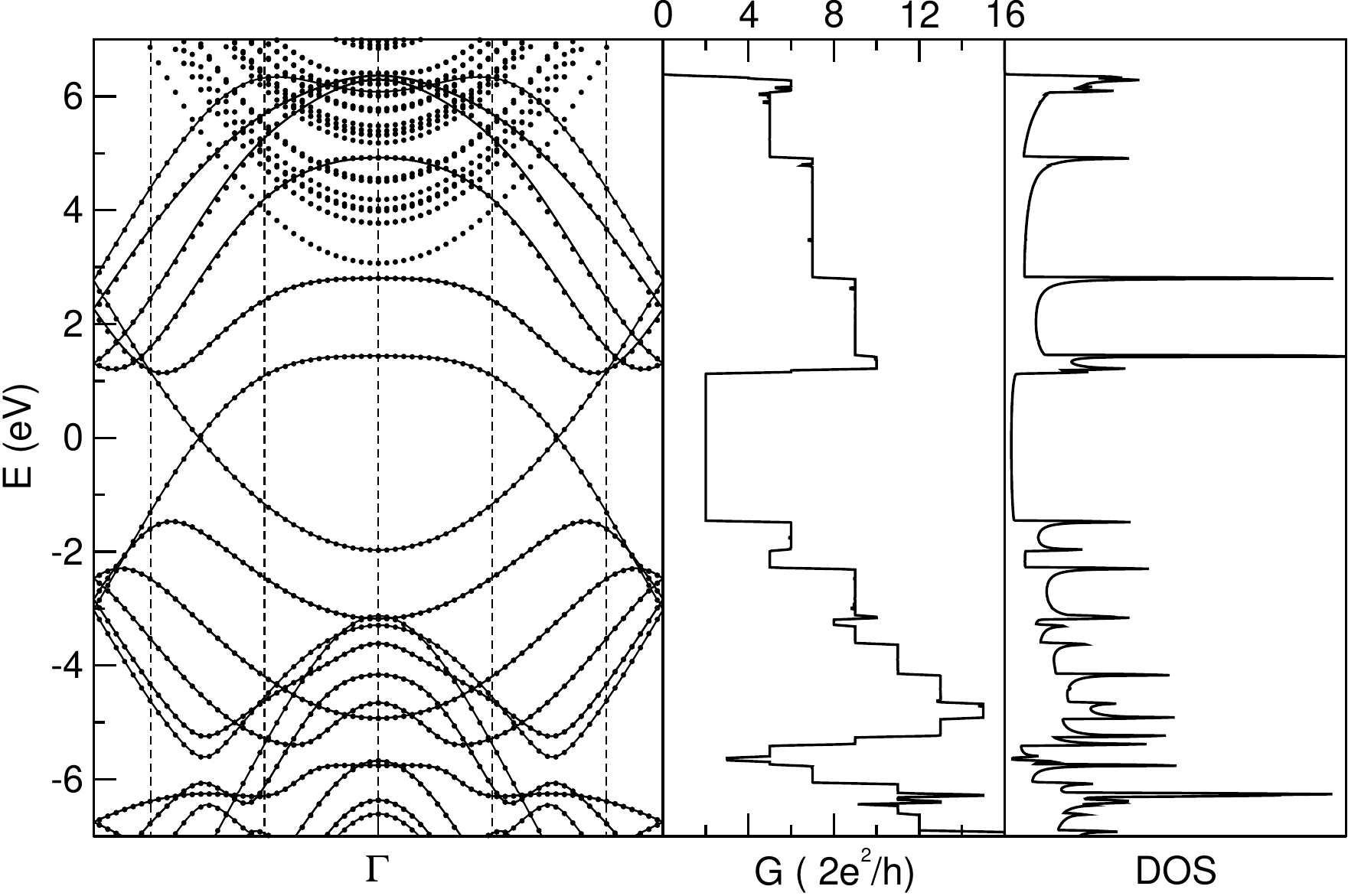}
\caption{Left panel: band structure of a (5,5) carbon nanotube,
  calculated by Wannier interpolation (solid lines), or from a full
  diagonalization in a planewave basis set (circles).  The five
  vertical dashed lines indicate the five $\kk$-points corresponding
  to the $\Gamma$ point in a 100-atom supercell.  The middle and right
  panels show the Wannier-based calculation of the quantum conductance
  and the density of states (see Sec. VII), with a perfect match of
  steps and peaks with respect to the exact band structure. From
  \textcite{lee-prl05}.}
\label{fig:nanotube-bands}
\end{center}
\end{figure}

As a second example, we consider Wannier interpolation in large
systems (such as nanostructures), that are often sampled only at the
zone center. We consider here a (5,5) carbon nanotube, studied in a
100-atom supercell (i.e.  five times the primitive unit cell) and with
$\Gamma$-point sampling.  The system is metallic, and the
disentanglement procedure is used to generate well-localized WFs,
resulting in either bond-centered combinations of $sp^2$ atomic
orbitals, or atom-centered $p_z$ orbitals.  The energy bands at any
other $\kk$-points are calculated by diagonalizing \equ{h_kw}, noting
that the size of the supercell has been chosen so that the Hamiltonian
matrix elements on the right-hand-side of this equation are non
negligible only for WFs up to neighboring supercells $\RR^{(\pm)}$ on
either side of $\RR=0$.  Fig.~\ref{fig:nanotube-bands} shows as solid
lines the interpolated bands, unfolded onto the 20-atom primitive
cell.  Even if with this sampling the system has a pseudogap of ~2eV,
the metallic character of the bands is perfectly reproduced, and these
are in excellent agreement with the bands calculated directly on the
primitive cell by direct diagonalization of the Kohn-Sham Hamiltonian
in the full plane-wave basis set (open circles). The vertical dashed
lines indicate the equivalent first-principles mesh obtained by
unfolding the $\Gamma$-point \footnote{When $\Gamma$ sampling is 
used, special care should be used in calculating matrix elements
between WFs, since the center of a periodic image of, e.g., the ket
could be closer to the bra that the actual state considered. Similar
considerations apply for transport calculations, and might require 
calculation of the matrix elements in real-space\cite{lee-phd06}.}.

\subsubsection{GW quasiparticle bands 
}
\label{sec:gw-bands}

In the two examples above the WFs were generated from Kohn-Sham Bloch
functions, and the eigenvalues used in \equ{ham-R} were the
corresponding Kohn-Sham eigenvalues. Many of the deficiencies of the
Kohn-Sham energy bands, such as the underestimation of the energy gaps
of insulators and semiconductors, can be corrected using many-body
perturbation theory in the form of the GW approximation
(for a review, see \textcite{aryasetiawan-ropp98}).

One practical difficulty in generating GW band structure plots is that
the evaluation of the quasiparticle (QP) corrections to the
eigenenergies along different symmetry lines in the BZ is
computationally very demanding. At variance with the DFT formalism,
where the eigenenergies at an arbitrary $\kk$ can be found starting
from the self-consistent charge density, the evaluation of the QP
corrections at a given $\kk$ requires a knowledge of the Kohn-Sham
eigenenergies and wavefunctions on a homogeneous grid of points
containing the wavevector of interest.  What is often done instead is
to perform the GW calculation at selected $\kk$-points only, and then
deduce a ``scissors correction,'' i.e., a constant shift to be applied
to the conduction-band Kohn-Sham eigenvalues elsewhere in the
Brillouin zone.

As already mentioned briefly in Sec.~\ref{sec:manybody},
\textcite{hamann-prb09} proposed using Wannier interpolation to
determine the GW QP bands very efficiently and accurately at arbitrary
points in the BZ. The wannierization and interpolation procedures are
identical to the DFT case. The only difference is that the starting
eigenenergies and overlaps matrices over the uniform first-principles
mesh are now calculated at the GW level.  (In the simplest $G_0W_0$
approximation, where only the eigenenergies, not the eigenfunctions,
are corrected, the wannierization is done at the DFT level, and the
resulting transformation matrices are then applied to the corrected QP
eigenenergies.)


\begin{figure}
\begin{center}
\includegraphics[width=3.2in]{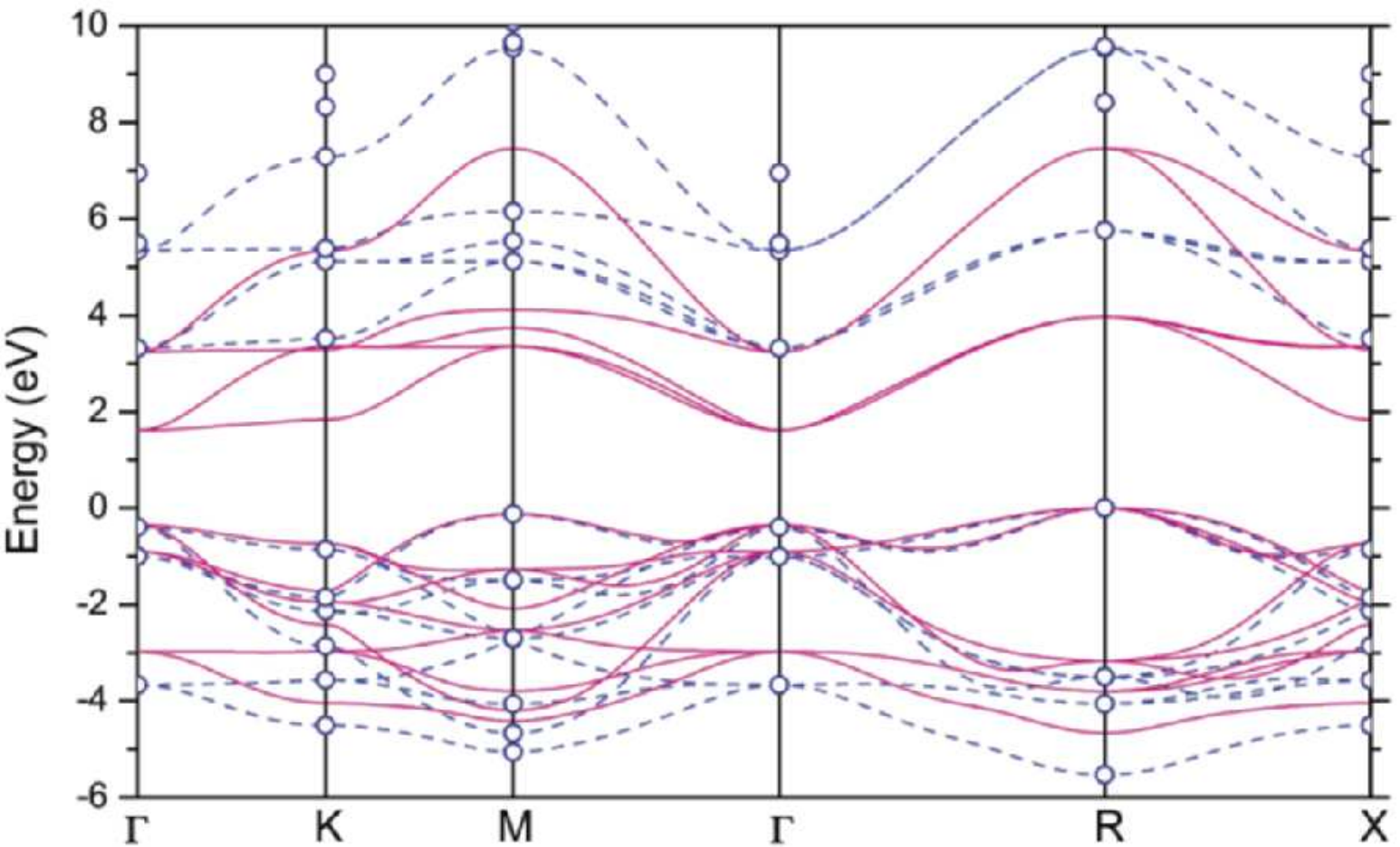}
\caption{(Color online) Wannier-interpolated upper valence and lower conduction bands of 
SrTiO$_3$ from LDA (solid red) and GW (dashed blue) calculations. The open
circles at the symmetry points denote the exact GW results taken directly
from the first-principles calculation. From \textcite{hamann-prb09}.}
\label{fig:srtio3-gw}
\end{center}
\end{figure}

Figure~\ref{fig:srtio3-gw} shows a comparison between the interpolated
GW (dashed lines) and DFT-LDA (solid lines) bands of
SrTiO$_3$~\cite{hamann-prb09}. Note that the dashed lines pass through
the open circles at the symmetry points, which denote exact (non-interpolated)
GW results.

Among the recent applications of the GW+Wannier method, we mention the
study of the energy bands of zircon and hafnon~\cite{shaltaf-prb09},
and a detailed comparative study between the DFT-LDA,
scissors-shifted, and QP $G_0W_0$ bands of Si and Ge
nanowires~\cite{peelaers-prb11}. In the latter study the authors found
that the simple scissors correction to the DFT-LDA bands is accurate
near the $\Gamma$ point only, and only for bands close to the highest
valence and lowest conduction band. \textcite{kioupakis-prb10} used
the method to elucidate the mechanisms responsible for free-carrier
absorption in GaN and in the In$_{0.25}$Ga$_{0.75}$N alloy.
\textcite{yazyev-prb12} investigated the quasiparticle effects on
the band structure of the topological insulators Bi$_2$Se$_3$ and
Bi$_2$Te$_3$, {and \textcite{aberg-arxiv12} studied in
detail the electronic structure of LaBr$_3$.}

\subsubsection{Surface bands of topological insulators}
\label{sec:surface-bands}

Topological insulators (TIs) were briefly discussed in
Sec.~\ref{sec:top} (see \textcite{hasan-rmp10,hasan-arcmp11} for
useful reviews).  Here we focus on the non-magnetic variety, the ${\cal
  Z}_2$-odd TIs. The recent flurry of activity on this class of
materials has been sustained in part by the experimental confirmation
of the ${\cal Z}_2$-odd character of certain quantum-well structures
and of a rapidly increasing number of bulk crystals.

In the case of 3D TIs, the clearest experimental signature of the
${\cal Z}_2$-odd character is at present provided by ARPES
measurements of the surface electron bands. If time-reversal symmetry
is preserved at the surface, ${\cal Z}_2$-odd materials possess
topologically-protected surface states which straddle the bulk gap,
crossing the Fermi level an odd number of times. These surface states
are doubly-degenerate at the time-reversal-invariant momenta of the
surface BZ, and in the vicinity thereof they disperse linearly,
forming Dirac cones.

First-principles calculations of the surface states for known and
candidate TI materials are obviously of great interest for comparing
with ARPES measurements.  While it is possible to carry out a direct
first-principles calculation for a thick slab in order to study the
topologically protected surface states, as done by
\textcite{yazyev-prl10}, such an approach is computationally
expensive. \textcite{zhang-njp10} used a simplified Wannier-based
approach which succeeds in capturing the essential features of the
topological surface states at a greatly reduced computational cost.
%
%
Their procedure is as follows.  First, an inexpensive calculation is
done for the bulk crystal without spin-orbit interaction.
Next, disentangled WFs spanning the upper valence and low-lying
conduction bands are generated, and the corresponding TB
Hamiltonian matrix is constructed.
The TB Hamiltonian is then augmented with spin-orbit couplings
$\lambda\bf L\cdot S$, where $\lambda$ is taken
from atomic data; this is possible because the WFs have been
constructed so as to have specified $p$-like characters.
The augmented TB parameters are then used to construct
sufficiently thick free-standing ``tight-binding slabs'' by a simple
truncation of the effective TB model, and the dispersion relation is
efficiently calculated by interpolation as a function of the
wavevector $\kk_\parallel$ in the surface BZ.

It should be noted that
this approach contains no surface-specific information, being based
exclusively on the bulk WFs. Even if its accuracy is questionable,
however, this method is useful for illustrating the ``topologically
protected'' surface states that arise as a manifestation of the bulk
electronic structure~\cite{hasan-rmp10}.

Instead of applying the naive truncation,
it is possible to refine the procedure so as to incorporate the
changes to the TB parameters near the surface. To do so, the bulk
calculation must now be complemented by a calculation on a thin slab,
again followed by wannierization. Upon aligning the on-site energies
in the interior of this slab to the bulk values, the changes to the TB
parameters near the surface can be inferred.  However,
\textcite{zhang-prl11b} found that the topological surface states are
essentially the same with and without this surface correction.

\begin{figure}
\begin{center}
\includegraphics[width=6.5cm]{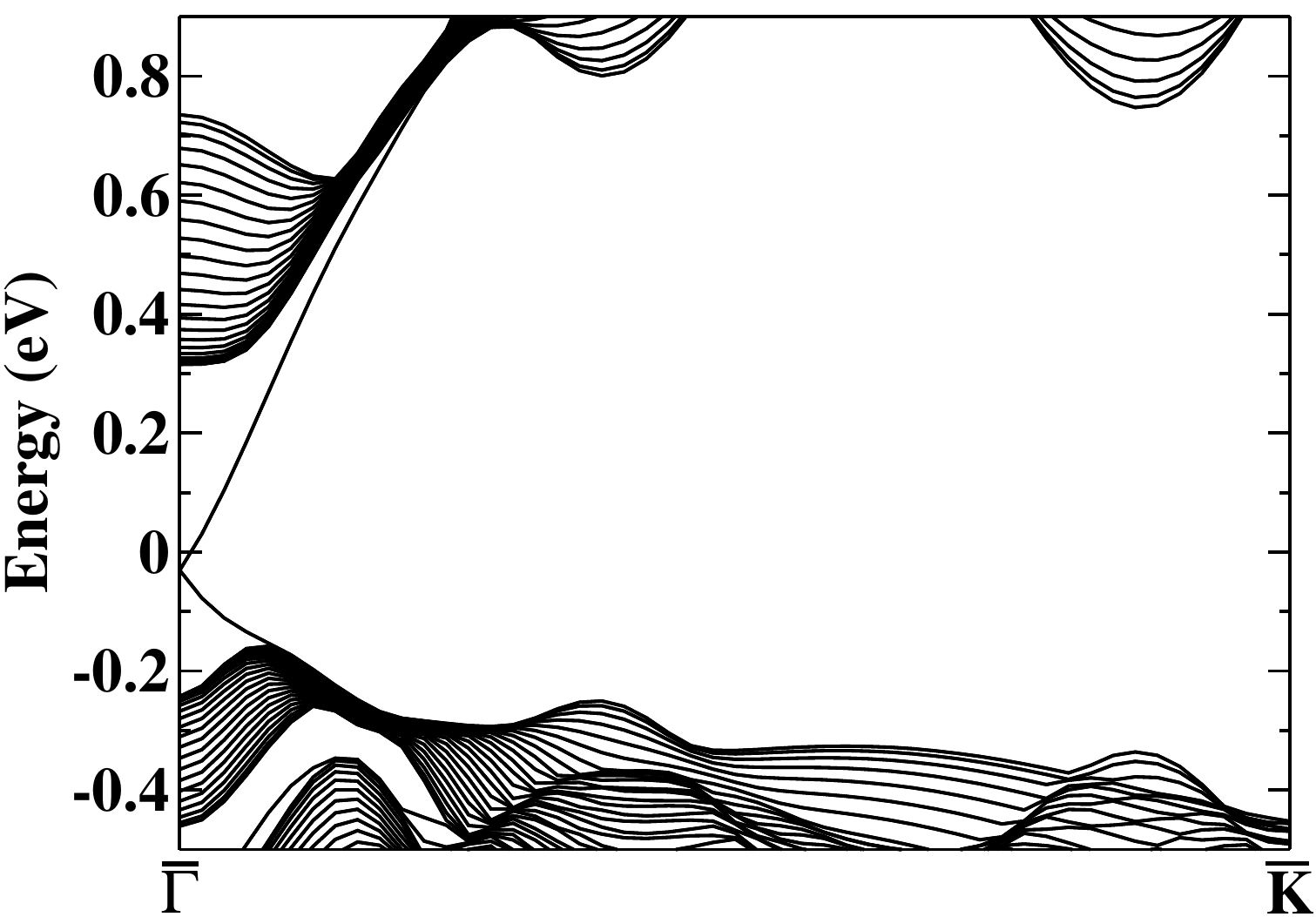}
\caption{Wannier-interpolated energy bands of a free-standing (111)
  slab containing 25 quintuple layers of
  Bi$_2$Se$_3$~\cite{zhang-njp10}, plotted along the
  $\overline{\Gamma}$--$\overline{\rm K}$ line in the surface BZ.  A
  pair of topologically-protected surface bands can be seen emerging
  from the dense set of projected valence and conduction bulk-like
  bands and crossing at the time-reversal-invariant point
  $\overline{\Gamma}$. Adapted from \textcite{zhang-njp10}.}
\label{fig:BiSe-slab}
\end{center}
\end{figure}

The truncated-slab approach was applied by \textcite{zhang-njp10} to
the stoichiometric three-dimensional TIs Sb$_2$Te$_3$, Bi$_2$Se$_3$,
and Bi$_2$Te$_3$.  The calculations on Bi$_2$Se$_3$ revealed the
existence of a single Dirac cone at the $\overline{\Gamma}$ point as
shown in Fig.~\ref{fig:BiSe-slab}, in agreement with ARPES
measurements~\cite{xia-np09}.

An alternative strategy for calculating the surface bands was used
earlier by the same authors~\cite{zhang-np09}. Instead of explicitly
diagonalizing the Wannier-based Hamiltonian $H(\kk_\parallel)$ of a
thick slab, the Green's function for the semi-infinite crystal as a
function of atomic plane is obtained via iterative
methods~\cite{lopezsancho-jpf84,lopezsancho-jpf85}, using the
approach of \textcite{lee-prl05}.  Here the
localized Wannier representation is used to break down the
semi-infinite crystal into a stack of ``principal layers'' consisting
of a number of atomic planes, such that only nearest-neighbor
interactions between principal layers exist (see Ch.~\ref{sec:basis}
for more details).

Within each principal layer one forms, starting from the
fully-localized WFs, a set of hybrid WFs
which are extended (Bloch-like) along the surface but remain localized
(Wannier-like) in the surface-normal direction (see
Secs.~\ref{sec:hybrid} and \ref{sec:layer}).  This is achieved by
carrying out a partial Bloch sum over the in-plane lattice vectors,
\beq 
\label{eq:hybrid-blochsum}
\ket{l,n \kk_\parallel}=\sum_{\RR_\parallel}\,
e^{i\kk_\parallel\cdot\RR_\parallel}\ket{\RR n}, 
\eeq
where $l$ labels the principal layer, $\kk_\parallel$ is the in-plane
wavevector, and $\RR_\parallel$ is the in-plane component of
$\RR$. The matrix elements of the Green's function in this basis are
\beq
G_{ll'}^{nn'}(\kk_\parallel,\epsilon)=\bra{\kk_\parallel ln}
\frac{1}{\epsilon-H}\ket{\kk_\parallel l'n'}.
\eeq
The nearest-neighbor coupling between principal layers means that for
each $\kk_\parallel$ the Hamiltonian has a block tri-diagonal form
(the dependence of the Hamiltonian matrix on $\kk_\parallel$ is given
by the usual Fourier sum expression).  This feature can be exploited
to calculate the diagonal elements of the Green's function matrix very
efficiently using iterative
schemes~\cite{lopezsancho-jpf84,lopezsancho-jpf85}.\footnote{ A
  pedagogical discussion, where a continued-fractions expansion is
  used to evaluate the Green's function of a semi-infinite linear
  chain with nearest-neighbor interactions, is given by
  \textcite{grosso-book00}.}
From these, the density of states (DOS) projected onto a given atomic
plane ${\cal P}$ can be obtained\cite{grosso-book00} as

\beq 
N_l^{\cal P}(\kk_\parallel,\epsilon)=-\frac{1}{\pi}\,{\rm
  Im}\sum_{n\in {\cal P}}\, G_{ll}^{nn}(\kk_\parallel,\epsilon+i\eta), 
\eeq
where the sum over $n$ is restricted to the orbitals ascribed to the
chosen plane and $\eta$ is a positive infinitesimal.

\begin{figure}
\begin{center}
\includegraphics[width=8.0cm]{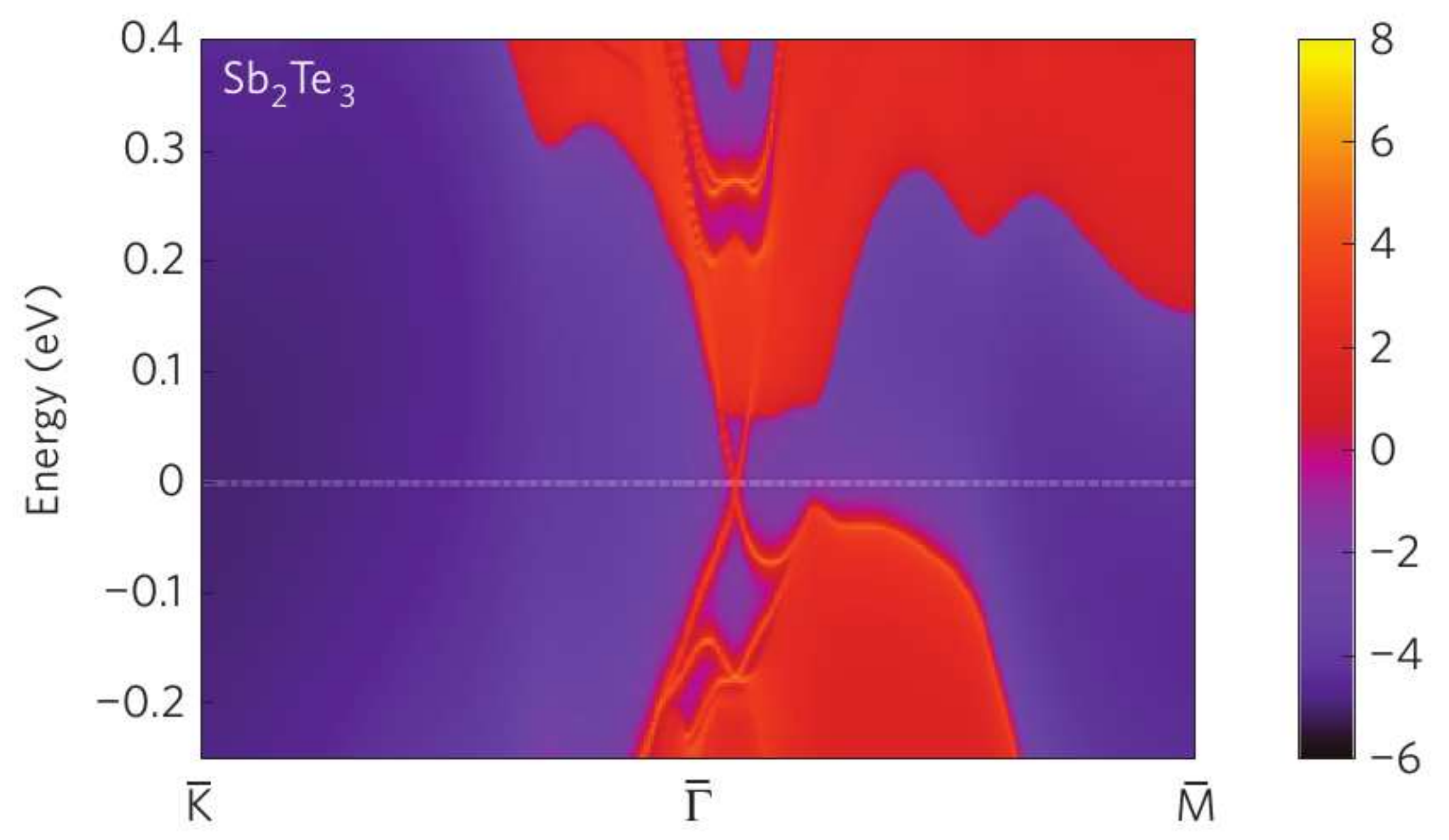}
\caption{(Color online) Surface density-of-states (SDOS) of a
  semi-infinite crystal of Sb$_2$Te$_3$ terminated with a [111]
  surface~\cite{zhang-np09}. Warmer colors represent a higher
  SDOS. The surface states can be clearly seen around
  $\overline{\Gamma}$ as red lines dispersing in the bulk gap.  From
  \textcite{zhang-np09}.}
\label{fig:SbTe-SDOS}
\end{center}
\end{figure}

The projection of the DOS onto the outermost atomic plane is shown in
Fig.~\ref{fig:SbTe-SDOS} as a function of energy $\epsilon$ and
momentum $\kk_\parallel$ for the (111) surface of Sb$_2$Te$_3$.  The
same method has been used to find the dispersion of the surface bands
in the TI alloy Bi$_{1-x}$Sb$_x$~\cite{zhang-prb09} and in ternary
compounds with a honeycomb lattice~\cite{zhang-prl11b}.

\subsection{Band derivatives}
\label{sec:band-der}

The first and second derivatives of the energy eigenvalues with
respect to $\kk$ (band velocities and inverse effective masses) appear
in a variety of contexts, such as the calculation of transport
coefficients~\cite{ashcroft-book76,grosso-book00}.  There is therefore
considerable interest in developing simple and accurate procedures for
extracting these parameters from a first-principles band structure
calculation.

A direct numerical differentiation of the eigenenergies calculated on
a grid is cumbersome and becomes unreliable near band crossings.  It
is also very expensive if a Brillouin zone integration is to be
carried out, as in transport calculations. A number of efficient
interpolation schemes, such as the method implemented in the {\sc
  BoltzTraP} package~\cite{madsen-ccp06}, have been developed for this
purpose, but they are still prone to numerical instabilities near band
degeneracies~\cite{uehara-prb00}.  Such instabilities can be avoided
by using a tight-binding parametrization to fit the first-principles
band structure~\cite{schulz-prb92,mazin-prb00}.  As shown by
\textcite{graf-prb92} and \textcite{boykin-prb95}, both the first and
the second derivatives are easily computed within tight-binding
methods, even in the presence of band degeneracies, and the same can
be done in a first-principles context using WFs.

Let us illustrate the procedure by calculating the band gradient away
from points of degeneracy; the treatment of degeneracies and second
derivatives is given in \textcite{yates-prb07}. The first step is to
take analytically the derivative $\partial_\alpha=\partial/\partial
k_\alpha$ of \equ{h_kw},
\beq
\label{eq:H_alpha}
H^\ww_{\kk,\alpha}
\equiv \partial_\alpha H^\ww_\kk
=\sum_\RR\,e^{i\kk\cdot\RR}iR_\alpha\bra{\0}H\ket{\RR}.
\eeq
The actual band gradients are given by the diagonal elements of the
rotated matrix,
\beq
\label{eq:band-grad}
\partial_\alpha\enkh=
\left[
  U_\kk^\dagger H_{\kk,\alpha}^\ww U_\kk
\right]_{nn}
\eeq
where $U_\kk$ is the same unitary matrix as in \equ{h_k-diag}.

It is instructive to view the columns of $U_\kk$ as orthonormal state
vectors in the $\nwann$-dimensional ``tight-binding space'' defined by
the WFs.  According to Eq.~(\ref{eq:h_k-diag}) the $n$-th column
vector, which we shall denote by $\vt\phi_{n\kk}\ra$, satisfies the
eigenvalue equation
$H_\kk^\ww\vt\phi_{n\kk}\ra=\enkh\vt\phi_{n\kk}\ra$.  Armed with this
insight, we now recognize in Eq.~(\ref{eq:band-grad}) the
Hellmann-Feynman result $\partial_\alpha\enkh= \la \phi_{n\kk}\vt
\partial_\alpha H_\kk^\ww
\vt \phi_{n\kk}\ra$.

\subsubsection{Application to transport coefficients
}
\label{sec:low-field-hall}

Within the semiclassical theory of
transport, the electrical and
thermal conductivities of metals and doped semiconductors can be
calculated from a knowledge of the band derivatives and relaxation
times $\tau_{n\kk}$ on the Fermi surface.  An example is the low-field
Hall conductivity $\sigma_{xy}$ of non-magnetic cubic metals, which in
the constant relaxation-time approximation is independent of $\tau$
and takes the form of a Fermi-surface integral containing the first and
second band derivatives~\cite{hurd-book72}.

Previous first-principles calculations of $\sigma_{xy}$ using various
interpolation schemes encountered difficulties for materials such as
Pd, where band crossings occur at the Fermi level~\cite{uehara-prb00}.
A Wannier-based calculation free from such numerical instabilities was
carried out by \textcite{yates-prb07}, who obtained
carefully-converged values for $\sigma_{xy}$ in Pd and other cubic
metals.

A more general formalism to calculate the electrical conductivity
tensor in the presence of a uniform magnetic field involves
integrating the equations of motion of a wavepacket under the field to
find its trajectory on the Fermi surface~\cite{ashcroft-book76}.  A
numerical implementation of this approach starting from the
Wannier-interpolated first-principles bands was carried out by
\textcite{liu-prb09}. This formalism is not restricted to cubic
crystals, and the authors used it to calculate the Hall conductivity
of hcp Mg~\cite{liu-prb09} and the magnetoconductivity of
MgB$_2$~\cite{yang-prl08}.

Wannier interpolation has also been used to determine the Seebeck
coefficient in hole-doped LaRhO$_3$ and CuRhO$_2$~\cite{usui-jpcm09},
in electron-doped SrTiO$_3$~\cite{usui-prb10}, in 
SiGe nanowires~\cite{shelley-epl11}, and in ternary 
skutterudites~\cite{volja-arxiv11}.

\subsection{Berry curvature and anomalous Hall conductivity}
\label{sec:curv-ahc}

The velocity matrix elements between Bloch eigenstates take the
form~\cite{blount-ssp62}
\beq
\label{eq:velocity}
\bra{\psi_{n\kk}}\hbar v_\alpha\ket{\psi_{m\kk}}=
\delta_{nm}
\partial_\alpha\enk
-i(\epsilon_{m\kk}-\enk)\left[A_{\kk,\alpha}\right]_{nm},
\eeq
where 
\beq
\label{eq:berry-connection-matrix}
\left[A_{\kk,\alpha}\right]_{nm}=
i\bra{u_{n\kk}}\partial_\alpha u_{m\kk}\rangle
\eeq
is the matrix generalization of the Berry
connection of \equ{berry-connection}.

In the examples discussed in the previous section the static transport
coefficients could be calculated from the first term in
\equ{velocity}, the intraband velocity.
The second term describes vertical interband transitions, which
dominate the optical spectrum of crystals over a wide frequency
range. Interestingly, under certain conditions, {\it virtual}
interband transitions also contribute to the dc Hall
conductivity. This so-called {\it anomalous Hall effect} occurs in
ferromagnets from the combination of exchange splitting and spin-orbit
interaction. For a recent review, see \textcite{nagaosa-rmp10}.

In the same way that WFs proved helpful for evaluating $\partial_\kk
\enk$, they can be useful for calculating quantities containing
$\kk$-derivatives of the cell-periodic Bloch states, such as the Berry
connection of \equ{berry-connection-matrix}. A number of properties
are naturally expressed in this form. In addition to the interband
optical conductivity and the anomalous Hall conductivity (AHC), other
examples include the electric polarization (Sec.~\ref{sec:pol-theo})
as well as the orbital magnetization and magnetoelectric coupling
(Sec.~\ref{sec:magnetism}).

Let us focus on the Berry curvature ${\bm{\mathcal{F}}}_{n\kk}$
[\equ{curv-pot}], a quantity with profound effects on the dynamics of
electrons in crystals~\cite{xiao-rmp10}.  ${\bm{\mathcal{F}}}_{n\kk}$
can be nonzero if either spatial inversion or time-reversal symmetries
are broken in the crystal, and when present acts as a kind of
``magnetic field'' in $\kk$-space, with the Berry connection ${\bf
  A}_{n\kk}$ playing the role of the vector potential. This effective
field gives rise to a Hall effect in ferromagnets even in the absence
of an actual applied ${\bf B}$-field (hence the name
``anomalous''). The 
AHC is given by\footnote{\equa{ahc} gives the so-called ``intrinsic''
  contribution to the AHC, while the measured effect also contains
  ``extrinsic'' contributions associated with scattering from
  impurities~\cite{nagaosa-rmp10}.}
\beq
\label{eq:ahc}
\sigma_{\alpha\beta}^{\rm AH}=-\frac{e^2}{\hbar}\int_{\rm BZ}
\frac{d\kk}{(2\pi)^3}\Omega^{\rm tot}_{\kk,\alpha\beta},
\eeq
where
$\Omega^{\rm tot}_{\kk,\alpha\beta}=\sum_n\,f_{n\kk}\Omega_{n\kk,\alpha\beta}$
($f_{n\kk}$ is the Fermi-Dirac distribution function), and we have
rewritten the pseudovector ${\bm{\mathcal{F}}}_{n\kk}$ as an
antisymmetric tensor. 

\begin{figure}
\begin{center}
\includegraphics[width=3.2in]{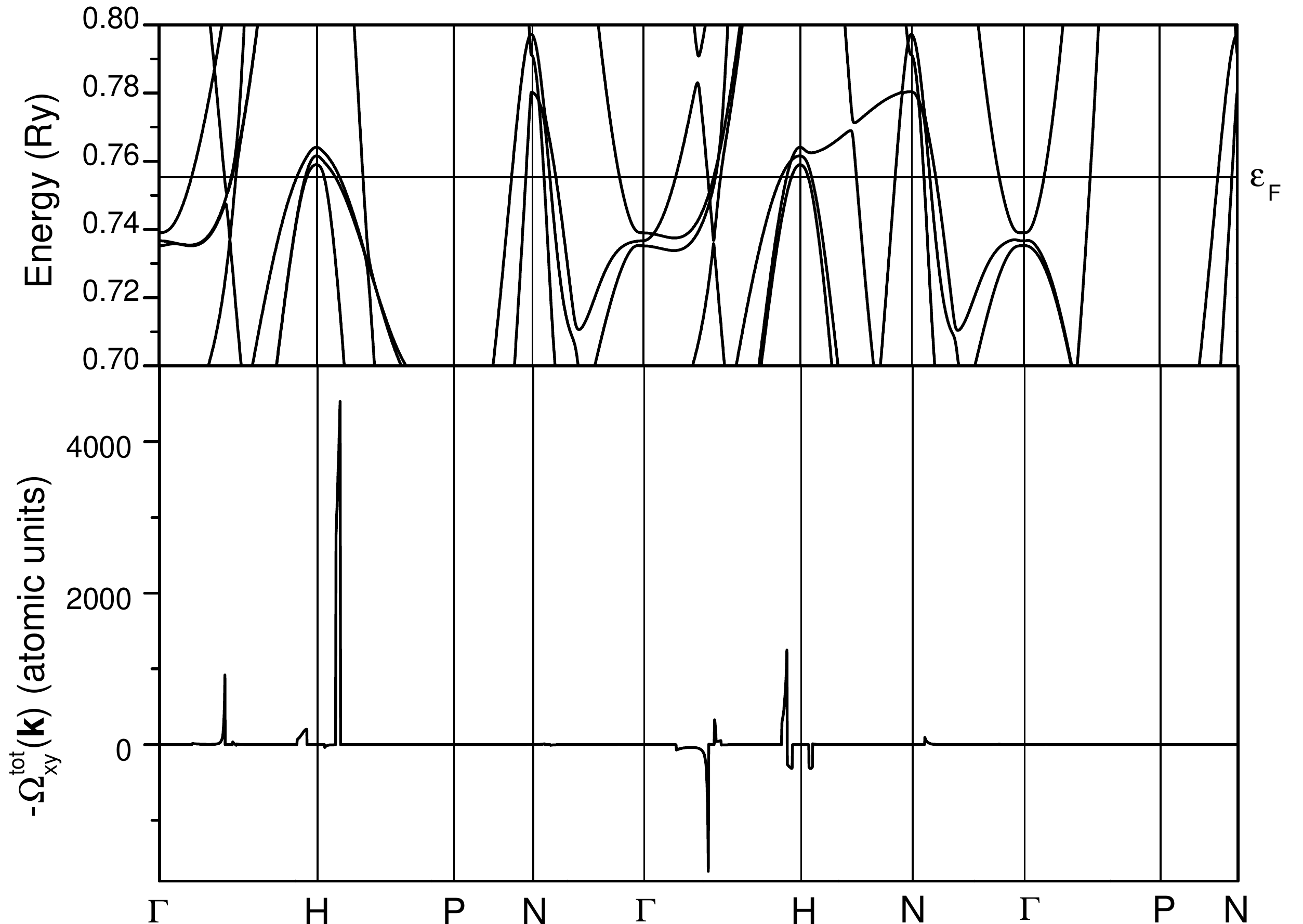}
\caption{Upper panel: band structure near the Fermi level of bcc Fe
  with the spontaneous magnetization along $\hat{\bf z}$. Lower panel:
  the Berry curvature summed over the occupied bands
  [\equ{berry-curv-sum}], plotted along the same symmetry lines. The
  sharp spikes occur when two spin-orbit-coupled bands are separated
  by a small energy across the Fermi level, producing a resonance
  enhancement. Adapted from \textcite{yao-prl04}.}
\label{fig:Fe-curv-spikes}
\end{center}
\end{figure}

The interband character of the intrinsic AHC can be seen by using
$\kk\cdot{\bf p}$ perturbation theory to write the $\kk$-derivatives in
\equ{curv-pot}, leading to the ``sum-over-states'' formula
\beq
\label{eq:berry-curv-sum}
\Omega^{\rm tot}_{\alpha\beta}=i\sum_{n,m}\,(f_n-f_m)
\frac{\bra{\psi_n}\hbar v_\alpha\ket{\psi_m}
      \bra{\psi_m}\hbar v_\beta\ket{\psi_n}
     }
     {(\epsilon_m-\epsilon_n)^2}.
\eeq
The AHC of bcc Fe and SrRuO$_3$ was evaluated from the previous two
equations by \textcite{yao-prl04} and \textcite{fang-science03}
respectively. These pioneering first-principles calculations revealed
that in ferromagnetic metals the Berry curvature displays strong and
rapid variations in $\kk$-space (Fig.~\ref{fig:Fe-curv-spikes}). As a
result, an ultradense BZ mesh containing millions of $\kk$-points is
often needed in order to converge the calculation.  This is the kind
of situation where the use of Wannier interpolation can be most
beneficial.

The strategy for interpolating the Berry curvature is similar to that
used in Sec.~\ref{sec:band-der} for the band gradient.  One first
evaluates certain objects in the Wannier gauge using Bloch sums, and
then transform to the Hamiltonian gauge. Because the gauge
transformation mixes the bands, it is convenient to introduce a
generalization of \equ{curv-pot} having two band indices instead of
one. To this end we start from \equ{berry-connection-matrix} and
define the matrices
\beq
\label{eq:berry-curv-mat}
\Omega_{\alpha\beta}=
\partial_\alpha A_\beta-\partial_\beta A_\alpha=
i\bra{\partial_\alpha u}\partial_\beta u\rangle-
i\bra{\partial_\beta u}\partial_\alpha u\rangle,
\eeq
where every object in this expression should consistently carry either
an H or W label.  Provided that the chosen WFs correctly span all
occupied states, the integrand of \equ{ahc} can now be expressed as
$\Omega^{\rm tot}_{\alpha\beta}=
\sum_{n=1}^\nwann\,f_n\Omega^\hh_{\alpha\beta,nn}$.

A useful expression for $\Omega^\hh_{\alpha\beta}$ can be obtained
with the help of the gauge-transformation law for the Bloch states,
$\ket{u_\kk^\hh}=\ket{u_\kk^\ww}U_{\kk}$
[\equ{psik-h}]. Differentiating both sides with respect to $k_\alpha$
and then inserting into Eq.~(\ref{eq:berry-curv-mat}) yields, after a
few manipulations,
\beq
\label{eq:omega-h}
\Omega^\hh_{\alpha\beta}=\overline{\Omega}_{\alpha\beta}
-\left[ D_\alpha,\overline{A}_\beta \right]
+\left[ D_\beta,\overline{A}_\alpha \right]
-i\left[ D_\alpha,D_\beta \right],
\eeq
where $D_\alpha=U^\dagger\partial_\alpha U$, and
$\overline{A}_\alpha$, $\overline{\Omega}_{\alpha\beta}$ are related
to the connection and curvature matrices in the Wannier gauge through
the definition $\overline{O}_\kk=U^\dagger_\kk O_\kk^\ww U_\kk$. Using
the band-diagonal elements of \equ{omega-h} in the expression for
$\Omega^{\rm tot}_{\alpha\beta}$ eventually leads to
\beq
\begin{split}
\label{eq:ahc-wanint}
\Omega^{\rm tot}_{\alpha\beta}=\sum_n^\nwann\, 
& f_n\overline{\Omega}_{\alpha\beta,nn}
+\sum_{mn}^\nwann\,(f_m-f_n)
\big(
        D_{\alpha,nm}\overline{A}_{\beta,mn} \\&
       -D_{\beta,nm}\overline{A}_{\alpha,mn}
       +iD_{\alpha,nm}D_{\beta,mn}
\big).
\end{split}
\eeq
This is the desired expression, which in the Wannier interpolation
scheme takes the place of the sum-over-states formula. In contrast to
\equ{berry-curv-sum}, note that the summations over bands now run over
the small set of Wannier-projected bands.
{(Alternatively, it is possible to recast
  \equ{ahc-wanint} in a manifestly gauge-invariant form such that the
  trace can be carried out directly in the Wannier gauge; this
  formulation was used by
  \textcite{lopez-prb12} to compute both the AHC and the orbital
  magnetization of ferromagnets.)}

The basic ingredients going into \equ{ahc-wanint} are the Wannier
matrix elements of the Hamiltonian and of the position operator. From
a knowledge of $\bra{\0}H\ket{\RR}$ the energy eigenvalues and
occupation factors, as well as the matrices $U$ and $D_\alpha$, can be
found using band-structure interpolation
(Sec.~\ref{sec:band-interpolation}). The information about
$A^\ww_\alpha$ and $\Omega^\ww_{\alpha\beta}$ is instead encoded in
the matrix elements $\bra{\0}\rr\ket{\RR}$, as can be seen by
inverting \equ{rmatel},
\beq
\label{eq:A-w}
A^\ww_{\alpha}
=\sum_\RR\,e^{i\kk\cdot\RR}\bra{\0}r_\alpha\ket{\RR}.
\eeq
As for $\Omega^\ww_{\alpha\beta}$, according to \equ{berry-curv-mat}
it is given by the curl of this expression, which can be taken
analytically.

The strategy outlined above was demonstrated by \textcite{wang-prb06}
in calculating the AHC of bcc Fe, using the spinor WFs of
Sec.~\ref{sec:fe-bands}. Both the $\kk$-space distribution of the Berry
curvature and the integrated AHC were found to be in excellent
agreement with the sum-over-states calculation of
\textcite{yao-prl04}.

Table~\ref{table:ahc} lists the AHC of the ferromagnetic transition
metal elements, calculated with the magnetization along the respective
easy axes. The magnetic anisotropy of the AHC was investigated by
\textcite{roman-prl09}.  While the AHC of the cubic metals Fe and Ni
is fairly isotropic, that of hcp Co was found to decrease by a factor
of four as the magnetization is rotated from the $c$-axis to the basal
plane.  The Wannier method has also been used to calculate the AHC in
FePt and FePd ordered alloys~\cite{seeman-prl10,zhang-prl11}, and the
spin-Hall conductivity in a number of metals~\cite{freimuth-prl10}.

\begin{table}
  \caption{
    Anomalous Hall conductivity in S/cm of the ferromagnetic
    transition metals, calculated from first-principles with the 
    magnetization along the respective easy axes.
    The first two rows show values obtained using the Wannier interpolation
    scheme
    to either integrate the Berry curvature over the Fermi 
    sea (see main text), or to
    evaluate the Berry phases of planar loops around the Fermi 
    surface~\cite{wang-prb07}. 
    Results obtained using the sum-over-states formula, \equ{berry-curv-sum},
    are included for comparison, as well as representative experimental
    values. Adapted from \textcite{wang-prb07}.}
\begin{ruledtabular}
\begin{tabular}{lrrr}
& bcc Fe & fcc Ni & hcp Co \cr
\hline
Berry curvature & 753\,\, & $-$2203\,\, & 477\,\, \cr
Berry phase & 750\,\, & $-$2275\,\, & 478\,\, \cr
Sum-over-states & 751\footnotemark[1]
    & $-$2073\footnotemark[2] & 492\footnotemark[2] \cr
Experiment& 1032  & $-$646 &480 \cr
\end{tabular}
\end{ruledtabular}
\footnotetext[1]{\textcite{yao-prl04}.}
\footnotetext[2]{Y. Yao, private communication.}
\label{table:ahc}
\end{table}

As already mentioned, for certain applications the Berry connection
matrix [\equ{berry-connection-matrix}] is the object of direct
interest. The interpolation procedure described above can be directly
applied to the off-diagonal elements describing vertical interband
transitions, and the magnetic circular dichroism spectrum of bcc Fe
has been determined in this way~\cite{yates-prb07}.
 
The treatment of the diagonal elements of the Berry connection matrix
is more subtle, as they are locally gauge-dependent. Nevertheless, the
Berry phase obtained by integrating over a closed loop in $\kk$-space,
$\varphi_n=\oint {\bf A}_{n\kk}\cdot d{\bf l}$, is
gauge-invariant~\cite{xiao-rmp10}.  Recalling that
${\bm{\mathcal{F}}}_{n\kk}={\nabla}_\kk\times {\bf A}_{n\kk}$
[\equ{curv-pot}] and using Stokes' theorem, \equ{ahc} for the AHC can
be recast in terms of the Berry phases of Fermi loops on planar slices
of the Fermi surface. This approach has been implemented by
\textcite{wang-prb07}, using Wannier interpolation to sample
efficiently the orbits with the very high density required near
band-crossings.  Table~\ref{table:ahc} lists values for the AHC
calculated using both the Berry curvature (``Fermi sea'') and
Berry-phase (``Fermi surface'') approaches. 

{It should be possible to devise similar Wannier interpolation
  strategies for other properties requiring dense BZ sampling, such as
  the magnetic shielding tensors of metals~\cite{davezac-prb07}. In
  the following we discuss electron-phonon coupling, for which
  Wannier-based methods have already proven to be of great utility.}

\subsection{Electron-phonon coupling}
\label{sec:el-ph}

The electron-phonon interaction~\cite{grimvall-book81} plays a key
role in a number of phenomena, from superconductivity to the
resistivity of metals and the temperature dependence of the optical
spectra of semiconductors.  The matrix element for scattering an
electron from state $\psi_{n\kk}$ to state $\psi_{m,\kk+\qq}$ while
absorbing a phonon $\qq\nu$ is proportional to the electron-phonon
vertex
\beq
\label{eq:g_kq}
g_{\nu,mn}(\kk,\qq)=\bra{\psi_{m,\kk+\qq}}\partial_{\qq\nu} V\ket{\psi_{n\kk}}.
\eeq
Here $\partial_{\qq\nu} V$ is the derivative of the self-consistent
potential with respect to the amplitude of the phonon with branch
index $\nu$ and momentum $\qq$.  Evaluating this vertex is a key task
for a first-principles treatment of electron-phonon couplings.

State-of-the-art calculations using first-principles linear-response
techniques~\cite{baroni-rmp01} have been successfully applied to a
number of problems, starting with the works of
  \textcite{savrasov-prl94} and \textcite{mauri-prl96b}, who used
respectively the LMTO and planewave pseudopotential methods.  The cost of evaluating
\equ{g_kq} from first-principles over a large number of
$(\kk,\qq)$-points is quite high, however, and this has placed a
serious limitation on the scope and accuracy of first-principles
techniques for electron-phonon problems.

The similarity between the Wannier interpolation of energy bands and
the Fourier interpolation of phonon dispersions was already noted.  It
suggests the possibility of interpolating the electron-phonon vertex
in both the electron and the phonon momenta, once \equ{g_kq} has been
calculated on a relatively coarse uniform $(\kk,\qq)$-mesh.  Different
electron-phonon interpolation schemes have been put forth in the
literature~\cite{giustino-prl07,eiguren-prb08,calandra-prb10} In the
following we describe the approach first developed by
\textcite{giustino-prb07} and implemented in the software package {\sc
  EPW}~\cite{noffsinger-cpc10}.
To begin, let us set the notation for lattice
dynamics~\cite{maradudin-rmp68}. We write the instantaneous nuclear
positions as
%
$\RR+{\boldsymbol\tau}_s+{\bf u}_{\RR s}(t)$,
%
where $\RR$ is the lattice vector, ${\boldsymbol\tau}_s$ is the
equilibrium intracell coordinate of ion $s=1,\ldots,p$, and ${\bf
  u}_{\RR s}(t)$ denotes the instantaneous displacement.  The normal
modes of vibration take the form
\beq
\label{eq:normal_mode}
{\bf u}^{\qq\nu}_{\RR s}(t)={\bf u}^{\qq\nu}_s
e^{i(\qq\cdot\RR-\omega_{\qq\nu}t)}.
\eeq
The eigenfrequencies $\omega_{\qq\nu}$ and mode amplitudes ${\bf
  u}^{\qq\nu}_s$ are obtained by diagonalizing the dynamical matrix
$\left[D_\qq^{\rm ph}\right]_{st}^{\alpha\beta}$, where $\alpha$ and
$\beta$ denote spatial directions.
It is expedient to introduce composite indices $\mu=(s,\alpha)$ and
$\nu=(t,\beta)$, and write $D^{\rm ph}_{\qq,\mu\nu}$ for the dynamical
matrix. With this notation, the eigenvalue equation becomes
\beq
\left[e_\qq^\dagger D^{\rm ph}_\qq e_\qq\right]_{\mu\nu}=
\delta_{\mu\nu}\omega^2_{\qq\nu},
\eeq
where $e_\qq$ is a $3p\times 3p$ unitary matrix. In analogy with the
tight-binding eigenvectors $\vt \phi_{n\kk}\ra$ of
Sec.~\ref{sec:band-der}, we can view the columns of $e_{\qq,\mu\nu}$
as orthonormal phonon eigenvectors ${\bf e}^{\qq\nu}_s$.  They are
related to the complex phonon amplitudes by ${\bf
  u}^{\qq\nu}_s=(m_0/m_s)^{1/2}{\bf e}^{\qq\nu}_s$ ($m_0$ is a
reference mass), which we write in matrix form as $U^{\rm
  ph}_{\qq,\mu\nu}$.

\begin{figure}
\begin{center}
\includegraphics[width=6.0cm]{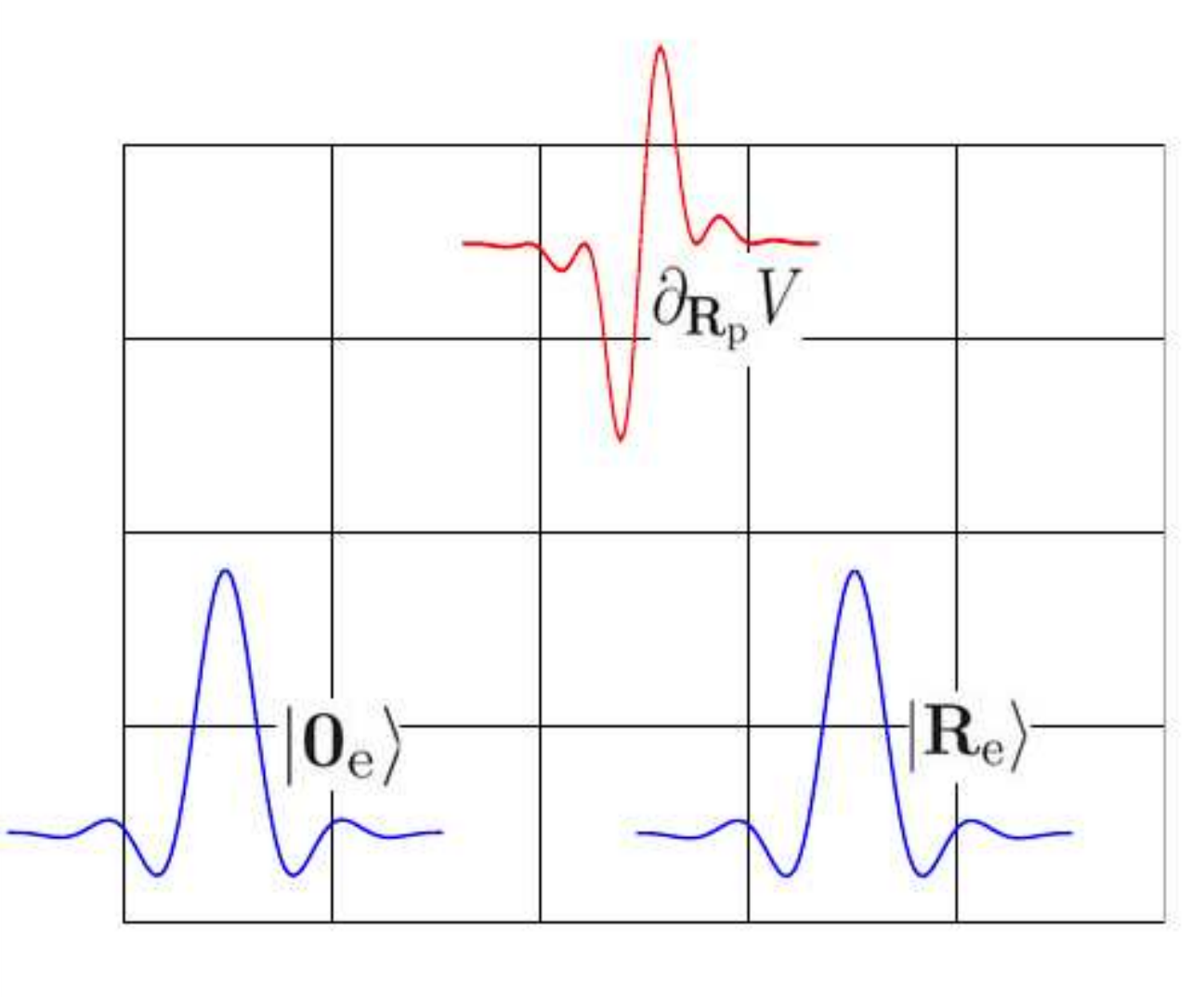}
\caption{(Color online) Real-space representation of
  $\bra{\0_e}\partial_{\RR_p\mu}V\ket{\RR_e}$, the electron-phonon
  vertex in the Wannier basis. The black squares denote the crystal
  lattice, the blue lines the electron Wannier functions $\ket{{\bf
      0}_e}$ and $\ket{\RR}_e$, and the red line the phonon
  perturbation in the lattice Wannier representation,
  $\partial_{\RR_p}V(\rr)$. Whenever two of these functions are
  centered on distant unit cells, the vertex becomes vanishingly
  small. From \textcite{giustino-prb07}.}
\label{fig:vertex-w}
\end{center}
\end{figure}

Returning to the electron-phonon vertex, \equ{g_kq}, we can now write
explicitly the quantity $\partial_{\qq\nu} V$ therein as
\beq
\label{eq:derV-q}
\begin{split}
\partial_{\qq\nu} V(\rr)&=\frac{\partial}{\partial\eta}
V(\rr;\{\RR+{\boldsymbol\tau}_s
+\eta{\bf u}^{\qq\nu}_{\RR s} \})\\
&=\sum_{\RR,\mu}\,e^{i\qq\cdot\RR}
\partial_{\RR\mu}V(\rr)
U^{\rm ph}_{\qq,\mu\nu},
\end{split}
\eeq
where $\partial_{\RR\mu}V(\rr)$ is the derivative of the
self-consistent potential with respect to $u_{\RR s,\alpha}$.  As will
be discussed in Sec.~\ref{sec:other}, it is possible to view these
single-atom displacements as maximally-localized ``lattice Wannier
functions.''  With this interpretation in mind we define the
Wannier-gauge counterpart of $\partial_{\qq\nu} V(\rr)$
as
\beq
\label{eq:g-w}
\partial^\ww_{\qq\mu}V(\rr)=\sum_\RR\,e^{i\qq\cdot\RR}
\partial_{\RR\mu}V(\rr),
\eeq
related to the ``eigenmode-gauge''  quantity $\partial_{\qq\nu} V(\rr)$ by
\beq
\label{eq:latt-gauge-transf}
\partial_{\qq\nu}V(\rr)=\sum_\mu\,\partial^\ww_{\qq\mu}V(\rr)
U^{\rm ph}_{\qq,\mu\nu}.
\eeq
Next we introduce the Wannier-gauge vertex $g^\ww_\mu(\kk,\qq)=
\bra{\psi^\ww_{\kk+\qq}}\partial^\ww_{\qq\mu} V\ket{\psi^\ww_{\kk}}$,
which can be readily interpolated onto an arbitrary point
$(\kk',\qq')$ using Eqs.~(\ref{eq:bloch-w-gauge}) and (\ref{eq:g-w}),
\beq
\label{eq:g-intp-w}
g^\ww_\mu(\kk',\qq')=\sum_{\RR_e,\RR_p}\,
e^{i(\kk'\cdot\RR_e+\qq'\cdot\RR_p)}
\bra{\0_e}\partial_{\RR_p\mu}V\ket{\RR_e}.
\eeq
The object $\bra{\0_e}\partial_{\RR_p\mu}V\ket{\RR_e}$, the
electron-phonon vertex in the Wannier representation, is depicted
schematically in Fig.~\ref{fig:vertex-w}. Its localization in
real-space ensures that \equ{g-intp-w} smoothly interpolates $g^\ww$
in the electron and phonon momenta. Finally, we transform the
interpolated vertex back to the Hamiltonian/eigenmode gauge,
\beq
\label{eq:g-intp-h}
\begin{split}
g^\hh_\nu(\kk',\qq')&=
\bra{\psi^\hh_{\kk'+\qq'}}\partial_{\qq'\nu} V\ket{\psi^\hh_{\kk'}}\\
&=U^\dagger_{\kk'+\qq'}
\Big[
  \sum_\mu g^\ww_\mu(\kk',\qq')
  U^{\rm ph}_{\qq',\mu\nu}
\Big]
U_{\kk'},
\end{split}
\eeq
where we used Eqs.~(\ref{eq:psik-h}) and (\ref{eq:latt-gauge-transf}).

Once the matrix elements $\bra{\0_e}\partial_{\RR_p\mu}V\ket{\RR_e}$
are known, the electron-phonon vertex can be evaluated at arbitrary
$(\kk',\qq')$ from the previous two equations.  The procedure for
evaluating those matrix elements is similar to that leading up to
\equ{ham-R} for $\bra{\0}H\ket{\RR}$: invert \equ{g-intp-w} over the
first-principles mesh, and then use Eqs.~(\ref{eq:psik-w}) and
(\ref{eq:latt-gauge-transf}).  
%
%

The above interpolation scheme has been applied to a number of
problems, including the estimation of $T_c$ in
superconductors~\cite{giustino-prl07,noffsinger-prb08,noffsinger-prb09};
the phonon renormalization of energy bands near the Fermi level in
graphene~\cite{park-prl07} and copper oxide
superconductors~\cite{giustino-n08}; the phonon renormalization of the
band gap of diamond~\cite{giustino-prl10}; the vibrational
lifetimes~\cite{park-nl08} and electron linewidths~\cite{park-prl09}
arising from electron-phonon interactions in graphene; {and the
  phonon-assisted optical absorption in
  silicon~\cite{noffsinger-prl12} (in this last application both the
  velocity and the electron-phonon matrix elements were treated by
  Wannier interpolation).}

We mention in closing the work of \textcite{calandra-prb10}, where the
linear-response calculation of the deformation potential
$\partial_{\qq\nu} V$ is carried out taking into account nonadiabatic
effects (that is, going beyond the usual approximation of static ionic
displacements). Using the electron-phonon interpolation scheme of
\textcite{giustino-prb07} to perform the BZ integrations, and a
Wannier interpolation of the dynamical matrix to capture Kohn anomalies, 
the authors
found significant nonadiabatic corrections to the phonon frequencies
and electron-phonon coupling matrix elements in MgB$_2$ and CaC$_6$.

\section{WANNIER FUNCTIONS AS BASIS FUNCTIONS}
\label{sec:basis}


In Ch.~\ref{sec:winterp}, we described the use of Wannier functions as
a compact tight-binding basis that represents a given set of energy
bands exactly, and that can be used to calculate a variety of
properties efficiently, and with very high accuracy. This is possible
because (a) WFs and bands are related by unitary transformations, and
(b) WFs are sufficiently localized that any resulting tight-binding
representation may be truncated with little loss of accuracy. In this
Chapter, we review two further general approaches to the use of WFs as
optimal and compact basis functions for electronic-structure
calculations that exploit their localization and transferability.

The first category includes methods in which WFs are used to go up in the
length scale of the simulations, using the results of electronic-structure 
calculations on small systems in order to construct accurate
models of larger, often meso-scale, systems. Examples include using
WFs to construct tight-binding Hamiltonians for large, structurally
complex nanostructures (in particular for studying quantum transport
properties), to parametrize semi-empirical force-fields, and to
improve the system-size scaling of quantum Monte Carlo (QMC)
and GW calculations, and the evaluation of exact-exchange integrals.

The second category includes methods in which WFs are used to
identify and focus on a desired, physically
relevant subspace of the electronic degrees of freedom that is singled
out (``downfolded'') for further detailed analysis or special treatment with a more
accurate level of electronic structure theory, an approach that is
particularly suited to the study of strongly-correlated electron
systems, such as materials containing transition metal, lanthanoid, or
even actinoid ions.

\subsection{WFs as a basis for large-scale calculations}
\label{sec:wfs-large-scale}

Some of the first works on linear-scaling electronic structure algorithms
\cite{yang-prl91,baroni-el92,galli-prl92,hierse-prb94} highlighted the connection 
between locality in electronic structure, which underpins linear-scaling algorithms
(Sec.~\ref{sec:linear-scaling}), and the
transferability of information across length-scales. In particular,
\textcite{hierse-prb94} 
considered explicitly two large systems A and B, different globally but which
have a certain similar local feature, such as a particular chemical
functionalization and its associated local environment, which we will
call C. They argued that the local electronic structure information
associated with C should be similar whether calculated from system A
or system B and, therefore,
that it should be possible to transfer this information from a
calculation on A in order to construct a very good
approximation to the electronic structure in the locality of feature C
in system B. In this way, large computational savings could be made on
the self-consistency cycle, enabling larger-scale calculations. 

The units of electronic structure information that
\textcite{hierse-prb94} used were localized non-orthogonal orbitals,
optimized in order to variationally minimize the total energy 
of the system. These orbitals are also referred to in the literature as ``support
functions''~\cite{hernandez-prb95} or ``non-orthogonal generalized
Wannier functions''~\cite{skylaris-prb02}. 

\subsubsection{{MLWFs as electronic-structure building blocks}}
\label{sec:building-blocks}

{Since MLWFs encode chemically accurate, local (and thus transferable)
information, they can act as building blocks to construct the electronic structure 
of very large-scale systems~\cite{lee-prl05}. 
In this approach the Hamiltonian matrix of a
large nanostructure for which a full, conventional DFT calculation would be intractable, is
built using first-principles calculations performed on smaller, typically periodic fragments.
The matrix elements in the basis of MLWFs that are obtained from
the calculations on the fragments can be used to construct the entire 
Hamiltonian matrix of the desired system,
with the size of the fragments a
systematically controllable determinant of the accuracy of the
method~\cite{shelley-cpc11}. 
Such an approach has been
applied to complex systems containing tens of thousands of
atoms~\cite{lee-prl06,cantele-nl09,shelley-cpc11,shelley-epl11,li-acn11,li-acn11b}. }


An issue that arises when combining matrix elements from more than one 
DFT calculation into a single tight-binding Hamiltonian is that a common
reference potential must be defined.
For example, consider combining the results of a calculation on a 
perfect bulk material and one on the same material with an isolated 
structural defect. In the latter case, the diagonal (on-site) matrix
elements $\matel{w_n}{H}{w_n}$
in the system with the defect should converge to the same value as 
those in the pristine system as one goes further away from the location
of the defect.
With periodic boundary conditions, however, this is not guaranteed:
the difference between the matrix elements in the respective calculations
will, in general, tend to a non-zero constant, this being the difference
in reference potential between the two calculations.
In other words, the difference in value between the on-site matrix elements from 
the bulk calculation and 
those from the calculation with the defect, but far away from the location 
of the defect, give a direct measure of the potential offset between the 
two calculations. This offset can then be used to align all of the diagonal
matrix elements from one calculation with those of the other, prior to 
them being combined in a tight-binding Hamiltonian for a larger nanostructure.
The potential alignment approach described above has been used 
for transport calculations (see next subsection), and for the
determination of band offsets and Schottky 
barriers~\cite{singhmiller-phd09},
for calculating formation energies of point
defects~\cite{corsetti-prb11}, and for developing tight-binding models of 
the surfaces of topological insulators~\cite{zhang-njp10} (see 
Sec.~\ref{sec:surface-bands}).

{
Last, it should be noted that charge self-consistency could play an important role
when trying to build the electronic structure of large or complex nanostructures, and one 
might have to resort to more sophisticated approaches. 
As a suggestion, electrostatic consistency could be
attained solving the Poisson equation for the entire system~\cite{leonard-prl99}, 
then using the updated electrostatic potential to shift appropriately
the diagonal elements of the Hamiltonian. In a more general fashion, the 
electronic charge density and the Hartree and exchange-correlation potentials 
could be updated and optimized self-consistently in a basis of disentangled, frozen
MLWFs, using a generalized occupation matrix~\cite{marzari-prl97}.
}

\subsubsection{Quantum transport}

A local representation of electronic structure is particularly
suited to the study of quantum transport, as we illustrate here in the case of
quasi-one-dimensional systems.
\begin{figure}
\begin{center}
\includegraphics[width=3.2in]{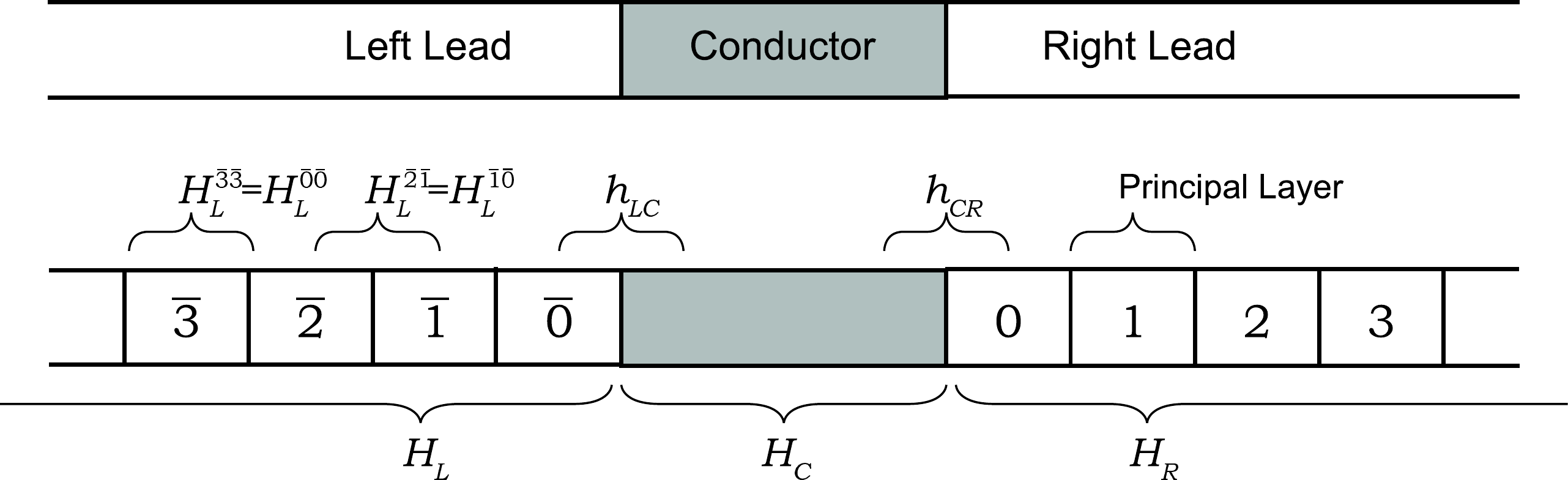}
\caption{Top: Illustration of the lead-conductor-lead geometry
  used in quantum transport calculations. The conductor ($C$), left lead
  ($L$) and right lead ($R$) are described by Hamiltonians $H_C$, $H_L$
  and $H_R$, respectively. The coupling between adjacent regions is
  described by matrices $h_{LC}$ and $h_{CR}$. Bottom: The leads are
  split into principal layers corresponding to the sub-matrices in
  \equ{hamloc}. From \textcite{shelley-cpc11}.}
\label{fig:LCR}
\end{center}
\end{figure}
We consider a system composed of a conductor $C$ connected to two
semi-infinite leads, $R$ and $L$, as shown in Fig.~\ref{fig:LCR}. 
The conductance ${\cal G}$ through
a region of interacting electrons is related to the scattering
properties of the region itself via the Landauer
formula~\cite{landauer-pm70} 
\beq
{\cal G}(E) = \frac{2 e^2}{h} {\cal T}(E),
\label{eq:qc}
\eeq
where the transmission function ${\cal T}(E)$ is the probability that
an electron with energy $E$ injected at one end of the conductor will 
transmit to the other end. 
This transmission function can be expressed in terms of the Green's 
functions of the conductors and the coupling of the conductor to the
leads \cite{datta-95,fisher-prb91},
\beq
{\cal T}(E) = {\rm Tr}(\Gamma_{L} G_{C}^{r} \Gamma_{R} G_{C}^{a}),
\eeq
where $G_{C}^{\{r,a\}}$ are the retarded (r) and advanced (a) Green's
functions of the conductor, and $\Gamma_{\{L,R\}}$ are functions that
describe the coupling of the conductor to the left ($L$) and right ($R$)
leads. Since $G^{a}=(G^{r})^{\dagger}$, we consider $G^{r}$ only and drop
the superscript.

Expressing the Hamiltonian $H$ of the system in terms of a localized,
real-space basis set enables it to be partitioned without ambiguity
into sub-matrices that correspond to the individual subsystems. 
A concept that is particularly useful is that of a \emph{principal
  layer}~\cite{lee-prb81} (PL), which is a section of lead that is
sufficiently long such that $\matel{\chi_i^n}{H}{\chi_j^m}\simeq
0$ if $|m-n| \ge 2$, where $H$ is the Hamiltonian operator of
the entire system and $\ket{\chi_i^n}$ is the $i^{\rm th}$ basis
function in the $n^{\rm th}$ PL. Truncating the matrix elements of the
Hamiltonian in this way 
incurs a small error which is systematically controlled by increasing
the size of the PL. The Hamiltonian matrix in this basis then takes
the block diagonal form (see also Fig.~\ref{fig:LCR})
\beq
H=\left( 
\begin{array}{ccccccc}
\ddots & \vdots & \vdots & \vdots & \vdots & \vdots & \iddots \\
\cdots & H_L^{\bar{0}\bar{0}} & H_L^{\bar{1}\bar{0}} & 0 & 0 & 0 &
\cdots \\
\cdots & H_L^{\bar{1}\bar{0}\dag}	& H_L^{\bar{0}\bar{0}}& h_{LC}
& 0 & 0 & \cdots \\
\cdots & 0 & h_{LC}^\dag & H_C & h_{CR} & 0 & \cdots \\
\cdots & 0 & 0 & h_{CR}^{\dag}	& H_R^{00} & H_R^{01} & \cdots \\
\cdots & 0 & 0 & 0 & H_R^{01\dag} & H_R^{00} & \cdots \\
\iddots & \vdots & \vdots & \vdots & \vdots &\vdots & \ddots \\
\end{array}\right),
\label{eq:hamloc}
\eeq
where $H_C$ represents the Hamiltonian matrix of the
conductor region, $H_L^{\bar{0}\bar{0}}$ and $H_R^{00}$ 
are those of each PL of the left and right leads, respectively,
$H_L^{\bar{1}\bar{0}}$ and $H_R^{01}$ are couplings between adjacent
PLs of lead, and $h_{LC}$
and $h_{CR}$ give the coupling between the conductor and the leads.

In order to compute the Green's function of the conductor one starts
from the equation satisfied by the Green's function $G$ of the whole
system, 
\beq
(\epsilon - H)G = \mathbb{I}
\label{eq:green}
\eeq
where $\mathbb{I}$ is the identity matrix, and $\epsilon = E + i\eta$,  where
$\eta$ is an arbitrarily small, real constant.

From \eqs{green}{hamloc}, it can be shown that
the Green's function of the conductor is then given by~\cite{datta-95}
\beq
G_{C} = (\epsilon -H_{C} -\Sigma_{L} -\Sigma_{R})^{-1} ,
\label{eq:gconduct}
\eeq
where we define $\Sigma_{L} = h_{LC}^\dagger g_{L} h_{LC}$ and $\Sigma_{R}
= h_{RC} g_{R} h_{RC}^\dagger$, the self-energy terms due to the
semi-infinite leads, and $g_{\{L,R\}}=(\epsilon-H_{\{L,R\}})^{-1}$,
the surface Green's functions of the leads. 

The self-energy terms can be viewed as effective Hamiltonians that
arise from the coupling of the conductor with the leads. Once the
Green's functions of the leads are known, the coupling functions
$\Gamma_{\{L,R\}}$ can be easily obtained as~\cite{datta-95} 
\beq
\Gamma_{\{L,R\}} = i [\Sigma_{\{L,R\}}^{r} - \Sigma_{\{L,R\}}^{a}] ,
\eeq
where $\Sigma_{\{L,R\}}^{a}=(\Sigma_{\{L,R\}}^{r})^{\dagger}$. 

As for the surface Green's functions $g_{\{L,R\}}$ of the
semi-infinite leads, it is well-known that any solid (or surface) can
be viewed as an infinite (semi-infinite in the case of surfaces) stack
of principal layers with nearest-neighbor
interactions~\cite{lee-prb81}. This corresponds to transforming the
original system into a linear chain of PLs. Within this
approach, the matrix elements of \equ{green} between layer orbitals
will yield a set of coupled equations for the Green's functions which
can be solved using an efficient iterative scheme due to
\textcite{lopezsancho-jpf84,lopezsancho-jpf85}. Knowledge of the
finite Hamiltonian sub-matrices in \equ{hamloc}, therefore, is
sufficient to calculate the conductance of the open
lead-conductor-lead system given by \equ{qc}.


There are a number of possibilities for the choice of localized basis
$\ket{\chi}$. Early work used model tight-binding
Hamiltonians~\cite{chico-prb96, saito-prb96, anantram-prb98,
  nardelli-prb99}, but the increasing sophistication of computational
methods meant that more realistic first-principles approaches could be
adopted~\cite{nardelli-prb01,fattebert-book03}.  Maximally-localized
Wannier functions were first used in this context by
\textcite{calzolari-prb04}, who studied Al and C chains and a (5,0)
carbon nanotube with a single Si substitutional defect, and by
\textcite{lee-prl05}, who studied covalent functionalizations of
metallic nanotubes - capabilities now encoded in the open-source
packages Wannier90 \cite{mostofi-cpc08} and WanT \cite{want}. This was
quickly followed by a number of applications to ever more realistic
systems, studying transport through molecular
junctions~\cite{thygesen-cp05, strange-jcp08}, decorated carbon
nanotubes and nanoribbons~\cite{lee-prl06, cantele-nl09, rasuli-prb10,
  li-acn11}, organic monolayers~\cite{bonferroni-n08}, and silicon
nanowires~\cite{shelley-epl11}, as well as more methodological work on
improving the description of electron correlations within this
formalism~\cite{calzolari-n07,bonferroni-n08,
  ferretti-prl05,ferretti-prb05}.

The formulation described above relies on a localized description of 
the electronic-structure problem, and it should be noted that several 
approaches to calculating electronic transport properties have been developed 
using localized basis sets rather than MLWFs, ranging from Gaussians~\cite{hod-jcp06} to 
numerical atomic orbitals~\cite{brandbyge-prb02, markussen-prb06, dasilva-prl08}.

{In addition, in the Keldysh formalism one can add
more complex interaction terms to the self-energies, such as
electron-phonon or (for molecular conductors) electron-vibration
interactions~\cite{frederiksen-prb07}. These latter can also be
conveniently expressed in a MLWFs representation, and a natural extension
of the previous quantum-transport formalism to the case of inelastic
scattering channels due to electron-vibration interactions has recently
been developed in a MLWFs basis~\cite{kim-arxiv12}.
}

\subsubsection{Semi-empirical potentials}

First-principles molecular dynamics simulations of large-scale
(thousands of atoms) systems for long (nanoseconds) time-scales are
computationally costly, if not intractable. Molecular dynamics
simulations with empirical interatomic potentials are a feasible
alternative and there is an ongoing effort in developing potentials
that are more accurate, more transferable and, therefore, more
predictive.
One approach in this direction is to fit the parameters that appear
within empirical potentials so that they reproduce target properties,
such as forces and stresses, obtained from accurate DFT
calculations on a large number of atomic 
configurations~\cite{ercolessi-el94}. 
In the particular class of ionic condensed matter systems, e.g.,
first and second row metal oxides, it is well-known that the
electronic properties of an ion can be significantly affected by its
coordination environment and, therefore, that it is also important to 
include an accurate description of polarization effects in any
interatomic potential. 
While simple potentials may attempt to account for these many-body
effects in an average manner, the result is that the potential loses
transferability and, hence, predictive power. 
As a result, there has been some effort in developing potentials such
that they are also fitted to information from DFT calculations 
regarding the electron distribution around each ion, in particular,
dipoles and quadrupoles. 

In this vein, \textcite{aguado-fd03} introduced the use of MLWFs in
order to calculate dipole and quadrupole moments in MgO and used these
to construct an interatomic potential based on the aspherical ion
method (AIM) potential~\cite{rowley-jcp98}. In an ionic system such as
this, MLWFs are found to be well-localized close to the ions and each can
therefore be associated unambiguously with a particular ion. 
The dipole moment ${\bf \mu}^{I}$ of each ion $I$ is
then calculated as 
\beq
\mu^{I}_{\alpha} = -2 \sum_{n\in I} \bar{r}_{n\alpha} + Z_I R^{I}_{\alpha},
\eeq
where $\alpha$ is a Cartesian component, $Z_I$ is the charge of ion
$I$ at position $\RR_I$, $\bar{r}_{n\alpha}$ is the center of
the $n^{\mathrm{th}}$ MLWF and is given by
\equ{r} or \equ{rcenter}, and the factor of two accounts for spin degeneracy.

For the quadrupole moments $\theta^{I}_{\alpha\beta}$, the fact that
the MLWFs are localized within the simulation cell is exploited in
order to explicitly compute the real-space integral
\beq
\theta^{I}_{\alpha\beta} = -2\sum_{n\in I} \int_{V^{I}_{c}} 
\! d\rr \; \left|w_{n}(\rr)\right|^{2} \left[ 3
  r^{I}_{\alpha}r^{I}_{\beta} - (r^{I})^{2}\delta_{\alpha\beta}\right]/2,
\eeq
where $\rr^{I}=\rr - \RR^{I}$, and the integral is performed over a
sphere $V^{I}_{c}$, centered on $R^{I}$, such that the integral of the
electron density associated with the MLWF within this sphere is greater
than some threshold. The potential obtained
exhibits good transferability and the method has been used
to parametrize similar potentials for other alkaline earth
oxides~\cite{aguado-jcp03}, Al$_2$O$_3$~\cite{jahn-prb06}, and the
CaO-MgO-Al$_2$O$_3$-SiO$_2$ (CMAS) mineral
system~\cite{jahn-poteapi07}. 

The use of MLWFs for attempting to obtain better interatomic potentials
has not been limited to the 
solid state. In biomolecular simulations, an important factor
in developing accurate force-fields is having an accurate description
of the electrostatic potential. Starting from DFT calculations on
isolated molecules, \textcite{sagui-jcp04} 
partition the electronic charge density into contributions from
individual MLWFs and calculate the multipoles of each using an
order-by-order expansion of gauge-invariant
cumulants~\cite{resta-prl98, souza-prb00b} (the reader is referred to
\textcite{sagui-jcp04} for full details). Using fast particle mesh
Ewald and multigrid methods, these multipoles can then be 
used to generate the electrostatic potential. \textcite{sagui-jcp04} show that higher
order multipoles, e.g., up to hexadecapole, may be incorporated without
computational or numerical difficulty and that agreement with the
`exact' potential obtained from DFT is very good. 
{
The idea of partitioning the charge density according to individual
MLWFs was also employed by \textcite{kirchner-jcp04} in order to
determine atomic charges in dimethyl sulfoxide,
showing that there can be significant deviations between the gaseous and
aqueous forms and, therefore, underlining the importance of using
polarizable force-fields for describing solvated systems.
}

Finally, we note that recently
\textcite{rotenberg-prl10} have proposed force-fields whose
parametrization is based entirely on a partitioning of the
electronic density in terms of MLWFs. Their method has been applied
successfully to water and molten salts. It
remains to be seen whether the approach is extensible to more complex
or anisotropic systems. 
%

\subsubsection{Improving system-size scaling}

The localized nature of MLWFs in real-space makes them a natural and
appealing choice of basis for electronic structure calculations as the
sparsity exhibited by operators that are 
represented in a localized basis may
be exploited in order to achieve computational efficiencies and
improved scaling of numerical algorithms with respect to system
size. Recently, therefore, MLWFs have been used for this very purpose in a
number of contexts and we mention, in brief, some of them here.

In quantum Monte Carlo (QMC) calculations~\cite{foulkes-rmp01}, a
significant computational 
effort is expended in evaluating the Slater determinant for a given
electronic configuration of $N$ electrons. These Slater determinants are usually
constructed from a set of extended single-particle states, obtained
from, e.g., a DFT or Hartree-Fock
calculation, represented in a basis set of, e.g., extended
plane-waves. This gives rise to $\mathcal{O}(N^3)$ scaling of the
computational cost of evaluating such a determinant.
\textcite{williamson-prl01} suggested, instead, to use
MLWFs that were smoothly truncated to zero beyond a certain cut-off
radius that is independent of system size. This ensures that each
electron falls only within the localization region of a fixed number
of MLWFs, thus reducing the asymptotic scaling by one factor of
$N$. Furthermore, by representing the MLWFs in a basis of localized spline
functions, rather than plane-waves or even Gaussian functions, the
evaluation of each orbital is rendered independent of system size,
thereby reducing the overall cost of computing the determinant of the
given configuration to $\mathcal{O}(N)$. 
More recently, rather than truncated MLWFs, the use of non-orthogonal
orbitals obtained by projection~\cite{reboredo-prb05} or other
localization criteria~\cite{alfe-jpcm04} has also been suggested.

In another development, \textcite{wu-prb09} use MLWFs in order to
compute efficiently Hartree-Fock exact-exchange integrals in extended
systems. Hybrid exchange-and-correlation
functionals~\cite{becke-jcp93} for DFT calculations, in which some
proportion of Hartree-Fock exchange is included in order to alleviate
the well-known problem of self-interaction that exists in local and semi-local
functionals such as the local-density approximation and its
generalized gradient-dependent variants, have been used relatively
little in extended systems.
This is predominantly due to the computational
cost associated with evaluating the exchange integrals between
extended eigenstates that are represented in a plane-wave basis
set. \textcite{wu-prb09} show that by performing a unitary
transformation of the eigenstates to a basis of MLWFs, and working in
real-space in order to exploit the fact that spatially distant MLWFs
have exponentially vanishing overlap, the number of such overlaps that
needs to be calculated scales linearly, in the limit of large system-size, 
with the number of orbitals (as opposed to
quadratically), which is a sufficient improvement to enable Car-Parrinello
molecular dynamics simulations with hybrid functionals.

Similar ideas that exploit the locality of
MLWFs have been applied  to many-body perturbation theory approaches
for going beyond DFT and Hartree-Fock calculations, for example, in the
contexts of the GW approximation~\cite{umari-prb09}, the evaluation
of local correlation in extended systems~\cite{buth-prb05,
  pisani-jcp05}, and the Bethe-Salpeter
equation~\cite{sasioglu-prb10}. The improved scaling and efficiency of
these approaches open the possibility of such calculations on larger
systems than previously accessible. 

{
Finally, we note that MLWFs have been used recently to compute van der
Waals (vdW) interactions in an approximate but efficient
manner~\cite{silvestrelli-prl08, silvestrelli-jpca09,
  andrinopoulos-jcp11}. 
The method is based on an expression due to
\textcite{andersson-prl96} for the vdW energy in terms of pairwise
interactions between fragments of charge density. MLWFs provide 
a localized decomposition of the electronic charge density of a system
and can be used as the basis for computing the vdW contribution to the
total energy in a post-processing (i.e., non-self-consistent)
fashion. In order to render tractable the necessary multi-dimensional 
integrals, the
true MLWFs of the system are substituted by analytic hydrogenic
orbitals that have the same centers and spreads as the true MLWFs. The 
approach has been applied to a variety of systems, including molecular
dimers, molecules physisorbed on surfaces, water clusters and
weakly-bound solids~\cite{silvestrelli-prl08, silvestrelli-jpca09,
  silvestrelli-jcp09, silvestrelli-cpl09, andrinopoulos-jcp11,
  espejo-cpc12}. Recently, \textcite{ambrosetti-prb12} have suggested
an alternative, simpler formulation that is based on London's
expression for the van der Waals energy of two interacting
atoms~\cite{london-zp30}.
}


\subsection{WFs as a basis for strongly-correlated systems}
\label{sec:wfs-min-models}

For many strongly-correlated electron problems, the essential physics
of the system can be explained by considering only a subset of the
electronic states. A recent example is understanding the behavior
of unconventional (high-$T_{\rm c}$) superconductors, in which
a great deal of insight can be gained by considering only the MLWFs of
$p$ and $d$ character on Cu and O, respectively, for
cuprates~\cite{sakakibara-prl10}, and those on As and Fe,
respectively, for  the iron pnictides~\cite{kuroki-prl08,
  cao-prb08, hu-jpcc10, suzuki-jpsj11}. 
Other strongly-correlated materials for which MLWFs have been used to
construct minimal models to help understand the physics include
manganites~\cite{kovacik-prb10}, topological
insulators~\cite{zhang-np09, zhang-prb09} (see also
Sec.~\ref{sec:surface-bands}), and polyphenylene vinylene (PPV), in
particular relating to electron-hole excitations~\cite{karabunarliev-jcp03a,
  karabunarliev-jcp03b}.  

Below we outline some of the general principles behind the
construction and solution of such minimal models. 

\subsubsection{First-principles model Hamiltonians}
\label{sec:mod-ham}
Consider a strongly-correlated electron system described by a 
Hamiltonian of the form
\beq
\label{eq:scham1}
\hat{H} = \hat{H}_{0} + \hat{H}_{\rm int},
\eeq
where $\hat{H}_{0}$ contains the one-particle terms and $\hat{H}_{\rm
  int}$ the interaction (e.g., Coulomb repulsion) terms. 
In second-quantized notation and expressed in terms of a complete
tight-binding basis, this may be expressed as
\beq
\begin{split}
\hat{H} =& \sum_{ij}\sum_{\RR,\RR'} h_{ij}^{\RR\RR'}
            (\hat{c}_{i\RR}^{\dagger} \hat{c}_{j\RR'} + {\rm H.c.})
            \\& + \frac{1}{2}\sum_{ijkl} \sum_{\RR\RR'\RR''\RR'''}
            U_{ijkl}^{\RR\RR'\RR''\RR'''}
            \hat{c}_{i\RR}^{\dagger} \hat{c}_{j\RR'}^{\dagger}
            \hat{c}_{k\RR''}\hat{c}_{l\RR'''},  
\label{eq:scham2}
\end{split}
\eeq
where $\RR$ labels the correlated site, lower-case indices (such as
$i$ and $j$) refer to the orbital and spin degrees of freedom,
$\hat{c}_{i\RR}^{\dagger}$ ($\hat{c}_{i\RR}$) creates (annihilates) an
electron in the orbital $w_{i\RR}(\rr)$, and ${\bf h}$ and ${\bf U}$
contain the matrix elements of the single-particle and (screened)
interaction parts of the Hamiltonian, respectively.

Usually, a complete tight-binding representation of all the states of the
system is not required, and the essential physics can be described by
a smaller set of physically relevant orbitals. For example, those
states close to the Fermi level, or those of a particularly symmetry,
or those localized on specific sites, may be sufficient. In this way,
the size of the basis used to represent the many-body Hamiltonian is
greatly reduced, rendering \equ{scham2} more amenable to solution
(see, e.g., \textcite{solovyev-jpcm08}).

In this spirit, MLWFs obtained from DFT calculations have been used as
the orbital basis for these minimal models.\footnote{An alternative
  approach is to obtain the orbitals via the downfolding method,
  discussed in Ch.~\ref{sec:locorb}.}
Advantages of using MLWFs include the fact that they can be
chosen to span precisely the required energy range (using the
disentanglement procedure outlined in Sec.~\ref{sec:disentanglement}),
and that they naturally incorporate hybridization 
and bonding appropriate to their local environment.

The single-particle hopping parameters of the model Hamiltonian are
obtained easily from the matrix elements of the DFT Hamiltonian
represented in the basis of MLWFs, using \equ{ham-R}.
{
The interaction parameters of the model Hamiltonian can be calculated,
for example, from either constrained DFT~\cite{dederichs-prl84,anisimov-prb91,
nakamura-prb06,mcmahan-prb90}, within the random phase
approximation~\cite{springer-prb98, aryasetiawan-prb04,
  solovyev-prb05, miyake-prb09},
}
or by direct calculation of the matrix elements of
a suitable screened Coulomb interaction between, for example,
MLWFs~\cite{miyake-prb06, nakamura-prb06}. It is interesting to note
that numerical evidence suggests that on-site Coulomb interactions
(both screened and bare), are maximized when calculated with a basis
of MLWFs~\cite{miyake-prb08} and, therefore, that MLWFs may be an
optimally localized basis for this purpose. This is perhaps not
surprising given the broad similarities between MLWFs and WFs obtained
via the Edmiston-Ruedenberg localization scheme~\cite{edmiston-rmp63},
discussed in Sec.~\ref{sec:alt-loc}, which maximizes the electronic 
Coulomb self-interaction of each orbital.

Once the parameters of the model have been determined, the model
Hamiltonian is then said to be ``from first-principles'', in the sense
that the parameters of the model are determined from DFT rather than
by fitting to experiments. 
The many-body Hamiltonian in the minimal basis of MLWFs may be then
solved by direct diagonalization, or one of a number of other
approaches that are too numerous to review here but which include, for
example, the dynamical mean-field theory (DMFT) approach. 
DMFT maps
the many-body problem on to an Anderson impurity 
model~\cite{anderson-pr61} in which on-site correlation is treated
non-perturbatively and the correlated sites are coupled to a
self-consistent energy bath that represents the rest of the system.
The impurity sites, also known as the ``correlated subspaces'', are
defined by localized projector functions and MLWFs are a common
choice~\cite{lechermann-prb06, trimarchi-jpcm08, weber-prb10}. In
particular, one would typically choose orbitals of $d$ or $f$
character associated with transition metal, lanthanoid or actinoid ions.
The Green's function for the impurity site is calculated
self-consistently, for example, by a numerical functional integration
(which constitutes the bulk of the computation). Further
self-consistency with the DFT ground-state may also be attained by
using the solution to the impurity problem to update the 
electronic density that is then used to construct an updated Kohn-Sham
potential, which determines a new set of eigenstates, MLWFs and,
hence, model Hamiltonian parameters that can then be fed back in to
the DMFT cycle. The reader is referred to \textcite{kotliar-romp06}
and \textcite{held-aip07} for further details; other examples that use
localized Wannier functions or generalized tight-binding models to
address correlated electrons problems can be found in  
\textcite{amadon-prb08, anisimov-prb05a, held-jpcm08, korotin-epjb08,
  korshunov-prb05, ku-prl02}.

\subsubsection{{Self-interaction and DFT + Hubbard $U$}}

{
In the approaches just described, the results of a DFT
calculation are used to parametrize the model Hamiltonian of a
strongly correlated electron system. In contrast, in a DFT+$U$ 
formulation~\cite{anisimov-prb91, anisimov-prb93}
the energy functional is explicitly augmented with a 
Hubbard term $U$~\cite{hubbard-prsa63}
aimed at improving the description of strong interactions,
such as those associated with localized $d$ and $f$ electronic states,
and at repairing the mean-field underestimation of on-site Coulomb repulsions. 
}

{In DFT+$U$ the Hubbard manifold
is defined by a set of projectors that are typically
atomic-like orbitals of $d$ or $f$ character.
Localization of this manifold plays a key role, since 
DFT+$U$ effectively corrects for 
the lack of piecewise linearity in approximate energy 
functionals~\cite{perdew-prl82,cococcioni-prb05}
and thus greatly reduces self-interaction 
errors~\cite{perdew-prb81,kulik-prl06,morisanchez-prl08}.
Since strongly localized orbitals are those that suffer most from
self-interaction, MLWFs can become an appealing choice to define
Hubbard manifolds adapted to the chemistry of the
local environment. In fact, MLWFs have been successfully used
as Hubbard projectors~\cite{fabris-prb05, miyake-prb08, anisimov-jpcm07}, 
and it has been argued that
their shape can constitute an additional degree of
freedom in the calculations~\cite{oregan-prb10}, provided 
their localized, atomic character is maintained. 
It should also be pointed out that the value of 
$U$ entering the calculations should not be considered 
universal, as it depends strongly on the manifold chosen
(e.g. for pseudopotential calculations on the oxidation state of the reference 
atomic calculation~\cite{kulik-jcp08}), or on the structure or electronic-structure
of the problem studied.
}

{
Last, it should be pointed out that functionals that aim to correct
directly for some effects of self-interaction -- such as the Perdew-Zunger
correction~\cite{perdew-prb81} or Koopmans-compliant
functionals~\cite{dabo-prb10} -- can lead naturally in a periodic
system to Wannier-like localized orbitals that minimize the total 
energy~\cite{stengel-prb08,park-arxiv11}, while the canonical orbitals 
that diagonalize the Hamiltonian still preserve Bloch periodicity.
}

\section{WANNIER FUNCTIONS IN OTHER CONTEXTS}
\label{sec:other}

As described in Sec.~\ref{sec:bloch-wannier}, Wannier
functions provide an alternative, localized, description of a
manifold of states spanned by the eigenstates (energy bands) of a
single-particle Hamiltonian that describes electrons in a periodic
potential.
The equivalence of the Wannier representation and the
eigenstate representation may be expressed in terms of the
band projection operator $\hat{P}$, see \equ{mbandproj}.
This operator satisfies the idempotency condition
$P^2=P$, which embodies simultaneously the
requirements of orthogonality and Pauli exclusion.

From their conception, and until relatively recently, Wannier
functions have been used almost exclusively in this context, namely to
represent a manifold of single-particle orbitals for
electrons. Furthermore, as discussed in Sec.~\ref{sec:theory}, we need
not restrict ourselves to an isolated group of states, such as the
occupied manifold: the disentanglement procedure enables a subspace of
a larger manifold, e.g., of occupied and unoccupied states, to be
selected which may then be wannierized. This has, for example,
opened up areas of application in which Wannier functions are used as
tight-binding basis functions for electronic structure and transport
calculations, as described in Sec.~\ref{sec:winterp} and
Sec.~\ref{sec:basis}.

From a general mathematical point of view, however, the set of
orthogonal eigenfunctions of any self-adjoint (Hermitian) operator may
be ``rotated'' by unitary transformation to another orthogonal 
basis that spans the same space.
As we have seen, the unitary transformation is arbitrary and may be
chosen to render the new basis set maximally-localized, which has
computational advantages when it comes to representing physical
quantities that are short-ranged. When the operator in question
has translational symmetry, the maximally-localized functions thus
obtained are reminiscent of the Wannier functions familiar from
electronic structure theory. Often, such a basis is also preferable to
using another localized basis because information regarding the
symmetries of the self-adjoint operator from which the basis is
derived is encoded within it.

These ideas have led to the appropriation of the MLWF formalism
described in Sec.~\ref{sec:theory} for contexts other than the
description of electrons: the single-particle electronic Hamiltonian
is replaced by another suitable periodic self-adjoint operator, and
the Bloch eigenstates are replaced by the eigenfunctions of that
operator, which may then be transformed to give a MLWF-like
representation that may be used as an optimal and compact basis for
the desired calculation, for example, to analyze the eigenmodes of the
static dielectric matrix in ice and liquid water~\cite{lu-prl08}.

Below we review the three most prominent of these applications, namely
to the study of phonons, photonic crystals, and cold atom lattices.

\subsection{Phonons}
\label{sec:phonons}

Lattice vibrations in periodic crystals are usually described in
terms of normal modes, which constitute a delocalized orthonormal
basis for the space of vibrations of the lattice such that an arbitrary
displacement of the atoms in the crystal may be expressed in terms of
a linear combination of its normal modes.
By analogy with the electronic case, \textcite{kohn-prb73} first
showed (for isolated phonon branches in one dimension) that a similar
approach could be used for constructing a localized orthonormal basis
for lattice vibrations that span the same space as the delocalized
normal modes. The approach was subsequently generalized to isolated
manifolds in three-dimensions by
\textcite{tindemansvaneijndhoven-jpcsp75}. The localized modes are now
generally referred to as \emph{lattice Wannier functions}
(LWFs)~\cite{rabe-prb95, iniguez-prb00}.

Following the notation of Sec.~\ref{sec:el-ph}, we denote by 
$\qq$ the phonon wavevector, and by $e_\qq$ the matrix whose 
columns are the eigenvectors of the dynamical matrix. As in
case of electronic Wannier functions, the phases of these eigenvectors
are undetermined. A unitary transformation of the form
\beq
\left[ \tilde{e}_{\qq} \right]_{\mu\nu} 
 = \left[ M_{\qq} e_{\qq}\right]_{\mu\nu},
\label{eq:lwf-gauge}
\eeq
performed within a subspace of dispersion branches that is
invariant with respect to the space group of the crystal, results in
an equivalent representation of generalized extended modes
$\left[\tilde{e}_{\qq}\right]_{\mu\nu}$ that are also orthonormal. 
LWFs may then be defined by
\beq
\left[w_{\RR}\right]_{\mu\nu} = \frac{1}{N_p} \sum_{\qq}
e^{-i\qq\cdot\RR} \left[\tilde{e}_{\qq}\right]_{\mu\nu}, 
\label{eq:lwf-transform}
\eeq
with the associated inverse transform
\beq
\left[\tilde{e}_{\qq}\right]_{\mu\nu} = \sum_{\RR} e^{i\qq\cdot\RR}
\left[w_{\RR}\right]_{\mu\nu}.
\eeq
By construction, the LWFs are periodic according to $w_{\RR+\bf{t}} =
w_{\RR}$, where $\bf{t}$ is a translation vector of the Born-von
K\'{a}rm\'{a}n supercell.

The freedom inherent in \equ{lwf-gauge} allows very localized LWFs to
be constructed, by suitable choice of the transformation matrix $M_{\qq}$. 
As noted by \textcite{kohn-prb73}, the proof of exponential
localization of LWFs follows the same reasoning as for electronic
Wannier functions (Sec.~\ref{sec:exp-loc}).

The formal existence of LWFs was first invoked in order to justify
the construction of approximate so-called \emph{local modes} of
vibration which were used in effective Hamiltonians for the study of
systems exhibiting strong coupling between electronic states and
lattice instabilities, such as perovskite
ferroelectrics~\cite{thomas-prl68, pytte-pr69}. 

\textcite{zhong-prl94} used first-principles methods in order to
calculate the eigenvector associated with a soft mode at $\qq=\0$ in
BaTiO$_3$. A localized displacement pattern, or local mode, of the
atoms in the cell was then parametrized, taking account of the
symmetries associated with the soft mode, and the parameters were
fitted to reproduce the calculated soft mode eigenvector at
$\qq=\0$. The degree of localization of the local mode was determined
by setting to zero all displacement parameters beyond the second shell
of atoms surrounding the central atom. Although this spatial
truncation results in the local modes being non-orthogonal, it does
not hamper the accuracy of practical calculations. As the local modes
are constructed using information only from the eigenvector at
$\qq=\0$, they do not correspond to a particular phonon branch in the
Brillouin zone.
\textcite{rabe-prb95} generalized the approach to allow fitting to
more than just $\qq=\0$, but rather to a small set of, usually
high-symmetry, $\qq$-points. The phase-indeterminacy of the
eigenvectors is exploited in order to achieve optimally rapid decay of
the local modes. 
Another approach, introduced by \textcite{iniguez-prb00}, constructs
local modes via a projection method that preserves the correct
symmetry. The procedure is initiated from simple atomic displacements
as trial functions. The quality of the local modes thus obtained may
be improved by systematically densifying the $\qq$-point mesh that is
used in \equ{lwf-transform}. Although there is no formal criterion of
maximal-localization in the approach, it also results in
non-orthogonal local modes that decay exponentially.

These ideas for generating local modes from first-principles
calculations have been particularly successful for the study of
structural phase transitions in ferroelectrics such as
BaTiO$_3$~\cite{zhong-prl94, zhong-prb95},
PbTiO$_3$~\cite{waghmare-prb97}, KbNiO3~\cite{krakauer-jpm99},
Pb$_3$GeTe$_4$~\cite{cockayne-prb97} and perovskite
superlattices~\cite{lee-jap08}.

The use of maximal-localization as an exclusive criterion for
determining LWFs was first introduced by \textcite{giustino-prl06}. In
this work, a real-space periodic position operator for non-interacting
phonons was defined, by analogy with the periodic position operator for
non-interacting electrons (\equ{rcenter}). 

The problem of minimizing the total spread of a set of WFs in
real-space is equivalent to the problem of simultaneously 
diagonalizing the three non-commuting matrices corresponding to the
three components of the position operator represented in the WF
basis, and \textcite{giustino-prl06} use the method outlined by
\textcite{gygi-cpc03} to achieve this. It is worth noting that
\textcite{giustino-prl06} furthermore define a generalized spread
functional that, with a single parameter, allows a trade-off between
localization in energy (the eigenstate or Bloch limit) and
localization in space (the Wannier limit), resulting in so-called
\emph{mixed Wannier-Bloch functions} which may be obtained for
the electrons as well as the phonons.

Finally, as first pointed out by
\textcite{kohn-prb73}, and subsequently by \textcite{giustino-prb07},
maximally-localized lattice Wannier functions correspond to
displacements of individual atoms. This may be
seen by considering a vibrational eigenmode, $\hat{\bf e}^{\nu}_{\qq
s} \equiv {\bf e}^{\nu}_{\qq s} e^{i\qq\cdot\RR}$, and noting that it
may be expressed as 
\beq
\label{eq:eigmode}
\hat{\bf e}^{\nu}_{\qq s} = 
    \sum_{s'\RR'}\,e^{i\qq\cdot\RR'}
\delta_{\RR\RR'}\delta_{s s'} 
    {\bf e}^{\nu}_{\qq s'}.
\eeq
\equ{eigmode} stands in direct correspondence to the electronic
analogue given by the inverse of \equ{wannier}, from which we conclude
that the LWFs do indeed correspond to individual atomic displacements
$\delta_{\RR\RR'}\delta_{ss'}$ and, furthermore, that the required
unitary transformation is the matrix of eigenvectors
$\left[e_{\qq}\right]_{\mu\nu}$.
As discussed in Sec.~\ref{sec:el-ph}, \textcite{giustino-prb07}
exploit this property for the efficient interpolation of dynamical
matrices and calculation of electron-phonon couplings.

\subsection{Photonic crystals}
\label{sec:phot-crys}

Photonic crystals are periodic arrangements of dielectric materials
that are designed and fabricated in order to control the flow of
light~\cite{john-prl87, yablonovitch-prl87}. They are very much to
light what semiconductors are to electrons and, like semiconductors
that exhibit an electronic band gap in which an electron may not
propagate, photonic crystals can be engineered to exhibit photonic
band gaps: ranges of frequencies in which light is forbidden to
propagate in the crystal. In the electronic case, a band gap results
from scattering from the periodic potential due to the ions in the
crystal; in the photonic case, it arises from scattering from the
periodic dielectric interfaces of the crystal. Again by analogy with
electronic materials, localized defect states can arise in the gap by
the deliberate introduction of defects into a perfect photonic crystal
structure. The ability to control the nature of these states promises
to lead to entirely light-based integrated circuits, which would have
a number of advantages over their electronic counterparts, including
greater speeds of propagation, greater bandwidth, and smaller energy
losses~\cite{joannopoulos-n97}.

In SI units, Maxwell's equations in source-free regions of space are
\begin{align}
&\nabla\cdot{\bf E} = 0,& &\nabla\cdot\BB = 0,& \\
&\nabla\times{\bf E} = -\frac{\partial\BB}{\partial t},&
&\nabla\times\HH = \frac{\partial\DD}{\partial t},& 
\end{align}
where the constitutive relations between the fields are
\beq
\DD = \eps_{\rm r}\eps_0{\bf E}, \quad
\BB = \mu_{\rm r}\mu_0\HH.
\eeq
Considering non-magnetic materials ($\mu_{\rm r}=1$) with a position
dependent dielectric constant $\eps_{\rm r}(\rr)$, and fields that
vary with a sinusoidal dependence $e^{-i\omega t}$, it is
straightforward to derive electromagnetic wave equations in terms of
either the electric field ${\bf E}$ or the magnetic field $\HH$, 
\bea
\nabla\times\left(\nabla\times{\bf E}(\rr)\right) &=&
      \frac{\omega^2}{c^2}\eps_{\rm r}(\rr){\bf E}(\rr) ,
      \label{eq:wave-eqn-E}\\
\nabla\times\left(\eps^{-1}_{\rm r}(\rr) \nabla\times\HH(\rr)\right) &=&
      \frac{\omega^2}{c^2}\HH(\rr), 
      \label{eq:wave-eqn-H}
\eea
where $c=(\mu_0\eps_0)^{-1/2}$ is the speed of light.

For a perfect periodic dielectric structure, $\eps_{\rm
r}(\rr)=\eps_{\rm r}(\rr + \RR)$, where $\RR$ is a lattice
vector. Application of Bloch's theorem leads to solutions that are
indexed by wavevector $\kk$, which may be chosen to lie in the first
Brillouin zone, and a band index $n$. For example 
$$
\HH_{\bnk}(\rr)=e^{i\kk\cdot\rr}{\bf u}_{\bnk}(\rr) ,
$$
where ${\bf u}_{\bnk}(\rr)={\bf u}_{\bnk}(\rr+\RR)$ is the periodic
part of the magnetic field Bloch function. The electromagnetic wave
equations can be solved, and hence the Bloch functions obtained, by a
number of methods including finite-difference time
domain~\cite{yee-itoap66, taflove-book00}, transfer
matrix~\cite{pendry-prl92, pendry-jpm96}, empirical tight-binding
methods~\cite{lidorikis-prl98, yariv-ol99}, and Galerkin techniques in
which the field is expanded in a set of orthogonal basis
functions~\cite{mogilevtsev-jlt99}. Within the latter
class, use of a plane-wave basis set is particularly
common~\cite{ho-prl90, johnson-oe01}.

The operators $\nabla\times\nabla$ and $\nabla\times\eps^{-1}_{\rm
r}(\rr)\nabla$ are self-adjoint and, therefore, the fields satisfy 
orthogonality relations given by\footnote{The notation 
${\bf{A}}\cdot{\bf{B}} = \sum_{i=1}^{3}A_i B_i$, and denotes the 
scalar product of the vectors $\bf{A}$ and $\bf{B}$, with 
Cartesian components $\{A_{i}\}$ and $\{B_{i}\}$, respectively.}
\bea
\int d\rr\, \HH^{\ast}_{n\kk}(\rr) \cdot \HH_{n'\kk'}(\rr) &=&
         \delta_{nn'}\delta(\kk-\kk'), \\
\int d\rr\, \epsilon_{\rm r}(\rr)\: {\bf E}^{\ast}_{n\kk}(\rr)\cdot  
         {\bf E}_{n'\kk'}(\rr) &=&
         \delta_{nn'}\delta(\kk-\kk').
\eea

\textcite{leung-josabp93} first suggested that transforming to a basis
of Wannier functions localized in real space would be advantageous for
computational efficiency, especially when dealing with defects in
photonic crystals which, using conventional methods, require very
large supercells for convergence. Although of great formal importance
for justifying the existence of a suitable localized basis, and hence
the tight-binding approach, the non-uniqueness of the transformation
between Bloch states and Wannier functions caused difficulties. As a
result, early work was limited to the case of single, isolated
bands~\cite{leung-josabp93, konotop-josabp97} or composite bands in
which the matrix elements $U_{mn}^{(\kk)}$ were treated as parameters
to fit the tight-binding band structure to the plane-wave result.

The formalism for obtaining maximally-localized Wannier functions,
however, removed this obstacle and several applications of MLWFs to
calculating the properties of photonic crystals have been reported
since, in both
two-dimensional~\cite{garcia-martin-n03, whittaker-prb03, jiao-ijqe06}
and three-dimensional~\cite{takeda-prb06} photonic crystal structures,
as well as for the case of entangled bands~\cite{hermann-josabp08}
(see \textcite{busch-jpcm03} for an early review).

Typically one chooses to solve for either the electric field ${\bf E}_{\bnk}$
or the magnetic field $\HH_{\bnk}$. Once the Bloch states for the
periodic crystal are obtained, a basis of magnetic or electric field
Wannier functions may be constructed using the usual definition, e.g.,
for the magnetic field 
\beq
\WW_{n\RR}^{(\HH)}(\rr) = 
         \ibz d\kk\,e^{-i\kk\cdot\RR}
\sum_{m} U_{mn}^{(\kk)}
         \HH_{m\kk}(\rr), 
\eeq
satisfying orthogonality relations
\beq
\int d\rr\, \WW_{n\RR}^{(\HH)\ast}\cdot\WW_{n'\RR'}^{(\HH)} 
   = \delta_{nn'}\delta_{\RR\RR'} ,
\eeq
where the unitary transformation $U_{mn}^{(\kk)}$ is chosen in the
same way described in Sec.~\ref{sec:theory} such that the sum of the
quadratic spreads of the Wannier functions is minimized, i.e., such
that the Wannier functions are maximally localized. 

Concentrating on the magnetic field, it may be expanded in the basis of
Wannier functions with some expansion coefficients $c_{n\RR}$,
\beq
\HH(\rr) = \sum_{n\RR} c_{n\RR} \WW_{n\RR}^{(\HH)}(\rr),
\label{eq:H-expansion}
\eeq
which on substitution into \equ{wave-eqn-H} gives the
tight-binding representation of the wave-equation for the magnetic
field in the Wannier function basis. 

\begin{figure}[h!]
\begin{center}
\includegraphics[width=1.6in]{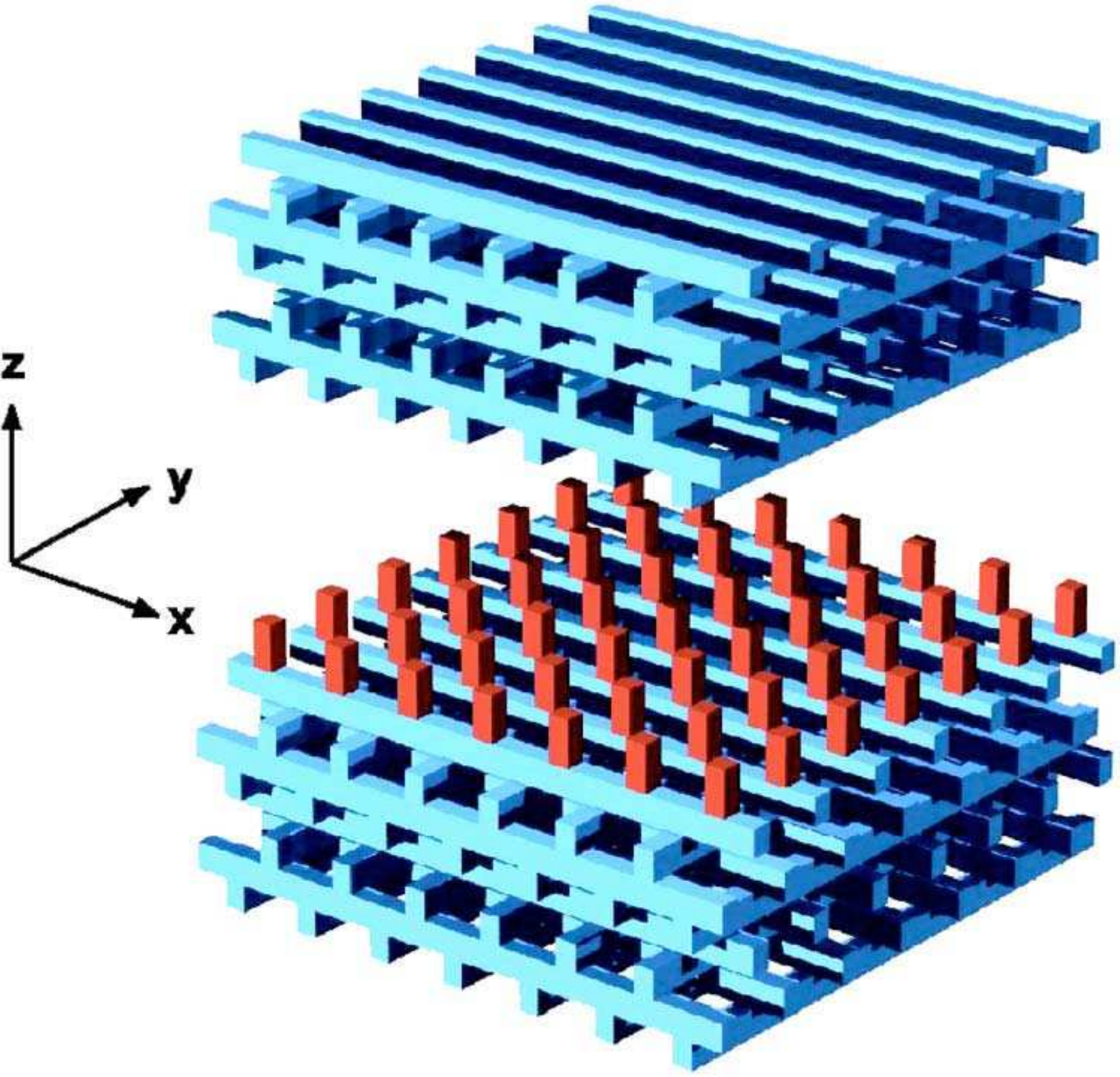}
\includegraphics[width=1.6in]{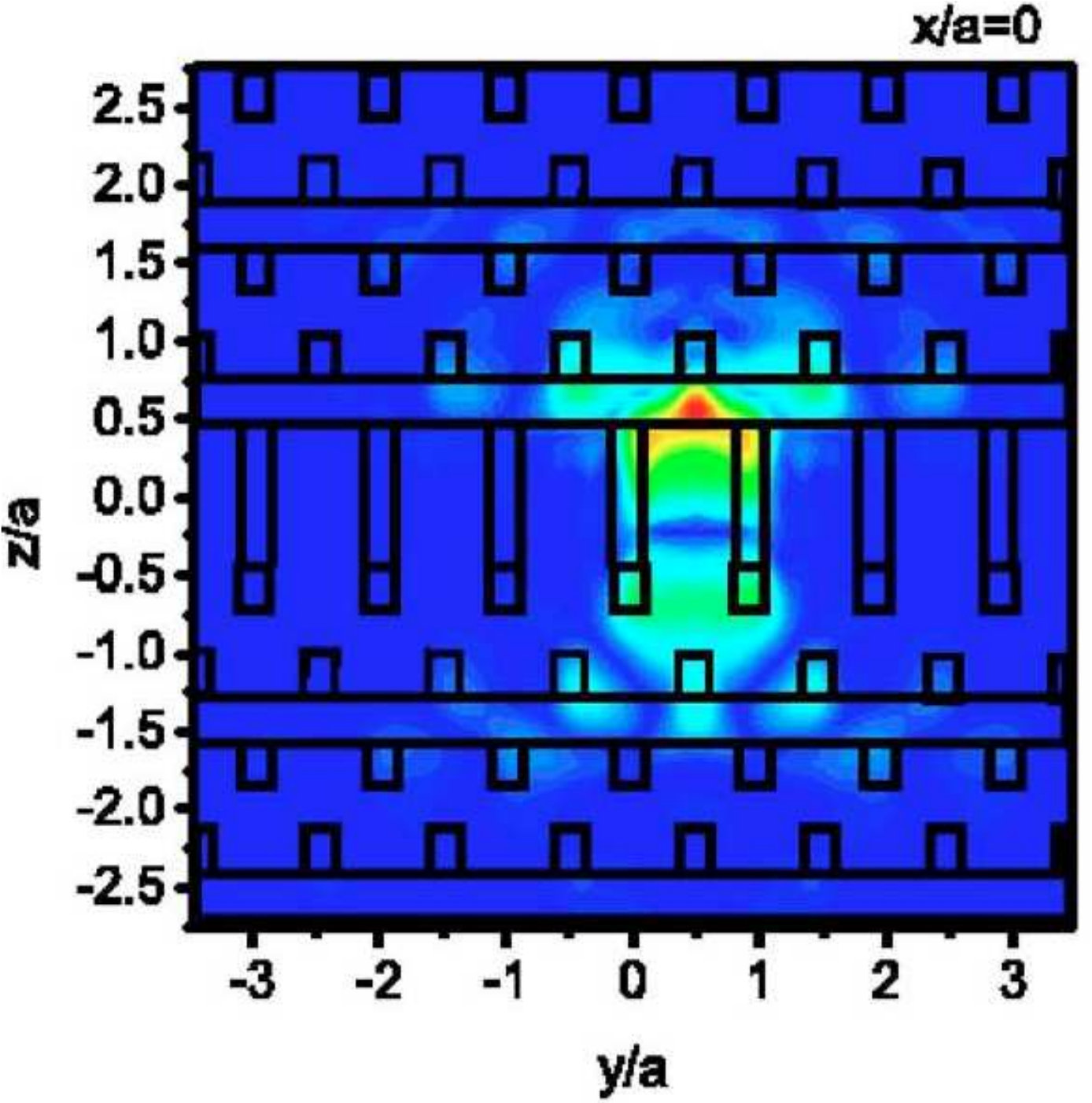}\\
\vspace{3mm}
\includegraphics[angle=90,width=2.5in]{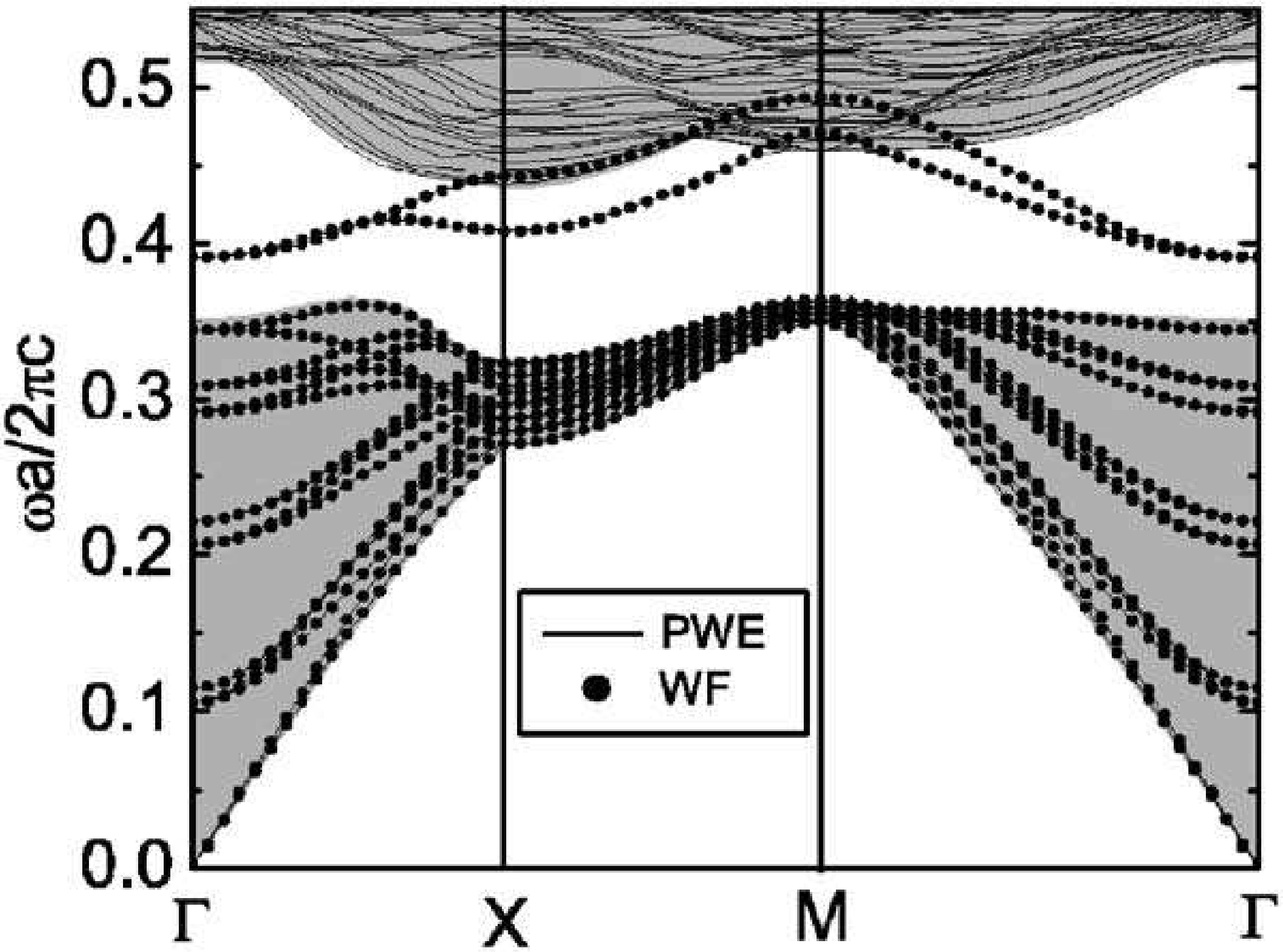}
\caption{(Color online) Photonic band structure (bottom) of the 3D Si woodpile structure
intercalated with a 2D layer consisting of a square lattice of
square rods (top left). Solid lines indicate the photonic band structure
calculated by the plane-wave expansion (PWE) method, and black points
indicate that reproduced by the MLWFs. Shaded regions indicate
the photonic band structure of the woodpile projected onto the 2D
$\kk_\parallel$ space. The square rods in the 2D layer are chosen
to structurally match the woodpile. The thickness of the layer is $0.8a$, 
where $a$ is the lattice parameter of the woodpile structure. Top right: absolute
value of the 17th MLWF of the magnetic field in the $yz$ plane.
Adapted from \textcite{takeda-prb06}.}
\label{fig:photonic}
\end{center}
\end{figure}

The utility of the approach becomes evident when considering the
presence of a defect in the dielectric lattice such that $\eps_{\rm
r}(\rr)\rightarrow \eps_{\rm r}(\rr) + \delta\eps(\rr)$. The magnetic
field wave equations become
\beq
\nabla\times\left( \left[\eps^{-1}_{\rm r}(\rr)+\Delta^{-1}(\rr)\right]
      \nabla\times\HH(\rr)\right) = \frac{\omega^2}{c^2}\HH(\rr),
\eeq
where
\beq
\Delta(\rr) = \frac{-\delta\eps(\rr)}
        {\eps_{\rm r}(\rr)[\eps_{\rm r}(\rr)+\delta\eps(\rr)]}.
\eeq
Using the Wannier functions from the defect-free
calculation as a basis in which to expand $\HH(\rr)$, as in
\equ{H-expansion}, the wave equations may be written in matrix form,
\beq
\sum_{n'\RR'}\left(A_{nn'}^{\RR\RR'} + B_{nn'}^{\RR\RR'} \right)
   c_{n'\RR'} = \frac{\omega^2}{c^2}c_{n\RR},
\eeq
where 
\beq
A_{nn'}^{\RR\RR'} = \ibz d\kk\, e^{i\kk\cdot(\RR-\RR')}\sum_{m}
   U_{nm}^{(\kk)\dagger} \left(\frac{\omega_{m\kk}}{c}\right)^2
   U_{mn'}^{(\kk)},
\eeq
and
\beq
B_{nn'}^{\RR\RR'}=\int d\rr\,\Delta(\rr)
                  \left[\nabla\times\WW_{n\RR}(\rr)\right]^{\ast}
             \cdot\left[\nabla\times\WW_{n'\RR'}(\rr)\right].
\eeq
Due to the localization and compactness of the basis, these matrix
equations may be solved efficiently to find frequencies of localized
cavity modes, dispersion relations for waveguides, and the
transmission and reflection properties of complex waveguide
structures. Fig.~\ref{fig:photonic}, for example, shows the photonic
band structure for a three-dimensional photonic crystal structure 
with a two-dimensional defect. 

\subsection{Cold atoms in optical lattices}
\label{sec:cold-atoms}

A good 70 years after Albert Einstein predicted that a system of
non-interacting bosons would undergo a phase transition to a new state
of matter in which there is macroscopic occupation of the lowest
energy quantum state, major achievements in laser cooling and
evaporative cooling of atoms enabled the first experimental
realizations of Bose-Einstein condensation~\cite{anderson-s95,
davis-prl95, bradley-prl95} and the award of Nobel prizes in 1997 and
2001. Since then, the study of cold atoms trapped in optical lattices
has flourished. For a reviews see \textcite{morsch-rmp06,
bloch-rmp08}.

Ultracold atoms trapped in optical lattices provide a versatile
alternative to electrons in crystal lattices for the study of quantum
phenomena. Indeed, they have a number of advantages over
the solid state in this respect, such as the absence of lattice
defects, the absence of a counterpart to the electron-phonon
interaction, and the ability to control precisely both the nature of
the inter-atomic interactions and the depth and periodicity of the
optical lattice potential.

The second quantized Hamiltonian for a system of $N$
weakly interacting bosons of zero spin and mass $m$ in a (periodic)
external potential $V_{0}(\rr)=V_{0}(\rr + \RR)$ is
given by~\cite{yukalov-lp09}
\bea
\hat{\mathcal{H}} &=& \intr \hat{\Psi}^{\dagger}(\rr) \left[
-\frac{\hbar^2}{2m}\nabla^2 + V_{0}(\rr) \right]
\hat{\Psi}(\rr) \nonumber \\ 
& & + \: \frac{g}{2} \intr
\hat{\Psi}^{\dagger}(\rr) \hat{\Psi}^{\dagger}(\rr)
\hat{\Psi}(\rr)\hat{\Psi}(\rr),
\label{eq:cold-atom-ham}
\eea
where $g=4\pi a_s \hbar^2 / m$, it is assumed that the atoms
interact via a short-range pseudopotential with $a_s$ as the
$s$-wave scattering length, and $\hat{\Psi}(\rr)$ and
$\hat{\Psi}^{\dagger}(\rr)$ are bosonic field operators obeying 
canonical commutation relations~\cite{fetter-book-ch1}.

In a Bose-Einstein condensate, wherein the condensate particle
densities are typically of the order of $10^{14}$\:cm$^{-3}$ or more,
the mean-field limit of this Hamiltonian is usually taken, which leads
to the Gross-Pitaevskii equation, also known as the non-linear
Schr\"{o}dinger equation (NLSE),
\beq
i\hbar \frac{\partial}{\partial t} \varphi (\rr, t) = 
\left[ -\frac{\hbar^2}{2m} + V_{0}(\rr) + 
g \left|\varphi(\rr,t)\right|^2 \right] \varphi(\rr,t).
\label{eq:gp}
\eeq
$\varphi(\rr,t)$ is the condensate wavefunction, the squared
norm of which gives the condensate density. The Gross-Pitaevskii
equation has been used with remarkable success in the study of
BEC~\cite{leggett-rmp01}.

As shown, for example, by \textcite{alfimov-pre02}, a basis of Wannier
functions, localized to each site $\alpha$ of the optical lattice, may
be used to expand the condensate wavefunction,
\beq
\varphi(x,t) = \sum_{n,\alpha} c_{n\alpha}(t)w_{n\alpha}(x).
\label{eq:condensate-exp}
\eeq
The Wannier functions are related to the Bloch eigenstates
$\psi_{nk}(x)$ of the eigenvalue equation
\beq
\left[ -\frac{d^2}{dx^2} + V_{0}(x)\right] \psi_{nk}(x) =
\epsilon_{nk}\psi_{nk}(x)
\eeq
by the usual Wannier transformation 
\beq
w_{n\alpha}(x) = \frac{L}{2\pi} \int_{\rm BZ} dk\,
e^{i\phi_{n}(k)} e^{-ik\alpha L}\psi_{nk}(x)
.
\eeq
Substituting \equ{condensate-exp} into \equ{gp}, leads to a tight-binding
formulation known as the discrete non-linear Schr\"{o}dinger equation
(DNLSE),
\beq
i\frac{d}{d t}c_{n\alpha} = \sum_{\beta} c_{n\beta}
\tilde{\epsilon}_{n,\alpha-\beta} 
 + \: \sum_{\beta,\gamma,\eta} \sum_{i, j, k}
c^{\ast}_{i\beta} c_{j\gamma} c_{k\eta} U_{nijk}^{\alpha\beta\gamma\eta},
\label{eq:dnlse}
\eeq
where
\beq
\tilde{\epsilon}_{n\alpha} = \frac{L}{2\pi} 
\int_{\rm BZ} dk\, e^{-ik\alpha L}
\epsilon_{nk},
\eeq
and the interaction matrix is given by
\beq
U_{nijk}^{\alpha\beta\gamma\eta} = g \int dx\,
w_{n\alpha}(x) w_{i\beta}(x)
w_{j\gamma}(x) w_{k\eta}(x).
\eeq

Truncating the first term on the right-hand side of \equ{dnlse} to
nearest-neighbors only, and the second term to on-site
($\alpha=\beta=\gamma=\eta$) terms within a single band ($n=i=j=k$)
results in the usual tight-binding description~\cite{chiofalo-epjd00,
trombettoni-prl01}, 
\beq
i\frac{d}{d t}c_{n\alpha} = c_{n\alpha} \tilde{\epsilon}_{n0} 
+ \tilde{\epsilon}_{n1}(c_{n,\alpha-1} + c_{n,\alpha+1})
+ U_{nnnn}^{\alpha\alpha\alpha\alpha}
\left|c_{n\alpha}\right|^2 c_{n\alpha}.
\label{eq:dnlse-simple}
\eeq

As pointed out by \textcite{alfimov-pre02}, using a WF basis enables
the range and type of interactions encapsulated in the DNLSE to be
systematically controlled and improved. On the most part, however, WFs
have been used in the context of the NLSE in order to carry out
formal derivations and to justify the use of empirical or
semi-empirical tight-binding models.

An interesting analogy with electrons in atomic lattices manifests
itself when the filling of sites in the optical lattice is low and
hence particle correlations need to be accounted for more
rigorously. This is done via the Bose-Hubbard model, developed by
\textcite{fisher-prb89} in the context of He-4, and first applied to
cold atoms in optical lattices by \textcite{jaksch-prl98}. The
Bose-Hubbard Hamiltonian is derived from \equ{cold-atom-ham} by
expanding the boson field operator in terms of WFs of a single band,
localized at the lattice sites,
\beq
\hat{\Psi} = \sum_\alpha \hat{b}_\alpha w_{\alpha}(\rr),
\eeq
where the bosonic particle creation and annihilation operators,
$\hat{b}_\alpha$ and $\hat{b}^{\dagger}_{\alpha}$, respectively, satisfy
the usual commutation rules~\cite{fetter-book-ch1}.
This, on approximation to nearest-neighbor coupling,
and on-site only interactions, results in the standard Bose-Hubbard
Hamiltonian~\cite{jaksch-prl98}
\beq
\hat{H}_{\rm BH} = -J \sum_{<\alpha,\beta>} \hat{b}^{\dagger}_{\alpha}
\hat{b}_{\beta} + 
\frac{U}{2} \sum_\alpha \hat{n}_\alpha (\hat{n}_\alpha - 1),
\eeq
where 
$\hat{n}_\alpha=\hat{b}^{\dagger}_{\alpha}\hat{b}_{\alpha}$ is the
number operator for lattice site $\alpha$, and the nearest-neighbor
hopping and on-site repulsion parameters are given by
\beq
J = -\intr w_{0}(\rr) 
\left[ -\frac{\hbar^2}{2m}\nabla^2 + V_{0}(\rr) \right] 
w_{1}(\rr),
\eeq
and
\beq
U = g \int d\rr\, \left|w(\rr)\right|^4,
\eeq
which may be calculated explicitly using WFs constructed from Bloch
eigenstates~\cite{vaucher-njp07, shotter-pra08}. The Bose-Hubbard
model is the bosonic analogue to the Hubbard model for fermions. 
As in the latter case, the behavior of the model
depends on the competition between hopping ($J$) and on-site
($U$) energies which determines whether the system is in a superfluid or
a Mott insulator phase.

Finally, we note that in work that is closely related to, and combines
elements from both ideas developed in photonic crystals and cold
atoms, WFs have also been used to represent polaritons in coupled
cavity arrays, a class of systems that serves as another experimental
realization of the Bose-Hubbard model~\cite{hartmann-np06,
hartmann-lpr08}.

\section{SUMMARY AND PROSPECTS}
\label{sec:summary}

In this review, we have summarized methods for constructing
WFs to represent electrons in periodic solids
or other extended systems.  While several methods
have been surveyed, our emphasis has been on the one of
\textcite{marzari-prb97}, essentially the generalization
of the approach of Foster and Boys
\cite{boys-rmp60,foster-rmp60a,foster-rmp60b,boys-66} to
periodic systems, in which the gauge freedom is resolved
by minimizing the sum of the quadratic spreads of the WFs.
The widespread adoption of this approach is reflected in the fact
that it has been incorporated as a feature into a large number
of modern first-principles electronic-structure code packages
including
{\sc Quantum ESPRESSO} \cite{gianozzi-jpcm09},
{\sc Abinit} \cite{gonze-cpc09},
{\sc FLEUR} \cite{freimuth-prb08},
{\sc WIEN2k} \cite{schwarz-cpc02,kunes-cpc10},
{\sc Siesta} {\cite{soler-jpcm02,korytar-jpcm10}},
and {\sc VASP} \cite{kresse-prb96,franchini-arxiv11}.
In the above cases
this has been done by providing an interface to the {\sc Wannier90}
package \cite{mostofi-cpc08}, an open-source post-processing code
developed by the Authors, offering most of the capabilities described
in this review. Other efforts have also seen the implementation of
MLWFs in {\sc CPMD} \cite{cpmd}, {\sc GPAW} \cite{enkovaara-jpcm10},
{\sc OpenMX} \cite{openmx}, and {\sc WanT} \cite{calzolari-prb04,want}
- this latter, and {\sc Wannier90}, also allowing for
quantum-transport calculations.

After an initial wave of applications increased the visibility of
WFs in the community and demonstrated their utility for a variety
of applications, other methods for constructing WFs were also
developed, as discussed in Secs.~\ref{sec:theory} and
\ref{sec:locorb}.  For some purposes, e.g., for many plane-wave
based LDA+U and DMFT calculations, methods based on simple
projection onto trial orbitals proved sufficient.  Methods tuned
specifically to $\Gamma$-point sampling of the BZ for supercell
calculations also became popular.  And, as surveyed briefly
in Sec.~\ref{sec:other}, the construction and application of WFs
was also extended to periodic systems outside the electronic-structure
context, e.g., to phonons, photonic crystals, and cold-atom optical
lattices.

Still, the vast majority of applications of WF methods have been
to electronic structure problems.  The breadth of such applications
can be appreciated by reviewing the topics covered in
Secs.~\ref{sec:bonding}-\ref{sec:basis}.  Very broadly, these
fall into three categories: investigations into the nature of chemical
bonding (and, in complex systems such as liquids, the statistics
of chemical bonding), as discussed in Sec.~\ref{sec:bonding};
applications that take advantage of the natural ability of
WFs and WF charge centers to describe dipolar and orbital
magnetization phenomena in dielectric, ferroelectric, magnetic,
and magnetoelectric materials, as reviewed in
Sec.~\ref{sec:pol}; and the use of WF for basis functions, as
surveyed in Secs.~\ref{sec:winterp}-\ref{sec:basis}.

Today these methods find applications in many topical areas
including investigations into novel superconductors, multiferroics,
and topological insulators.
The importance of WFs is likely to
grow in response to future trends in computing, which are clearly
moving in the direction of more massive parallelization based on
algorithms that can take advantage of real-space partitioning.
This feature of WFs should also facilitate their adoption in
formulating new beyond-DFT methods in which many-body interactions
are included in a real-space framework.  Thus, the growing pressures
for increased efficiency and accuracy are likely to elevate the
importance of WF-based methods in coming years.
Overall, one can look forward
to continued innovation in the development and application of
WF-based methods to a wide variety of problems in modern
condensed-matter theory.

\bibliographystyle{apsrmp}
\bibliography{bib/wannier}

\end{document}